\documentclass[12pt]{book}
\usepackage[cp1251]{inputenc}
\usepackage{hyperref}
\usepackage{graphicx}
\usepackage{amsmath}
\usepackage{amssymb}
\usepackage{dcolumn}
\usepackage{bm}
\usepackage{amsmath}
\usepackage{enumitem}
\usepackage{epsfig}
\usepackage{pdfpages}

\begin{titlepage}

\title{Dynamics of the Universe in Problems}

\author{{\Large A thousand problems in Cosmology}\\ \\ \\
Yu.L.~Bolotin${}^{1}{}^*$ , V.A.~Cherkaskiy${}^{1}$,
G.I.~Ivashkevych${}^{1}$,\\ O.A.~Lemets${}^{1}$, I.V.~Tanatarov${}^{1,2}$,\\ D.A.~Yerokhin${}^{1}$\\
{ }\\
\it ${}^1$A.I.Akhiezer Institute for Theoretical Physics,\\ \it
National Science Center "Kharkov Institute of Physics and
Technology",\\
 \it Akademicheskaya Str. 1, 61108 Kharkov, Ukraine\\
\it ${}^{2}$V.N. Karazin Kharkov National University,\\
\it Svobody Sq. 4, 61077 Kharkov, Ukraine\\ {}\\ ${}^*$ybolotin@gmail.com}
\date{}
\end{titlepage}

\begin{document}
\maketitle
\tableofcontents
\chapter*{Abstract}
To our best knowledge, there are no problem books on cosmology yet, that would include its spectacular recent achievements. We believe there is a strong need for such now, when cosmology is swiftly becoming a strict and vast science, and the book would be extremely useful for the youth pouring in this area of research. Indeed, the only way to rise over the popular level in any science is to master its alphabet, that is, to learn to solve problems.

Of course, most of modern textbooks on cosmology include problems. However, a reader, exhausted by high theory, may often be thwarted by the lack of time and strength to solve them. Might it be worth sometimes to change the tactics and just throw those who wish to learn to swim into the water?

We present an updated version of the ''Dynamics of the Universe in Problems'' We have the following new sections , 'Gravitational Waves', ''Interactions in the Dark Sector'', ''Horizons'' and ''Quantum Cosmology'' .    A number of new problems have been added to almost every section. The total number of problems exceeds fifteen hundred. Solutions to all the problems can be found at \href{http://universeinproblems.com/}{Dynamics of the Universe in Problems} 

\chapter{Cosmo-warm-up}
\section{Astronomy ''before the Common Era''}
\begin{enumerate}
\item \label{razm-AD0} Aristotle knew there were stars, that one can see in Egypt but not in Greece. What conclusions could be made from this observation?

\item \label{razm-AD1} How could Eratosthenes in 250 BCE determine the Earth's radius with the help of a well?

\item \label{razm-AD2} How could Aristarchus (310-230 BCE) compute the distance to the Moon, knowing that it sets in about two minutes, and the maximum duration of a lunar eclipse is three hours?

\item \label{razm-AD3} How can one calculate the size and distance to the Sun, while observing the phases and eclipses of the Moon? 

\item \label{razm-AD4} Can one estimate how farther is the Sun than the Moon using only naked eye observations?
\item\label{razm-AD5} How to determine the distance to an Earth's satellite (Moon, for example) or to the Sun, using only a chronometer? 

\item \label{razm-AD6} Ancient Babylonians knew, that apparent motion of Mars relative to the Earth is periodic with the orbital period 780 days (the synodic period of Mars). Tycho Brahe, in XVI century, measured the apperent Mars' trajectory with great precision. How could Johannes Kepler, using Brahe's data, accurately calculate the orbits of both Earth \emph{and} Mars relative to the Sun and discover the laws named after him, which in time enabled Newton to biuld the classical theory of gravity?

\item \label{razm-AD7} The orbital period of Io, one of the satellites of Jupiter, is equal to $42.5$ hours. In the 1670-ties Ole R\o mer was measuring the time between two successive eclipses of Io when Earth moved on its orbit towards Jupiter and when it moved in the opposite direction. The noticed difference allowed him to estimate the speed of light. Try to reproduce the reasoning and estimates of R\o mer.

\item \label{razm-AD8} In 1982 Doppler predicted the effect of change in the percieved frequency of oscillations when there is relative motion of emitter and detector. Doppler assumed that this effect can cause the difference in the color of stars: a star moving towards the Earth seems ``bluer'', while the one moving away ``reddens''. Explain why the Doppler effect cannot substantially change the color of a star.

\item \label{razm-AD9} In the beginning of the XXth century J.F.~Hartmann, a German astronomer, was studying the spectra of double stars. The wavelengths of their spectral lines shifted periodically due to their relative motion, with the period equal to the orbital one. In the spectra of some of the binaries he also noticed there were absorption lines with wavelengths that did not change with time. What discovery did Hartmann make due to this observation?

\item \label{razm-AD10} Originally the interstellar gas was discovered by its absorption of spectral lines of calcium. Does that mean that calcium is the dominating component of the interstellar medium?

\item \label{razm-AD11} One of the creators of the Theory of Relativity, Henry Poincare, when speaking in 1904 (!) of the fact that the speed of light $c$ enters all the equations of electro\-mag\-ne\-tism, compared the situation with the Ptolemy's geocentric theory of epicycles, in which the Earth's year enters all the relations for the relative motion of celestial bodies. Poincare expressed hope that a future Copernicus would rid electrodynamics of $c$. Recall other examples of blunders of geniuses.

\end{enumerate}

\section{Quantities large and small}
\begin{enumerate}[resume]
\item \label{razm2} From what distance will one astronomical unit length have visible size of one angular second?

\item \label{razm3} What is the angular dimension of our Galaxy for an observer in the Andromeda galaxy, if the distance to it is about $700 kpc$? Compare it with the angular size of the Sun viewed from the Earth.

\item \label{razm6} A glance on the night sky makes the impression of invariability of the Universe. Why do the stars seem to us practically static?

\item \label{razm4} * A supernova outburst in the Andromeda galaxy has been observed on Earth. Estimate the time since the star's explosion.

\item \label{razm5} A galaxy at distance $R$ from us at the moment of observation recedes with velocity $V$. At what distance was it situated at the moment of emission of the observed light?

\item \label{razm26} Suppose that we have concentrated the whole cosmic history (14 billion years) in one day. Display the main events in the history of the Universe using the logarithmic time scale. Start from the Planck's time to avoid singularities.

\item \label{razm30} According to the Big Bang model the initial ratio of the uranium isotopes' abundances was $U^{235}/U^{238}\approx 1.65$, while  the presently observed one is $U^{235}/U^{238}\approx 0.0072$. Taking into account that the half--value periods of the isotopes are equal to $t_{1/2}(U^{235})=1.03\cdot 10^9 $ years and $t_{1/2}(U^{238})=6.67\cdot 10^9 $ years, determine the age of the Universe.


\item \label{razm28} Estimate the mass $M_G$ of Milky Way and the number of stars in it, if the Sun is an avarage star of mass $M_\odot$, situated almost at the edge of our Galaxy and it orbits its center with the period $T_\odot=250$ millions years at the distance $R_G=30$ thousands light years.

\item \label{razm29} Estimate the density of luminous matter in the Universe assuming that the Milky Way containing $\sim10^{11}$ stars of solar type is a typical galaxy, and average intergalactic distance is of order of $L=1 Mpc$.

\item \label{razm41} Assume that the space is infinite and on average uniformly filled with matter. Estimate the distance from our observable part of the Universe to the part of the Universe with identical distribution of galaxies and the same Earth.

\item \label{razm34}  Show that in the hydrogen atom the ratio of electrical forces to gravitational ones is close to the ratio of the size of the Universe to the size of an electron (this fact was first noted by P.Dirac).

\item \label{razm37} Express the Bohr radius through the fine structure constant and Compton wavelength.

\item \label{razm38} Estimate the free path of a hydrogen atom in the intergalactic space.

\item \label{razm39} ** Protons accelerated at LHC ($E=7 TeV$) and photons are participants of a cosmic Earth--Sun race. How much will the protons lose in time and distance?

\item \label{razm49} Estimate the total amount of energy collected by optical telescopes during the past XX century and compare it with the energy needed to turn over a page of a book.

\item \label{razm25n1} * Estimate your own weight on the surface of white dwarf, neutron star, black hole.

\item \label{razm25n2} Densities of astrophysical objects vary in a wide range. Estimate the ratio of a neutron star's density to the average density of Milky Way.

\item \label{razm52} What cosmological process releases the maximum amount of energy simultaneosly since the Big Bang?
\item \label{razm52n1} Demonstrate, that for any Standard Model particle quantum gravity effects are completely negligible at the particle level.
\end{enumerate}

\section{Geometric warm-up}
\begin{enumerate}[resume]
\item \label{razm7} What is maximum sum of angles in a triangle on a sphere?

\item \label{razm8} * Consider the sphere of radius $R$. A circle is drawn on the sphere which has radius $r$ as measured along the sphere. Find the circumference of the circle as a function of $r$.

\item \label{razm9} Suppose that galaxies are distributed evenly on a two-dimensional sphere of radius $R$  with number density $n$  per unit area. Determine the total number $N$  of galaxies inside a radius $r$. Do you see more or fewer galaxies out to the same radius, compared to the flat case?

\item \label{razm10} An object of size $A$ is situated at distance $B$. Determine the angle at which the object is viewed in flat space and in spaces of constant (positive and negative) curvature.
\end{enumerate}

\section{Astrophysical warm-up}
\begin{enumerate}[resume]
\item \label{razm-Dop0} Obtain the non--relativistic approximation (to the first order in $V/c$) for the Doppler effect.

\item \label{razm-Dop1} Obtain the relativistic formula for the Doppler effect.

\item \label{razm-Dop2} What are the differences between the Doppler effect for light and the  ''analogous'' effect for sound?

\item \label{razm-Dop3} Using the Doppler effect, how could we demonstrate that time is running differently for observers which move relative to each other?

\item \label{razm11} Every second about $1400 J$ of solar energy falls onto one square meter of the Earth. Estimate the  absolute emittance of the Sun.

\item \label{razm12} Assuming that the constant emittance stage for the Sun is of order of $10^{10}$ years, find the portion of solar mass lost due to radiation.

\item \label{razm13} Why was the connection between the emittance of variable stars (Cepheids) and the period of their brightness variation discovered from observation of stars in Great Magellan cloud rather than in our Galaxy?

\item \label{razm14} A supernova in its maximum brightness reaches the absolute stellar magnitude of $M=-21$. How often will the supernova outbursts be registered if observation is carried out on the whole sky up to the limiting magnitude $m=14$? Assume that in a typical galaxy a supernova bursts on average once per 100 years, and galaxies are distributed with spatial density of one galaxy per $10 Mpc^3$.

\item \label{razm-Astro1} Calculate the average rate of star formation in our Galaxy.

\item \label{razm13n2} * Determine the wavelengths of hydrogen resonant lines.

\item \label{razm13n3} The main method for investigation of interstellar neutral hydrogen are observations in unltraviolet  band. The strongest interstellar absobtion line is $\alpha$--Lyman hydrogen line ($\lambda=121.6\mbox{nm}$). This line corresponds to transition of electron from state with quantum number $n=1$ to the state with $n=2$. At the same time, the Balmer series characterized by electron transition from excited $n=2$ state are not observed. Why does this happen?

\item \label{razm61} Hydrogen burning (hydrogen to helium transformation) in stars is realized in the so-called $p$--$p$ cycle, which starts from the reaction of deuterium formation \(p+p\to d+e^+ +\nu.\) To support such a reaction the colliding protons have to overcome Coulomb barrier (in order to enter the region where nuclear forces act: $r_{nf}\approx10^{-13}\mbox{\it cm}$.) It requires energy as high as \(E_{c}=e^2/r_{nf}\approx1.2\mbox{\it MeV.}\) Typical solar temperature is only $T_\odot=10^7 \mbox{\it K}\approx0.9\mbox{\it keV}$. The Coulomb barrier can be overcome due to the quantum tunneling effect (classical probability to overcome the barrier for the protons in the tail of Maxwell distribution is too low). Estimate the probability of tunneling through the Coulomb barrier for protons on the Sun.

\item \label{razm63} At present hydrogen in the Sun burns (transforms into helium) at temperature $1.5\cdot10^7\ K$, but much higher temperature will be  required to synthesize carbon from helium (when hydrogen is exhausted) due to higher Coulomb barrier. What physical mechanism could provide the increase of the Sun's temperature at the later stages of its evolution?

\item \label{razm46} Accretion is the process of gravitational capture of matter and its subsequent precipitation on a cosmic body, i.e. a star. In such a process the kinetic energy of the falling mass $m$ transforms with some efficiency into radiation energy, which leads to additional contribution to the brightness of the accreting system. Determine the limiting brightness due to accretion (the Eddington limit).

\item \label{razm61_1} Masses of stars vary in the limits \[0.08{{M}_{\odot }}<{{M}_{star}}<100{{M}_{\odot }}.\] How could this be explained?

\item \label{razm61_2} How many quantities completely determine the orbit of a component of a double star?

\end{enumerate}

\section{Planck scales and fundamental constants $c$, $G$, $\hbar$}
The problems \ref{Okun1} and \ref{Okun2} are inspired by a paper by Okun'\footnote{Okun L B ''The theory of relativity and the Pythagorean theorem'' Phys. Usp. 51 622–631 (2008);}.

\begin{enumerate}[resume]
\item \label{Okun1} Consider some physical quantity $A$. The multiplication of $A$ by any power of arbitrary fundamental constant certainly changes it's dimensionality, but not its physical meaning. For example, the quantity $e\equiv E/c^2$ is energy, despite the fact that it has the dimensionality of mass. Why, then, do we refer to the quantity $E/\hbar$ as frequency, but not as energy, although the Planck constant $\hbar$ is a fundamental constant as well as $c$?

\item \label{Okun2} In special relativity mass is determined by the relation
\[m^{2}=e^{2}-p^{2},\qquad e=E/c^{2}.\]
This expression presents the simpliest possible relation between energy, momentum and mass. Why the relation between these quantities could not be linear?

\item \label{razm15} Construct the quantities with dimensionalities of length, time, mass, temperature, density from fundamental constants $c, G, \hbar$ and calculate their values (the corresponding quantities are called the Planck units).

\item \label{razm15n} Perform the same procedure for just $c,G$. Considered quantities are called Newton units. Construct, in particular, the Newton force unit and Newton power unit. What is the physical meaning of these quantities? Why is there no Newton length scale?

\item \label{razm16} * Compare the delay of reception of an object located at $1~\mbox{m}$ from flat mirror, with Planck time.

\item \label{razm17} Demonstrate that gravitational radius of a particle with Planck mass coincides with its Compton  wavelength. Note that gravitational radius in General Relativity is a radius of the spherically symmetric mass, for which the escape velocity at the surface is equal to the speed of light.

\item \label{razm18} Demonstrate that in  the units $c=\hbar=1$
\[1\,GeV\approx 1.8\cdot 10^{-24}\, g;\quad
	1\, GeV^{-1}\approx 0.7\cdot 10^{-24}\,c
	 \approx 2\cdot 10^{-14}\,cm.\]

\item \label{razm18new} Estimate the energy scale, which correspond to current age of the Universe in units $\hbar =c=1$ .

\item \label{razm19} Express Planck mass in terms of $K$, $cm^{-1}$, $s^{-1}$.

\item \label{razm19n} Express Newton's constant $G$ in units $c=1$.

\item \label{razm20} Show that the fine structure constant $\alpha=e^2/\hbar c$ is dimensionless only in a space of dimension $D=3$.

\item \label{razm21} Construct a dimensionless combination from the constants $c,\ \hbar,\ e,\ G$ in the space of arbitrary dimension.

\item \label{razm35} * Compare the constants of strong, weak, electromagnetic and gravitational interactions.

\item \label{razm36} * Estimate the order of magnitude of the temperature of Great Unification, i.e the temperature at which intensity of gravitation comes up to intensities of three other interactions.

	\item Construct planck units in a space of arbitrary dimension.
\end{enumerate}

\section{Gravity}
\begin{enumerate}[resume]
 \item \label{razm29ngr_1} Consider two observers with constant distance $L$ between them, moving far from any mass, i.e. in the absence of gravitational field, with constant acceleration $a.$ At $t_0$ the rear observer emits a photon with wavelength $\lambda$. Calculate the redshift that the leading observer will detect.

 \item \label{razm29ngr_2}  Let's suppose that an elevator's rope breaks and elevator enters the state of free fall. Is it possible to determine experimentally, been inside, that the elevator is falling near the Earth's surface?

\item \label{razm29ngr}  Richard Feynman wrote: ''The striking similarity of electrical and gravitational forces $\ldots$ has made some people conclude that it would be nice if antimatter repelled matter.'' What arguments did Feynman use to demonstrate the inconsistence of this assumption (at least in our world)? 

\item \label{razm48} What is the difference (quantitative and qualitative) between the gravitational waves and the electromagnetic ones?

\item \label{razm50} Find the probability that transition between two atomic states occurs due to gravitation rather than electromagnetic forces.

\item \label{razm50n2} In his Lectures on Gravitation Feynman asks: ''$\ldots$maybe nature is trying to tell us something new here, maybe we should not try to quantize gravity. Is it possible perhaps that we should not insist on a uniformity of nature that would make everything quantized?''. And answers this question. Try to reproduce his arguments.  

\item \label{razm54}  Evidently the role of gravitation grows with the mass of a body. Show that gravitation dominates if the number of atoms in the body exceeds the critical value \(N_{cr}\simeq(\alpha/\alpha_G)^{3/2}\simeq10^{54},\) where $\alpha=e^2/(\hbar c)$ is the fine structure constant and $\alpha_G\equiv Gm_p^2/(\hbar c)$ is the ''fine structure constant'' for gravitation, $m_p$ is proton's mass.

\item \label{razm56n1} Stars form from gas and dust due to gravitational instability, which forces gas clouds to compress. This process is known as Jeans instability after the  famous English cosmologist James Jeans (1877 -- 1946). What is the physical cause of the Jeans instability?

\item \label{razm56n2} Observations show that stars form not individually, but in large groups. Young stars are detected in clusters, which contain, usually, several hundreds of stars, which were formed at the same time. Theoretical calculations show that formation of individual stars is almost impossible. How could this claim be justified?

\item \label{razm57} A gas cloud of mass $M$ consisting of molecules of mass $\mu$ is unstable if the gravitational energy exceeds the kinetic energy of thermal motion. Derive the stability condition for the spherically symmetric homogeneous cloud of radius $R$ (the Jeans criterion).

\item \label{razm58} Estimate the critical density for a hydrogen cloud of solar mass at temperature $T=1\ 000\ K$.

\item \label{razm60} Compare the gravitational pressure in the centers of the Sun ($\rho=1.4\ g/cm^3$) and the Earth ($\rho=5.5\ g/cm^3$).

\item \label{razm60n1}  Gravitational forces on the Sun are balanced by gas pressure. If pressure ''switches of'' at some moment, Sun would collapse. The time of gravitational collapse is called the dynamic time. Calculate this time for the Sun.

\item \label{razm60n2} Show that for any star its dynamic time is approximately equal to the ratio of star's radius to the escape velocity on its surface.  Compare this estimate for the Sun with the exact value obtained in the previous problem.

\item \label{razm62} Show that stars (and star clusters) have the amazing property of negative thermal capacity: the more the star looses energy due to radiation from its surface, the higher is the temperature at its center.

\item \label{razm62n} Along with cosmological redshift, there is the gravitational redshift (the effect of General Relativity), which consists in change in clock under varying gravitational potential. How could these effects be distinguished qualitatively?
\end{enumerate}

The following four problems are based on the paper\footnote{Okun L B “The theory of relativity and the Pythagorean theorem” Phys. Usp. 51 622–631 (2008)}. Here we try to understand the effect of gravity on matter (particles and photons) in terms of Newtonian and relativistic physics, but without relying on the General Theory of Relativity (GTR). Thus the terminology differs from that of GTR, and one should not be surprised to see that i.e. $c$ is not constant. For readers familiar with GTR it would be instructive to reformulate the solutions in its terms and establish the relations between the notions used in both approaches.

\begin{enumerate}[resume]
\item \label{Okun3} Demonstrate that a photon emitted at lower floor of a building due to the transition between two nuclear (atomic) levels could not induce the reverse transition in the same nucleus (atom) at higher floor.

\item \label{Okun4} Consider a photon in a static gravitational field. Which characteristics of this photon (energy, frequency, momentum, wavelength) change and which remain the same?

\item \label{Okun5} Describe the mechanism of light deflection in the gravitational fields of the Sun and galaxies (see ***).

\item \label{Okun6} Trace the photon from problem \ref{Okun3} with clocks, located along the photon's trajectory (at each floor).

\end{enumerate}

\section{Forest for the trees} 
\begin{enumerate}[resume]
\item \label{razm42} If the Universe was infinitely old and infinitely extended, and stars could shine eternally, then in  whatever direction you look the line of your sight should cross the surface of a star, and therefore all the sky should be as bright as the Sun's surface. This notion is known under the name of Olbers' paradox. Formulate the Olbers' paradox quantitatively\footnote{L.~Anchordoqui, arXiv:0706.1988, physics.ed-ph}.

\item \label{razm43} There are $n$ trees per hectare in a forest. Diameter of each one is $D$. Starting from what distance will we be unable to see the forest for the trees? How is this question connected to the Olbers' paradox\footnote{Ryden, Introduction to cosmology, ADDISON-Wesley.}?

\item \label{razm44} In a more romantic formulation this problem looks as following. Suppose that in the Sherwood Forest the average radius of a tree is $30cm$ and average number of trees per unit area is $n=0.005m^{-2}$. If Robin Hood shoots an arrow in a random direction, how far, on average, will it travel before it strikes a tree?

\item \label{razm45} The same problem in the cosmological formulation looks this way. Suppose you are in a infinitely large, infinitely old Universe in which the average number density of stars is $n_{st}=10^{11} Mpc^{-3}$ and the average stellar radius is equal to the Sun's radius: $R_{st}=R_{\bigodot}=7\cdot10^8m$. How far, on average, could you see in any direction before your line of sight strucks a star?

\item \label{razm45n} Demonstrate that stars in galaxies can be considered as collisionless medium.
\end{enumerate}

\section{Life on Mars?}
\begin{enumerate}[resume]
\item \label{Mars1} Consider a highly developed alien civilization actively exploring the Earth in search of intelligent life. How would citizens of the opposite part of Milky Way see their ''brothers in mind'' from the Earth? And what about citizens of Andromeda galaxy?

\item \label{Mars2} Estimate the number of stars, which had already received the signal of homo sapiens'existence on the Earth.

\item \label{Mars3} The inhabited planets in our Galaxy are likely to be located within the so--called galactical habitable zone, which has the form of narrow ring with radius greater or approximately equal to half the radius of Galaxy. Explain, why genesis of life is unlikely to happen at too small or too large galactocentric distances.

\item \label{Mars4} Why does the width of galactic habitable zone gradually increase in time?

\item \label{Mars5} One of the possible formulations of the Fermi paradox, or the ''great silence'' paradox, is the following: if there is intelligent life in the Universe, why does it not emit any signal into space and generally manifest itself in any way? This paradox is related to the name of Fermi, because once having listened to the arguments of his colleagues stating that there exist a great number of highly developed technological civilizations, he asked after some pause: ''So, where are they?'' Give arguments to support or disprove the paradox.
\end{enumerate}

\subsection{Fine tuning of the Universe}
\begin{enumerate}[resume]
\item \label{razm50n} * Imagine for a moment, that electron has the mass several times larger that aclual.  If all other conditions were the same, how would our Universe change?

\item \label{razm50nn} Our Sun emits energy during billions of years and thus supports life development on the Earth. It is possible due to very slow reaction of deuteron creation from two interacting protons. Find the limiting value of electron mass providing the necessarily low rate of such reaction.
\end{enumerate}

\section{Thermo warm-up}
\begin{flushright}
\parindent=7cm
 {\it If someone points out to you that your pet theory of the universe is in disagreement with Maxwell's equations---then so much the worse for Maxwell's equations. If it is found to be contradicted by observation---well these experimentalists do bungle things sometimes. But if your theory is found to be against the second law of thermodynamics I can give you no hope; there is nothing for it but to collapse in deepest humiliation. \\
Sir Arthur Stanley Eddington}
\end{flushright}

\begin{enumerate}[resume]
\item \label{razm-therm0} When designing a suit for open space, what should ingeneers be  more careful of -- heating of heat extraction?

\item \label{razm-therm1} Estimate the number of photons in gas oven at room temperature and at maximum heat.

\item \label{razm-therm2} Estimate the temperature at the surface of the Sun, assuming that the Earth with mean temperature at its surface $15\,C^\circ$ is in thermal equilibrium with the Sun.

\item \label{razm-therm3} What is the difference between the entropy of gravitational degrees of freedom and ordinary entropy (e.g., entropy of ideal gas)?

\item \label{razm-therm4} One of the most used classifications divide physical systems into ''open'' and ''isolated''. The entropy in an isolated system could only increase, eventually reaching the thermal equilibrium.  In contrast, due to external interactions, entropy  in open systems could decrease, for example, through absorption of a component with low entropy. Explain why the Sun is a source of low entropy for the Earth.
    
\item
Estimate how many infrared photons Earth radiates per one ultraviolet photon it absorbs from the Sub.

\item
A human during his lifetime increases the entropy of the Universe by converting chemical energy contained in the food to heat. Estimate the corresponding entropy gain.

\item
Estimate the contribution of Earth during its existence to the entropy production.

\item
Show that the Sun has contributed the entropy increase of about $10^{40} J/K$

\section{Play with Numbers after Sivaram}
\item\label{Siv_1}
(after C.Sivaram, Dark Energy may link the numbers of Rees, arXiv: 0710.4993)
Given $\Lambda$-dominated Universe, the requirement that for various large scale structures (held together by self gravity) to form a variety of length scales, their gravitational self energy density should at least match the ambient vacuum energy repulsion, as was shown to imply \cite{siv3}
a scale invariant mass-radius relationship to the form (for the various structures):
\[\frac M{R^2}\approx\sqrt\Lambda\frac{c^2} G.\]
This equation  predicts a universality of $M/R^2$ for a large variety of structures. Check this statement for such structures as a galaxy, a globular cluster, a galaxy cluster.

\item\label{Siv_2}
(after C.Sivaram, Scaling Relations for self-Similar Structures and the Cosmological Constant, arXiv: 0801.1218)
In recent papers \cite{siv1}, it was pointed out that the surface gravities of a whole hierarchy
of astronomical objects (i.e. globular clusters, galaxies, clusters, super clusters,
GMC's etc.) are more or less given by a universal value $a_0\approx cH_0\approx 10^{-8} cm\ s^{-2}$. Thus
\[a=\frac{GM}{R^2}\approx a_0\]
for all these objects, $M$ being their typical mass and $R$ their typical radius. Also interestingly enough it was also pointed out \cite{siv2} that the gravitational self energy of a typical elementary particle (hadron) was shown to be
\[E_G\approx\frac{Gm^3 c}{\hbar}\approx\hbar H_0\]
implying the same surface gravity value for the particle
\[a_h=\frac{GM}{r^2}\approx \frac{Gm^3 c}{\hbar}\times\frac c\hbar\approx cH_0\approx a_0.\]
Calculate actual value of the ratio
\[\frac M{R^2}\approx\sqrt\Lambda\frac{c^2} G\sim1\]
for such examples as a galaxy, whole Universe, globular cluster, a GMC, a supercluster, nuclei, an electron, Solar system, planetary nebula.

\end{enumerate}

\chapter{Dynamics of Expanding Universe}
\begin{flushright}
\parindent=7cm
 {\it Matter tells Specetime how to curve, and\\
Spacetime tells matter how to move.\\
John Wheeler}
\end{flushright}

\section[Homogeneous and isotropic Universe, Hubble's law]{Homogeneous and isotropic Universe, \\ Hubble's law}
\begin{enumerate}
\item \label{equ1} Most cosmological models are based on the assumption that the Universe is spatially homogeneous and isotropic. Give examples to show that the two properties do not automatically follow one from the other.

\item \label{equ2} Show that if some spatial distribution is everywhere isotropic then it is also homogeneous. Is the opposite true?

\item \label{equ3} What three-dimensional geometrical objects are both homogeneous and isotropic?

\item \label{equ4} Why the notion of ''Big Bang'' regarding the early evolution of the Universe should not be treated too literally?

\item \label{equ5} Show that the Hubble's law is invariant with respect to Galilean transformations.

\item \label{equ6} Show that the Hubble's law represents the only form of expansion compatible with homogeneity and isotropy of the Universe.

\item \label{equ7} Show that if expansion of the Universe obeys the Hubble's law then the initial homogeneity is conserved for all its subsequent evolution.

\item \label{equ9} In the 1940-ties Bondi, Gold and Hoyle proposed a stationary model of the Universe basing on the generalized cosmological principle, according to which there is no privileged position either in space or in time. The model describes a Universe, in which all global properties and characteristics (density, Hubble parameter and others) remain constant in time. Estimate the rate of matter creation in this model.

\item \label{equ10} Galaxies typically have peculiar (individual) velocities of the order of $V_p \approx 100~\mbox{\it km/s}.$ Estimate how distant a galaxy should be for its peculiar velocity to be negligible compared to the velocity of Hubble flow $V_H=H_{0}R$.

\item \label{equ11} Estimate the age of the Universe basing on the observed value of the Hubble's constant (the Hubble time $t_H$).

\item \label{equ13} Show that the model of the expanding Universe allows one to eliminate the Olbers' paradox.
\end{enumerate}

\section{Equations of General Relativity}
\begin{flushright}\begin{minipage}{0.6\textwidth}
{\it
I have thought seriously about this question,\linebreak
 and have come to the conclusion that\linebreak
 what I have to say cannot reasonably be conveyed\linebreak
 without a certain amount of mathematical notation\linebreak
 and the exploration of genuine mathematical concepts.
\begin{flushright} Roger Penrose\\
	The Road to Reality\end{flushright}}
\end{minipage}\end{flushright}

\begin{enumerate}[resume]
\item \label{equ_oto1} Consider a spacetime with diagonal metric
\[ds^2=g_{00}(dx^{0})^2 + g_{11}(dx^{1})^2
    +g_{22}(dx^{2})^{2}+g_{33}(dx^{3})^{2}.\]
Find the explicit expressions for the intervals of proper time and spatial length, and for the 4-volume. Show that the invariant $4$-volume is given by
\[\sqrt{-g}\;d^{4}x\equiv
    \sqrt{-g}\;dx^{0}dx^{1}dx^{2}dx^{3},\]
where $g=\det(g_{\mu\nu})$.

\item \label{equ_oto1a} Let there be an observer with $4$-velocity $u^{\mu}$. Show that the energy of a photon with $4$-wave vector $k^\mu$ that he registers is $u^{\mu}k_{\mu}$, and the energy of a massive particle with $4$-momentum $p^{\mu}$ is $u^{\mu}p_{\mu}$.

\item \label{equ_oto2} The \emph{covariant derivative (or
connection)} $\nabla_{\mu}$ is a tensorial generalization of partial derivative of a vector field $A^{\mu}(x)$ in the curved space-time. It's action on vectors is defined as
\begin{equation}\label{nabla}
    \nabla_{\mu}A^{\nu}=\partial_{\mu}A^{\nu}+
    {\Gamma^{\nu}}_{\lambda\mu}A^{\lambda};\qquad
    \nabla_{\mu}A_{\nu}=\partial_{\mu}A_{\nu}-
    {\Gamma^{\rho}}_{\nu\mu}A_{\rho},
\end{equation}
where matrices ${\Gamma^{i}}_{jk}$ are called the connection coefficients, so that $\nabla_{\mu}A^{\nu}$ and $\nabla_{\mu}A_{\nu}$ are tensors. The connection used in GR is symmetric in lower indices (${\Gamma^{\lambda}}_{\mu\nu}=    {\Gamma^{\lambda}}_{\nu\mu}$) and compatible with the metric $\nabla_{\lambda}g_{\mu\nu}=0$. It is called the Levi-Civita's connection, and the corresponding coefficients ${\Gamma^{\lambda}}_{\mu\nu}$ the Christoffel symbols. The action on tensors is defined through linearity and Leibniz rule. Express the Christoffel symbols through the metric tensor.

\item \label{equ_oto3} Derive the transformation rule for matrices ${\Gamma^{\lambda}}_{\mu\nu}$ under coordinate transformations. Show that for any given point of spacetime there is a coordinate frame, in which ${\Gamma^{\lambda}}_{\mu\nu}$ are equal to zero in this point. It is called a locally inertial, or locally geodesic frame.

\item \label{equ_oto4} Free falling particles' worldlines in General Relativity are \textit{geodesics} of the spacetime--- i.e the curves $x^{\mu}(\lambda)$ with tangent vector $u^{\mu}=dx^{\mu}/d\lambda$, such that covariant derivative of the latter along the curve equals to zero:
\[u^{\mu}\nabla_{\mu}u^{\nu}=0.\]
In a (pseudo-)Euclidean space the geodesics are straight lines. Obtain the general equation of geodesics in terms of the connection coefficients. Show that the quantity $u^{\mu}u_{\mu}$ is conserved along the geodesic.

\item \label{equ_oto5} Consider the action for a massive particle of the form
\[S_{AB}=-mc\int_{A}^{B} ds,\quad\text{where}\quad
    ds=\sqrt{g_{\mu\nu}dx^{\mu}dx^{\nu}}\]
and derive the geodesic equation from the principle of least action. Find the canonical $4$-momentum of a massive particle and the energy of a photon.

\item \label{equ_oto-kill1} A Killing vector field, or just Killing vector, is a vector field $K^{\mu}(x)$, such that infinitesimal coordinate transformation $x\to x'+\varepsilon K $ (where $\varepsilon\to 0$) leaves the metric invariant in the sense\footnote{That is, let $g_{\mu\nu}(x)$ be the components of the metric at some point $A$ in the original frame. Then $g'_{\mu\nu}(x)$ are the components of the metric in the new frame, taken at point $A'$, which has the same coordinates in the new frame as $A$ had in the old frame.}
\[g_{\mu\nu}(x)=g'_{\mu\nu}(x).\]
A Killing vector defines a one-parametric symmetry group of the metric tensor, called isometry.

Show that a Killing vector obeys the equation
\[\nabla_{\mu}K_{\nu}+\nabla_{\nu}K_{\mu}=0,\]
called the Killing equation.

\item \label{equ_oto-kill2} Suppose there is a coordinate frame, in which the metric does not depend on one of the coordinates $x^{k}$. Show that in this case the vectors $\partial_{k}$ constitute the Killing vector field, and that the inverse is also true: if there is a Killing vector, we can construct such a coordinate frame.

\item \label{equ_oto-kill3} Prove that if $K^{\mu}$ is a Killing vector, the quantity $K^{\mu}u_{\mu}$ is conserved along a geodesic with tangent vector $u^{\mu}$.

\item \label{equ_oto6} The Riemann curvature tensor ${R^{i}}_{klm}$ can be defined through the so-called Ricci identity, written for arbitrary $4$-vector $A^i$:
\[\nabla_{m}\nabla_{l}A^{i}-\nabla_{l}\nabla_{m}A^{i}
    ={R^{i}}_{klm}A^{k}.\]
Express ${R^{i}}_{klm}$ in terms of the Christoffel symbols. Show that the Ricci tensor
\[R_{km}={R^{l}}_{klm}\]
is symmetric.
\item\label{equ_oto6a} Prove the differential Bianchi identity for the curvature tensor:
\begin{equation}\label{BianchiId}
    \nabla_{i}{R^{j}}_{klm}+\nabla_{l}{R^{j}}_{kmi}
    +\nabla_{m}{R^{j}}_{kil}=0,
\end{equation}
and show that
\[\nabla_{i}R^{i}_{j}=\tfrac{1}{2}\partial_{j}R.\]

\item \label{equ_oto6b} The energy-momentum tensor in General Relativity is defined through the variational derivative of the action for matter
\[S_{m}[g^{\mu\nu},\psi]
	=\frac{1}{c}\int d^{4}x\sqrt{-g}
	\;L_{m}(g^{\mu\nu},\psi)\]
with respect to metric $g^{\mu\nu}$:
\[\delta_{g}S=\tfrac{1}{2}\int d^{4}x\sqrt{-g}\;
    \delta g^{\mu\nu}T_{\mu\nu}.\]
Here $L_{m}$ is the Lagrange function for the matter fields $\psi$. Show that for the cases of a massless scalar field and electromagnetic field the above definition reduces to the usual one.

\item \label{equ_oto7} The full action consists of the action for matter, discussed in the previous problem, and the action for the gravitational field $S_{g}$:
\[S=S_{g}+S_{m};\qquad\text{where}\quad
S_{g}=-\frac{c^{3}}{16\pi G}\int d^{4}x\sqrt{-g}\;R,\]
$R=R^{i}_{i}=g^{ik}R_{ik}$ is scalar curvature, $G$ is the gravitational constant. Starting from the variational principle, derive the Einstein-Hilbert equations\footnote{Also called Einstein field equations, or just Einstein equation (either in singular or plural); the action $S_{g}$ is referred to as Hilbert or Einstein-Hilbert action.} for the
gravitational field
\begin{equation}\label{EinsteinHilbetEq}
     R_{\mu\nu}-\tfrac{1}{2}Rg_{\mu\nu}
    =\frac{8\pi G}{c^4}T_{\mu\nu}.
\end{equation}

\item \label{equ_oto8} Show that the Einstein's equation can be presented in the following form
\[R_{\mu\nu} = \frac{8\pi G}{c^4}
     \left( T_{\mu\nu}-\tfrac{1}{2}T g_{\mu\nu}\right),
    \quad\text{where}\quad T=T^{\mu}_{\mu}.\]
Show that it leads to the ''energy-momentum conservation law'' for matter
\[\nabla_{\mu}T^{\mu\nu}=0.\]
Does it mean the energy and momentum of matter are actually conserved in general?
\end{enumerate}

\section{Friedman--Lema\^{i}tre--Robertson--Walker (FLRW) metric}
\begin{enumerate}[resume]
\item \label{equ16} Consider two points $A$ and $B$ on a two-dimensional sphere with radius $a(t)$ depending on time. Find the distance between the points $r_{AB}$, as measured along the surface of the sphere, and their relative velocity $v_{AB}={dr_{AB}}/{dt}$.

\item \label{equ17} The comoving reference frame is defined so that matter is at rest in it, and the distance $\chi_{AB}$ between any two points $A$ and $B$ is constant. Show that in a homogeneous and isotropic Universe the proper (physical) distance $r_{AB}$ between two points is
related to the comoving one as
\[r_{AB}=a(t)\cdot \chi_{AB},\]
where quantity $a$ is called the scale factor and it can depend on time only. Integrate the Hubble's law and find  $a(t)$.

\item \label{equ21} Consider a spacetime with  homogeneous and isotropic spatial section of constant time $dt=0$. Show that in the comoving coordinates its metric necessarily has the form of the Friedman--Lema\^{i}tre--Robertson--Walker (FLRW)\footnote{Depending on geographical or historical preferences, named after a subset of the four scientists: Alexander Friedmann, Georges Lema\^{i}tre, Howard Percy Robertson and Arthur Geoffrey Walker. Thus abbreviations FRW, RW or FL are also used.} metric:
\begin{equation}\label{FLRW1}
    ds^2=dt^2-a^2(t)
    \left\{ d\chi^2+\Sigma^2(\chi)
    (d\theta^2+\sin^2\theta d\varphi^2)\right\},
\end{equation}
where
\[\Sigma^2(\chi)=
    \left\{\begin{array}{lcl}
        \sin^2\chi \\
        \chi^2 \\
        \sinh^2 \chi. \\
\end{array}\right.\]
The time coordinate $t$, which is the proper time for the comoving matter, is referred to as cosmic (or cosmological) time.

\item \label{equ20} Show that the FLRW metric (\ref{FLRW1}) can be presented in the form
\begin{equation}\label{FLRW2}
    ds^2=dt^2-a^2(t) \left\{
    \frac{dr^2}{1-kr^2}+r^2
    (d\theta^2+\sin^2\theta d\varphi^2) \right\},
\end{equation}
where $k=0,\pm1$ is the sign of spatial curvature (see problem
 \ref{equ28n}).

\item \label{equ57} Show that only the sign of spatial curvature has physical meaning, as renormalization of the scale factor rescales the curvature.

\item \label{equ56} Why is the normalization of the scale factor not fixed for a spatially flat Universe, for which $k=0$?

\item \label{equ22} Consider a closed Universe (with $k=+1$) and find the length of equator and full volume of its spatial section $dt=0$.

\item \label{equ23} Present arguments in favor of the affirmation that the electric charge of a closed Universe should be exactly zero.

\item \label{equ24} Using the FLRW metric, derive the Hubble's law.

\item \label{equ64} Conformal time $\eta$ is defined as
\[dt=a(\eta)d\eta.\]
It can be interpreted as the time measured by a clock that decelerates along with the expansion of the Universe. Rewrite the FLRW metric in conformal time. Show that the logarithmic derivative of the scale factor with respect to conformal time determines its evolution in the physical time.

\item \label{equ71} Express the FLRW metric in
comoving coordinates and conformal time. Show that in the case $k=0$ it is conformally flat, i.e. it can be made flat
(pseudo-Euclidean) by means of global stretching.

\item \label{equ66n} Consider an arbitrary function of time $f(t)$ and express $\dot{f}$ and $\ddot{f}$ in terms of derivatives with respect to conformal time.

\item \label{equ65} Obtain the equation of a photon's  worldline in terms of conformal time for the case of the isotropic and spatially flat Universe.

\item \label{equGeo1} Derive the equations of geodesics in terms of conformal time and comoving coordinates for the case of radial motion in the FLRW metric.

\item \label{equGeo2} A comoving observer is the one that is at rest in the comoving coordinates. He sees the Universe as isotropic, and can also be called an isotropic observer. Show that the frequency of a photon and velocity of a free particle, as measured by a comoving observer\footnote{We will refer to these quantities as to the ``physical'' energy and momentum of a particle, to stress that they are the ones directly measured in the most natural way.} at time $t$, are proportional to $1/a(t)$.

\item \label{equ70} Express the detected redshift of a photon as a function of the cosmic time $t$ at the moment of its emission and vice versa: express the time $t$ and conformal time $\eta$ at the moment of its emission in terms of its registered redshift.

\item \label{equ69} Obtain the relation between the scale factor and conformal time using the properties of conformal time interval.

\item \label{equ72} Is it possible for an open Universe to evolve into a closed one or vice versa?

\item \label{equ27} Calculate all connection coefficients (Christoffel symbols) for the FLRW metric.

\item \label{equ28} Derive the components of Ricci tensor, scalar curvature and the trace of energy-momentum tensor for the FLRW metric.

\item \label{equ28n} Obtain the components of the Ricci tensor and scalar curvature ${}^{(3)}R$ of the spatial section $t=const$ of the FLRW metric. Show that $k=sign^{(3)}R$ if ${}^{(3)}R\neq 0$.

\item \label{equ29n} Derive the components and trace of the energy-momentum tensor which satisfies the cosmological principle.
\end{enumerate}

\section[Expanding Universe: ordinarity, difficulties and paradoxes]{Expanding Universe: \\ordinarity, difficulties and paradoxes}
\begin{flushright}\begin{minipage}{0.6\textwidth}
 {\it
\ldots how is it possible for space,
 which is utterly empty, to expand?
 How can nothing expand?
 The answer is: space does not expand.
 Cosmologists sometimes talk about expanding space,
 but they should know better.
\begin{flushright} Stephen Weinberg\end{flushright}}
\end{minipage}\end{flushright}

\subsection{Warm-up}
\begin{enumerate}[resume]
\item\label{equ_exp1} An elastic rubber cord of $1$ meter length $1$ is attached to a wall. A spider sits on it at the junction to the wall, and a man holds the other end. The man starts moving away from the wall with velocity $1\, m/s$, and at the same time the spider starts to run along the cord with velocity $1\, cm/s$. Will the spider come up with the man?

\item\label{equ_b0} Does the law of inertia hold in an expanding Universe?

\item\label{equ_exp2} Suppose a particle's mean free path in an expanding Universe is small enough. Show that its momentum decreases as\footnote{This is a generalization of the previous problem to relativistic case, but still a simplification of the general formulation \ref{equGeo2}.} $p(t)\propto a(t)^{-1}$.

\item \label{equ_exp3} Show that the comoving phase volume equals to the physical one.

\item \label{equ72nn} Show that the twins' paradox is resolved in the FLRW Universe in the same way as in the Minkowski spacetime: the twin who experienced non-zero acceleration appears to be younger than his brother\footnote{O.Gron, S. Braeck, arXiv:0909.5364}.

\item\label{equ_b7} Show that the Hubble's time $H_{0}^{-1}$ gives the characteristic time scale for any stage of evolution of the Universe.

\item\label{equ_b8} Show that in a Universe which expands with acceleration the Hubble's radius decreases.

\item\label{equ_b9} Show that the surface of Hubble's sphere recedes with velocity $V=c(1+q)$, where $q=-a\ddot{a}/(\dot{a})^2$ is the deceleration parameter.

\item \label{equ63} Show that the standard definition of  redshift is valid only inside the Hubble's sphere.
\end{enumerate}

\subsection{The tethered galaxy problem}
Inspired by\footnote{T. Davis, C. Lineweaver, J. Webb, 2001. Solutions to the tethered galaxy problem (arXiv:astro-ph/0104349v3)}
\begin{enumerate}[resume]
\item\label{equ_b1} Let us consider radial motion in the uniform and homogeneous Universe. For this case the FLRW metric reduces to
\[ds^2 =c^2 dt^2 -a^{2}(t)d\chi^{2}.\]
Proper (physical) distance is defined as the distance (measured along the constant time section $dt=0$) between an observer and a galaxy with given comoving coordinate. Let us define the total velocity of a test galaxy as the time derivative of the proper distance
\[v_{tot}=\dot{D},\quad
    \dot{D}=\dot{a}\chi+a\dot{\chi},\quad
    v_{tot}=v_{rec}+v_{pec}.\]
Here $v_{rec}$ is the recession velocity of the test galaxy and $v_{pec}$ is its peculiar velocity. What can be said of the possible values of these velocities?

\item \label{equ_b1a} Determine the distance to a galaxy which, due to the Hubble's expansion, recedes from us with the speed of light.

\item \label{equ_b1b} Is it possible for cosmological objects to recede from us with superluminal speeds?

\item \label{equ_b1c} Is it possible to observe galaxies receding with superluminal speeds?

\item\label{equ_b2} Imagine that we separate a small test galaxy from the Hubble flow by tethering it to an observer such that the proper distance between them remains constant. We can think of the tethered galaxy as one that has received a peculiar velocity boost toward the observer that exactly matches its recession velocity. We then remove the tether (or turn off the boosting rocket) to establish the initial condition of constant proper distance $\dot{D}_{0}=0$. Determine the future fate of the test galaxy: will it approach the observer, recede from him or remain at constant distance?

\item\label{equ_b3} Show that, provided the Universe expands forever, the test galaxy considered in the previous problem asymptotically joins the Hubble flow.

\item\label{equ_b4} Obtain the analogue of the Hubble's law for acceleration in presence of radial peculiar velocity.

\item\label{equ_b5} Derive the result of the previous problem by direct differentiation of the Hubble's law.

\item\label{equ_b6} In the context of special relativity (Minkowski space), objects at rest with respect to an observer have zero redshift. However, in an expanding universe special relativistic concepts do not generally apply. ``At rest'' is defined to be ``at constant proper distance'' ($v_{tot}=\dot{D}= 0$), so our untethered galaxy with $\dot{D}=0$ satisfies the condition
for being at rest. Will it therefore have zero redshift?
That is, are $z_{tot}=0$ and $v_{tot}=0$ equivalent?

\item\label{equ_b7} Show that, although radial recession and peculiar velocities add vectorially, their corresponding redshifts combine as
\[1+z_{tot}=(1+z_{rec})(1+z_{pec}).\]
\end{enumerate}

\subsection{Cosmological redshift}
Inspired by\footnote{E. Bunn, D. Hogg. The kinematic origin of the cosmological redshift. \textit{Am. J. Phys.} \textbf{77}:688-694, (2009); arXiv:0808.1081.}
\begin{enumerate}[resume]
\item \label{equ_red1} Derive the cosmological redshift as the result of addition of infinitesimal Dopper shifts due to relative velocities of galaxies along the worldline of a photon.

\item \label{equ_red2} Suppose the source galaxy $A$ and detector galaxy $B$ are moving with the Hubble flow. Imagine a family of comoving observers situated along the trajectory of the photon. Let the observer 1, closest to the source galaxy, measure his velocity $v_1$ relative to the galaxy and send this information along with the photon to the next closest to him observer 2. Observer 2 measures his velocity $u$ relative to observer 1 and calculates his velocity relative to the galaxy $v_2$ according to the special relativistic formula
\[v_{2}=\frac{v_1 +u}{1+v_1 u}.\]
He sends this information along. What will be the velocity  $v_{rel}$ of the observers relative to the galaxy, defined this way, in terms of scale factors at the moment of emission and at the moment of detection?

\item \label{equ_red3} Show that the registered cosmological redshift corresponds to Doppler effect with this very velocity $v_{rel}$.

\item \label{equ_red4} Find the relative physical velocity of two particles with $4$-velocities $u_{1}^{\mu}$ and $u_{2}^{\mu}$. Let $u_{1}^{\mu}$ be the $4$-velocity of the comoving detector at the moment of detection, and let $u_{2}^{\mu}$ be the $4$-velocity of the source at the moment of emission, parallel transported to the detector along the worldline of the photon\footnote{A vector $a$ is parallel transported along a curve with tangent vector $u^{\mu}$, if $u^{\mu}\nabla_{\mu}a^{\nu}=0$.}.

\item\label{equ_kill4} A Killing tensor $K_{\mu\nu}$ is a tensor field, which obeys the generalization of the Killing equation
\[\nabla_{(\mu}K_{\nu\lambda)}=0,\]
where parenthesis denote symmetrization over all indices. Prove that the quantity $K_{\mu\nu}u^{\mu}u^{\nu}$ is conserved along a geodesics with tangent vector $u^{\mu}$.

\item\label{equ_kill5} Verify that the tensor
\begin{equation}\label{FLRWKillingTensor}
    K_{\mu\nu}
	=a^{2}\big(u_{\mu}u_{\nu}-g_{\mu\nu}\big),
\end{equation}
where $u^{\mu}$ is the $4$-velocity of a comoving particle, is a Killing tensor for the FLRW metric.

\item\label{equ_kill6} Show that due to the Killing tensor (\ref{FLRWKillingTensor}), the physical momentum of a particle, measured by comoving observers, changes with time as $p\sim 1/a$.
\end{enumerate}

\section{Friedman Equations}
\begin{enumerate}[resume]
\item \label{equ29} Starting from the Einstein equations, derive the equations for the scale factor $a(t)$ of the FLRW metric---the Friedman equations:
\begin{align}\label{FriedmanEqI}
    \Big(\frac{\dot a}{a}\Big)^2& =\;
    \frac{8\pi G}{3}\rho -\frac{k}{a^2};\\
    \label{FriedmanEqII}
    \frac{\ddot a}{a} &=- \frac{4\pi G}3\big(\rho+3p).
\end{align}
Consider matter an ideal fluid with $T^{\mu}_{\nu}=diag(\rho,-p,-p,-p)$ (see problem \ref{equ29n}).

\item \label{equ66} Derive the Friedman equations in terms of conformal time.

\item \label{equ40} Show that the first Friedman equation is the first integral of the second one.

\item \label{equ53} Show that in the weak field limit of General Relativity the source of gravity is the quantity $(\rho + 3p)$.

\item \label{equ54} How does the magnitude of pressure affect the expansion rate?

\item \label{equ41} Consider the case $k = 0$ and show that the second Friedman equation can be presented in the form
\[HH' =  - 4\pi G\left(\rho  + p\right),\]
where $H' \equiv \frac{dH}{d\ln a}.$

\item \label{equ29nn} Are solutions of Friedman equations Lorentz-invariant?

\item \label{equ31} The critical density corresponds to the case of spatially-flat Universe $k=0$. Determine its actual value.

\item \label{equ33} Show that the first Friedman equation can be presented in the form
\[\sum\limits_i\Omega_i=1,\]
where $\Omega_i$ are relative densities of the components,
\[\Omega_i\equiv\frac{\rho_{i}}{\rho_{cr}},
    \quad \rho_{cr}=\frac{3H^{2}}{8\pi G},
    \quad \rho_{curv}=-\frac{3}{8\pi G}\frac{k}{a^2},\]
and $\rho_{curv}$ describes the contribution to the total density of the spatial curvature.

\item \label{equ34n} Express the scale factor $a$ in a non-flat Universe through the Hubble's radius and total relative density $\rho/\rho_{cr}$.

\item \label{equ34} Show that the relative curvature density $\rho_{curv}/\rho_{cr}$ in a given region can be interpreted as a measure of difference between the average potential and kinetic energies in the region.

\item  \label{equ35} Prove that in the case of spatially flat Universe
\[\dot{H} = - 4\pi G (\rho + p).\]

\item \label{equ36} Obtain the Raychadhuri equation
\[H^2 + \dot H =  - \frac{4\pi G}{3}(\rho  + 3p).\]

\item \label{equ37} Starting from the Friedman equations, obtain the conservation equation for matter in an expanding Universe:
\[\dot{\rho}+3H(\rho+p)=0.\]
Show that it can be presented in the form
\[\frac{d\ln\rho}{d\ln a}+3(1+w)=0,\]
where $w=p/\rho$ is the state parameter for matter.

\item \label{equ68} Obtain the conservation equation in terms of the conformal time.

\item \label{equ38} Starting from the energy momentum  conservation law $\nabla_{\mu}T^{\mu\nu}=0$, obtain the conservation equation for the expanding Universe.

\item \label{equ46} Show that among the three equations (two Friedman equations and the conservation equation) only two are independent, i.e. any of the three can be obtained from the two others.

\item \label{equ_46e} For a spatially flat Universe rewrite the Friedman equations in terms of the $e$-foldings number for the scale factor
\[N(t)=\ln\frac{a(t)}{a_0}.\]

\item \label{equ47} Express the conservation equation in terms of $N$ and find $\rho(N)$ for a substance with equation of state $p=w\rho$.

\item \label{equ43} For the case of spatially flat Universe express pressure in terms of the Hubble parameter and its time derivative.

\item \label{equ44} Express the state parameter $w = p/\rho$ in terms of the Hubble parameter and its time derivative.

\item\label{equ_duality} Show that for a spatially flat Universe the Friedman equations are invariant\footnote{V. Faraoni. A symmetry of the spatially flat Friedmann equations with barotropic fluids. arXiv:1108.2102v1.} under the change of variables to a new scale factor
\[a \to \alpha=\frac{1}{a}\]
and a the new equation of state:
\[(w+1)\to (\omega+1)=-(w+1).\]

\item \label{equ68n} Evaluate the derivatives of the state parameter $w$ with respect to cosmological and conformal times.

 \item \label{equ44_1} Consider an FLRW Universe dominated by a substance, such that Hubble parameter depends on time as $H=f(t)$, where $f(t)$ is an arbitrary differentiable function. Find the state equation for the considered substance.

\item \label{equ45} Find the upper bound for the state parameter $w$.

\item \label{equ52} Show that for non-relativistic particles the state parameter $w$ is much less unity.
\end{enumerate}

\section{Newtonian cosmology}
\begin{enumerate}[resume]
\item \label{equ48} Obtain the first Friedman equation basing only on Newtonian mechanics.

\item \label{equ49} Derive the analogue of the second Friedman equation in the Newtonian mechanics.

\item \label{equ51} Obtain the conservation equation for non-relativistic matter from the continuity equation for the ideal fluid.

\item \label{equ55} Show that equations of Newtonian hydrodynamics and gravity prohibit the existence of a  uniform, isotropic and static cosmological model, i.e. a Universe constant in time, uniformly filled by ideal fluid.

\item\label{equ_peeb1} Find the generalization of the Newtonian energy conservation equation to an expanding cosmological background\footnote{P.J.E. Peebles, Principles of Physical Cosmology, Princeton University Press, 1993}.

\item\label{equ_peeb2} Using the Layzer-Irvene equation, discussed in the previous problem, recover the Newtonian virial theorem
\end{enumerate}

\section{Energy balance in an expanding Universe}
\begin{enumerate}[resume]
\item \label{equ_en1} Show that the first Friedman equation can be treated as the energy conservation law in Newtonian mechanics. Use this equation to classify the solutions describing different dynamics of the Universe.

\item \label{equ_en2} Obtain the conservation equation for the expanding Universe using only thermodynamical considerations.

\item\label{equ_en3} A photon's wavelength is redshifted due to the Universe' expansion. Estimate the rate of change of the energy of the Universe due to this process.
\end{enumerate}

\section{Cosmography}
{\it In problems \ref{equ60}-\ref{equ63} we use an approach to the description of the evolution of the Universe's, which is called ``cosmography''\footnote{See Weinberg, Gravitation and Cosmology, chapter 14}. It is based entirely on the cosmological principle and on some consequences of the equivalence principle. The term ''cosmography'' is a synonym for ''cosmo\-kinematics''. Let us recall that the kinematics represent the part of mechanics which describes motion of bodies regardless of the forces responsible for it. In this sense cosmography represents nothing else than the kinematics of cosmological expansion.

In order to construct the key cosmological quantity $a(t)$ one needs the equations of motion (the Einstein's equation) and some assumptions an the material composition of the Universe, which enable one to obtain the energy-momentum tensor. The efficiency of cosmography lies in the ability to test cosmological models of any kind, that are compatible with the cosmological principle. Modifications of General Relativity or introduction of new components (such as dark matter or dark energy) certainly change the dependence $a(t)$, but they absolutely do not affect the kinematics of the expanding Universe.

The rate of Universe's expansion, determined by Hubble parameter $H(t)$, depends on time. The deceleration parameter $q(t)$ is used to quantify this dependence. Let us define it through the expansion of the scale factor $a(t)$ in a Taylor series in the vicinity of current time ${{t}_{0}}$:
\[a(t)=a\left( {{t}_{0}} \right)
	+\dot{a}\left( {{t}_{0}} \right)\left[ t-{{t}_{0}} \right]
	+\frac{1}{2}\ddot{a}({t}_{0})
	{{\left[ t-{{t}_{0}} \right]}^{2}}+\cdots\]
Let us present this in the form
\[\frac{a(t)}{a\left( {{t}_{0}} \right)}
	=1+{{H}_{0}}\left[ t-{{t}_{0}} \right]
	-\frac{{{q}_{0}}}{2}H_{0}^{2}
	{{\left[ t-{{t}_{0}} \right]}^{2}}+\cdots\]
where the deceleration parameter is
\[q(t)\equiv -\frac{\ddot{a}(t)a(t)}{{{{\dot{a}}}^{2}}(t)}
	=-\frac{\ddot{a}(t)}{a(t)}\frac{1}{{{H}^{2}}(t)}.\]

Note that the accelerated growth of scale factor takes place for $q<0$. When the sign of the deceleration parameter was originally defined, it seemed evident that gravity is the only force that governs the dynamics of Universe and it should slow down its expansion. The choice of the sign was determined then by natural wish to deal with positive quantities. This choice turned out to contradict the observable dynamics and became an example of historical curiosity.

In order to describe the kinematics of the cosmological expansion in more details it is useful to consider the extended set of the parameters:}
\begin{align}
	 H(t)\equiv &\frac{1}{a}\frac{da}{dt}\\
q(t)\equiv& -\frac{1}{a}\frac{{{d}^{2}}a}{d{{t}^{2}}}{{\left[ \frac{1}{a}\frac{da}{dt} \right]}^{-2}}\\
j(t)\equiv &\frac{1}{a}\frac{{{d}^{3}}a}{d{{t}^{3}}}{{\left[ \frac{1}{a}\frac{da}{dt} \right]}^{-3}}\\
s(t)\equiv &\frac{1}{a}\frac{{{d}^{4}}a}{d{{t}^{4}}}{{\left[ \frac{1}{a}\frac{da}{dt} \right]}^{-4}}\\
l(t)\equiv&\frac{1}{a}\frac{{{d}^{5}}a}{d{{t}^{5}}}{{\left[ \frac{1}{a}\frac{da}{dt} \right]}^{-5}}
\end{align}

\begin{enumerate}[resume]
\item \label{equ60} Using the cosmographic parameters introduced above, expand the scale factor into a Taylor series in time.

\item \label{equ61} Using these cosmographic parameters, expand the redshift into a Taylor series in time.

\item \label{equ61_1}\label{q(h)} Obtain the following relations between the deceleration parameter and Hubble's parameter
\[q(t)=\frac{d}{dt}\left( \frac{1}{H} \right)-1;\quad
    q(z)=\frac{1+z}{H}\frac{dH}{dz}-1;\quad
     q(z)=\frac{d\ln H}{dz}(1+z)-1.\]

 \item \label{equ61_2} Show that  for the deceleration parameter the following relation holds:
\[q\left( a \right)=-\left( 1+
    \frac{\frac{dH}{dt}}{{{H}^{2}}} \right)
    -\left( 1+\frac{a\frac{dH}{da}}{H} \right).\]

\item \label{equ61_3} Show that derivatives of lower cosmographic parameters can expressed through the higher ones.

\item \label{equ61_4} Prove that
\[\frac{dq}{d\ln (1+z)}=j-q(2q+1).\]

\item \label{equ61_5} Show that the derivatives $\frac{dH}{dz}$ and $\frac{{{d}^{2}}H}{d{{z}^{2}}}$ can be expressed through the parameters $q$ and $j$.

\item \label{equ61_6} Show that the time derivatives of the Hubble's parameter can be expressed through the cosmographic parameters as follows:
\begin{eqnarray}
\dot{H} &=& -{{H}^{2}}(1+q); \\
\ddot{H} &=& {{H}^{3}}\left( j+3q+2 \right); \\
\dddot{H}&=& {{H}^{4}}\left[ s-4j-3q(q+4)-6 \right];  \\
\ddddot{H}&=& {{H}^{5}}\left[ l-5s+10\left( q+2 \right)j+30(q+2)q+24\right].
\end{eqnarray}

\item \label{equ61_7} Consider the case of spatially flat Universe and express the scalar (Ricci) curvature and its time derivatives in terms of the cosmographic parameters $q,j,s,l$.

\item \label{equ61_7n} Show that the accelerated growth of expansion rate $\dot{H}>0$ takes place on the condition $q<-1$.

\item \label{equ61_2_1} Obtain the relation
\[H(z)=-\frac{1}{1+z}\frac{dz}{dt}.\]

\item\label{dyn23} Let $t_a$ be the moment in the history of the Universe when the decelerated expansion turned to the accelerated one, i.e. $q(t_a)=0$, and let $t_{1}<t_{a}$ and $t_{2}>t_{a}$ be two moments in the vicinity of $t_{a}$. Show that
\[\Delta t\equiv t_1-t_2 =\frac{1}{H_1}-\frac{1}{H_2}.\]

\item\label{dyn27} Obtain the following integral relation between the Hubble's parameter and the deceleration parameter
\[H=H_0\exp\left[   \int\limits_0^z
    [q(z^\prime)+1] d\ln(1+z^\prime)\right].\]

\item \label{equ62} Reformulate the Hubble's law in terms of redshift for close galaxies $(z\ll 1)$.

\item \label{equ62_1} Show that
\[\frac{d}{dt}=-(1+z)H\frac{d}{dz}.\]

\item \label{equ62_21} Obtain the transformation law from the higher time derivatives to the derivatives with respect to redshift:
\[\frac{d^{(i)}}{dt}\to \frac{d^{(i)}}{dz}.\]

\item \label{equ62_3} Calculate the derivatives of Hubble's parameter squared with respect to redshift \[\frac{d^{(i)}H^2}{dz^{(i)}},\quad i=1,2,3,4\]
 and express them in terms of the cosmographic parameters.

\item\label{dyn72} Show that the deceleration parameter $q$ can be presented in the form
\[q(x) = \frac{H'(x)}{H(x)}x - 1;\; x = 1 + z.\]

\item\label{dyn72n} Show that
\[q(z)=\frac{1}{2}\frac{d\ln {{H}^{2}}}{d\ln (1+z)}.\]

\item\label{dyn73} Express the derivatives $dH/dz$ and $d^2H/dz^2$ throgh the parameters $q$ and \[r \equiv \frac{\dddot a}{aH^3}.\]

\item \label{equ62_2} Consider the time average of the deceleration parameter
\[\bar{q}\left( {{t}_{0}} \right)
    =\frac{1}{{{t}_{0}}}\int_{0}^{{{t}_{0}}}{q(t)dt}\] and show that it can be evaluated without integration of equation of motion for the scale factor.

\item \label{equ62_3b} Show that the current age of the Universe is proportional to $H_{0}^{-1}$ and the proportionality coefficient is determined by the average value of the deceleration parameter.

\item \label{equ62_4m} Show that the proper distance to an object with redshift $z$ is related to the current deceleration parameter $q_0$ as
\[R=\frac{c}{H_{0}q_{0}^{2}}\,\frac{1}{1+z}\,
	\Big[q_{0}z +(q_{0}-1)\big(\sqrt{1+2q_0}-1\big)\Big].\]

\item
Express the current values of deceleration parameter
$q_0  \equiv \left. { - \frac{1}{a}\frac{d^2 a}{dt^2}\left[ {\frac{1}{a}\frac{da}{dt}} \right]^{ - 2} } \right|_{t = t_9 }$
and the jerk parameter
$j \equiv \left. {\frac{1}{a}\frac{d^3 a}{dt^3}\left[ {\frac{1}{a}\frac{da}{dt}} \right]^{ - 3} } \right|_{t = t_0 }$
in terms of $N \equiv  - \ln \left( {1 + z} \right)$

\item
Using $d_l (z)=a_0 (1+z)f(\chi)$, where
\[\chi=\frac1{a_0}\int\limits_0^z\frac{du}{H(u)},\quad f(\chi)=\left\{\begin{array}{rcl}
    \chi & - & flat\ case\\
    \sinh(\chi) & - & open\ case\\
    \sin(\chi) & - & closed\ case.
    \end{array}\right.\]
find the standard luminosity distance-versus-redshift relation up to the second order in $z$:
\[d_L=\frac z{H_0}\left[1+\left(\frac{1-q_0}2\right)z+O(z^2)\right].\]

\item
Many supernovae give data in the $z>1$ redshift range. Why is this a problem for the above formula for the $z$-redshift? (Problems 2) - 10) are all from the Visser convergence article)

\item
Give physical reasons for the above divergence at $z=-1$.

\item
Sometimes a "pivot" is used:
\[z=z_{pivot}+\Delta z,\]
\[\frac1{1+z_{pivot}+\Delta z}=\frac{a(t)}{a_0}=1+H_0(t-t_0)-\frac1{2!}{q_0 H_0^2}(t-t_0)^2+\frac1{3!}{j_0 H_0^3}(t-t_0)^3+O(|t-t_0|^4).\]
What is the convergence radius now?

\item
The most commonly used definition of redshift is
\[z=\frac{\lambda_0-\lambda_e}\lambda_e=\frac{\Delta\lambda}\lambda_e.\]
Let's introduce a new redshift:
\[z=\frac{\lambda_0-\lambda_e}\lambda_0=\frac{\Delta\lambda}\lambda_0.\]

\item
Argue why, on physical grounds, we can't extrapolate beyond $y=1$.

\item\label{cg_7}
The most commonly used distance is the luminosity distance, and it is related to the distance modulus in the following way:
\[\mu_D=5\log_10[d_L/(10\ pc)]=5\log_10[d_L/(1\ Mpc)]+25.\]
However, alternative distances are also used (for a variety of mathematical purposes):

1) The "photon flux distance": \[d_F=\frac{d_L}{(1+z)^{1/2}}.\]

2) The "photon count distance": \[d_P=\frac{d_L}{(1+z)}.\]

3) The "deceleration distance": \[d_Q=\frac{d_L}{(1+z)^{3/2}}.\]

4) The "angular diameter distance": \[d_A=\frac{d_L}{(1+z)^{2}}.\]

Obtain the Hubble law for these distances (in terms of $z$-redshift, up to the second power by z)

\item
Do the same for the distance modulus directly.

\item
Do problem \ref{cg_7} but for $y$-redshift.

\item\label{cg_10}
Obtain the THIRD-order luminosity distance expansion in terms of $z$-redshift

\item
Alternative (if less physically evident) redshifts are also viable. One promising redshift is the $y_4$ redshift: $y_4 = \arctan(y)$. Obtain the second order redshift formula for $d_L$ in terms of $y_4$.

\item
$H_0a_0/c\gg1$ is a generic prediction of inflationary cosmology. Why is this an obstacle in proving/measuring the curvature of pace based on cosmographic methods?

\item
When looking  at the various series formulas, you might be tempted to just take the highest-power formulas you can get and work with them.  Why is this not a good idea?

\item
Using the definition of the redshift, find the ratio \[\frac{\Delta z}{\Delta t_{obs}},\] called "redshift drift" through the Hubble Parameter for a fixed/commoving observer and emitter:
\[\int\limits_{t_s}^{t_o}\frac{dt}{a(t)}=\int\limits_{t_s+\Delta t_s}^{t_o+\Delta t_o}\frac{dt}{a(t)}.\]

\item
A photon's physical distance traveled is \[D=c\int dt=c(t_0-t_s).\] Using the definition of the redshift, construct a power series for $z$-redshift based on this physical distance.

\item
Invert result of the previous problem (up to $z^3$).

\item
Obtain a power series (in $z$) breakdown of redshift drift up to $z^3$ (hint: use $(dH)/(dz)$ formulas)

\item
Using the Friedmann equations, continuity equations, and the standard definitions for heat capacity(that is, \[C_V=\frac{\partial U}{\partial T},\quad C_P=\frac{\partial h}{\partial T},\] where  $U=V_0\rho_ta^3$, $h=V_0(\rho_t+P)a^3$, show that
\[C_P=\frac{V_0}{4\pi G}\frac{H^2}{T'}\frac{j-1}{(1+z)^4},\]
\[C_V=\frac{V_0}{8\pi G}\frac{H^2}{T'}\frac{2q-1}{(1+z)^4}.\]

\item
What signs of $C_p$ and $C_v$ are predicted by the $\Lambda-CDM$ model? ($q_{0\Lambda CDM}=-1+\frac32\Omega_m$, $j_{0\Lambda CDM}=1$ and experimentally, $\Omega_m=0.274\pm0.015$)

\item\label{cg_20}
In cosmology, the scale factor is sometimes presented as a power series
\[a(t)=c_0|t-t_\odot|^{\eta_0}+c_1|t-t_\odot|^{\eta_1}+c_2|t-t_\odot|^{\eta_2}+c_3|t-t_\odot|^{\eta_3}+\ldots\]. Get analogous series for $H$ and $q$.

\item
What values of the powers in the power series are required for the following singularities?

a) Big bang/crunch (scale factor $a=0$)

b) Big Rip (scale factor is infinite)

c) Sudden singularity ($n$th derivative of the scale factor is infinite)

d) Extremality event (derivative of scale factor $= 1$)

\item
Analyze the possible behavior of the Hubble parameter around the cosmological milestones (hint: use your solution of problem \ref{cg_20}).

\item
Going outside of the bounds of regular cosmography, let's assume the validity of the Friedmann equations. It is sometimes useful to expand the Equation of State as a series (like $p=p_0+\kappa_0(\rho-\rho_0)+O[(\rho-\rho_0)^2],$) and describe the EoS parameter ($w9t)=p/\rho$) at arbitrary times through the cosmographic parameters.

\item
Analogously to the previous problem, analyze the slope parameter \[\kappa=\frac{dp}{d\rho}.\] Specifically, we are interested in the slope parameter at the present time, since that is the value that is seen in the series expansion.

\item
Continuing the previous problem, let's look at the third order term - $d^2 p/d\rho^2$.

\end{enumerate}

\chapter{Dynamics of the Universe in the Big Bang model}
\begin{flushright}
{\it At present the Big Bang model does not have any significant flaws.\\ I would even say that it is confirmed to be true to the same extent\\ that we know that the Earth revolves around the Sun.\\
Yakov B. Zel'dovich}
\end{flushright}

\section{General questions}
\begin{enumerate}
\item\label{dyn2} What main observations underlie the Big Bang model?
\end{enumerate}

\section{Solutions of Friedman equations in the Big Bang model}
\begin{enumerate}[resume]
\item\label{dyn5} Derive $\rho(a)$, $\rho(t)$ and $ a(t)$ for a spatially flat\footnote{The ``spatial'' part is often omitted when there cannot be any confusion (and even when there can be), generally and hereatfer.} ($k=0$) Universe \label{r_m} that consists of only
\begin{description}
    \item[a)] radiation,
    \item[b)] non-relativistic matter\footnote{In this context and below also quite often called just ``matter'' or dust.}.
\end{description}

\item\label{dyn6} Consider two spatially flat Universes. One is filled with radiation, the other with dust. The current energy density is the same. Compare energy densities when both of them are twice as old.

\item\label{dyn11} Find the scale factor and density of each component as functions of time in a flat Universe which consists of dust and radiation, for the case one of the components is dominating. Present the results graphically.

\item\label{dyn17} Derive the time dependence of the Hubble parameter for a flat Universe in which either matter or radiation is dominating.

\item\label{dyn12} Derive the exact solutions of the Friedman equations for the Universe filled with matter and radiation. Plot the graphs of scale factor and both densities as functions of time.

\item\label{dyn20} \label{m_r_eq} At what moment after the Big Bang did matter's density exceed that of radiation for the first time?

\item\label{dyn21} Determine the age of the Universe in which either matter or radiation has always been dominating.

\item\label{dyn14} Derive the dependence  $a(t)$ for a spatially flat Universe \label{a(w)} that consists of matter with equation of state $p=w\rho$, assuming that the parameter $w$ does not change throughout the evolution.

\item\label{dyn18} Find the Hubble parameter as function of time for the previous problem \label{H(w)}. 

\item\label{dyn15} Using the first Friedman equation, construct the effective potential \label{V(a)} $V(a)$, which governs the one-dimensional motion of a fictitious particle with coordinate $a(t)$, for the Universe filled with several non-interacting components.

\item\label{dyn16} Construct the effective one-dimensional potential $V(a)$ (using the notation of the previous problem) for the Universe consisting of non-relativistic matter and radiation. Show that motion with $\dot{a}>0$ in such a potential can only be decelerating.

\item\label{dyn39} Derive the exact solutions of the Friedman equations for the Universe with arbitrary curvature, filled with radiation and matter.

\item\label{dyn41} Show that for a spatially flat Universe consisting of one component with equation of state $p = w\rho$ the deceleration parameter $q\equiv -\ddot{a}/(aH^2)$ is equal to $\frac{1}{2}(1 + 3w)$.

\item\label{dyn21_1} Express the age of the Universe through the deceleration parameter $q=-\ddot{a}/(aH^2)$ for a spatially flat Universe filled with single component with equation of state $p=w\rho$.

\item\label{dyn41nn} Find the generalization of relation $q = \frac{1}{2}(1 + 3w)$ for the non-flat case.

\item\label{dyn49nn} Show that for a one-component Universe filled with ideal fluid of density $\rho$
\[q=-1-\frac{1}{2}\,\frac{d\ln\rho}{d\ln a}.\]

\item\label{dyn42} Show that for a Universe consisting of several components with equations of state $p_{i}  = w_{i} \rho_{i}$ the deceleration parameter is
\[q = \frac{\Omega }{2} +
	\frac{3}{2}\sum\limits_i
	{w_i \Omega_i },\]
where $\Omega$ is the total relative density.

\item\label{dyn41n} For which values of state parameter $w$ the rate of expansion of a one-component flat Universe increases with time?

\item\label{dyn34} Show that for a spatially closed  ($k=1$) Universe that contains only non-relativistic matter the solution of the Friedman equations can be given in the form
\[a(\eta)=a_{\star}(1-\cos\eta);
	\qquad t(\eta)=a_{\star}(\eta -\sin\eta);
	\qquad a_{\star}=\frac{4\pi G\rho_0}{3};
	\quad 0<\eta<2\pi.\]

\item\label{dyn37} Find the relation between the maximum size and the total lifetime of a closed Universe filled with dust.

\item\label{dyn30} Suppose we know the current values of the Hubble constant $H_0$ and the deceleration parameter $q_0$ for a closed Universe filled with dust only. How many times larger will it ever become?

\item\label{dyn28} In a closed Universe filled with non-relativistic matter the current values of the Hubble constant is $H_0$, the deceleration parameter is $q_0$. Find the current age of this Universe.

\item\label{dyn31} Suppose in the same Universe radiation is dominating during a negligibly small fraction of total time of evolution. How many times will a photon travel around the Universe during the time from its ``birth'' to its ``death''?

\item\label{dyn57} In a closed Universe filled with dust the current value of the Hubble constant is $H_0$ and of the deceleration parameter $q_0$.
\begin{description}
\item[a)] What is the total proper volume of the Universe at present time?
\item[b)]  What is the total current proper volume of space occupied by matter which we are presently observing?
\item[c)] What is the total proper volume of space which we are directly observing?
\end{description}

\item\label{dyn_open} Find the solution of Friedman equations for spatially open ($k=-1$) Universe filled with dust in the parametric form $a(\eta)$, $t(\eta)$.

\item\label{dyn40} Suppose the density of some component in a spatially flat Universe depends on scale factor as  $\rho(t) \sim a^{-n}(t)$. How much time is needed for the density of this component to change from $\rho_1$ to $\rho_2$?

\item\label{dyn44} Using the expression for $H(t)$, calculate the deceleration parameter for the cases of domination of
\begin{description}
    \item[a)] radiation,
    \item[b)] matter.
\end{description}

\item\label{dyn77} Consider a Universe consisting of $n$ components, with equations of state $p_{i}=w_{i}\rho_{i}$, and find $w_{tot}$, the parameter of the equation of state $p_{tot}=w_{tot}\rho_{tot}$.

\item\label{dyn78} Derive the equations of motion for  relative densities $\Omega_{i}=\rho_{i}/\rho_{cr}$ of the two components comprising a spatially flat two-component Universe, if their equations of state are $p=w_i\rho$, $i=1,2$.

\item\label{dyn60} Suppose a Universe is initially filled with a gas of non-relativistic particles of mass density  $\rho_{0}$, pressure $p_{0}$, and $c_p/c_v=\gamma$. Construct the equation of state for such a system.

\item\label{dyn19} Derive the expression for the critical density $\rho_{cr}$ from the condition that Hubble's expansion velocity equals the second cosmic velocity (escape velocity) $v=\sqrt{2gR}$.

\item\label{dyn42nn} Suppose the Universe is filled with non-relativistic matter and some substance with equation of state $p_X=w\rho_X$. Find the evolution equation for the quantity $r \equiv \frac{\rho_m}{\rho _X}$.

\item\label{dyn43nn} Express the deceleration parameter through the ratio $r$ for the conditions of the previous problem.

\item\label{dyn46nn}  Let a spatially flat Universe be filled with non-relativistic dust and a substance with equation of state $p_{X}= w\rho_{X}$. Show that in case $\rho_{X}\propto H^2$, the ratio $r =\rho_{m}/\rho_{X}$ does not depend on time.\marginpar{BAD}

\item\label{dyn47nn}  Show that for the model of the Universe described in the previous problem the parameter $r$ is related with the deceleration parameter as\marginpar{BAD}
\[\dot{r}=-2H\frac{\Omega_{curv}}{\Omega_X}q.\]

\item\label{dyn48nn} Show that in the model of the Universe of problem \ref{dyn46nn}, in case $k=+1$ and $q>0$ (decelerated expansion) $r$ increases with time, in case $k=+1$  and $q<0$ (accelerated expansion) $r$ decreases with time, and for $k=-1$ vice-versa.\marginpar{BAD}
\end{enumerate}

The following three problems on power-law cosmology are inspired by Kumar\footnote{S.Kumar arXiv:1109.6924.}
\begin{enumerate}[resume]
\item\label{dyn-Kumar2} Let us consider a general class of power-law cosmologies described by the scale factor
\[a(t) =a_{0}\Big(\frac{t}{t_0}\Big)^\alpha,\]
where $t_0$ is the present age of theUniverse and $\alpha$ is a dimensionless positive parameter. Show that:
\begin{enumerate}
\item the scale factor in terms of the deceleration parameter  may be written as
\[a(t) =a_{0}\Big(\frac{t}{t_0}\Big)^{1/1 + q},
	\quad\text{i.e.}\quad \alpha=\frac{1}{1+q}.\]
\item the expansion of the  Universe is described by Hubble parameter
\[H=\frac{1}{(1+q)\;t}\]
or in terms of redshift
\[H(z)=H_{0}(1+z)^{1+q}.\]
\end{enumerate}

\item\label{dyn-Kumar3} In the power-law cosmology find the age of the Universe at redshift $z$.

\item\label{dyn-Kumar4} For the power-law cosmology find  the luminosity distance between the observer and the object with redshift $z$.
\end{enumerate}

\section{The role of curvature in the dynamics of the Universe}
\begin{enumerate}[resume]
\item\label{dyn38} Derive $\rho(t)$ in a spatially open Universe filled with dust for the epoch when the curvature term in the first Friedman equation is dominating .

\item\label{dyn3} Show that in the early Universe the curvature term is negligibly small.

\item\label{dyn33} Show that $k =\text{sign}(\Omega-1)$ and express the current value of the scale factor $a_{0}$ through the observed quantities $\Omega_{0}$ and $H_{0}$.

\item\label{dyn_curv4} Find the lower bound for $a_{0}$, knowing that the Cosmic background (CMB) data combined with SSNIa data imply
\[-0.0178<(1-\Omega)<0.0063.\]

\item\label{dyn43} Fnd the time dependence \label{o-1} of $\left|\Omega-1\right|$ in a Universe with domination of
\begin{description}
    \item[a)] radiation,
    \item[b)]  matter.
\end{description}

\item\label{dyn4} Estimate the upper bound of the curvature term in the first Friedman equation during the electroweak epoch ($t\sim 10^{-10}$~s) and the nucleosynthesis epoch ($t\sim 1-200$~s).

\item\label{dyn-Kumar1} Derive and analyze the conditions of accelerated expansion for a one-component Universe of arbitrary curvature with the component's state parameter\footnote{S.Kumar arXiv: 1109.6924} $w$.
\end{enumerate}

\section{The Milne Universe}
\begin{enumerate}[resume]
\item\label{Milne1} Find the solutions of the Friedman equations for the Milne Universe: the expanding Universe with $\rho\to 0$ and $k=-1$. Why is it necessarily spatially open? What is the scalar curvature of this spacetime?

\item\label{Milne2} Let us consider the Minkowski spacetime in spherical spatial coordinates $(T,R,\theta,\phi)$. Let at some moment of initial explosion a cloud of particles emerge from the origin with all possible velocities $v<c$ in all directions, which stay constant. Their mass is considered negligible, so that they do not interact and do not affect the underlying spacetime. The larger is the velocity of a particle, the further away from the origin it is at a given moment of time, so the velocity of a particle $v$, or alternatively its ``rapidity''
\[r=\text{artanh}\, v
	\equiv\tfrac{1}{2}\ln\frac{1+v}{1-v}\]
serve as radial coordinates in the region $R<T$. Let $\tau$ be the proper time of the particle. Show that the region $R<T$ in coordinates $(\tau,r,\theta,\phi)$ \emph{is} the Milne Universe\footnote{In fact, this is the way Milne in his papers of 1935 and 1948 introduced this spacetime, trying to show that Big Bang can be described by pure kinematics and in the frame of Special Theory of Relativity only. This is in general not possible, but his renowned example is very instructive.}.

\item\label{Milne3} Let the density of matter in the Milne Universe (in the comoving frame) be small but finite. Find the dependence of (number) density on the distance to the horizon $R=T$ in the Minkowski spacetime (the laboratory frame with regard to the experiment of the Big Bang), if the distribution in the Milne Universe is homogeneous. What is the total number of particles (galaxies) in each of the frames of reference?
\end{enumerate}

\section{Cosmological horizons}
\begin{enumerate}[resume]
\item\label{dyn47} Calculate the particle horizon for a Universe with dominating
\begin{description}
    \item[a)] radiation,
    \item[b)] dust,
	\item[c)] matter with state equation $p=w\rho$.
\end{description}

\item\label{dyn48} Show the the comoving particle horizon equals to the age of the Universe in conformal time.

\item\label{dyn49} Show that, if ultrarelativistic matter is dominating in the matter content of a spatially flat Universe ($k=0$), its particle horizon coincides with the Hubble radius.

\item\label{dyn35} Find the comoving Hubble radius $R_{H}/a$ as function of the scale factor for a spatially flat Universe that consists of one component with equation of state $p=w\rho$.

\item\label{dyn50} Express the comoving particle horizon $L_{p}/a$ through the comoving Hubble radius $R_{H}/a$ for the case of domination of a substance with state parameter $w$.

\item\label{dyn52}  Show that in an open Universe filled with dust the number of observed galaxies increases with time.

\item\label{dyn56} Show that even in early Universe the scale of particle horizon is much less than the curvature radius, and thus curvature does not play significant role within the horizon. 

\item\label{dyn53} Estimate the ratio of the volume enclosed by the Hubble sphere to the total volume of the closed Universe.
\end{enumerate}

The following five problems are based on work by F. Melia\footnote{F. Melia, arXiv: 0711.4181, arXiv:0907.5394}

Standard cosmology is based on the FLRW metric for a spatially homogeneous and isotropic three-dimensional space, expanding or contracting with time. In the coordinates used for this metric, $t$ is the cosmic time, measured by a comoving observer (and is the same everywhere), $a(t)$ is the expansion factor, and $r$ is an appropriately scaled radial coordinate in the comoving frame.

F.Melia  demonstrated the usefulness of expressing the FRLW metric in terms of an observer-dependent coordinate $R=a(t)r$, which explicitly reveals the dependence of the observed intervals of distance, $dR$, and time on the curvature induced by the mass-energy content between the observer and $R$; in the metric, this effect is represented by the proximity of the physical radius $R$ to the cosmic horizon $R_{h}$, defined by the relation
\[R_{h}=2G\,M(R_h).\]
In this expression, $M(R_h)$ is the mass enclosed within $R_h$ (which terns out to be the Hubble sphere). This is the radius at which a sphere encloses sufficient mass-energy to create divergent time dilation for an observer at the surface relative to the origin of the coordinates.
\begin{enumerate}[resume]
\item\label{dyn-Melia1} Show that in a flat Universe $R_{h}=H^{-1}(t)$.

\item\label{dyn-Melia2} Represent the FLRW metric in terms of the observer-dependent coordinate $R=a(t)r$.

\item\label{dyn-Melia3} Show, that if we were to make a measurement at a fixed distance $R$ away from us, the time interval $dt$ corresponding to any measurable (non-zero) value of $ds$ must go to infinity as $r\to R_h$.

\item\label{dyn-Melia4} Show that $R_h$ is an increasing function of cosmic time $t$ for any cosmology with $w>-1$.

\item\label{dyn-Melia5} Using FLRW metric in terms of the observer-dependent coordinate $R=a(t)r$, consider specific cosmologies:
\begin{enumerate}
\item[a)] the De Sitter Universe ;
\item[b)] a cosmology with  $R_h =t$, ($w=-1/3$);
\item[c)] radiation dominated Universe  ($w=1/3$);
\item[d)] matter dominated Universe ($w=0$).
\end{enumerate}
\end{enumerate}

\section{Energy conditions and the Raychaudhuri equation}
\subsection{Energy conditions}
S.~Carroll \cite{Carroll} writes:
\begin{quote} Sometimes it is useful to think about Einstein's equation without spe\-ci\-fying the theory of matter from which $T^{\mu\nu}$ is derived. This leaves us with a great deal of arbitrariness; consider for example the question, What metrics obey Einstein's equation? In the absence of some constraints on $T^{\mu\nu}$, the answer is any metric at all; simply take the metric of your choice, compute the Einstein tensor $G^{\mu\nu}$ for this metric, and then demand that $T^{\mu\nu}$ be equal to $G^{\mu\nu}$. It will automatically be conserved, by the Bianchi identity. Our real concern is with the existence of solutions to Einstein's equation in the presence of ``realistic'' sources of energy and momentum, whatever that means. One strategy is to consider specific kinds of sources, such as scalar fields, dust, or electromagnetic fields. However, we occasionally wish to understand properties of Einstein's equations that hold for a variety of different sources. In this circumstance it is convenient to impose energy
conditions that limit the arbitrariness of $T^{\mu\nu}$.
\end{quote}
The energy conditions are formulated in coordinate-independent way, but in the context of cosmology they are most useful in application to the energy-momentum tensor of a perfect fluid\footnote{For more detailed discussion see textbooks: Carroll S. \textit{Spacetime and geometry: an introduction to General Relativity}. AW, 2003; ISBN 0805387323, 525p. (\textsection 4.6), and Poisson E. \textit{A relativist's toolkit}. CUP, 2004; ISBN 0521830915, 248p. (ch 2)}:
\[\begin{array}{lcc}
\text{Name}&\text{Statement}&\text{For perfect fluid}\\
\text{Weak}\phantom{\Big|}& T_{\mu\nu}v^{\mu}v^{\nu}\geq0&
	\rho\geq 0,\quad \rho+p>0;\\[0.2cm]
\text{Null}& T_{\mu\nu}k^{\mu}k^{\nu}\geq 0&
	\rho+p\geq 0;\\[0.2cm]
\text{Strong}&
	(T_{\mu\nu}-\tfrac{1}{2}Tg_{\mu\nu}) 	
		 v^{\nu}v^{\nu}\geq 0\quad&
	\quad\rho+p\geq 0,\quad \rho+3p\geq 0;\\[0.2cm]
\text{Dominant}& \quad T^{\mu}_{\nu}v^{\nu}\;
	\mbox{\begin{tabular}{c}is non-spacelike and\\ future-directed\end{tabular}}\quad &
	\rho\geq |p\,|.
\end{array}\]
The conditions are assumed to hold for arbitrary timelike vectors $v^{\mu}$ and arbitrary null vectors $k^{\mu}$.
\begin{enumerate}[resume]
\item\label{EnCond1} Derive the energy conditions for the perfect fluid, shown in the last column, from the coordinate-independent formulations.

\item\label{EnCond2} Does the weak energy condition follow from the strong one? Which of the energy conditions imply the others?

\item\label{EnCond3} Express the energy conditions in terms of scale factor and its derivatives.

\item\label{EnCond4} Express the null, weak and strong energy conditions in terms of the Hubble parameter and redshift.

\item\label{EnCond5} Find the restrictions that the energy conditions impose on the deceleration parameter in  a flat Universe with $\rho>0$.
\end{enumerate}

\subsection{Raychaudhuri equation}
\begin{enumerate}[resume]
\item\label{Ray1} Consider a timelike curve $x^{\mu}(\tau)$ and find the projection operators on its tangent vector and on its orthogonal complement.

\item\label{Ray2} A \textit{congruence} is a set of curves having the property that each point in a given region belongs to one and only one curve of the set. Consider a congruence of timelike geodesics. Let us mark two infinitely close geodesics and look at their relative evolution along their length. Let $\xi^{\mu}$ be the infinitesimal $4$-vector that is directed normal to one of the curves towards the other. Show that
\[\frac{d\xi_{\nu}}{d\tau}
	=B_{\nu\mu}\xi^{\mu},\]
where
\[B_{\nu\mu}=u_{\nu;\mu},\]
is a three-dimensional spacelike tensor orthogonal to $u^{\mu}$, and $\tau$ is the parameter along the geodesic.

\item\label{Ray3} Show that any tensor field of second rank defined on an $n-$dimensional Riemannian manifold with positive definite metric $g_{\mu\nu}$ can be uniquely decomposed into
\begin{equation}\label{TensorDecomposition}
	 B_{\mu\nu}=\frac{1}{n}\Theta g_{\mu\nu}
	+\sigma_{\mu\nu}+\omega_{\mu\nu},
\end{equation}
where $\Theta={B^{\mu}}_{\mu}$, $\sigma_{\mu\nu}$ is the symmetric traceless part of $B_{\mu\nu}$, and $\omega_{\mu\nu}$ is the antisymmetric part of $B_{\mu\nu}$.

\item\label{Ray4} Let there be a congruence in a three-dimensional Riemannian manifold. What is the geometric meaning of $\Theta$, $\sigma$ and $\omega$ for $B_{\mu\nu}=u_{\nu;\mu}$?

\item\label{Ray5} Derive the Raychaudhuri equation for a congruence of timelike geodesics in spacetime
\begin{equation}\label{Raychaudhuri}
	\frac{d\Theta}{d\tau}=
	-\frac{1}{3}\Theta^{2}
	-\sigma_{\mu\nu}\sigma^{\mu\nu}
	+\omega_{\mu\nu}\omega^{\mu\nu}
	-R_{\mu\nu}u^{\mu}u^{\mu}.
\end{equation}
Here $\Theta$, $\sigma_{\mu\nu}$ and $\omega_{\mu\nu}$ are the components (\ref{TensorDecomposition}) of decomposition of $B_{\mu\nu}=u_{\mu;\nu}$.

\item\label{Ray6} Show that for a congruence of geodesics orthogonal to a family of hypersurfaces $\omega_{\mu\nu}=0$. Prove further, that in case the strong energy condition $R_{\mu\nu}u^{\mu}u^{\nu}\geq0$ holds (see problem \ref{EnCond1} and the following), then the following (the focusing theorem) also is true:  if $\Theta=\Theta_{0}<0$ at some initial moment, then in a finite period of proper time $\Theta$ diverges and tends to $-\infty$.

\item\label{Ray7} Write out the Raychaudhuri equation for the geodesics of comoving matter in the FLRW Universe and show that it is reduced to the second Friedman equation.
\end{enumerate}

\subsection{Sudden Future Singularities}
The following problems are composed in the spirit of\footnote{John D. Barrow, Sudden Future Singularities, arXiv:0403084v3.}
\begin{enumerate}[resume]
\item\label{sing1} Let us consider the possibility of \textit{sudden future singularities}. The ``suddenness'' implies that they occur at some time in the future, while both the scale factor and the Hubble constant remain bounded and separated from zero:
\[a\to a_{s}\neq 0,\infty,
	\qquad	H\to H_{s}\neq 0,\infty.\]
What scalars can in principle become unbounded in this scenario?

\item\label{sing2} Consider a solution of Friedman equations of the form
\[a(t)=A+Bt^{q}+C(t_{s}-t)^{n},\]
where $A,B,q,n>0$ and $C$ are some free constants. What values of $q$ and $n$ are compatible with the sudden singularity of the previous problem?

\item\label{sing3} Is any energy condition violated by the solutions with the sudden future singularity? What physical constraint on matter can be introduced that would prevent it?
\end{enumerate}

\section{Influence of cosmological expansion on local systems}
Does the expansion of space mean that everything in it is stretched? Galaxies? Atoms? A shallow answer to this question is: ``bounded'' systems do not take part in the expansion. However, if space is stretched, then how can these systems not experience some, at least minimal, extension? Should bounded systems be stretched less intensively? The following several problems attempt to clarify the question by the example of a simple model: a classical atom, which consists of a negatively charged electron with negligible mass, rotating around a positively charged massive nucleus.

Let us place this atom in a homogeneous Universe which expands with scale factor $a(t)$. We will use two sets of spatial coordinates for its description, both spherical with the atom at the origin. The first set consists of physical coordinates $R,\theta,\varphi$, with $R$ being the distance between the electron and the nucleus at given time. The second set $r,\theta,\varphi$ is the comoving coordinates, the fixed points that partake in the cosmological expansion. The two sets are related through
\[R = a(t)r.\]
The angular coordinates are the same, as we assume that the cosmological expansion is radial.

\begin{enumerate}[resume]
\item\label{dyn7} How can we understand in terms of the physical and comoving coordinates whether the atom partakes in the cosmological (Hubble) expansion or not?

\item\label{dyn8} Derive the equation of motion for the atom's electron accounting for the cosmological expansion.

\item\label{dyn9} Write the effective potential for the electron for the case of exponential expansion $a(t) = e^{\beta t}$ and use it to analyze the dynamics for the case $L^2=C$, where $C$ is the constant of electrostatic interaction.

\item\label{dyn10} Why does the Solar system not expand despite of expansion of all the Universe? Give quantitative arguments.
\end{enumerate}

\section{Dynamics of the Universe in terms of redshift and conformal time}
\begin{enumerate}[resume]
\item\label{dyn13} Express the first Friedman equation in terms of redshift and analyze the contribution of different terms in different epochs.

\item\label{dyn63} Find the conformal time as function of the scale factor for a Universe with domination of a) radiation and b) non-relativistic matter.

\item\label{dyn64} Find the relation between time and redshift in the Universe with dominating matter.

\item\label{dyn65} Derive $a(\eta)$ for a spatially flat Universe with dominating radiation.

\item\label{dyn66} Express the cosmic time through the conformal time in a Universe with dominating radiation.

\item\label{dyn67} Derive $a(\eta)$ for a spatially flat Universe with dominating matter.

\item\label{dyn68} Find $a(\eta)$ for a spatially flat Universe filled with a mixture of radiation and matter.

\item\label{dyn69} Suppose a component's state parameter  $w_i=p_i/\rho_i$ is a function of time. Find its density as function of redshift.

\item\label{dyn70} Derive the Hubble parameter as a function of redshift in a Universe filled with non-relativistic matter.


\item\label{dyn74} The redshift of any object slowly changes with time \label{dot z} due to acceleration (or deceleration) of the Universe's expansion. Find the rate of change of redshift $\dot{z}$ for a Universe with dominating non-relativistic matter.

\item\label{dyn51} The Universe is known to have become transparent for electromagnetic waves at $z\approx 1100$ (in the process of formation of neutral hydrogen,  recombination), i.e when it was $1100$ times smaller than at present. Thus in practice the possibility of optical observation of the Universe optically is limited by the so-called optical horizon: the maximal distance that light travels since the moment of recombination. Find the ratio of the optical horizon to the particle one for a Universe dominated by matter.

\item\label{dyn54} Derive the particle horizon as function of redshift for a Universe filled with matter and radiation with relative densities $\Omega_{m0}$ and $\Omega_{r0}$.

\item\label{dyn55} Show that any signal emitted from the cosmological horizon will arrive to the observer with infinite redshift.

\section{Time Evolution of CMB}

\item Show that in the expanding Universe the quantity $aT$ is an approximate invariant.

\item
Show that the electromagnetic radiation frequency decreases with expansion of the Universe
as $\omega(t)\propto a(t)^{-1}$.

\item Show that if the radiation spectrum was equilibrium at some initial moment, then it will remain equilibrium during the following expansion.
\item Find the CMB temperature one second after the Big Bang.
\item Show that creation of the relic radiation (the photon decoupling) took place in the matter-dominated epoch.
\item What was the color of the sky at the recombination epoch?
\item When the night sky started to look black?
\item Estimate the moment of time when the CMB energy density was comparable to that in the microwave oven.
\item Estimate the moment of time when the CMB wavelength will be comparable to that in the microwave oven, which is $\lambda=12.6\ cm$.
\item When the relic radiation obtained formal right to be called CMB? And for what period of time?
\item Calculate the presently observed density of photons for the CMB and express it in Planck units.
\item Find the ratio of CMB photons' energy density to that of the neutrino background.
\item Determine the average energy of a CMB photon at present time.
\item Why, when calculating the energy density of electromagnetic radiation in the Universe, we can restrict ourselves to the CMB photons?
\item The relation $\rho_\gamma\propto a^{-4}$ assumes conservation of photon's number. Strictly speaking, this assumption is inaccurate. The Sun, for example, emits of the order of $10^{45}$ photons per second. Estimate the accuracy of this assumption regarding the photon's number conservation.
\item Can hydrogen burning in the thermonuclear reactions provide the observed energy density of the relic radiation?
\item Find the ratio of relic radiation energy density in the epoch of last scattering to the present one.
\item Find the ratio of average number densities of photons to baryons in the Universe.
\item Explain qualitatively why the temperature of photons at the surface of last scattering (0.3 eV) is considerably less than the ionization energy of the hydrogen atom (13.6 eV).
\item\label{cmb20} Estimate the moment of the beginning of recombination: transition from ionized plasma to gas of neutral atoms.
\item\label{cmb21} Determine the moment of time when the mean free path of photons became of the same order as the current observable size of Universe).
\item How will the \ref{cmb20} and the time when \ref{cmb21} change if one takes into account the possibility of creation of neutral hydrogen in excited states?
\item Why is the cosmic neutrino background (CNB) temperature at present lower than the one for CMB?

\end{enumerate}

\chapter{Black Holes. Problems}

\section{Technical warm-up}
The problems collected here introduce the tools needed for solving problems in the following sections. They can be solved in succession, or one can return to them if and when questions arise. For more background see the section ``Equations of General Relativity'' of chapter 2.

\subsection{Uniformly accelerated observer, Rindler metric}	
\emph{Einstein's equivalence principle states that locally a gravitational field cannot be dis\-tin\-guished from a non-inertial frame of reference. Therefore a number of effects of General Relativity, such as time dilation in a gravitational field and formation of horizons, can be studied in the frame of Special Theory of Relativity when considering uniformly accelerated observers.}
\begin{enumerate}
\item\label{BlackHole01}
Derive the equation of motion $x(t)$ of a charged particle in Minkowski space in a uniform electric field without initial velocity. Show that its acceleration is constant.
\item\label{BlackHole02}
What region of spacetime is unobservable for such an accelerated observer? In what region is this observer  unobservable?
\item\label{BlackHole03}
Consider the set of particles, which move with constant accelerations $a=const>0$ in Minkowski space, with initial conditions at time $t=0$ set as $x=\rho=c^{2}/a$. Let $\tau$ be the proper time of these particles in the units of $\rho/c$. What region of spacetime is parametrized by the pair of positive numbers $(\tau,\rho)$? Express the metric in this region in the coordinates $(\rho,\varphi)$, where $\varphi=c\tau/\rho$. This is the Rindler metric.
\end{enumerate}

\subsection{Metric in curved spacetime}
\emph{We see here, how, given an arbitrary metric tensor, to determine physical distance between points, local time and physical velocity of a particle in an arbitrary frame of reference.}

\emph{This problem, though fundamentally important,  is necessary in full form only for consideration of particle dynamics in the Kerr metric. In order to analyze the dynamics in the Schwarzschild metric, it suffices to answer all the questions with a substantially simplifying condition $g_{0i}=0$, where $i=1,2,3$ (see the last of the problems).}

Let the spacetime metric have the general form
\[ds^2=g_{\mu\nu}dx^{\mu}dx^{\nu}.\]
Coordinates are arbitrary and do not carry direct metrical meaning. An observer, stationary in a given coordinate frame, has 4-velocity $u^{\mu}=(u^{0},0,0,0)$, and the interval determines his proper ``local'' time
\[c^2 d\tau^{2}=ds^{2}=g_{00}(dx^{0})^2.\]
An observer in point A, with coordinates $x^\mu$, determines the physical ``radar'' distance to an infinitely close point $B$, with coordinates $x^{\mu}+dx^{\mu}$, in the following way. She sends a light beam to $B$ and measures the time it takes for the reflected beam to come back. Then distance to $B$ is half the proper time she waited from emission to detection times $c$. It is also natural for her to consider the event of the beam reflection in $B$ to be simultaneous with the middle of the infinitely small 4-distance between the events of emission and detection of light beam in $A$.

\begin{enumerate}[resume]
\item\label{BlackHole09}
Find the physical distance $dl$ between two events with coordinates $x^\mu$ and $x^\mu+dx^{\mu}$.
\item\label{BlackHole10}
Find the difference between coordinate times of two infinitely close simultaneous events.
\item\label{BlackHole11}
Let a particle's world line be $x^{\mu}(\lambda)$. What is the proper time interval $\delta\tau$ of a stationary observer, in which this particle covers distance from $x^{\mu}$ to $x^\mu+dx^\mu$?
\item\label{BlackHole12}
Physical velocity $v$ of the particle is defined as  $dl/\delta\tau$. Express it through the $4$-velocity of the particle and through its coordinate velocity $dx^{i}/dx^{0}$; find the interval along the world line $ds$ in terms of $v$ and local time $d\tau$.
\item\label{BlackHole13}
How are all the previous answers simplified if $g_{0i}=0$?
\end{enumerate}

\section{Schwarzschild black hole}
The spherically symmetric solution of Einstein's equations in vacuum for the spacetime metric has the form \cite{Schw}
\begin{align}\label{Schw}
ds^{2}=h(r)\,dt^2-h^{-1}(r)\,dr^2-r^2 d\Omega^{2},
	&\qquad\mbox{where}\quad
	h(r)=1-\frac{r_g}{r};\quad r_{g}=\frac{2GM}{c^{2}};\\
d\Omega^{2}=d\theta^{2}+\sin^{2}\theta\, d\varphi^{2}&\;\mbox{-- metric of unit sphere.}\nonumber
\end{align}
The Birkhoff's theorem (1923) \cite{Birkhoff,Jebsen} states, that this solution is unique up to coordinate transformations. The quantity $r_g$ is called the Schwarzschild radius, or gravitational radius, $M$ is the mass of the central body or black hole.

\subsection{Simple problems}
\begin{enumerate}[resume]
\item\label{BlackHole15}
Find the interval of local time (proper time of stationary observer) at a point $(r,\theta,\varphi)$ in terms of coordinate time $t$, and show that $t$ is the proper time of an observer at infinity. What happens when $r\to r_{g}$?

\item\label{BlackHole16}
What is the physical distance between two points with coordinates $(r_{1},\theta,\varphi)$ and $(r_{2},\theta,\varphi)$? Between $(r,\theta,\varphi_{1})$ and $(r,\theta,\varphi_{2})$? How do these distances behave in the limit $r_{1},r\to r_{g}$?

\item\label{BlackHole17}
What would be the answers to the previous two questions for $r<r_g$ and why\footnote{It is actually not a very simple problem}? Why the Schwarzschild metric cannot be imagined as a system of ``welded'' rigid rods in $r<r_g$, as it can be in the external region?

\item\label{BlackHole18}
Calculate the acceleration of a test particle with zero velocity.

\item\label{BlackHoleExtra1}
Show that Schwarzschild metric is a solution to Einstein's equations.
\end{enumerate}

\subsection{Symmetries and integrals of motion of Schwarzschild metric}
\begin{enumerate}[resume]
\item\label{BlackHole19}
What integral of motion arises due to existance of a timelike Killing vector? Express it through the physical velocity of the particle.

\item\label{BlackHole20}
Derive the Killing vectors for a sphere in Cartesian coordinate system; in spherical coordinates.
\item\label{BlackHole21}
Verify that in coordinates $(t,r,\theta,\varphi)$ vectors
\[ \begin{array}{l}
	\Omega^{\mu}=(1,0,0,0),\\
	R^{\mu}=(0,0,0,1),\\
	S^{\mu}=(0,0,\cos\varphi,-\cot\theta\sin\varphi),\\
	T^{\mu}=(0,0,-\sin\varphi,-\cot\theta\cos\varphi)
\end{array}\]
are the Killing vectors of the Schwarzschild metric.

\item\label{BlackHole22}
Show that existence of Killing vectors $S^\mu$ and $T^\mu$ leads to motion of particles in a plane.

\item\label{BlackHole23}
Show that the particles' motion in the plane is stable.

\item\label{BlackHole24}
Write down explicitly the conserved quantities  $p_{\mu}\Omega^{\mu}$ and $p_{\mu}R^{\mu}$ for movement in the plane $\theta=\pi/2$.

\item\label{BlackHole25}
What is the work needed to pull a particle from the horizon to infinity? Will a black hole's mass change if we drop a particle with zero initial velocity from immediate proximity of the horizon?

\end{enumerate}
\subsection{Radial motion in Schwarzchild metric}
Consider a particle's radial motion: $\dot{\varphi}=\dot{\theta}=0$. In this problem one is especially interested in asymptotes of all functions as $r\to r_{g}$.

\begin{enumerate}[resume]
\item\label{BlackHole26}
Derive the equation for null geodesics (worldlines of massless particles).

\item\label{BlackHole27}
Use energy conservation to derive $v(r)$, $\dot{r}(r)=dr/dt$, $r(t)$ for a massive particle. Initial conditions: $g_{00}|_{\dot{r}=0}=h_{0}$ (the limiting case $h_{0}\to 1$ is especially interesting and simple).

\item\label{BlackHole28}
Show that the equation of radial motion in terms of proper time of the particle is the same as in the non-relativistic Newtonian theory. Calculate the proper time of the fall from $r=r_0$ to the center. Derive the first correction in $r_{g}/r$ to the Newtonian result. Initial velocity is zero.

\item\label{BlackHole29}
Derive the equations of radial motion in the ultra-relativistic limit.

\item\label{BlackHole30}
A particle (observer) falling into a black hole is emitting photons, which are detected on the same radial line far away from the horizon (i.e. the photons travel from emitter to detector radially). Find  $r$, $v$ and $\dot{r}$ as functions of the signal's detection time in the limit  $r\to r_g$.
\end{enumerate}

\subsection{Blackness of black holes}
A source radiates photons of frequency $\omega_i$, its radial coordinate at the time of emission is $r=r_{em}$. Find the frequency of photons registered by a detector situated at $r=r_{det}$ on the same radial line in different situations described below. By stationary observers here, we mean stationary in the static Schwarzschild metric; ``radius'' is the radial coordinate $r$.
\begin{enumerate}[resume]
\item\label{BlackHole31}
The source and detector are stationary.

\item\label{BlackHole32}
The source is falling freely without initial velocity from radius $r_0$; it flies by the stationary detector at the moment of emission.

\item\label{BlackHole33}
The source is freely falling the same way, while the detector is stationary at $r_{det}>r_{em}$.

\item\label{BlackHole34}
The source is falling freely and emitting continuously photons with constant frequency, the detector is stationary far away from the horizon $r_{det}\gg r_{g}$. How does the detected light's intensity depend on $r_{em}$ at the moment of emission? On the proper time of detector?
\end{enumerate}

\subsection{Orbital motion, effective potential}
Due to high symmetry of the Schwarzschild metric, a particle's worldline is completely determined by the normalizing condition $u^{\mu}u_{\mu}=\epsilon$, where $\epsilon=1$ for a massive particle and $\epsilon=0$ for a massless one, plus two conservation laws---of energy and angular momentum.

\begin{enumerate}[resume]
\item\label{BlackHole35}
Show that the ratio of specific energy to specific angular momentum of a particle equals to $r_{g}/b$, where $b$ is the impact parameter at infinity (for unbounded motion).

\item\label{BlackHole36}
Derive the geodesics' equations; bring the equation for $r(\lambda)$ to the form
\[\frac{1}{2}\Big(\frac{dr}{d\lambda}\Big)^{2}
	+V_{\epsilon}(r)=\varepsilon,\]
where $V_{\epsilon}(r)$ is a function conventionally termed as effective potential.

\item\label{BlackHole37}
Plot and investigate the function $V(r)$. Find the radii of circular orbits and analyze their stability; find the regions of parameters $(E,L)$ corresponding to bound and unbound motion, fall into the black hole. Consider the cases of a) massless, b) massive particles.

\item\label{BlackHole38}
Derive the gravitational capture cross-section for a massless particle; the first correction to it for a massive particle ultra-relativistic at infinity. Find the cross-section for a non-relativistic particle to the first order in $v^2/c^2$.

\item\label{BlackHole39}
Find the minimal radius of stable circular orbit and its parameters. What is the maximum gravitational binding energy of a particle in the Schwarzschild spacetime?
\end{enumerate}

\subsection{Miscellaneous problems}
\begin{enumerate}[resume]
\item\label{BlackHole40}
Gravitational lensing is the effect of deflection of a light beam's (photon's) trajectory in the gravitational field. Derive the deflection of a photon's trajectory in Schwarzschild metric in the limit $L/r_{g}\gg 1$. Show that it is twice the value for a massive particle with velocity close to $c$ in the Newtonian theory.

\item\label{BlackHole41}
Show that the $4$-acceleration of a stationary particle in the Schwarzschild metric can be presented in the form
\[a_{\mu}=-\partial_{\mu}\Phi,\quad
	\text{where}\quad \Phi=\ln \sqrt{g_{00}}
		=\tfrac{1}{2}\ln g_{00}\]
is some generalization of the Newtonian gravitational potential.

\item\label{BlackHole42}
Let us reformulate the problem in a coordinate-independent manner. Suppose we have an arbitrary stationary metric with timelike Killing vector $\xi^\mu$, and we denote the $4$-velocity of a stationary observer by $u^{\mu}=\xi^{\mu}/V$. What is the $4$-force per unit mass that we need to apply to a test particle in order to make it stay stationary? Show in coordinate-independent way that the answer coincides with $\partial_{\mu}\Phi$ (up to the sign), and rewrite $\Phi$ in coordinate-independent form.

\item\label{BlackHole43}
Surface gravity $\kappa$ of the Schwarzschild horizon can be defined as acceleration of a stationary particle at the horizon, measured in the proper time of a stationary observer at infinity. Find $\kappa$.
\end{enumerate}

Solving Einstein's equations for a spherically symmetric metric of general form in vacuum (energy-momentum tensor equals to zero), one can reduce the metric to
\[ds^2=f(t)\Big(1-\frac{C}{r}\Big)dt^2
	-\Big(1-\frac{C}{r}\Big)^{-1}dr^2-r^2 d\Omega^2,\]
where $C$ is some integration constant, and $f(t)$ an arbitrary function of time $t$.

\begin{enumerate}[resume]
\item\label{BlackHole44}
Suppose all the matter is distributed around the center of symmetry, and its energy-momentum tensor is spherically symmetric, so that the form of $g_{\mu\nu}$ written above is correct. Show that the solution in the exterior region is reduced to the Schwarzschild metric and find the relation between $C$ and the system's mass $M$.

\item\label{BlackHole45}
Let there be a spherically symmetric void $r<r_{0}$ in the spherically symmetric matter distribution. Show that spacetime in the void is flat.

\item\label{BlackHole46}
Let the matter distribution be spherically symmetric and filling regions  $r<r_{0}$ and $r_{1}<r<r_{2}$ ($r_{0}<r_{1}$). Can one affirm, that the solution in the layer of empty space $r_{0}<r<r_{1}$ is also the Schwarzschild metric?
\end{enumerate}

\subsection{Different coordinates, maximal extension}
We saw that a particle's proper time of reaching the singularity is finite. However, the Schwarzschild metric has a (removable) coordinate singularity at $r=r_{g}$. In order to eliminate it and analyze the casual structure of the full solution, it is convenient to use other coordinate frames. Everywhere below we transform the coordinates $r$ and $t$, while leaving the angular part unchanged.

\begin{enumerate}[resume]
\item\label{BlackHole47}
Make coordinate transformation in the Schwarzschild metric near the horizon $(r-r_{g})\ll r_{g}$ by using physical distance to the horizon as a new radial coordinate instead of $r$, and show that in the new coordinates it reduces near the horizon to the Rindler metric.

\item\label{BlackHole48}
Derive the Schwarzschild metric in coordinates $t$ and  $r^\star=r+r_{g}\ln|r-r_g|$. How do the null geodesics falling to the center look like in $(t,r^\star)$? What range of values of $r^\star$ corresponds to the region $r>r_g$?

\item\label{BlackHole49}
Rewrite the metric in coordinates $r$ and $u=t-r^\star$, find the equations of null geodesics and the value of $g=det(g_{\mu\nu})$ at $r=r_{g}$. Likewise in coordinates $r$ and $v=t+r^\star$; in coordinates $(u,v)$. The coordinate frames $(v,r)$ and $(u,r)$ are called the ingoing and outgoing Eddington-Finkelstein coordinates.

\item\label{BlackHole50}
Rewrite the Schwarzschild metric in coordinates $(u',v')$ and in the Kruskal coordinates $(T,R)$ (Kruskal solution), defined as follows:
\[v'=e^{v/2r_g},\quad u'=-e^{-u/2r_g};\qquad
	T=\frac{u'+v'}{2},\quad R=\frac{v'-u'}{2}.\]
What are the equations of null geodesics, surfaces $r=const$ and $t=const$, of the horizon $r=r_{g}$, singularity $r=0$, in the coordinates $(T,R)$? What is the range space of $(T,R)$? Which regions in the Schwarzschild coordinates do the regions $\{\text{I}:\;R>|T|\}$, $\{\text{II}:\;T>|R|\}$, $\{\text{III}:\;R<-|T|\}$ and $\{\text{IV}:\;T<-|R|\}$ correspond to? Which of them are casually connected and which are not? What is the geometry of the spacelike slice $T=const$ and how does it evolve with time $T$?

\item\label{BlackHole51}
Pass to coordinates
\[v''=\arctan\frac{v'}{\sqrt{r_g}},\quad
	u''=\arctan\frac{u'}{\sqrt{r_g}}\]
and draw the spacetime diagram of the Kruskal solution in them.

\item\label{BlackHole52}
The Kruskal solution describes an eternal black hole. Suppose, for simplicity,  that some black hole is formed as a result of radial collapse of a spherically symmetric shell of massless particles. What part of the Kruskal solution will be realized, and what will not be? What is the casual structure of the resulting spacetime?
\end{enumerate}

\section{Kerr black hole}
Kerr solution is the solution of Einstein's equations in vacuum that describes a rotating black hole (or the metric outside of a rotating axially symmetric body) \cite{Kerr63}. In the Boyer-Lindquist coordinates \cite{BoyerLindquist67} it takes the form
\begin{align}\label{Kerr}
	&&ds^2=\bigg(1-\frac{2\mu r}{\rho^2}\bigg)dt^2
		+\frac{4\mu a \,r\sin^{2}\theta}{\rho^2}
				\;dt\,d\varphi
		-\frac{\rho^2}{\Delta}\;dr^2-\rho^2\, d\theta^2
	+\qquad\nonumber\\
	&&-\bigg(
		r^2+a^2+\frac{2\mu r\,a^2 \,\sin^{2}\theta}{\rho^2}
	\bigg) \sin^2 \theta\;d\varphi^2;\\
	\label{Kerr-RhoDelta}
&&\text{where}\quad
	\rho^2=r^2+a^2 \cos^2 \theta,\qquad
	\Delta=r^2-2\mu r+a^2.
\end{align}
Here $\mu$ is the black hole's mass, $J$ its angular momentum, $a=J/\mu$; $t$ and $\varphi$ are time and usual azimuth angle, while $r$ and $\theta$ are some coordinates that become the other two coordinates of the spherical coordinate system at $r\to\infty$.

\subsection{General axially symmetric metric}
\emph{A number of properties of the Kerr solution can be understood qualitatively without use of its specific form. In this problem we consider the axially symmetric metric of quite general kind}
\begin{equation}\label{AxiSimmMetric}
		ds^2=A dt^2-B(d\varphi-\omega dt)^{2}-
	C\,dr^2-D\,d\theta^{2},\end{equation}
\emph{where functions $A,B,C,D,\omega$ depend only on $r$ and $\theta$.}
\begin{enumerate}[resume]
\item\label{BlackHole53}
Find the components of metric tensor $g_{\mu\nu}$ and its inverse $g^{\mu\nu}$.
\item\label{BlackHole54}
Write down the integrals of motion corresponding to Killing vectors $\partial_t$ and $\partial_\varphi$.
\item\label{BlackHole55}
Find the coordinate angular velocity $\Omega=\tfrac{d\varphi}{dt}$ of a particle with zero angular momentum $u_{\mu}(\partial_{\varphi})^{\mu}=0$.
\item\label{BlackHole55plus}
Calculate $A,B,C,D,\omega$ for the Kerr metric.
\end{enumerate}

\subsection{Limiting cases}
\begin{enumerate}[resume]
\item\label{BlackHole56}
Show that in the limit $a\to 0$ the Kerr metric turns into Schwarzschild with $r_{g}=2\mu$.
\item\label{BlackHole57}
Show that in the limit $\mu\to 0$ the Kerr metric describes Minkowski space with the spatial part in coordinates that are related to Cartesian as
\begin{align*}
	&x=\sqrt{r^2+a^2}\;\sin\theta\cos\varphi,
		\nonumber\\
	&y=\sqrt{r^2+a^2}\;\sin\theta\sin\varphi,\\
	&z=r\;\cos\theta\nonumber,\\
	&\text{where}\quad
	r\in[0,\infty),\quad
	\theta\in[0,\pi],\quad \varphi\in[0,2\pi).\nonumber
\end{align*}
Find equations of surfaces $r=const$ and $\theta=const$ in coordinates $(x,y,z)$. What is the surface $r=0$?
\item\label{BlackHole58}
Write the Kerr metric in the limit $a/r \to 0$ up to linear terms.
\end{enumerate}

\subsection{Horizons and singularity}
Event horizon is a closed null surface. A null surface is a surface with null normal vector $n^\mu$:
\[n^{\mu}n_{\mu}=0.\]
This same notation means that $n^\mu$ belongs to the considered surface (which is not to be wondered at, as a null vector is always orthogonal to self). It can be shown further, that a null surface can be divided into a set of null geodesics. Thus the light cone touches it in each point: the future light cone turns out to be on one side of the surface and the past cone on the other side. This means that world lines of particles, directed in the future, can only cross the null surface in one direction, and the latter works as a one-way valve, -- ``event horizon''.
\begin{enumerate}[resume]
\item\label{BlackHole59}
Show that if a surface is defined by equation $f(r)=0$, and on it $g^{rr}=0$, it is a null surface.
\item\label{BlackHole60}
Find the surfaces $g^{rr}=0$ for the Kerr metric. Are they spheres?
\item\label{BlackHole61}
Calculate surface areas of the outer and inner horizons.
\item\label{BlackHole62}
What values of $a$ lead to existence of horizons?
\end{enumerate}
On calculating curvature invariants, one can see they are regular on the horizons and diverge only at $\rho^2 \to 0$. Thus only the latter surface is a genuine singularity.
\begin{enumerate}[resume]
\item\label{BlackHole63}
Derive the internal metric of the surface $r=0$ in Kerr solution.
\item\label{BlackHole64}
Show that the set of points $\rho=0$ is a circle. How it it situated relative to the inner horizon?
\end{enumerate}

\subsection{Stationary limit}
Stationary limit is a surface that delimits areas in which particles can be stationary and those in which they cannot. An infinite redshift surface is a surface such that a phonon emitted on it escapes to infinity with frequency tending to zero (and thus its redshift tends to infinity). The event horizon of the Schwarzschild solution is both a stationary limit and an infinite redshift surface (see problems \ref{BlackHole31}-\ref{BlackHole34}). In the general case the two do not necessarily have to coincide.
\begin{enumerate}[resume]
\item\label{BlackHole65}
Find the equations of surfaces $g_{tt}=0$ for the Kerr metric. How are they situated relative to the horizons? Are they spheres?
\item\label{BlackHole66}
Calculate the coordinate angular velocity of a massless particle moving along $\varphi$ in the general axially symmetric metric  (\ref{AxiSimmMetric}). There should be two solutions, corresponding to light traveling in two opposite directions. Show that both solutions have the same sign on the surface $g_{tt}=0$. What does it mean? Show that on the horizon $g^{rr}=0$ the two solutions merge into one. Which one?
\item\label{BlackHole67}
What values of angular velocity can be realized for a massive particle? In what region angular velocity cannot be zero? What can it be equal to near the horizon?
\item\label{BlackHole68}
A stationary source radiates light of frequency $\omega_{em}$. What frequency will a stationary detector register? What happens if the source is close to the surface $g_{tt}=0$? What happens if the detector is close to this surface?
\end{enumerate}

\subsection{Ergosphere and the Penrose process}
Ergosphere is the area between the outer stationary limit and the outer horizon. As it lies before the horizon, a particle can enter it and escape back to infinity, but $g_{tt}<0$ there. This leads to the possibility of a particle's energy in ergosphere to be also negative, which leads in turn to interesting effects.

\textit{All we need to know of the Kerr solution in this problem is that it \emph{has an ergosphere}, i.e. the outer horizon lies beyond the outer static limit, and that on the external side of the horizon all the parameters $A,B,C,D,\omega$ are positive (you can check!). Otherwise, it is enough to consider the axially symmetric metric of general form.}
\begin{enumerate}[resume]
\item\label{BlackHole69}
Let a massive particle move along the azimuth angle $\varphi$, with fixed $r$ and $\theta$. Express the first integral of motion $u_t$ through the second one\footnote{Relations problem \ref{BlackHole12} and  \ref{BlackHole19}) do not hold, as they were derived in assumption that $g_{00}>0$.} $u_{\varphi}$ (tip: use the normalizing condition $u^\mu u_{\mu}=1$).
\item\label{BlackHole70}
Under what condition a particle can have $u_{t}<0$? In what area can it be fulfilled? Can such a particle escape to infinity?
\item\label{BlackHole71}
What is the meaning of negative energy? Why in this case (and in GR in general) energy is \emph{not} defined up to an additive constant?
\item\label{BlackHole72}
Let a particle $A$ fall into the ergosphere, decay into two particles $B$ and $C$ there, and particle $C$ escape to infinity. Suppose $C$'s energy turns out to be greater than $A$'s. Find the bounds on energy and angular momentum carried by the particle $B$.
\end{enumerate}

\subsection{Integrals of motion}
\begin{enumerate}[resume]
\item\label{BlackHole73}
Find the integrals of motion for a massless particle moving along the azimuth angle $\varphi$ (i.e. $dr=d\theta=0$). What signs of energy $E$ and angular momentum $L$ are possible for particles in the exterior region and in ergosphere?
\item\label{BlackHole74}
Calculate the same integrals for massive particles. Derive the condition for negativity of energy in terms of its angular velocity $d\varphi/dt$. In what region can it be fulfilled? Show that it is equivalent to the condition on angular momentum found in problem \ref{BlackHole70}.
\item\label{BlackHole75}
Derive the integrals of motion for particles with arbitrary $4$-velocity $u^{\mu}$. What is the allowed interval of angular velocities $\Omega=d\varphi/dt$? Show that for any particle $(E-\tilde{\Omega} L )>0$ for any $\tilde{\Omega}\in(\Omega_{-},\Omega_{+})$.
\end{enumerate}

\subsection{The laws of mechanics of black holes}
If a Killing vector is null on some null hypersurface $\Sigma$, $\Sigma$ is called a Killing horizon.
\begin{enumerate}[resume]
\item\label{BlackHole76}
Show that vector  $K=\partial_{t}+\Omega_{H}\partial_{\varphi}$ is a Killing vector for the Kerr solution, and it is null on the outer horizon $r=r_{+}$. Here $\Omega_{H}=\omega\big|_{r=r_+}$ is the angular velocity of the horizon.

\item\label{BlackHole77}
Let us define the surface gravity for the Kerr black hole as the limit
\[\kappa=\lim\limits_{r\to r_{+}}
	\frac{\sqrt{a^{\mu}a_{\mu}}}{u^0}\]
for a particle near the horizon with $4$-velocity $\bm{u}=u^{t}(\partial_{t}+\omega\partial_{\varphi})$. In the particular case of Schwarzschild metric this definition reduces to the one given in problem \ref{BlackHole43}. Calculate $\kappa$ for particles with zero angular momentum in the Kerr metric. What is it for the critical black hole, with $a=\mu$?
\item\label{BlackHole78}
Find the change of (outer) horizon area of a black hole when a particle with energy $E$ and angular momentum $L$ falls into it. Show that it is always positive.
\item\label{BlackHole79}
Let us define the irreducible mass $M_{irr}$ of Kerr black hole as the mass of Schwarzschild black hole with the same horizon area. Find $M_{irr}(\mu,J)$ and $\mu(M_{irr},J)$. Which part of the total mass of a black hole can be extracted from it with the help of Penrose process?
\item\label{BlackHole80}
Show that an underextremal Kerr black hole (with $a<\mu$) cannot be turned into the extremal one in any continuous accretion process.
\end{enumerate}
This problem's results can be presented in the form that provides far-reaching analogy with the laws of thermodynamics.
\begin{itemize}
\item[0:] Surface gravity $\kappa$ is constant on the horizon of a stationary black hole. The zeroth law of thermodynamics: a system in thermodynamic equilibrium has constant temperature $T$.
\item[1:] The relation
\[\delta\mu=\frac{\kappa}{8\pi}\delta A_{+}
	+\Omega_{H}\delta J\]
gives an analogy of the first law of thermodynamics, energy conservation.
\item[2:] Horizon area $A_+$ is nondecreasing. This analogy with the second law of thermodynamics hints at a correspondence between the horizon area and entropy.
\item[3:] There exists no such continuous process, which can lead as a result to zero surface gravity. This is an analogy to the third law of thermodynamics: absolute zero is unreachable.
\end{itemize}

\subsection{Particles' motion in the equatorial plane}
The following questions refer to a particle's motion in the equatorial plane $\theta=\pi/2$ of the Kerr metric.
\begin{enumerate}[resume]
\item\label{BlackHole81}
Put down explicit expressions for the metric components and parameters $A,B,C,D,\omega$.
\item\label{BlackHole82}
What is the angular velocity of a particle with zero energy?
\item\label{BlackHole83}
Use the normalizing conditions for the $4$-velocity $u^{\mu}u_{\mu}=\epsilon^2$ and two conservation laws to derive geodesic equations for particles, determine the effective potential for radial motion.
\item\label{BlackHole84}
Integrate the equations of motion for null geodesics with $L=aE$, investigate the asymptotes close to the horizons, limits $a\to 0$ and $a\to \mu$.
\item\label{BlackHole85}
Find the minimal radii of circular geodesics for massless particles, the corresponding values of integrals of motion and angular velocities. Show that of the three solutions one lies beyond the horizon, one describes motion in positive direction and one in negative direction. Explore the limiting cases of Schwarzschild $a\to0$ and extreme Kerr $a\to\mu$.
\item\label{BlackHole86}
Find $L^2$ and $E^2$ as functions of radii for circular geodesics of the massive particles.
\item\label{BlackHole87}
Derive equation for the minimal radius of a stable circular orbit; find the energy and angular momentum of a particle on it, the minimal radius in the limiting cases $a/\mu\to 0,1$.
\end{enumerate}


\section{Physics in general black hole spacetimes}
In this section we use the $(-+++)$ signature, Greek letters for spacetime indices and Latin letters for spatial indices.

\subsection{Frames, time intervals and distances}
In the next several problems we again consider the procedure of measuring time and space intervals by different observers, but in a different, more formal and powerful approach.

\begin{enumerate}[resume]
\item\label{OZ01} Let a particle move with the four-velocity $U^{\mu }$. It can be viewed as
some observer carrying a frame attached to him. Locally, it defines the
hypersurface orthogonal to it. Show that
\begin{equation}
h_{\mu \nu }=g_{\mu \nu }+U_{\mu }U_{\nu }  \label{h}
\end{equation}
is (i) the projection operator onto this hypersurface, and at the same time
(ii) the induced metric of the hypersurface. This means that  (i) for any
vector projected at this hypersurface by means of $h^\mu_\nu$, only the components orthogonal to $U^{\mu}$ survive, (ii) the repeated application of the projection operation leaves the vector within the hypersurface unchanged. In other words, $h_{\mu \nu}$ satisfies
\begin{align}
&h^{\mu}_{\nu}U^{\nu }=0 ;  \label{1} \\
&h^{\mu}_{\nu }h^{\nu}_{\lambda}=h^{\mu }_{\lambda}.  \label{2}
\end{align}


\item\label{OZ02}

Let us consider a particle moving with the four-velocity $U^{\mu }$. The
interval $ds^{2}$ between two close events is defined in terms of
differentials of coordinates,%
\begin{equation}
ds^{2}=g_{\mu \nu }dx^{\mu }dx^{\nu }.
\end{equation}

For given $dx^{\mu }$, what is the value of the proper time $d\tau _{obs}$ between the corresponding events measured by this observer? How can one define
locally the notions of simultaneity and proper distance $dl$ for the observer in terms of its four-velocity and the corresponding projection operator $h^\mu_\nu$ ? How is the interval $ds^{2}$ related to $d\tau _{obs}$ and $dl$?

\item\label{OZ03}

Let our observer measure the velocity of some other particle passing in its
immediate vicinity. Relate the interval to $d\tau _{obs}$ and the particle's
velocity $w$.

\item\label{OZ04}

Analyze the formulas derived in the previous three problems applied to the case of flat spacetime (Minkovskii space) and compare them to
the known formulas of special relativity.

\item\label{OZ05}

Consider an observer being at rest with respect to a given coordinate frame:
$x^{i}=const$ ($i=1,2,3$). Find $h_{\mu \nu }$, $d\tau _{obs}$, the
condition of simultaneity and $dl^{2}$ for this case. Show that the
corresponding formulas are equivalent to eqs. (84.6), (84.7) of \cite{LL},
where they are derived in a different way.

\item\label{OZ06}

Consider two events at the same point of space but at different values of
time. Find the relation between $dx^{\mu}$ and $d\tau _{obs}$ for such an
observer.
\end{enumerate}

\subsection{Fiducial observers}
\begin{enumerate}[resume]

\item\label{OZ07}

Consider an observer with
\begin{equation}
U_{\mu }=-N\delta _{\mu }^{0}=-N(1,0,0,0)\text{.}  \label{uz}
\end{equation}%
We call it a fiducial observer (FidO) in accordance with \cite{mb}. This
notion is applied in \cite{mb} mainly to static or axially symmetric rotating
black holes. In the latter case it is usually called the ZAMO (zero angular
momentum observer). We will use FidO in a more general context.

Show that a FidO's world-line is orthogonal to hypersurfaces of constant time $t=const$.

\item\label{OZ08}

Find the explicit form of the metric coefficients in terms of the components
of the FidO's four-velocity. Analyze the specific case of axially symmetric metric in
coordinates $(t,\phi ,r,\theta )$ with $g_{0i}=g_{t\phi }\delta _{i}^{\phi }$.

\item\label{OZ09}

Consider a stationary metric with the time-like Killing vector field $\xi^\mu =(1,0,0,0)$. Relate the energy $E$ of a particle with four-velocity $u^\mu$ as measured at
infinity by a stationary observer to that measured by a local observer with 4-velocity $U^\mu$.

\item\label{OZ10}

Express $E_{rel}$ and $E$ in terms of the relative velocity $w$ between a
particle and the observer (i.e. velocity of the particle in the frame of the observer and vice versa).

\item\label{OZ11}

Show that in the flat spacetime the formulas derived in the previous problem are reduced to the usual ones
of the Lorentz transformation.

\item\label{OZ12}

Find the expression for $E$ for the case of a static observer ($U^{i}=0$).

\item\label{OZ13}

Find the expression for $E$ for the case of the ZAMO observer and, in particular, in case of axially symmetric metric.

\end{enumerate}

\subsection{Collision of particles: general relationships}

\begin{enumerate}[resume]

\item\label{OZ14}

Let two particles collide. Define the energy in the center of mass (CM)
frame $E_{c.m.}$ at the point of collision and relate it to $E_{rel}$ and
the Lorentz factor of relative motion of the two particles.

\item\label{OZ15}

Let us consider a collision of particles 1 and 2 viewed from the frame
attached to some other particle 0. How are different Lorentz factors related
to each other? Analyze the case when the laboratory frame coincides with
that of particle 0.

\item\label{OZ16}

When can the relative Lorentz factor of two particles $\gamma $ as a function of their individual Lorentz factors in some frame $\gamma _{1}$ and $\gamma _{2}$ grow
unbounded? How can the answer be interpreted in terms of relative velocities?

\item\label{OZ17}

A tetrad basis, or the orthonormal tetrad, is the set of four unit vectors $h_{(a)}^\mu$ (subscripts in parenthesis $a=0,1,2,3$ enumerate these vectors), of which one, $h_{(0)}^\mu$, is timelike, and three vectors $h_{(i)}^\mu$ ($i=1,2,3$) are spacelike, so that
\begin{equation}
g_{\mu\nu}h_{(a)}^\mu h_{(b)}^\nu =\eta_{ab},\qquad a,b=0,1,2,3.
\end{equation}
A vector's tetrad components are
\begin{equation}
u_{(a)}=u_\mu h_{(a)}^\mu,\qquad u^{(b)}=\eta^{ab}u_{(b)}.
\end{equation}

Define the local three-velocities with the help of the tetrad basis attached to the observer, which
would generalize the corresponding formulas of special relativity.

\item\label{OZ18}

Derive the analogues of the results of problems \ref{OZ09} and \ref{OZ10} for massless particles (photons). Analyze the cases of static and ZAMO observers.

\item\label{OZ19}

The ergosphere is a surface defined by equation $g_{00}=0$. Show that it is
the surface of infinite redshift for an (almost) static observer.

\item\label{OZ20}

Consider an observer orbiting with a constant angular velocity $\Omega $ in
the equatorial plane of the axially symmetric back hole. Analyze what
happens to redshift when the angular velocity approaches the minimum or
maximum values $\Omega _{\pm }$.

\item\label{OZ21}

Let two massive particle 1 and 2 collide. Express the energy of each particle
in the centre of mass (CM) frame in terms of their relative Lorentz factor $\gamma (1,2)$. Analyze the limiting cases of ultra-relativistic $\gamma
(1,2)\rightarrow \infty $ and non-relativistic $\gamma (1,2)\approx 1$ collisions.

\item\label{OZ22}

For a stationary observer in a stationary space-time the quantity $\alpha =(U^0)^{-1}$ is the redshifting factor: if this observer emits a ptoton with frequency $\omega_{em}$, it is detected at infinity by another stationary observer with frequency $\omega_{det}=\alpha \omega_{em}$. For a generic observer this interpretation is invalid, however, $\alpha =ds/dt$ still determines the time dilation for this observer, and thus can still be called the same way. Express the redshifting factor of the center of mass frame $\alpha _{c.m.}$ through the redshifting factors of the colliding particles $\alpha_{1}$ and $\alpha_2$.

\item\label{OZ23}

Relate the energy of a particle at infinity $E_{1}$, its energy at the point of collision in the C.M. frame $(E_{1})_{c.m.}$ and $\mu$.

\item\label{OZ24}

Solve the same problem when both particles are massless (photons). Write
down formulas for the ZAMO observer and for the C.M. frame.
\end{enumerate}

\section{Astrophysical black holes}

\subsection{Preliminary}
\begin{enumerate}[resume]

 \item\label{BlackHole88} Calculate in the frame of Newtonian mechanics the time of collapse of a uniformly distributed spherical mass with density $\rho_0$.

 \item\label{BlackHole89} Why are stars of a certain type called ``white dwarfs''?

  \item\label{BlackHole90} What is the physical reason for the stopping of thermonuclear reactions in the stars of white dwarf type?

  \item\label{BlackHole91} Estimate the radius and mass of a white dwarf.

  \item\label{BlackHole92} What is the average density of a white dwarf of one solar mass, luminosity one thousandth of solar luminosity and surface temperature twice that of the Sun?

  \item\label{BlackHole93} Explain the mechanism of explosion of massive enough white dwarfs, with masses close to the Chandrasekhar limit.

\item\label{BlackHole94} Thermonuclear explosions of white dwarfs with masses close to the Chandrasekhar limit lead to the phenomenon of supernova explosions of type I. Those have lines of helium and other relatively heavy elements in the spectrum, but no hydrogen lines. Why is that?

\item\label{BlackHole95} A supernova explosion of type II is related to the gravitational collapse of a neutron star. There are powerful hydrogen lines in their spectrum. Why?

\item\label{BlackHole96} Estimate the radius and mass of a neutron star.

\item\label{BlackHole97}	 Why do neutron stars have to possess strong magnetic fields?

\item\label{BlackHole98} Find the maximum redshift of a spectral line emitted from the surface of a neutron star.

 \item\label{BlackHole99} What is the gravitational radius of the Universe? Compare it with the size of the observable Universe.

 \item\label{BlackHole100} What is the time (the Salpeter time) needed for a black hole, radiating at its Eddington limit, to radiate away all of its mass?

\item\label{BlackHole101} The time scales of radiation variability of active galactic nuclei (AGNs) are from several days to several years. Estimate the linear sizes of AGNs.

\item\label{BlackHole102} What mechanisms can be responsible for the supermassive black hole (SMBH) in the center of a galaxy to acquire angular momentum?

\item\label{BlackHole103} The Galactic Center is so ``close'' to us, that one can discern individual stars there and examine  in detail their movement. Thus, observations carried out in 1992-2002 allowed one to reconstruct the orbit of motion of one of the stars (S2) around the hypothetical SMBH at the galactic center of the Milky Way. The parameters of the orbit are: period $15.2$ years, maximum distance from the black hole $120$ a.u., eccentricity $0.87$. Using this data, estimate the mass of the black hole.

\item\label{BlackHole104} Using the results of the previous problem, determine the density of the SMBH at the Galactic Center.

\item\label{BlackHole105} Show that for a black hole of mass $M$ the temperature of the surrounding hot gas in thermal equilibrium is proportional to
\[T\sim {{M}^{-1/4}}.\]

\item\label{BlackHole106} Show that luminosity of a compact object (neutron star or black hole) of several solar masses is mostly realized in the X-rays.

\item\label{BlackHole107} In order to remain bound while subject to the rebound from gigantic radiative power,
AGNs should have masses $M>{{10}^{6}}{{M}_{\odot }}$. Make estimates.

\item\label{BlackHole108} AGNs remain active for more than tens of millions of years. They must have tremendous masses to maintain the luminosity
\[L\sim {{10}^{47}}\text{erg/sec}\]
during such periods. Make estimates for the mass of an AGN.

\item\label{BlackHole109} What maximum energy can be released at the merger of two black holes with masses ${{M}_{1}}={{M}_{2}}=\frac{M}{2}$?

\item\label{BlackHole110} Show that it is impossible to divide a black hole into two black holes.

\item\label{BlackHole111} J.~Wheeler noticed that in the frame of classical theory of gravity the existence of black holes itself contradicts the law of entropy's increase. Why is that?

\item\label{BlackHole112} What is the reason we cannot attribute the observed entropy's decrease (see the previous problem) to the interior of the black hole?

\item\label{BlackHole113} Find the surface area of a stationary black hole as a function of its parameters: mass, angular momentum and charge.
\end{enumerate}

\subsection{Quantum effects}
\begin{enumerate}[resume]

\item\label{BlackHoleQ1} Estimate the maximum density of an astrophysical black hole, taking into account that black holes with masses $M<{{10}^{15}}g$ would not have lived to our time due to the quantum mechanism of evaporation.

\item\label{BlackHoleQ2} Determine the lifetime of a black hole with respect to thermal radiation.

\item\label{BlackHoleQ3} Determine the temperature of a black hole (Hawking temperature) of one solar mass, and the temperature of the supermassive black hole at the center of our Galaxy.

\item\label{BlackHoleQ4} Particles and antiparticles of given mass $m$ (neutrinos, electrons and so on) can be emitted only if the mass $M$ of the black hole is less than some critical mass ${M}_{cr}$. Estimate the critical mass of a black hole $M_{cr}(m)$.

\end{enumerate}

\chapter{Cosmic Microwave Background (CMB)}
\section{Thermodynamics of Black-Body Radiation}
\begin{enumerate}
\item \label{bbrazm1} Show that the photon gas in thermal equilibrium has zero chemical potential. 

\item \label{bbrazm2} Find number of photons in black-body radiation of temperature $T$ in volume $V$, which have frequencies in the interval $\left[ \omega ,\omega +d\omega  \right]$.

\item \label{bbrazm3} Find total photon number of black body radiation in volume $V$ at temperature $T$.

\item \label{cmb_td_1} Estimate number of photons in a gas oven at room temperature and at maximum heat.

\item \label{bbrazm4} Find energy of photons in black-body radiation of temperature $T$ in volume $V$, which have frequencies in the interval $\left[ \omega ,\omega +d\omega  \right]$.

\item \label{bbrazm5} Calculate free energy, entropy and total energy of black-body radiation.

\item \label{bbrazm6} Calculate thermal capacity of black-body radiation.

\item \label{bbrazm7} Find pressure of black-body radiation and construct its state equation.

\item \label{bbrazm8} Find adiabate equation for the photon gas of black-body radiation.

\item \label{cmb3} Why CMB cannot be used to warm up food like in the microwave oven?

\item \label{cmb6} The binding energy of electron in the hydrogen atom equals to $13.6\
eV$. What is the temperature of Planck distribution, with this
average photon energy?

\section{Time Evolution of CMB}

\item \label{cmb0} Show that in the expanding Universe the quantity $aT$ is an approximate invariant.

\item \label{cmb32} Show that the electromagnetic radiation frequency decreases with expansion of Universe
as $\omega(t)\propto a(t)^{-1}$.

\item \label{cmb34} Show that if the radiation spectrum was equilibrium at some initial
moment, then it will remain equilibrium during the following
expansion.

\item Find the CMB temperature one second after the Big Bang.

\item \label{cmb7} Show that creation of the relic radiation (the photon
decoupling) took place in the matter-dominated epoch.

\item \label{cmb_tmp_1} What color had the sky at the recombination epoch?

\item \label{cmb_tmp_2} When the night sky started to look black?

\item \label{cmb4} Estimate the moment of time when the CMB energy density was
comparable to that in the microwave oven.

\item \label{cmb5} Estimate the moment of time when the CMB wavelength will be
comparable to that in the microwave oven, which is $\lambda=12.6\ cm$.

\item \label{cmb_tmp_3} When the relic radiation obtained formal right to be called CMB? And for what period of time?

\item \label{cmb10} Calculate the presently observed density of photons for the
CMB and express it in Planck units.

\item \label{cmb11} Find the ratio of CMB photons' energy density to that of the neutrino background.

\item \label{cmb12} Determine the average energy of a CMB photon at present
time.

\item \label{cmb13} Why, when calculating the energy density of electromagnetic radiation in the Universe, we can restrict ourself to the CMB photons?

\item  \label{cmb44} The relation $\rho_\gamma\propto a^{-4}$ assumes conservation of
photon's number. Strictly speaking, this assumption is inaccurate.
The Sun, for example, emits of the order of $10^{45}$ photons per
second. Estimate the accuracy of this assumption regarding the
photon's number conservation.

\item \label{cmb14} Can hydrogen burning in the thermonuclear reactions provide
the observed energy density of the relic radiation?

\item  \label{cmb15} Find the ratio of relic radiation energy density in the epoch of last
scattering to the present one.

\item \label{cmb16} Find the ratio of average number densities of photons to baryons in the
Universe.

\item \label{cmb17} Explain qualitatively why the temperature of photons at the surface of last scattering (0.3
eV) is considerably less than the ionization energy of the hydrogen
atom (13.6 eV).

\item \label{cmb20} \label{item:time} Estimate the moment of the beginning of recombination---transition from ionized plasma to gas of neutral atoms.

\item \label{cmb21} \label{item:time2} Determine the moment of time when the mean free path of photons became of the same order as the current observable size of Universe).

\item \label{cmb22} How will the results of problems \ref{item:time} and \ref{item:time2} change if one takes into account the possibility of creation of neutral hydrogen in excited states?

\item \label{cmb35} Why is the cosmic neutrino background (CNB) temperature at present lower than the one for CMB?

\section{Statistical properties of CMB}

\item \label{cmb33} Construct the differential equation for the photons' distribution function
$\varphi(\omega,t)$ in a homogeneous and isotropic Universe.

\item \label{cmb18} The magnitude of dipole component of anisotropy generated by
the Solar system's motion relative to the relic radiation equals
$\Delta T_d\simeq3.35mK$. Determine the velocity of the Solar system
relative to the relic radiation.

\item \label{cmb19}Estimate the magnitude of annual variations of CMB anisotropy
produced by rotation of the Earth around the Sun.

\item \label{cmb27} Show that the angular resolution $\Delta\theta$ is related to
the maximum harmonic $l_{max}$ (in the spherical harmonics
decomposition) by \(\Delta\theta={180^\circ}/{l_{max}}.\)

\item  \label{cmb38} Does the measurement of velocity relative to the CMB mean violation of
the relativity principle and an attempt to introduce an absolute
reference frame?

\section{Primary anisotropy of CMB}

\item \label{cmb24} Determine the size of the sound horizon on the last scattering
surface.

\item \label{cmb25} Determine the radius of the surface of last scattering.

\item \label{cmb26} Estimate the thickness of the layer of last scattering.

\item \label{cmb28} Determine the position of the first acoustic peak in the CMB power
spectrum produced by baryon oscillations on the surface of last
scattering in the Einstein-de Sitter Universe.

\item \label{cmb45} Propose a scheme to use acoustic oscillations of CMB for
the determination of geometry of the Universe.

\item \label{cmb_lss1} Find relative perturbation in the CMB temperature due to the Doppler effect.

\item \label{cmb_lss2} Find relation between the relative perturbation in the CMB temperature and that of matter density for the case of adiabatic potential perturbations.

\item \label{cmb_lss3} How perturbations of gravitational potential affect the perturbations of the CMB temperature?

\item \label{cmb_lss4} Find mean square variation of the CMB temperature due to the potential perturbations (see problems \ref{cmb_lss1},\ref{cmb_lss2},\ref{cmb_lss3}).

\item \label{cmb_an1} Estimate angular resolution of CMB anisotropy measurements required to detect the seeds of large scale structures in form of density fluctuations on the last scattering surface.

\section{CMB interaction with other components}

\item \label{Cerazm1} Show that an isolated free electron can neither emit nor absorb a photon.

\item \label{Cerazm2} A photon of frequency $\omega$ impacts on an electron at rest and is scattered at angle $\vartheta $. Find the change of the photon frequency.

\item \label{cmb45n} \label{Cerazm3} When a relativistic charged particle is scattered on a photon, the
process is called the inverse Compton scattering of the photon.
Consider the inverse Compton scattering in the case when a charged
particle with rest mass $m$ and total energy $E\gg m$ in laboratory
frame impacts a photon with frequency $\nu$. What is the
maximum energy that the particle can transfer to the photon?

\item \label{cmb30}Estimate the probability of the fact that a photon observed on
Earth has already experienced Thomson scattering since the moment it
left the surface of last scattering.

\item \label{cmb46n}  Cosmic rays contain protons with energies up to $10^{20}eV$.
What energy can be transmitted from such a proton to a photon of
temperature $3K$?

\item \label{cmb46} Relic radiation passes through hot intergalactic gas and is scattered on the electrons. Estimate the CMB temperature variation due to the latter process (the Sunyaev-Zel'dovich effect).

\item \label{cmb47} A photon with energy $E\ll mc^2$ undergoes collisions and
Compton scattering in the electron gas with temperature $kT\ll
mc^2$. Show that to the leading order in $E$ and $T$ the average
energy lost by photons in collisions takes the form
\[\langle\Delta E\rangle=\frac{E}{mc^2}(E-4kT).\]

\item \label{cmb39} Find the force acting on an electron moving through the CMB with
velocity $v\ll c$.

\item \label{cmb40} Estimate the characteristic time of energy loss by high-energy
electrons with energy of order $100GeV$ passing through the CMB.

\item \label{cmb41} Find the energy limit above which the $\gamma$-rays interacting
with the CMB should not be observed.

\section{Extras}

\item \label{cmb2} In year 1953 the article "Extended Universe and creation of Galactics" by
G.A.~Gamov was published. 
In that paper Gamov took two numbers---the age of the world
and average density of matter in the Universe---and basing on them he
determined the third number: the temperature of the relic radiation (cosmic microwave background).
Try to repeat the scientific feat of Gamov.

\end{enumerate}

\chapter{Thermodynamics of Universe}
\begin{flushright}
\it  In the mid-forties G.A.~Gamov proposed the idea\\ of the "hot"
origin of the World. Therefore thermodynamics \\ was introduced into
cosmology, and nuclear physics too. \\Before him the science of
the evolution of Universe \\contained only dynamics and geometry of the World.\\
A.D.~Chernin.
\end{flushright}
\section{Thermodynamical Properties of Elementary Particles}
\begin{table}[h]
\centering
\begin{tabular}{|l|c | c| c |c|}
\hline
Particles & Mass &  \multicolumn{1}{c}{Number of } & states & $g$ (particles and \\
\cline{3-4}
 &  &  spin & color & anti-particles)\\
\hline
Photon($\gamma$) & 0 & 2 & 1 & 2\\

$W^+,\,W^-$ & $80.4\, \mbox{\it GeV}$ & 3 & 1 & 6\\

$Z$ & $91.2\, \mbox{\it GeV}$ & 3 & 1 & 3\\

Gluon ($g$) & 0&2&8&16\\

Higgs boson & $>114\, \mbox{\it GeV}$ &1&1&1\\
\hline
Bosons & \multicolumn{2}{c}{}& &28\\
\hline
$u,/,\bar{u}$&$3\,\mbox{\it MeV}$&2&3&12\\

$d,/,\bar{d}$&$6\,\mbox{\it MeV}$&2&3&12\\

$s,/,\bar{s}$&$100\,\mbox{\it MeV}$&2&3&12\\

$c,/,\bar{c}$&$1.2\,\mbox{\it GeV}$&2&3&12\\

$b,/,\bar{b}$&$4.2\,\mbox{\it GeV}$&2&3&12\\

$t,/,\bar{t}$&$175\,\mbox{\it GeV}$&2&3&12\\
\hline
$e^+,\,e^-$& $0.511\,\mbox{\it MeV}$&2&1&4\\

$\mu^+,\,\mu^-$& $105.7\,\mbox{\it MeV}$&2&1&4\\

$\tau^+,\,\tau^-$& $1.777\,\mbox{\it GeV}$&2&1&4\\

$\nu_e,\,\bar{\nu_e}$& $<3\,\mbox{\it eV}$&1&1&2\\

$\nu_\mu,\,\bar{\nu_\mu}$& $<0.19\,\mbox{\it MeV}$&1&1&2\\

$\nu_\tau,\,\bar{\nu_\tau}$& $<18.2\,\mbox{\it MeV}$&1&1&2\\
\hline
Fermions & \multicolumn{2}{c}{}& &90\\
\hline
\end{tabular}
\caption{The Standard Model particles and their properties.}
\label{tab:ter_sm}
\end{table}

\begin{enumerate}
\item \label{ter_1}  Find the energy and number densities for bosons and fermions in the relativistic limit.

\item \label{ter_2} Find the number densities and energy densities, normalized to the photon energy density at the same temperature, for neutrinos, electrons, positrons, pions and nucleons in the relativistic limit.

\item \label{ter_3} Calculate the average energy per particle in the relativistic and non-relativistic limits.

\item \label{ter_4} Find the number of internal degrees of freedom for a quark.

\item \label{ter_5n} Find the entropy density for bosons end fermions with zero chemical potential.

\item \label{ter_5} Find the effective number of internal degrees of freedom for a mixture of relativistic bosons and fermions.

\item \label{therm5} Generalize the results of the previous problem for the case when some $i$-components have temperature $T_i$ different from the CMB temperature $T$.

\item \label{ter_6} Find the effective number of internal degrees of freedom for the Standard Model
particles at temperature $T>1TeV$.

\item \label{ter_7} Find the change of the number of internal degrees of freedom due to the quark
hadronization process.

\item \label{ter_8} Find the relation between the energy density and temperature at $10^{10}\ K$.

\item \label{ter_9} Find the ratio of the energy density at temperature $10^{10}\ K$ to that at $10^8\ K$.

\item \label{razm40} Determine Fermi energy of cosmological neutrino background (CNB) (a gas).

\end{enumerate}

\section{Thermodynamics of Non-Relativistic Gas}
\begin{enumerate}[resume]
\item \label{therm26} Find the ratio of thermal capacities of matter in the form of monatomic gas and radiation.

\item  \label{therm27} Expansion of the Universe tends to violate thermal equilibrium
between the radiation ($T\propto a^{-1}$) and gas of
non-relativistic particles ($T\propto a^{-2}$). Which of these two
components determines the temperature of the Universe?

\item \label{therm28} Show that in the non-relativistic limit ($kT\ll mc^2$)
$p\ll\rho$.

\item \label{ter_n_19} Show that for a system of particles in thermal equilibrium the following holds
\[\frac{dp}{dT}=\frac1T(\rho+p).\]

\item \label{ter_n_20} Show that for a substance with the equation of state $p=w\rho$ the following holds
\[T\propto\rho^{\frac{w}{w+1}}.\]

\item \label{ter_n_21} Show that for a substance with the equation of state $p=w\rho$ the following holds
\[T\propto a^{-3w}.\]

\item \label{ter_n_22} Obtain a generalization of Stefan-Boltzmann law for cosmological fluid, described by generalized polytropic equation of state \[p=w\rho+k\rho^{1+1/n},\]
    assuming that $1+w+k\rho^{1/n}>0$.

\item \label{ter_n_22_1} Show that the velocity of sound for cosmological fluid described by generalized polytropic equation of state \(p=w\rho+k\rho^{1+1/n}\), where $1+w+k\rho^{1/n}>0$, vanishes at the point where the temperature is extremum.

\end{enumerate}

\section{Entropy of Expanding Universe}
\begin{enumerate}[resume]
\item \label{therm23_1} Transform the energy conditions for the flat Universe to
conditions for the entropy density (see arXiv:1009.4513).

\item \label{therm29} Find the entropy density for the photon gas.

\item \label{therm30} Use the result of the previous problem to derive an alternative proof of the fact that $aT=const$ in the adiabatically expanding Universe.

\item \label{therm31} Find the adiabatic index for the CMB.

\item \label{therm32} Show that the ratio of CMB entropy density to the baryon number
density $s_\gamma/n_b$ remains constant during the expansion of the
Universe.

\item \label{ter_6n} Estimate the current entropy density of the Universe.

\item \label{ter_7n} Estimate the entropy of the observable part of the Universe.

\item \label{ter_10} Why is the expansion of the Universe described by the Friedman equations adiabatic?

\item \label{ter_11} Show that entropy is conserved during the expansion of the Universe described by the Friedman equations.

\item \label{ter_12} Show that the entropy density behaves as $s\propto a^{-3}$.

\item \label{ter_13} Using only thermodynamical considerations, show that if the energy density of some component is $\rho=const$
than the state equation for that component reads $p=-\rho$.

\item \label{ter_14} Show that the product $aT$ is an approximate invariant in the Universe dominated by relativistic particles.

\end{enumerate}

\section{Connection between Temperature and Redshift}
\begin{enumerate}[resume]
\item \label{therm23} Find the dependence of radiation temperature on the redshift.

\item \label{therm24} Find the dependence of free non-relativistic gas' temperature on the redshift.

\item \label{therm25} Estimate the time moment when the recombination started, i.e. when ionized plasma transited to the gas of neutral atoms, and determine the corresponding redshift value. The recombination temperature equals to $T_{rec} \approx 0.3\,\mbox{\it eV}.$

\end{enumerate}

\section{Peculiarities of Thermodynamics in Early Universe}
\begin{enumerate}[resume]
\item \label{ter_15} \label{time1} Find the temperature dependence for the Hubble
parameter in the early flat Universe.

\item \label{ter_16n} \label{time2} Find the time dependence of the temperature of the early Universe by direct integration of the first Friedman equation.

\item \label{ter_17n} Prove that results of the problems \ref{time1} and \ref{time2} are equivalent.

\item \label{ter_16} What was the time dependence for temperature at the early
stages of evolution of the Universe?

\item \label{ter_17} Determine the energy density of the Universe at the Planck time.

\item \label{ter_18} Show that at Planck time the energy density of the Universe
corresponded to $10^{77}$ proton masses in one proton volume.

\item \label{ter_19} What was the temperature of radiation-dominated Universe at the Planck time?


\item  \label{ter_20} Determine the age of the Universe when its temperature was equal to $1\ MeV$.

\item \label{ter_21} In the first cyclic accelerator---the cyclotron (1931)---particles were accelerated up to energies of order $1MeV$. In the
next generation accelerators---the bevatrons---energy was risen
to $1GeV$. In the last generation accelerator---the LHC---protons are accelerated to energy of $1\ TeV$. What times in the Universe history do those energies allow to investigate?

\item \label{ter_23} Show that in the epoch when the energy density of the Universe was
determined by ultra-relativistic matter and effective number of
internal degrees of freedom did not change, held $\dot{T}/T\propto
-T^2$.

\item \label{therm37} Estimate the baryon-antibaryon asymmetry \(A\equiv(n_b-n_{\bar{b}})/n_{\bar{b}}\) in the early Universe.

\item \label{therm38n} Determine the monopoles' number density and their contribution to
the energy density of the Universe at the great Unification
temperature. Compare the latter with the photons' energy density at
the same temperature.

\item \label{therm39n} At what temperature and time does the contribution of
monopoles into the Universe energy density become comparable to the
contribution of photons?

\end{enumerate}

\section{The Saha equation}
{\it Degree of ionization of atomic hydrogen in thermal equilibrium
can be described by the Saha equation
\[\frac{1-X}{X^2}=n\lambda_{Te}^3e^{\frac{I}{kT}},\] where $X=n_e/n$ is the degree of ionization,
$n_e$ and $n$ are concentrations of electrons and atoms (both
neutral and ionized) respectively,
\[\lambda_{Te}^2=\frac{2\pi\hbar^2}{m_e kT}\] is the electron's
thermal de Broglie wave length and $I=13.6eV$ is the ionization
energy for hydrogen. It is often used in astrophysics for
description of stellar dynamics.}
\begin{enumerate}[resume]
\item \label{cmb23} Using the Saha equation, determine the hydrogen ionization
degree
\begin{description}
\item [a)] 100 seconds after the Big Bang;
\item [b)] at the epoch of recombination;
\item [c)] at present time.
\end{description}
Assume for simplicity $\Omega=1$.

\item \label{therm33} Assuming that the ionization degree at the last scattering was equal to $10\%$, determine the decoupling
temperature using the Saha equation.

\item \label{therm34} How many iterations in the Saha equation are needed in order to obtain the decoupling temperature with
accuracy $1K$? Write down analytically the approximate result.

\item \label{therm35} Estimate the duration of the epoch of recombination: how long did it take for hydrogen ionization
degree to change from $90\%$ to $10\%$ according to the Saha
equation?

\item  \label{therm36} Using the Saha equation, determine the hydrogen ionization degree in the center of the Sun ($\rho=100g/cm^3$, $T=1.5\cdot10^7K$).

\end{enumerate}

\section{Primary Nucleosynthesis}
\begin{enumerate}[resume]
\item \label{therm38} Find the ratio of neutrons to protons number
densities in the case of thermal equilibrium between them.

\item \label{therm39} Up to what temperature can the reaction $n\nu_e\leftrightarrow
pe^-$ support thermal equilibrium between protons and neutrons in
the expanding Universe?

\item \label{therm40} Determine the ratio $n_n/n_p$ at the temperature of freeze-out.

\item \label{therm41} Determine the age of Universe when it reached the temperature of freeze-out.

\item \label{therm41n} Determine the time period during which the synthesis of light elements
took place.

\item \label{therm42n} At what temperature and at what time did efficient deuterium
synthesis start?

\item \label{therm43n} Determine the ratio of neutrons to protons number
densities at temperature interval from the freeze-out to the
creation of deuterium.

\item \label{therm44n} Determine the relative abundance of ${}^4He$ in
the Universe.

\item \label{therm45n} How many helium atoms are there for each hydrogen atom?

\item \label{therm46n} What changes in relative ${}^4 He$ abundance would be caused
by
\begin{description}
  \item[a)] decreasing of average neutron lifetime $\tau_n$;
  \item[b)] decreasing or increasing of the temperature of
  freeze-out $T_f$?
\end{description}

\item \label{therm43} What nuclear reactions provided the ${}^4 He$ synthesis in the early Universe?

\item \label{therm44} Why is synthesis of elements heavier than ${}^7 Li$ suppressed in the early Universe?

\item \label{therm47n} In our Universe the neutron half-value period (the life-time)
approximately equals to 600 seconds. What would the relative helium
abundance be if the neutron life-time decreased down to 100 seconds?

\item \label{therm48nn} At what temperature in Universe did the synthesis reactions stop?

\end{enumerate}

\section*{Extras}
\begin{enumerate}[resume]
\item \label{razm39n} What is more important for the open space suit: heating or
cooling function?

\item \label{therm42} If the Universe is electrically neutral then how many
electrons are there for each baryon?

\item \label{therm45} Explain why the thermonuclear processes in the first stars considerably influenced the evolution of the Universe as a whole?

\item \label{therm46} Why is the present relative abundance of chemical elements approximately the
same as right after the creation of the Solar system?

\end{enumerate}

\chapter{Perturbation Theory}
\section{Non relativistic small perturbation theory}
{\it Perturbation theory in expanding Universe has a number of distinctive festures. Strictly speaking, this theory shloud be based within the framework of general relativity. However, if inhomogeneities are small one could neglect the effects of curvature and finite speed of interaction and use newtonian dynamics.

To describe the fluctuations of density in this approximation we need the continuity equation
$$
\frac{\partial \rho}
{\partial t} + \nabla  \cdot \left(\rho \vec v\right) = 0
$$
and Euler equation
$$
\frac{\partial \vec v}
{\partial t} + \left( \vec v\nabla \right)\vec v + \frac{1}
{\rho }\nabla P + \nabla \Phi  = 0,
$$
where Newtonian gravitational potential satisfies the Laplace equation
$$
\Delta \Phi  = 4\pi G\rho.
$$}
\begin{enumerate}

\item\label{per1} Express the deviation of expansion rate from Hubble law in terms of physical and comoving coordinates.

\item\label{per2} Obtain the equations for perturbations in linear approximation, assuming that unperturbed state is stationary gas. uniformly distributed in space.

\item\label{per3} Demonstrate, that perturbations depend exponentially on time if unperturbed solution is stationary.

\item\label{per4} Consider time dependent adiabatic perturbations and find the characterictic scale of instability (so--called Jeans instability).

\item\label{per5} Using the results of previous problem, consider the cases of
\begin{description}
    \item[a)] long--wave $\lambda  > \lambda _J$ and
    \item[b)] short--wave $\lambda < \lambda _J$
\end{description}
perturbations. Cosider also the limiting case of short waves ($\lambda \ll \lambda _J$).

\item\label{per6} Construct the equation for small relative fluctuations of density \[\delta  = \frac{\delta \rho }{\rho }\] in Newtonian approximation neglecting the entropy perturbations.

\item\label{per7} Rewrite equation from previous problem in terms of Fourier components, eliminating the Lagrangian coordinates. Estimate the order of ``physical'' Jeans wavelength for matter dominated Universe.

\item\label{per8n} Obtain the dependence of fluctuations on time in flat Universe when
a) matter,
b) radition
is dominating.

\item\label{per8} Assuming, that a particular solution to equation from prob. \ref{per6} has the form $\delta _1\left( t \right) \sim H\left( t \right)$, construct the general solution for $\delta (t)$. Consider the flat Universe filled with the substance with $p = w\rho.$

\item\label{per9} Demonstrate, that transverse or rotational mode in expanding Universe tends to decrease.

\end{enumerate}
\section{Introduction to relativistic theory of small perturbations}
In the following section we consider only the linear in perturbations theory.

\subsection{Perturbations on flat background}
\begin{enumerate}[resume]

\item\label{per15} Consider the stability of stationary Universe, filled with matter $(p=0)$ and by substance with $p = w\rho$.

 \item\label{per12}  Determine the dependence of density fluctuations on scale factor in the flat Universe when
\begin{description}
    \item[a)] radiation,
    \item[b)] matter
\end{description}
is dominating.

\item\label{per13} Determine the dependence of density fluctuations on time in the closed Universe with $k=1$.
\item\label{per14} Determine the dependence of density fluctuations on time in the open Universe with $k=-1$.
\end{enumerate}

\subsection{Metrics perturbations, coordinates transforms and perturbed energy--momentum tensor}
\begin{enumerate}[resume]
\item\label{per18} The inhomogeneities in matter distribution in the Universe generate perturbations of different kinds. In linear approximation these perturbations do not interact with each other and evolve independently. Construct the classifications of perturbations.

\item\label{per19} Determine the number of independent functions required for description of perturbations, considered in the previous problem.

\item\label{per21ntenzor} Demonstrate, that spatial velocity $v^i$ from previous problem is physical.

\item\label{per22ntenzor} Using the results of two previous problems, calculate the components of energy--momentum tensor perturbation $\delta T^\mu_\nu$, which is induced by the perturbations of density and pressure $$\widetilde{\rho}=\rho+\delta \rho, ~  \widetilde{p}=p+\delta p,$$ where $\rho,p$ are their background values.

\item\label{per23ntenzor} Obtain the equations of covariant convervation in linear approximation in components of tensor perturbations $h_{\mu\nu}$. Use the gauge $h_{0i}=0$ for simplicity.

\item\label{per20} In the first order of $h_{ik}$ calculate the components of energy--momentum tensor for the Universe. filled with ideal fluid with equation of state$p=w\rho$ and metrics \eqref{met_ten_per}

\item\label{per21} \label{item:ProbPL} In linear approximation determine the transformation of metrics $g_{\mu\nu}$, which is generated by the space transformation $x^\alpha\rightarrow \tilde{x}^\alpha=x^\alpha+\xi^\alpha$, where $\xi^\alpha$ is infinitesimal scalar function.

\item\label{per22} Using the results of previous problem determine the transformation of metrics, generated by the transformation $x^\alpha\rightarrow \tilde{x}^\alpha=x^\alpha+\xi^\alpha$. Here the four--vector  $\xi^\alpha=(\xi^0,\xi^i)$ satisfies the condition $\xi^i=\xi^i_\bot+\zeta^i$, $\xi^i_\bot$ is a three--vector with zero divergence ($\xi^i_{\bot,i}=0$) and $\zeta^i$ is a scalar function.

\item\label{per22_1} Demonstrate, that non--uniform flat Friedman metrics
  $$
  ds^2= \left(1-\frac{2}{\sqrt{\lambda}}\dot{f}(t,\vec{r})\right)dt^2-a^2(\delta_{ij}-2\mathcal{B}_{,ij})dx^idx^j,
  $$
     with arbitrary small function $f(t,\vec{r})$, can be trnsformed to the uniform one.
\end{enumerate}

\section{Expansion of cosmological perturbations in helicities}

{\it When dealing with perturbations in cosmology one usually doesn't distinguish between original functions and their Fourier transforms. Coordinate and momenetum representations are connected by Fourier transform:
\begin{equation*}
h_{\mu\nu}(\eta,\, \vec{x})= \int d^3ke^{i\vec{k}\vec{x}} h_{\mu\nu}(\vec{k}),
\end{equation*}
which is reduced to the replacement $\partial_i\longleftrightarrow ik_i$ (see problem \ref{per29nnn}), and $\vec{k}$ has a meaning of conformal momentum.

Due to isotropy, metrics is invariant under spatial rotations, while at fixed conformal momentum $\vec{k}$ it is invariant under the rotations around the direction of $\vec{k}$, i.e. posses posses $SO(2)$ symmetry. Arbitrary three--dimensional tensor can be expanded in irreducible representations of $SO(2)$, which have certain helicity (eigenvalues of a rotation operator $L_\alpha = -i\frac{\partial}{d \alpha}$ at angle $\alpha$ ). \\
For example, three--dimensional scalar has zero helicity, since it doesn't transform under rotations around $\vec{k}$. Since for three--dimensional vector $v_i \propto k_i$, it has unit helicity, or more precisely a superposition of helicities +1 and -1. Three--dimensional tensor of the form $h_{ij}\propto v_i v_j$  or $h_{ij}\propto \delta_{ij}$ has the same helicity. Symmetric transverse traceless tensor $h_{ij}^{TT}$ has double helicity. Mathematically this can be expressed as
\begin{equation}\label{hijTT}
    h_{ii}^{TT}=0,~ k_ih_{ij}^{TT}=k_jh_{ij}^{TT}=0.
\end{equation}
In linear theory, the Einstein's equations and equations of energy--momentum tensor covariance split into independent equations for helical components of cosmological perturbations. Thus, the perturbations are divided into tensor (double helicity), vector (unit helicity) and scalar (zero helicity). Expansion of perturbed metrics in helical components in general case has the form (see, for example \cite{Rubakov2}):
\begin{eqnarray}
  h_{00} &=& 2\Phi,\label{h00_gen} \\
  h_{0i} &=& i k_i Z+ Z_i^T, \label{h0i_gen}\\
  h_{ij} &=& -2\Psi \delta_{ij}-2k_i k_j E+i(k_i W_j^T+k_j W_i^T)+h_{ij}^{TT},\label{hij_gen}
\end{eqnarray}
where  $\Phi, Z, \Psi, E$ are scalar functions of coordinates,$Z_i^T, W_i^T$ are transverse vectors ($k_i Z_i^T=k_i W_i^T =0$) and $h_{ij}^{TT}$ is a transverse traceless tensor.

In the subsequent problems the all types of perturbations are considered: scalar, vecotr and tensor. Also, we will use the gauge $h_{0i}=0.$ }

\subsection{Scalar perturbations in conformal Newtonian gauge}

\begin{enumerate}[resume]

\item\label{per23} Calculate the Christoffel symbols in conformal Newtonian coordinate system with metrics
\begin{equation}
  \label{per_newton_conf} ds^2=a^2(\eta)[(1+2\Phi)d\eta^2-(1-2\Phi)\delta_{ij}dx^idx^j],
\end{equation}
where $\Phi$  is a scalar function.

\item\label{per24nnn}\label{cov_newt} In conformal Newtonian gauge and perturbed energy--momenetum tensor derive the covariant conservation equations

\item\label{per25nnn} Calculate the components of Ricci tensor in conformal Newtonian metrics \eqref{per_newton_conf}.

\item\label{per26nnn} Calculate the components of perturbed Einstein tensor $G_\mu^\nu$ using the components of Ricci tensor..

\item\label{per27nnn} Calculate the components of linearized Einstein tensor using the results of previous problems.

\item\label{per28nnn}\label{Fried_lin} Obtain the linearized Friedman equations in conformal Newtonian gauge using the results of previous problems.

\item\label{per29nnn} Rewrite the linearized Friedman equations in momentum representation.

\item\label{per30nnn} Using the linearized Friedman equations in momentum representation, obtain the equation, which contains only gravitational potential $\Phi.$

\item\label{per31nnn} Determine the evolution of relativistic matter perturbations in single component expansing Universe.

\item\label{per32nnn} Derive the continuity equation and Euler equation for single component relativistic medium.

\item\label{per34nnn} Calculate the fluctuations (perturbations) of density  and velocity of non--relativistic matter in the Universe dominated by non--relativistic matter.

\item\label{per35nnn} Determine the evolution of non--relativistic matter density perturbations in hypothetical expanding Universe, dominated by negative spatial curvature.

\item\label{per24} Consider the Universe filled with ideal fluid with the equation of state $p=w\rho$ with metrics from previous (?) problem. Derive the equations, corresponding to condition $T^\beta_{\alpha,\beta}=0$.

\item\label{per27} Using the metrics \eqref{per_newton_conf}, derive the equation of motion of scalar field $\varphi$ from variational princile.

\item\label{per30}\label{delta_phi} Derive the equation for fluctuations of scalar field $\delta \phi(t)$ using the metrics \eqref{per_newton_conf} and assuming linear approximation in scalar potential $\Phi$.

\item \label{Tmunu_inf} Using the conformal Newtonian gauge, determine the components of energy--momentum tensor of scalar field, which is represented as a sum of unifom background field $\varphi(t)$ and perturbation part $\psi(\vec{x},t)$.

\item \label{v_inf} Considering the scalar field from previous problem, determine which quantity plays a part of velocit of medium.

\item\label{per31} Using the conformal Newtonian gauge determine the scalar of three--dimensional curvature in co--moving reference frame.

\end{enumerate}

\subsection{Evolution of vector and tensor perturbations}

\begin{enumerate}[resume]

\item \label{per20h}Calculate the components of the Christoffel symbols in a coordinate system for which the metric is given in the form
\begin{equation}
  \label{per_gen_conf} ds^2=a^2(\eta)[(\eta_{\mu\nu}+h_{\mu\nu}(\vec{x}))dx^\mu dx^\nu],
\end{equation}
where $\eta_{\mu\nu}$ is a standard Minkowski metrics and $h_{\mu\nu}(\vec(x))$ -- function of the coordinates describing the perturbations over a flat background.

\item \label{per20R} Obtain the components of the Ricci tensor for perturbed metrics of the form \eqref{per_gen_conf}. Calculate also the Ricci scalar curvature.

\item \label{per20G}\label{Ein_gen_prob} Calculate the components of linearized Einstein tensor in metrics \eqref{per_gen_conf}.

\item \label{gr_wave_eq} Construct the equation describing the evolution of tensor perturbations in the expanding Universe.

\item \label{gr_wave_eq_action} Construct the expression for the action of gravitational waves in the expanding Universe.

\item \label{gr_wave_eq_Mimk} Construct  the equation for gravitational perturbations in the Minkowski background metrics.

\item Find the solution for the equation describing the evolution of tensor modes beyond the horizon.

\item Determine the amplitude of the tensor perturbations in the Universe filled with relativistic matter. Consider the case $\eta \to 0$.

\item Determine the amplitude of the tensor perturbations in the Universe filled with relativistic matter. Consider the case $\eta \to 0, ~\eta \to \infty.$

\end{enumerate}

\section{CMB anisotropy}

\begin{enumerate}[resume]
\item\label{per25} Obtain the equation of motion for photon in metrics \eqref{per_newton_conf} in linear approximation in $\Phi$.

\item\label{per26} For the Universe, dominated by a substance with equation of state $ p = w \rho $, connect in the first approximation the fluctuations of the gravitational potential of the CMB with $ \Phi$.

\item Estimate the spatial scale of Silk effect. Assume that at the temperatures we are interested in, photon changes direction randomly, and  its energy does not change when scattering on electrons.\\

\item Estimate the angular scale of the CMB anisotropy due to the Silk effect.\\

\end{enumerate}

\section{Initial perturbations in the Universe}
\subsection{Fluctuations power spectrum: non--relativistic approach}
\begin{enumerate}[resume]

 \item\label{per16} Construct the correlation function of the Fourier components of the relative density fluctuations, which satisfies the cosmological principle.
 \item\label{per17} Express the correlation function of the relative density fluctuations through the power spectrum of these fluctuations.
\end{enumerate}

\subsection{Quantum fluctuations of fields in inflationary Universe}
\begin{enumerate}[resume]
 \item \label{infl_vac_fluc0} Estimate the amplitude of vacuum fluctuations of the free massless scalar quantum field  $\varphi(\vec{x},t)$ with characteristic momenta $q$ and frequencies $w_q=q$ with background Minkowski metrics.

\item \label{infl_vac_fluc1} Considering the free massless scalar quantum field as a set of quantum harmonic oscillators, refine the estimate obtained in the problem \ref{infl_vac_fluc0}.\\

\item Representing the inflanton field as a superposition of uniform scalar field $\varphi_b(t)$ and small perturbation $\psi(\vec{x},t)$ on the background of unperturbed FRW metrics, obtain the equations of motion of small perturbation $\psi(\vec{x},t),$, assuming, that action for perturbation is quadratic.\\

\item Perform a qualitative analysis of the equation for small perturbations of the inflaton from previous problemin the different modes of inflation.

\item \label{infl_vac_per_app} Estimate the amplitude of vacuum fluctuations at the moment of exit of cosmological perturbations beyond the on the horizon.

\item Demonstrate that the inflationary stage provides amplification of vacuum fluctuations of inflaton field.\\

\item Consider the difference in the sequence of events during the evolution of cosmological perturbations in the radiation--dominated stage, or during the stage of domination nonrelativistic matter and inflationary cosmology.

\item \label{field_inf_Mink} Demonstrate that in the slow--roll regime at the beginning of inflation the inflaton field behaves like a massless scalar field in the Minkowski space.\\

\item What is the initial state of the inflaton quantum field towards the creation and annihilation operators?

\item For modes beyond the horizon at the inflationary stage, obtain a qualitative solution to the equation, obtained in problem \ref{field_inf_Mink}.

\item \label{inf_chi1} Obtain the exact solution to the equation from problem \ref{field_inf_Mink} at the inflationary stage. Consider the case of modes below and beyondthe horizon.

\item \label{infl_vac_fluc_hor-p} Demonstrate, that power spectrum $\mathcal{P}_k(\varphi)$ of the modes, which cross the horizon is the same as for free massless scalar field (see problem \ref{infl_vac_fluc1}.)

\item \label{eq_infPhi} Obtain the equation, connecting the gravitational potential $\Phi$, background inflanton scalar field  $\varphi(t)$ and its perturbation $\psi(\vec{x},t).$

\item \label{inf-z-u} Obtain the equations of evolution of scalar perturbations generated by the inflanton field perturbation in the case, when there is no other fields of matter in the Universe. What form does its solution has in the case for the mode under the horizon and what its implies?

\item \label{inf-z-u1} Construct the solution of the equation obtained in the previous problem for the case of inflation in the slow--roll approximation. Find the spatial curvature of hypersurfaces of constant inflaton field $\mathcal{R}$ in this regime.

\item Find the power spectrum of $ \mathcal {P}_\varphi (k) $ of  spatial curvature of hypersurfaces generated by fluctuations of the inflaton field in the comoving reference frame.

\item Express the amplitude of the scalar perturbations $ \Delta_ {\mathcal {P}} \equiv \sqrt {| \mathcal {P} _ \varphi (k) |} $ generated by fluctuations of the inflaton field through the potential of this field.

\item Express the amplitude of the scalar perturbations $\Delta_{\mathcal{P}}$  generated by fluctuations of the inflaton field for the potential $V(\varphi)=g\varphi^n$.

\item Find the power spectrum of initial perturbations $ \mathcal {P}_h ^{(T)}$, generated at the inflationary stage.

\item Construct the relation between the power amplitudes of the primary gravitational and scalar perturbations generated at the inflationary stage.

\end{enumerate}

\chapter{Inflationary Universe \label{inf}}
\begin{flushright}
{\it ''...some researchers to question whether\\ inflationary cosmology is a branch of science at all.''

\textit{J.Barrow}

Although inflation is remarkably successful as a phenomenological model for
the dynamics of the very early universe, a detailed understanding of the physical
origin of the inflationary expansion has remained elusive.

\textit{D. Baumann, L. McAllister}}

\end{flushright}
\section{Problems of the Hot Universe (the Big Bang Model)}
\label{prob}
\begin{flushright}
{\it ''Sad the week without a paradox,\\
a difficulty, an apparent contradiction!\\
For how can one then make progress?''\\}
\textit{John Wheeler}
\end{flushright}

\subsection{The Horizon Problem}
\begin{enumerate}
\item \label{prob_of_mod_1} Determine the size of the Universe at Planck temperature (the problem of the size of the Universe).

\item  \label{prob_of_mod_5} Determine the number of causally disconnected regions at the redshift $z$, represented in our causal volume today.

\item \label{prob_of_mod_6} What is the expected angular scale of CMB isotropy (the horizon problem)?

\item \label{prob_of_mod_7} Show that in the radiation-dominated Universe there are causally disconnected regions at any moment in the past.

\end{enumerate}

\subsection{The Flatness Problem}
\begin{enumerate}[resume]
\item   \label{prob_of_mod_2} If at present time the deviation of density from the critical one is $\Delta$, then what was the deviation at $t\sim t_{Pl}$ (the problem of the flatness of the Universe)?

\item \label{prob_of_mod_3} Show that both in the radiation--dominated and matter--dominated epochs the combination $a^2 H^2$ is a decreasing function of time. Relate this result to the problem of flatness of the Universe.

\item {\bf  \label{prob_of_mod_4} Show that both in the radiation--dominated and matter--dominated cases $x=0$ is an unstable fixed point for the quantity \[x\equiv\frac{\Omega-1}{\Omega}.\]}
\end{enumerate}

\subsection{The Entropy Problem}
\begin{enumerate}[resume]
\item  \label{prob_of_mod_9n} Formulate the horizon problem in terms of the entropy of the Universe.

\item  \label{prob_of_mod_18} Show that the standard model of Big Bang must include the huge dimensionless parameter--the initial entropy of the Universe--as an initial condition.

\end{enumerate}

\subsection{The Primary Inhomogeneities Problem}
\begin{enumerate}[resume]
\item   \label{prob_of_mod_X} Show that any mechanism of generation of the primary inhomogeneities generation in the Big Bang model violates the causality principle.

\item  \label{prob_of_mod_X+1} How should the early Universe evolve in order to make the characteristic size $\lambda_p$ of primary perturbations decrease faster than the Hubble radius $l_H$, if one moves backward in time?

\item   \label{prob_of_mod_8} Suppose that in some initial moment the homogeneity scale in our Universe was greater than the causality scale. Show that in the gravitation--dominated Universe this scale relation will be preserved in all future times.

\item  \label{prob_of_mod_9} If the presently observed CMB was strictly homogeneous, then in what number of causally independent regions would constant temperature be maintained at Planck time?

\item   \label{prob_of_mod_10} Suppose an initial homogeneous matter distribution in the Universe is given. The initial velocities must obey Hubble law (otherwise the initially homogeneous matter distribution will be quickly destroyed). What should the accuracy of the initial velocity field homogeneity be in order to preserve the homogeneous matter distribution until present time?

\item   \label{prob_of_mod_12} Estimate the present density of relict monopoles in the framework of the model of the Hot Universe.

\item   \label{prob_of_mod_19} The cyclic model of the Universe is interesting because it avoids the intrinsic problem of the Big Bang model--the ''initial singularity'' problem. However, as it often happens, avoiding old problems, the model produces new ones. Try to determine the main problems of the cyclic model of the Universe.

\item   \label{prob_of_mod_20} In the Big Bang model the Universe is homogeneous and isotropic. In this model the momentum of a particle decreases as $p(t) \propto a(t)^{ - 1} $ as the Universe expands. At first sight it seems that due to homogeneity of the Universe the translational invariance must ensure the conservation of the momentum. Explain this seeming contradiction.
\end{enumerate}

\section{Cosmological Inflation: The Canonic Theory}
\begin{flushright}
{\it ''Inflation hasn't won the race,\\
But so far it's the only horse''

\textit{Andrei Linde}}
\end{flushright}
\subsection{Scalar Field In Cosmology}
\begin{enumerate}[resume]
\item  \label{inf1} A scalar field $\varphi(\vec r,t)$ in a potential $V(\varphi)$ is described by Lagrangian
\[L=\frac12\left(\dot\varphi^2-\nabla\varphi\cdot\nabla\varphi\right)-V(\varphi)\]
Obtain the equation of motion (evolution) from the least action principle for that field.

\item  \label{inf2} Using the action for free scalar field minimally coupled to gravitation
\[S_\varphi=\int d^4x\sqrt{-g}\left(\frac12 g^{\mu\nu}\partial_\mu\varphi\partial_\nu\varphi\right)\]
obtain action for this field in the FRW metric.

\item  \label{inf3} Using the action obtained in the previous problem, obtain evolution equation for the scalar field in the expanding Universe.

\item  \label{inf4} Calculate the density and pressure of homogeneous scalar field $\varphi(t)$ in potential $V(\varphi)$ in the FRW metric.

\item  \label{inf5} Starting from the scalar field's action in the form
\[
S = \int {d^4 x\sqrt { - g} \left[ {{1 \over 2}(\nabla \varphi )^2  - V(\varphi )} \right]}
\]
obtain the equation of motion for this field for the case of FRW metric.

\item  \label{inf5n} Construct the Lagrange function describing the dynamics of the Universe filled with a scalar field in potential $V(\varphi)$. Using the obtained Lagrangian, obtain the Friedman equations and the Klein--Gordon equation.

\item \label{inf6} Obtain the equation of motion for a homogeneous scalar field $\varphi(t)$ in potential $V(\varphi)$ starting from the conservation equation \[\dot\rho+3\frac{\dot a}{a}(\rho+p)=0.\]

\item \label{inf13} Obtain the equation of motion for the homogeneous scalar field $\varphi(t)$ in potential $V(\varphi)$ using the analogy with Newtonian dynamics.

\item \label{inf14n} Express $V(\varphi)$ and $\varphi$ through the Hubble parameter $H$ and its derivative
$\dot{H}$ for the Universe filled with quintessence.

\item \label{inf14} Show that Friedman equations for the scalar field $\varphi(t)$ in potential $V(\varphi)$ can be presented in the form
\[H^2=\frac{8\pi G}{3}\left(\frac12 \dot\varphi^2+V(\varphi)\right),\]
\[\dot H=-4\pi G\dot\varphi^2.\]

\item \label{inf15} Provided that the scalar field $\varphi(t)$ is a single--valued function of time, transform the second order equation for the scalar field into a system of first order equations.

\item  \label{inf16} Express the equations for the scalar field in terms of conformal time.

\item \label{inf17} Show that condition $\dot H>0$ cannot be realized for the scalar field with positively defined kinetic energy.

\item \label{inf7} Show that the Klein--Gordon equation could be rewritten in dimensionless form
$$
\varphi '' + \left( {2 - q} \right)\varphi ' = \chi ;\quad \chi  \equiv  - \frac{1}{H^2 }\frac{dV}{d\varphi },
$$
where prime denotes the derivative by $\ln a$, and $q =  - {{a\ddot a} / {\dot a^2 }}$ is the deceleration parameter.

\item \label{inf7n} Represent the equation of motion for the scalar field in the form
$$
\pm \frac{V_{,\varphi}}{V} = \sqrt {\frac{3(1 + w)}{\Omega
_\varphi(a)}} \left[ 1 + \frac{1}{6}\frac{d\ln \left( x_{\varphi}
\right)}{d\ln a} \right],
$$
where
$$
x_{\varphi}=\frac{1+w_{\varphi}}{1-w_{\varphi}},
~w_{\varphi}=\frac{\dot{\varphi}^2+2V(\varphi)}{\dot{\varphi}^2-2V(\varphi)},
$$ in the system of units such that $8\pi G=1.$

\item \label{inf8} The term $3H\dot{\varphi}$ in the equation for the scalar field formally acts as friction that damps the inflation evolution. Show that, nonetheless, this term does not lead to dissipative energy production.

\item \label{inf9} Obtain the system of equations describing the scalar field dynamics in the expanding Universe
containing radiation and matter in the conformal time.

\item \label{inf10} Calculate pressure of homogeneous scalar field in the potential $V(\varphi)$ using the obtained above energy density of the field and its equation of motion.

\item \label{inf11} What condition should the homogeneous scalar field $\varphi(t)$ in potential $V(\varphi)$ satisfy in order to provide accelerated expansion of the Universe?

\item \label{inf19} What conditions should the scalar field satisfy in order to provide expansion of the Universe close to the exponential one?

\end{enumerate}

\subsection{Inflationary Introduction}
\begin{enumerate}[resume]

\item \label{inf12} What considerations led A.Guth to name his theory describing the early Universe dynamics as ''inflation theory''?

\item \label{inf12_new} A. Vilenkin in his cosmological bestseller ''Many world in one'' remembers: ''On a Wednesday afternoon, in the winter of 1980, I was sitting in a fully packed Harvard auditorium, listening to the most fascinating talk I had heard in many years. The speaker was Alan Guth, a young physicist from Stanford, and the topic was a new theory for the origin of the universe$\ldots$ The beauty of the idea was that in a single shot inflation explained why the universe is so big, why it is expanding, and why it was so hot at the beginning. A huge expanding universe was produced from almost nothing. All that was needed was a microscopic chunk of repulsive gravity material. Guth admitted he did not know where the initial chunk came from, but that detail could be worked out later. ''It's often said that you cannot get something for nothing.'' he said, ''but the universe may be the ultimate free lunch'' ''. Explain, why can this be.

\item \label{inf13_new} Is energy conservation violated during the inflation?

\item \label{inf18} The inflation is defined as any epoch for which the scale factor of the Universe has accelerated growth, i.e. $\ddot{a}>0$. Show that the condition is equivalent to the requirement of decreasing of the comoving Hubble radius with time.

\item \label{inf20} Show that in the process of inflation the curvature term in the Friedman equation becomes negligible. Even if that condition was not initially satisfied, the inflation quickly realizes it.

\item \label{inf20_1} It is sometimes said, that the choice of $k = 0$ is motivated by observations: the density of curvature is close to zero. Is this claim correct?

\end{enumerate}

\subsection{Inflation in the Slow-Roll Regime}
\begin{enumerate}[resume]
\item \label{inf21} Obtain the evolution equations for the scalar field in expanding Universe in the inflationary slow-roll regime.

\item \label{inf22} Find the time dependence of scale factor in the slow--roll regime for the case $V(\varphi)={m^2 \varphi^2 / 2}$.

\item \label{inf23} Find the dependence of scale factor on the scalar field in the slow-roll regime.

\item \label{inf25} Show that the conditions for realization of the slow--roll limit can be presented in the form:
    \[\varepsilon(\varphi)\equiv\frac{M^{*2}_{Pl}}{2}\left(\frac{V^\prime}{V}\right)^2\ll1;
    \ |\eta(\varphi)|\equiv\left|M^{*2}_{Pl}\frac{V^{\prime\prime}}{V}\right|\ll1;
    \ M^*_{Pl}\equiv(8\pi G)^{-1/2}.\]

\item \label{inf26} Show that the condition $\varepsilon\ll1$ for the realization of the slow--roll limit obtained in the previous problem is also sufficient condition for the inflation.

\item \label{inf27} Find the slow--roll condition for power law potentials.

\item \label{inf28} Show that the condition $\varepsilon \ll\eta$ is satisfied in the vicinity of inflection point of the inflationary potential $V(\varphi)$.

\item \label{inf29} Show that the inflation parameter $\varepsilon$ can be expressed through the parameter $w$ in the state equation for the scalar field.

\item \label{inf30} Show that the second Friedman equation \[\frac{\ddot{a}}{a}=-\frac{4\pi G}{3}(\rho+3p)\] can be presented in the form \[\frac{\ddot{a}}{a}=H^2(1-\varepsilon).\]

\item \label{inf31} Show that in the slow--roll regime $\varepsilon_H\rightarrow\varepsilon$ and
$\eta_H\rightarrow\eta-\varepsilon$.

\item \label{inf32} Show that the inflation parameters $\varepsilon_H,\ \eta_H$ can be presented in the following symmetric form \[\varepsilon_H=-\frac{d\ln H}{d\ln a};\ \eta_H=-\frac{d\ln H'}{d\ln a}.\]

\item \label{inf33} Prove that the definition of inflation as the regime for which $\ddot a>0$  is equivalent to condition $\varepsilon_H<1$.

\item \label{inf34} Show that inflation appears every time when the scalar field's value exceeds the Planck mass.

\item \label{inf51} Find the energy momentum tensor for homogeneous scalar field in the slow--roll regime.

\end{enumerate}

\subsection{Solution of the Hot Big Bang Theory Problems}
\begin{enumerate}[resume]
\item \label{inf37} Show that in the inflation epoch the relative density $\Omega$ exponentially tends to unity.

\item \label{inf38} Estimate the temperature of the Universe at the end of inflation.

\item \label{inf39} Estimate the size of the Universe at the end of inflation.

\item \label{inf40} Find the number $N_e$ of $e$-foldings of the scale factor in the inflation epoch.

\item \label{inf42} Find the number $N_e$ of $e$-foldings of the scale factor for the inflation process near the inflection point.

\item \label{inf43} Show that inflation transforms the unstable fixed point $x=0$ for the quantity \[x\equiv\frac{\Omega-1}{\Omega}\] into the stable one, therefore solving the problem of the flatness of the Universe.

\item \label{inf24} Find the particle horizon in the inflationary regime, assuming $H\approx const$.

\item \label{inf44} Find the solution of the horizon problem in the framework of inflation theory.

\item \label{inf56} Did entropy change during the inflation period? If yes, then estimate what its change was.

\item \label{inf57} Does the inflation theory explain the modern value of entropy?

\item \label{inf45} Find the solution of the monopole problem in frame of inflation theory.
\end{enumerate}

\subsection{Different Models of Inflation}
There is a number of inflationary models. All of them deal with potentials of scalar fields which realize the slow-roll regime during sufficiently long period of evolution, then the inflation terminates and Universe enters the hot stage. It is worth noting that the models considered below are among the simplest ones and they do not exhaust all the possibilities, however they give main idea about possible features of the evolution of inflanton and scale factor in the slow-roll regime.

\subsubsection{Chaotic Inflation (The Inflation with Power Law Potential)}
\emph{\textbf{The chaotic inflation} or the inflation with high field, is considered as a rule with the power law potentials of the form
$$V=g\varphi^n,$$
where $g$ is a dimensional constant of interaction:
$$[g]=(\mbox{mass})^{4-n}$$
It should be noted that the sow-roll conditions for the given potential are always satisfied for sufficiently high values of the inflaton field
$$\varphi\gg\frac{nM_{Pl}}{4\sqrt{3\pi}},$$
therefore the slow-roll takes place at field values which are great compared to Planck units.}

\begin{enumerate}[resume]
\item \label{inf_new_o1} Consider an inflaton with simple power law potential
      $$V=g\phi^n,$$
      where $g = (\mbox{mass}^{4-n})$ is the interaction constant, and show that there is wide range of scalar field values where classical Einstein equations are applicable and the slow-roll regime is realized too. Assume that the interaction constant $g$ is sufficiently small in Planck units.

\item \label{inf_new_o2} Estimate total duration of the chaotic inflation in the case of power law potentials of second and forth order.

\item \label{inf_new_o3} Express the slow-roll parameters for power law potentials in terms of $e$-folding number $N_e$ till the end of inflation.

\item \label{inf41} Obtain number $N$ of $e$-fold increase of the scale factor in the model \[V\left( \varphi \right) = \lambda \varphi ^4 \left( {\lambda = 10^{ - 10} } \right).\]

\item \label{inf35} Estimate the range of scalar field values corresponding to inflation epoch in the model \[V(\varphi ) = \lambda \varphi ^4 \left( {\lambda \ll 1} \right).\]

\item \label{inf36} Show that the classical analysis of the evolution of the Universe is applicable for the scalar field value $\varphi\gg M_{Pl}$, which allows the inflation to start.

\item \label{inf48} Find the time dependence for the scale factor in the inflation regime for potential $(1/n)\lambda\varphi^n$, assuming $\varphi\gg M_{Pl}$.

\item \label{inf49} The inflation conditions definitely break down near the minimum of the inflaton potential and the Universe leaves the inflation regime. The scalar field starts to oscillate near the minimum. Assuming that the oscillations' period is much smaller than the cosmological time scale, determine the effective state equation near minimum of the inflaton potential.

\item \label{inf50} Show that effective state equation for the scalar field, obtained in previous problem for potential $V\propto\varphi^n$, in the case $n=2$ corresponds to non--relativistic matter and for $n=4$---to the ultra--relativistic component (radiation).

\item \label{inf52} Obtain the time dependence of scalar field near the minimum of the potential.

\item \label{inf53} Find the energy-momentum tensor of a homogeneous scalar field in the regime of fast oscillations near the potential minimum.

\item \label{ch_inf_1} Check wether the chaotic inflation model agrees with the experimental data, which give the value $r=\mathcal{P}_\mathcal{T}/\mathcal{P}_\mathcal{R}<0.2$ for the tensor perturbation amplitude and $n_s=0.94\div 0.99$ for the spectrum slope. For the inflaton potential take $V(\varphi)=m^2\varphi^2/2$.

\item \label{ch_inf_2} Consider chaotic inflation with potential $V(\varphi)=m^2\varphi^2/2$ and obtain difference between the spectrum slopes for the waves corresponding to cosmological perturbations of sizes $100\ kpc$ and $10\ Gpc$.

\item \label{inf59} What is the difference between the chaotic inflation model by Linde and its original version by Starobinsky--Guth?

\end{enumerate}

\subsubsection{The Novel Inflation (the Inflation Near Minimum of the Potntial)}
\emph{As the reader might notice in the previous subsection, the chaotic inflation requires to include the super-Planck field values, however it is worth noting that there is an inflation model free of such requirement. Conditions of possible inflation start in this model considerably differ from the chaotic initial data. At the same time the flatness requirement for the inflaton potential is present in this model too.}

\begin{figure}[h]
\centering
  \includegraphics[width=250pt]{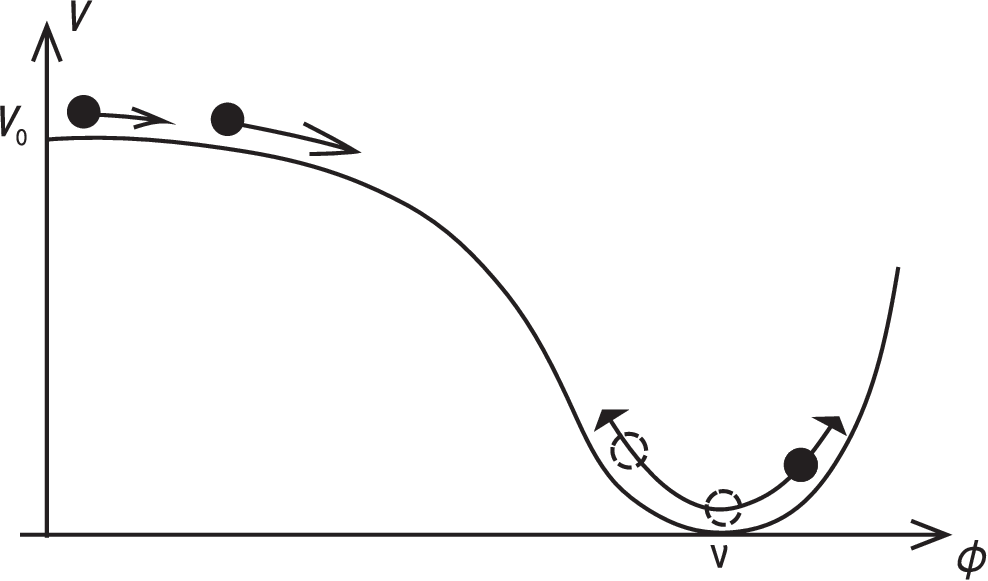}
  \caption{The inflaton potential in the inflation model with small field values.\label{new_inf_pic}}
\end{figure}

\emph{Consider the inflaton potential (see fig. \ref{new_inf_pic}) and suppose that near the origin it takes the form
$$V(\phi)=V_0-g\phi^n,$$
where $n\geq 3$. The most often considered potential is the forth order one
$$V(\phi)=V_0-\frac{\lambda}{4}\phi^4.$$
The inflation model with such potential is called the \textbf{novel inflation}.}

\begin{enumerate}[resume]
\item \label{new_inf_V1} Show that the inflation regime can start without super-Planck values of the inflaton field. Consider potential of the form \[V(\varphi)=V_0-\frac{\lambda}{4}\varphi^4.\]

\item \label{new_inf_V2} Consider an inflation model in which in inflaton potential near the origin takes the form $$V(\varphi)=V_0-\frac\lambda 4\varphi^4.$$ Assume that the slow-roll regime terminates at comparably small value of the inflaton field, so that $\lambda \varphi^4_e\ll V_0$, to obtain the ratio $\epsilon/|\eta |$. Determine which of the two parameters reaches unity first.

\item \label{new_inf_V2_1} Using the results of the previous problem and the condition $\lambda \varphi^4_e\ll V_0$, obtain the inequality giving the condition for realization of the novel inflation scenario.

\item \label{new_inf_V3} For the model of problem \ref{new_inf_V2} obtain the post-inflation heat up temperature $T_{reh}$.

\item \label{new_inf_V4} For the model of problem \ref{new_inf_V2} determine relation between the inflaton field value on the inflation stage and the $e$-folding number till the end of inflation.

\item \label{new_inf_V5} For the model of problem \ref{new_inf_V2} determine relation between the slow-roll parameters and the $e$-folding number till the end of inflation.

\item \label{new_inf_V10} Obtain estimate of the novel inflation duration.

\item \label{new_inf_V11} Obtain the relation $$\eta=-\frac{n-1}{n-2}\frac{1}{N_e}$$ for the novel inflation model with potential of the form $V(\varphi)=V_0-g\varphi^n,$ where $n\geq 3$.
\end{enumerate}

\subsubsection{Inflation with Exponential Potential (the Power Law Inflation)}
\begin{enumerate}[resume]
\item  \label{inf45n} Find the scalar field potential giving rise the power law for the scalar factor growth \[a(t)\propto {{t}^{p}}.\]

\item   \label{inf46} Find the exact particular solution of the system of equations for the scalar field in potential $V(\varphi)=g\exp(-\lambda\varphi)$.

\item  \label{inf47} Compare the solution obtained in the previous problem with the solution of evolution equations for the scalar field in the expanding Universe in the inflation limit.

\item   \label{inf55} Show that the dependence
\[H(\varphi)\propto\exp\left(-\sqrt{\frac{1}{2p}}\frac{\varphi}{M_{Pl}}\right)\]
leads to the power law inflation $a(t)=a_0 t^p$.

\item  \label{inf54} Show that the dependence $H(\varphi)=\varphi^{-\beta}$ leads to the so-called intermediate inflation (the Universe expansion goes faster than any power law and slower than the exponential one), at which
\[a(t)\propto\exp(At^f),\ 0<f<1,\quad A>0,\quad f=(1+\beta/2)^{-1}.\]

\item  \label{exp_inf_1} Consider the inflation model with the potential $V(\varphi)=\Lambda\exp (-\varphi/\varphi_0)$ and obtain the field values at which the slow-roll conditions are satisfied. Assume that the inflation terminates at $\varphi=\varphi_1$.

\item  \label{exp_inf_2} Consider the inflation model of the previous problem and obtain the initial value of the field required to prolong the inflation for $N_e\gtrsim 60$ $e$-foldings. What value of $\Lambda$ gives the correct amplitude $\delta\rho/\rho\simeq\sqrt{\mathcal{P}_\mathcal{R}}\simeq 5\cdot 10^{-5}$ of the scalar perturbations?
\end{enumerate}

\subsubsection{The Hybrid Inflation}
\begin{figure}[h]
\centering
  \includegraphics[width=250pt]{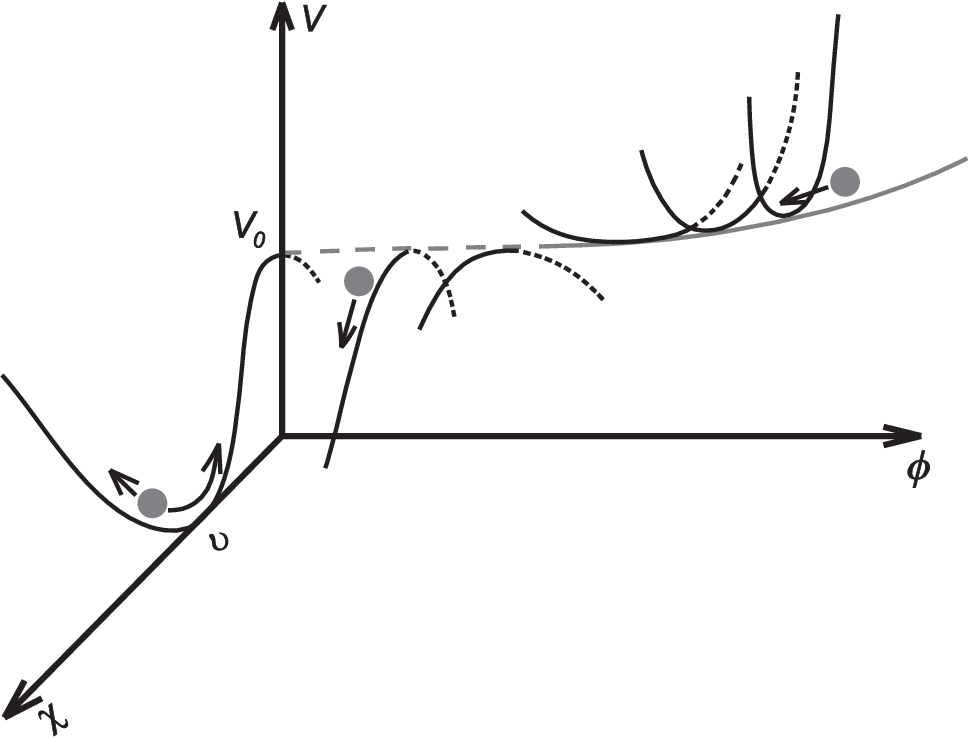}
  \caption{The inflaton potential in the hybrid inflation model.}\label{gibrid_inf_pic}
\end{figure}

\emph{One more inflation model free of super-Planck fields is called the \textbf{hybrid inflation}. In the simplest version this model contains two scalar fields---the inflaton field $\varphi$ and an additional field $\chi$ (see Fig. \ref{gibrid_inf_pic}). During the inflation stage the field $\varphi$ is high, and the system slowly rolls along the valley $\chi =0$. After that the valley $\chi =0$ transforms into saddle, the fast roll takes place in perpendicular direction, the inflation terminates, and the oscillations near the minimum $\varphi =0,$ $\chi=\nu$ lead to heat up of the Universe. }
\emph{
The potential of inflaton field for the hybrid inflation takes the form
\begin{equation}\label{gibrid_inf_v}
V(\varphi,\chi)=\frac12\left(g^2\varphi^2-\mu^2\right)\chi^2+\frac h4\chi^4+U(\varphi)+V_0,
\end{equation}
where $g$ and $h$ are positive dimensionless constants, $\mu$ is a parameter of dimension of mass, $U(\varphi)$ is the monotonically growing inflaton potantial.}

\begin{enumerate}[resume]
\item  \label{inf_gibrid_1} Consider the hybrid inflation model with potential of the form \ref{gibrid_inf_v}, where $U(\varphi=0)=0$ and $U(\varphi)=m^2\varphi^2/2$. The potential $V(\varphi,\chi)$ has maximum at $\varphi=0$, $\chi=\nu=\mu/\sqrt{h}.$ The constant $V_0$ is determined from the requirement that the potential value at minimum is equal to zero, resulting in
    $$V_0=\frac{\mu^4}{4h}.$$
    The line $\chi=0$ represents the valley of the potential $V(\varphi, \chi)$ at $\varphi > \varphi_c,$ where $\varphi_c=\mu/g.$ Consider the case when the contribution of the inflaton field into the total energy at critical value of the former is small
    \begin{equation}\label{gibrid_inf_full_energy}
        U_c\equiv U(\varphi_c)\ll V_0.
    \end{equation}
    Obtain the slow-roll parameters under the requirement that the inflation endures up to the moment when the inflaton rolls down the critical value $\varphi_c$ ($\varphi\gg\varphi_c$), and show that the inflation can take place at sub-Planck values of the inflaton field.

\item   \label{inf_gibrid_2}
Consider the inflation model of problem \ref{inf_gibrid_1} and obtain the potential value $V_{eff}(\chi)$ corresponding to the field value $\chi$, at which it starts to roll down to the potential minimum $\varphi=0$ after the Hubble time $\Delta t \sim H^{-1}(\varphi_c)$ and $\chi = \nu$ after the inflaton reaches the critical value $\varphi_c$.

\item  \label{inf_gibrid_2_1} Obtain the value $\eta_\chi$ corresponding to the fast roll down of the field $\chi$ along the potential $V_{eff}(\chi)$ (see the previous problem).

\item   \label{inf_gibrid_2_2} Use the results of the problems \ref{inf_gibrid_1} -- \ref{inf_gibrid_2_1}, the condition $U_c\equiv U(\varphi_c)\ll V_0$ and assume that the critical value of the inflaton field is small compared to the Planck mass to obtain the chain of inequalities giving the conditions to realize the hybrid inflation scenario.

\item   \label{inf_gibrid_3} Obtain dependence of the $e$-folding number till the inflation end on the inflaton field value.

\item  \label{inf_gibrid_4} Obtain relation between $\eta$ and $N_e$ for the hybrid inflation model in the case of quadratic potential of the inflaton field.

\item   \label{inf_gibrid_5} Consider the hybrid inflation model with $U_c\sim V_0$. Show that in this case $\varphi_c\gtrsim M_{Pl}.$ What region of the $(\epsilon,\eta)$ plane corresponds to the models of such type?
\end{enumerate}

\subsubsection{Other Types of Inflation}
\begin{enumerate}[resume]
\item   \label{inf58} Formulate the differences between the models of cold and warm inflation.

\item  \label{inf60} The Standard Model of particles allows instability of baryons, which can decay to leptons within the framework of this model. But as the rate of the processes violating the baryon number conservation is very low (for example, the proton lifetime $\tau_p\ge10^{30}years$), it is extremely difficult to find experimental proofs of the proton instability. Try to formulate the "inflationary proof" of the proton instability.

\item   \label{inf61} In what way is the inflation theory related to the solution of the problem: ''why is mathematics so efficient in description of our world and prediction of its evolution?''

\end{enumerate}

\chapter{Dark Energy \label{DE}}

\footnotesize {\it The observed accelerated expansion of the Universe requires either modification of General Relativity or existence within the framework of the latter of a smooth energy component with negative pressure, called the ''dark energy''. This component is usually described with the help of the state equation $p=w\rho$. As it follows from Friedman equation,
\[\frac{\ddot a}{a}=-\frac{4\pi G}{3}(\rho+3p).\]
For cosmological acceleration it is needed that $w<-1/3$. The allowed range of values of $w$ can be split into three intervals. The first interval $-1<w<-1/3$ includes scalar fields named the quintessence. The substance with the state equation $p=-\rho\ (w=-1)$ was named the cosmological constant, because $\rho=const$: energy density does not depend on time and is spacially homogeneous in this case. Finally, scalar fields with $w<-1$ were called the phantom fields. Presently there is no evidence for dynamical evolution of the dark energy. All available data agree with the simplest possibility - the  cosmological constant. However the situation can change in future with improved accuracy of observations. That is why one should consider other cases of dark energy alternative to the cosmological constant.}

\section{The Cosmological Constant}
\begin{enumerate}
\item \label{DE01} Derive the Einstein equations in the presence of the cosmological constant by variation of the gravitational field action with the additional term \[S_\Lambda=-\frac{\Lambda}{8\pi G}
	\int\sqrt{-g}\;d^4x.\]

\item \label{DE02} Show that the $\Lambda$-term is a constant (the cosmological constant): $\partial_\mu\Lambda=0$.

\item \label{DE03} Derive Friedman equations in the presence of the cosmological constant.

\item \label{DE03_1} Consider the case of two-component Universe with arbitrary curvature filled by non-relativistic matter and dark energy in the form of cosmological constant. Show that in such case the first Friedman equation can be presented in terms of dimensionless variables $\bar a\equiv a/a_0$, $\Omega_m,\Omega_\Lambda$ and $\tau\equiv H_0 t)$ in the following way:
\[\left(\frac{d\bar a}{d\tau}\right)^2=1+\Omega_{m0}\left(\frac{1}{\bar a}-1\right)+\Omega_{\Lambda0}({\bar a}^2-1).\]

\item \label{DE03_2} 
 Represent the first Friedman equation in terms of intrinsic Gaussian curvature of the three-space
\[K(t)=K_0/a^2(t).\]

\item \label{DE03_3} Analyze the contribution of different energy components to the intrinsic Gaussian curvature of the three-space.

\item \label{DE04}  Find equation for the scale factor in a two-component Universe filled with matter with equation of state $p=w\rho$ and cosmological constant.

\item \label{DE05}  Find the natural scale of length and time appearing due to introduction of cosmological constant into General Relativity.

\item \label{DE06} Show that the relativity principle results in the state equation $p=-\rho$ for dark energy in form of cosmological constant if it is treated as the vacuum energy.

\item \label{DE07} Show that the cosmological constant's equation of state $p=-\rho$ ensures Lorentz-invariance of the vacuum energy-momentum tensor.

\item \label{DE08} Show that the equation of state $p=-\rho$ is the only form which ensures Lorentz-invariance of the vacuum energy-momentum tensor.

\item \label{DE09} Show explicitly that the state equation $p=-\rho$ is Lorenz--invariant.

\item \label{DE10} Consider the observer that moves with constant velocity $V$ in the Universe described by FLRW metrics and filled with substance with equation of state $p=w\rho$. Calculate the energy density, which the observer will register. Consider the cases of decelerated and accelerated expansion of the Universe.

\item \label{DE11} Does the energy conservation law hold in the presence of dark energy in the form of cosmological constant?

\item \label{DE12} Show that by assigning energy to vacuum we do not revive the notion of "ether", i.e. we do not violate the relativity principle or in other words we do not introduce the notions of absolute rest and motion relative to vacuum.

\item \label{DE13} Suppose that density of the dark energy as cosmological constant is equal to the present critical density, $\rho_{\Lambda}=\rho_{cr}$. What is then the total amount of dark energy inside the Solar System? Compare this number with $M_\odot c^2$.

\item \label{DE14} Estimate the upper limit for the cosmological constant. Can an upper or lower limit be derived from the observed rate of growth of cosmological structures?

\item \label{DE15} Knowing the age of the oldest objects in the Universe, determine the lower physical limit of the physical vacuum density.

\item \label{DE16} Find time dependence of the scale factor in the case of flat Universe filled by dark energy in the form of cosmological constant and non-relativistic matter with current relative densities $\Omega_{\Lambda0}$ and $\Omega_{m0}$ respectively (see Chapter~11 for more detailed analysis).

\item \label{DE17} Consider flat Universe filled by matter and cosmological constant with $\Lambda<0$ and show that it collapses in time period
 \[t_{col}=\frac{2\pi}{\sqrt{3|\Lambda|}}.\]

\item \label{DE18} Find the value of redshift in the cosmological constant dominated flat Universe, for which a source of linear size $d$ has the minimum visible angular dimension.

\item \label{DE_24} Find (in the Newton's approximation) a critical distance $r_0$ around a point mass $m$, embedded into medium which is the cosmological constant $\Lambda>0$, where the gravity vanishes: it is attractive if $r<r_0$ and repulsive if $r>r_0$.

\section{Geometry and Destiny}
(see L.Krauss, M. Turner, arXiv:9904020; B.Ryden, Introduction to cosmology, Addison Wesley; J.E.Felten,R.Isaacman, Rev.Mod.Phys. 58,689 , 1986)

{\it Presence of the cosmological constant reevaluates standard notions about the connection between geometry and the fate of the Universe. The traditional philosophy of General Relativity is that ''Geometry is Destiny'', with ''geometry'' in this context implying openness or closure of the three-space of constant cosmological time.  If energy content is provided by ''ordinary'' matter (nonrelativistic matter or radiation) this slogan transforms into ''Density is Destiny''. If the density of matter is less or equal than the critical value (and the Universe is open), then the destiny of Universe is eternal expanding; if the density is greater, and the Universe is closed, then the destiny is recollapse. If the Universe contains cosmological constant (or any energy component with $w<-1/3$) the situation changes radically: an open Universe can recollapse, while a closed Universe can expand forever. Geometry no longer determines the fate of the Universe.}

\item \label{DE18_1} As $\Omega_{tot}-1=k/(H^2a^2)$ the sign of $k$ is determined by whether $\Omega_{tot}$ is greater or less than $1$.  A measurement of $\Omega_{tot}$ at any epoch---including the present---determines the geometry of the Universe. However, as opposed to situation with only non-relativistic matter, we can no longer claim that the magnitude of $\Omega_{tot}$ uniquely determines the fate of the Universe. Explain decoupling between $\Omega_{tot}$ and destiny using deceleration parameter.

\item \label{DE18_2} Show that decoupling between $\Omega_{tot}$ and destiny of the Universe is due to violation of strong energy condition.

\item \label{DE18_3} Find the maximum value of scale factor for a hypothetical flat Universe with $\Omega_{\Lambda0}<0$.

\item \label{DE18_4} For the Universe considered in the previous problem, find the analytical solution $t(a)$ and time of the collapse from $a=a_{\max}$ back down to $a=0$.

\item \label{DE18_5} In the flat Universe with $\Omega_{m0}<1$ and positive cosmological constant find the late time asymptotic for the scale factor.

\item \label{DE18_6} Show that eternal expansion is inevitable if and only if \[\Omega_\Lambda>4\Omega_m\left\{\cos\left[\frac13\arccos(\Omega_m^{-1}-1)+\frac{4\pi}{3}\right]\right\}.\]

\item \label{DE18_7} Show that if $\Omega_{m0}+\Omega_{\Lambda0}>1$ (positive spatial curvature) and $\Lambda>0$, then it is possible to have a Universe that expands at late times, but without the initial Big Bang ($a=0,\ t=0$).

\item \label{DE18_8} Show that collapse of the Universe is possible if the following conditions are satisfied:

1. Closed ($k>0$) Universe: $\rho_m>2\rho_\Lambda$ when $H=0$.

2. Open, flat ($k\le0$) or closed ($k>0$) Universe: $\rho_\Lambda\le0$.

\item \label{DE18_9} Show that eternal expansion of the Universe is possible if the following conditions are satisfied

1. Closed ($k>0$) Universe: $\rho_m<2\rho_\Lambda$ before $H=0$.

2. Open or flat ($k\le0$) Universe: $\rho_\Lambda\ge0$.

\section{Time-dependent Cosmological Constant}
(Inspired by I.Shapiro,J.Sola, H.Stefancic, hep-ph/0410095)

\item \label{DE19} Obtain the analogue of the conservation equation $\dot\rho+3H(\rho+p)=0$ for the case when the gravitation constant $G$ and the cosmological "constant" $\Lambda$ depend on time.

\item \label{DE20}  Show that when $G$ is constant, $\Lambda$ is also a constant if and only if the ordinary energy-momentum tensor $T_{\mu\nu}$ is also conserved.

\item \label{DE21} Show that in case the cosmological constant depends on time, the energy density related to the latter can be converted into matter.

\item \label{DE22} Derive the time dependence of the scale factor for a flat Universe with the $\Lambda$--dynamical constant $\Lambda=\Lambda_0(1+\alpha t)$.

\item \label{DE23} Construct the dynamics of the Universe in the cosmological model with $\Lambda=\sigma H ,\; \sigma>0.$

\item \label{DE24_0} Construct the dynamics of the Universe in the cosmological model with $\Lambda(H)=\sigma H +3\beta H^2$ and $\rho=\rho_\Lambda+\rho_m$.

\section{Static Einstein's Universe}
\item \label{DE24} Find the static solution of Friedman equations with cosmological constant and non-relativistic matter (static Einstein's Universe).

\item \label{DE25} Show that the static Einstein's Universe must be closed. Find the total volume and mass of this Universe.

\item \label{DE27} Find the parameters of static Einstein's Universe under the assumption that both matter and radiation are absent.

\item \label{DE28} Estimate the radius of static Einstein's Universe if the zero-point energy of electromagnetic field is cut off at the classical electron radius.

\item \label{DE29} Show that Einstein's Universe is unstable.

\item \label{DE30}  What are the concrete mechanisms that drive the instability of the static Einstein's Universe?

\item \label{DE31}  What is the most unsatisfactory peculiarity of the static Einstein's model of the Universe (besides the instability)?

\item \label{DE32}   Construct the effective one-dimensional potential $V(a)$ for the case of flat Universe filled with non-relativistic matter and dark energy in the form of cosmological constant.

\item \label{DE33}   Show that the static Einstein's Universe may be realized only in the maximum of the effective potential $V(a)$ of the previous problem.

\item \label{DE34}  Problem \ref{DE32} can be considered in more general setup.
Assuming arbitrary values for the contributions of the cosmological constant $\lambda$, matter $\mu$, radiation $\gamma$ and curvature $\kappa$ respectively, present the first Friedman equation
\begin{equation}\label{Friedman4k+}
H^2\equiv\left(\frac{\dot a}{a}\right)^2=\lambda-\frac{\kappa}{a^2}+\frac{\mu}{a^3}+\frac{\gamma}{a^4}
\end{equation}
in the form of the energy conservation law \[p^2/2 + U(q) = E.\]

\section{Dynamical Forms of Dark Energy}
\subsection{The Quintessence}

{\it The cosmological constant represents nothing but the simplest realization of the dark energy---the hypothetical substance introduced to explain the accelerated expansion of the Universe. There is a dynamical alternative to the cosmological constant---the scalar fields, formed in the post-inflation epoch. The most popular version is the scalar field $\varphi$ evolving in a properly designed potential $V(\varphi)$. Numerous models of such type differ by choice of the scalar field Lagrangian. The simplest model is the so-called quintessence. In antique and medieval philosophy this term (literally 'the fifth essence', after the earth, water, air and fire) meant the concentrated extract, the creative force, penetrating all the material world. We shall understand the quintessence as the scalar field in a potential, minimally coupled to gravity, i.e. feeling only the influence of space-time curvature. Besides that we restrict ourselves to the canonic form of the kinetic energy. The action for fields of such type takes the form
\[S=\int d^4x \sqrt{-g}\; L=\int d^4x \sqrt{-g}\left[\frac12g^{\mu\nu}\frac{\partial\varphi}{\partial x^\mu} \frac{\partial\varphi}{\partial x^\nu}-V(\varphi)\right].\]
The equations of motion for the scalar field are obtained as usual, by variation of the action with respect to the field (see Chapter 'Inflation').}

\item \label{DE35_0}   Obtain the Friedman equations for the case of flat Universe filled with quintessence.

\item \label{DE35}  Obtain the general solution of the Friedman equations for the Universe filled with free scalar field, $V(\varphi)=0$.

\item \label{DE47_1}   Show that in the case of Universe filled with non-relativistic matter and quintessence the following relation holds: \(\dot H=-4\pi G(\rho_m+\dot\varphi^2).\)

\item \label{DE47_2}   Show that in the case of Universe filled with non-relativistic matter and quintessence the Friedman equations
    \[H^2=\frac{8\pi G}{3}\left[\rho_m+\frac12\dot\varphi^2+V(\varphi)\right],\]
\[\dot H =-4\pi G(\rho_m+\dot\varphi^2)\]
can be transformed to the form
\[\frac{8\pi G}{3H_0^2}\left(\frac{d\varphi}{dx}\right)^2=\frac{2}{3H_0^2x}\frac{d\ln H}{dx}-\frac{\Omega_{m0}x}{H^2};\]
\[\frac{8\pi G}{3H_0^2}V(x)=\frac{H^2}{H_0^2}-\frac{x}{6H_0^2}\frac{d H^2}{dx}-\frac12\Omega_{m0}x^3;\]
\[x\equiv1+z.\]

\item \label{DE36}   Show that the conservation equation for quintessence can be obtained from the Klein-Gordon equation \[\ddot\varphi+3H\dot\varphi+\frac{dV}{d\varphi}=0.\]

\item \label{DE36_2}   Find the explicit form of Lagrangian describing the dynamics of the Universe filled with the scalar field in potential $V(\varphi)$. Use it to obtain the equations of motion for the scale factor and the scalar field.

\item \label{DE36_3}   In the flat Universe filled with scalar field $\varphi$ obtain the isolated equation for $\varphi$ only. [see S.Downes, B.Dutta, K.Sinha, arXiv:1203.6892 ]

\item \label{DE37}   What is the reason for the requirement that the scalar field's evolution in the quintessence model is slow enough?

\item \label{DE38}   Find the potential and kinetic energies for quintessence with the given state parameter $w$.

\item \label{DE39}   Find the dependence of state equation parameter $w$ for scalar field on the quantity \[x=\frac{\dot\varphi^2}{2V(\varphi)}\] and determine the ranges of $x$ corresponding to inflation in the slow-roll regime, matter-dominated epoch and the rigid state equation ($p\sim\rho$) limit correspondingly.

\item \label{DE40}   Show that if kinetic energy $K=\dot\varphi^2/2$ of a scalar field is initially much greater than its potential energy $V(\varphi)$, then it will decrease as $a^{-6}$.

\item \label{DE41}   Show that the energy density of a scalar field $\varphi$ behaves as $\rho_\varphi\propto
a^{-n}$, $0\le n\le6$.

\item \label{DE42}   Show that dark energy density with the state equation $p=w(a)\rho(a)$ can be presented as a function of scale factor in the form
\[\rho=\rho_0 a^{-3[1+\bar w(a)]},\]
where $\bar w(a)$ is the parameter $w$ averaged in the logarithmic scale
$$
\bar w(a) \equiv \frac{\int w(a)d\ln a}{\int {d\ln a} }.
$$

\item \label{DE55_1}   Consider the case of Universe filled with non-relativistic matter and quintessence with the state equation $p=w\rho$ and show that the first Friedman equation can be presented in the form
    \[H^2(z)=H_0^2\left[\Omega_{m0}(1+z)^3+(1-\Omega_{m0})e^{3\int_0^z\frac{dz'}{1+z'}(1+w(z'))}\right].\]

\item \label{DE55_2}   Show that for the model of the Universe considered in the previous problem the state equation parameter $w(z)$ can be presented in the form
    \[w(z)=\frac{\frac23(1+z)\frac{d\ln H}{dz}-1}{1-\frac{H_0^2}{H^2}\Omega_{m0}(1+z)^3}.\]

\item \label{DE55_3}   Show that the result of the previous problem can be presented in the form
    \[w(z)=-1+(1+z)\frac{2/3E(z)E'(z)-\Omega_{m0}(1+z)^2}{E^2(z)-\Omega_{m0}(1+z)^3},\quad E(z)\equiv\frac{H(z)}{H_0}.\]
 
\item \label{DE43}   Show that decreasing of the scalar field's energy density with increasing of the scale factor slows down as the scalar field's potential energy $V(\varphi)$ starts to dominate over the kinetic energy density $\dot\varphi^2/2$.

\item \label{DE44}   Express the time derivative $\dot\varphi$ through the quintessence' density $\rho_\varphi$ and the state equation parameter $w_\varphi$.

\item \label{DE45}   Estimate the magnitude of the scalar field variation $\Delta\varphi$ during time $\Delta t$.

\item \label{DE46}   Show that in the radiation-dominated or matter-dominated epoch the variation of the scalar field is small, and the measure of its smallness is given by the relative density of the scalar field.

\item \label{DE47}   Show that in the quintessence $(w>-1)$ dominated Universe the condition $\dot{H}<0$ always holds.

\item \label{DE48}   Consider simple bouncing solution of Friedman equations that avoid singularity. This solution requires positive spatial curvature $k=+1$, negative cosmological constant $\Lambda<0$ and a "matter" source with equation of state $p=w\rho$ with $w$ in the range
	\[-1<w<-\frac13.\]
In the special case $w=-2/3$ Friedman equations describe a constrained harmonic oscillator (a simple harmonic Universe). Find the corresponding solutions.
(Inspired by P.Graham et al. arXiv:1109.0282/hp-th)

\item \label{DE49}   Derive the equation for the simple harmonic Universe (see previous problem), using the results of problem \ref{DE04}. 

\item \label{DE50}   Barotropic liquid is a substance for which pressure is a single--valued function of density. Is quintessence generally barotropic?

\item \label{DE51}   Show that a scalar field oscillating near the minimum of potential is not a barotropic substance.

\item \label{DE52}   For a scalar field $\varphi$ with state equation $p=w\rho$ and relative energy density $\Omega_\varphi$ calculate the derivative \[w'=\frac{dw}{d\ln a}.\]

\item \label{DE53}   Calculate the sound speed in the quintessence field $\varphi(t)$ with potential $V(\varphi)$.

\item \label{DE54}  Find the dependence of quintessence energy density on redshift for the state equation $p_{DE}=w(z)\rho_{DE}$.

\item \label{DE55}   The equation of state $p=w(a)\rho$ for quintessence is often parameterized as $w(a)=w_0 + w_1(1-a)$. Show that in this parametrization energy density and pressure of the
scalar field take the form:
$$
\rho(a) \propto a^{-3[1+w_{eff}(a)]},\quad p(a) \propto
 (1+w_{eff}(a))\rho(a),
$$
where
$$
w_{eff}(a)=(w_0+w_1)+(1-a)w_1/\ln a.
$$

\item \label{DE56}   Find the dependence of Hubble parameter on redshift in a flat Universe filled with non-relativistic matter with current relative density $\Omega_{m0}$ and dark energy with the state equation $p_{DE}=w(z)\rho_{DE}$.

\item \label{DE57}   Show that  in a flat Universe filled with non--relativistic matter and arbitrary component with the state equation $p=w(z)\rho$ the first Friedman equation can be presented in the form:
\[w(z)=-1+\frac13\frac{d\ln(\delta H^2/H_0^2)}{d\ln(1+z)},\]
where
\[\delta H^2 = H^2 - \frac{8\pi G}{3}\rho_m\]
describes the contribution into the Universe's expansion rate of all components other than matter.

\item \label{DE58}   Express the time derivative of a scalar field through its derivative with respect to redshift \(d\varphi/dz.\)

\item \label{DE59}   Show that the particle horizon does not exist for the case of quintessence because the corresponding integral diverges (see Chapter 2(3)).

\item \label{DE60}   Show that in a Universe filled with quintessence the number of observed galaxies decreases with time.

\item \label{DE61}   Let $t$ be some time in the distant past $t\ll t_0$. Show that in a Universe dominated by a substance with state parameter $w>-1$ the current cosmic horizon (see Chapter 3) is
	\[R_h(t_0)\approx\frac32(1+\langle w\rangle)t_0,\]
where $\langle w\rangle$ is the time-averaged value of $w$ from $t$ to the present time
\[\langle w\rangle\equiv\frac{1}{t_0}\int\limits_t^{t_0} w(t)dt.\]

\item \label{DE62}    From WMAP\footnote{Wilkinson Microwave Anisotropy Probe is a spacecraft which measures differences in the temperature of the Big Bang's remnant radiant heat---the cosmic microwave background radiation---across the full sky.} observations we infer that the age of the Universe  is  $t_0\approx13.7\cdot10^9$ years and cosmic horizon equals to $R_h(t_0)=H_0^{-1}\approx13.5\cdot10^9$ light-years. Show that these data imply existence of some substance with equation of state $w<-1/3$,---''dark energy''. 

\item \label{DE63}    The age of the Universe today depends upon the equation-of-state of the dark energy. Show that the more negative parameter $w$ is, the older Universe is today.

\item \label{DE64}    \label{g} Consider a Universe filled with dark energy with state equation depending on the Hubble parameter and its derivatives,
\[p=w\rho+g(H,\dot H, \ddot H,\ldots,;t).\]
What equation does the Hubble parameter satisfy in this case?

\item \label{DE65}   Show that taking function $g$ (see the previous problem) in the form
\[g(H,\dot H, \ddot H)=-\frac{2}{\kappa^2}\left(\ddot H +
\dot H + \omega_0^2 H + \frac32(1+w)H^2-H_0\right),\ \kappa^2=8\pi
G\]
leads to the equation for the Hubble parameter identical to the one for the harmonic oscillator, and find its solution.

\item \label{DE66}    Find the time dependence of the Hubble parameter in the case of function $g$ (see problem \ref{g}) in the form \[g(H;t)= -\frac{2\dot f(t)}{\kappa^2f(t)}H,\ \kappa^2=8\pi G\] where $f(t)=-\ln(H_1+H_0\sin\omega_0t)$, $H_1>H_0$ is arbitrary function of time.

\item \label{DE68}   Show that in an open Universe the scalar field potential $V[\varphi(\eta)]$ depends monotonically on the conformal time $\eta$.

\item \label{DE69}   Reconstruct the dependence of the scalar field potential $V(a)$ on the scale factor basing on given dependencies for the field's energy density $\rho_\varphi(a)$ and state equation parameter $w(a)$.

\item \label{DE70}   Find the quintessence potential providing the power law growth of the scale factor $a\propto t^p$, where the accelerated expansion requires $p>1$.

\item \label{DE70_1}   Let $a(t)$, $\rho(t)$, $p(t)$ be solutions of Friedman equations. Show that for the case $k=0$ the function $\psi_n\equiv a^n$ is the solution of "Schr\"odinger equation" $\ddot\psi_n=U_n\psi_n$ with potential [see A.V.Yurov, arXiv:0905.1393] \[U_n=n^2\rho-\frac{3n}{2}(\rho+p).\]

\item \label{DE70_2}   Consider flat FLRW Universe filled with a scalar field $\varphi$. Show that in the case when $\varphi=\varphi(t)$, the Einstein equations with the cosmological term are reduced to the ''Schr\"odinger equation'' \[\ddot\psi=3(V+\Lambda)\psi\] with $\psi=a^3$. Derive the equation for $\varphi(t)$ [see A.V.Yurov, arXiv:0305019].

\item \label{DE70_3}  Consider FLRW space-time filled with non-interacting matter and dark energy components. Assume the following forms for the equation of state parameters of matter and dark energy
	\[w_m=\frac{1}{3(x^\alpha+1)},\quad w_{DE}=\frac{\bar{w}x^\alpha}{x^\alpha+1},\]
where $x=a/a_*$ with $a_*$ being some reference value of $a$, $\alpha$ is some positive constant and $\bar{w}$ is a negative constant. Analyze the dynamics of the Universe in this model. [see S.Kumar,L.Xu, arXiv:1207.5582]

\subsection{Tracker Fields}
{\it A special type of scalar fields---the so-called tracker fields---was discovered at the end of the nineties. The term reflects the fact that a wide range of initial values for the fields of such type rapidly converges to the common evolutionary track. The initial values of energy density for such fields may vary by many orders of magnitude without considerable effect on the long-time asymptote. The peculiar property of tracker solutions is the fact that the state equation parameter for such a field is determined by the dominant component of the cosmological background.

It should be stressed that, unlike the standard attractor, the tracker solution is not a fixed point (in the sense of a solution corresponding to the fixed point in a system of autonomous differential equations ): the ratio of the scalar field energy density to that of background component (matter or radiation) continuously changes as the quantity $\varphi$ descends along the potential. It is well desirable feature because we want the energy density $\varphi$ to exceed ultimately the background density and to transfer the Universe into the observed phase of the accelerated expansion.

Below we consider a number of concrete realizations of the tracker fields.}

\item \label{DE71}   Show that initial value of the tracker field should obey the condition $\varphi_0=M_{Pl}$.

\item \label{DE72}   Show that densities of kinetic and potential energy of the scalar field $\varphi$ in the potential of the form \[V(\varphi)=M^4\exp(-\alpha\varphi M),\quad M\equiv\frac{M_{PL}^2}{16\pi}\] are proportional to the density of the concomitant component (matter or radiation) and therefore it realizes the tracker solution.

\item \label{DE73}   Consider a scalar field potential \[V(\varphi)=\frac A n\varphi^{-n},\] where $A$ is a dimensional parameter and $n>2$. Show that the solution $\varphi(t)\propto t^{2/(n+2)}$ is a tracker field under condition $a(t)\propto t^m$, $m=1/2$ or $2/3$ (either radiation or non-relativistic matter dominates).
    
\item \label{DE74}  Show that the scalar field energy density corresponding to the tracker solution in the potential \[V(\varphi)=\frac A n\varphi^{-n}\] (see the previous problem \ref{DE73}) decreases slower than the energy density of radiation or non-relativistic matter.

\item \label{DE75}   Find the equation of state parameter $w_\varphi\equiv p_\varphi/\rho_\varphi$ for the scalar field of problem \ref{DE73}.

\item \label{DE76}   Use explicit form of the tracker field in the potential of problem \ref{DE73} to verify the value of $w_\varphi$ obtained in the previous problem.

\subsection{The K-essence}
{\it Let us introduce the quantity $$X\equiv \frac{1}{2}{{g}^{\mu \nu }}\frac{\partial \varphi }{\partial {{x}^{\mu }}}\frac{\partial \varphi }{\partial {{x}^{\nu }}}$$ and consider action for the scalar field in the form
$$
S=\int{{{d}^{4}}x\sqrt{-g}}\; L\left( \varphi ,X \right),
$$
where Lagrangian $L$ is generally speaking an arbitrary function of variables $\varphi$ and $X.$ The dark energy model realized due to modification of the kinetic term with the scalar field, is called the $k$-essence. The traditional action for the scalar field corresponds to
$$
L\left( \varphi ,X \right)=X-V(\varphi ).
$$
In the problems proposed below we restrict ourselves to the subset of Lagrangians of the form
$$
L\left( \varphi ,X \right)=K(X)-V(\varphi ),
$$
where $K(X)$ is a positively defined function of kinetic energy $X$. In order to describe a homogeneous Universe we should choose
$$
X=\frac{1}{2}{{\dot{\varphi }}^{2}}.
$$
}

\item \label{DE77}   Find the density and pressure of the $k$-essence.

\item \label{DE78}   Construct the equation of state for the $k$-essence.

\item \label{DE80}  The sound speed $c_s$ in any medium must satisfy two fundamental requirements: first, the sound waves must be stable and second, its velocity value should be low enough to preserve the causality condition. Therefore \[0\le c_s^2\le1.\] Reformulate the latter condition in terms of scale factor dependence for the equation of state parameter $w(a)$ for the case of the $k$-essence.

\item \label{DE81}   Find the state equation for the simplified $k$-essence model with Lagrangian $L=F(X)$ (the so-called pure kinetic $k$-essence).

\item \label{DE82}   Find the equation of motion for the scalar field in the pure kinetic $k$-essence.

\item \label{DE83}   Show that the scalar field equation of motion for the pure kinetic $k$-essence model gives the tracker solution.

\subsection{Phantom Energy}
{\it The full amount of available cosmological observational data shows that the state equation parameter $w$ for dark energy lies in a narrow range near the value $w=-1$. In the previous subsections we considered the region $-1\le w\le-1/3$. The lower bound $w=-1$ of the interval corresponds to the cosmological constant, and all the remainder can be covered by the scalar fields with canonic Lagrangians. Recall that the upper bound $w=-1/3$ appears due to the necessity to provide the observed accelerated expansion of Universe. What other values of parameter $w$ can be used? The question is very hard to answer for the energy component we know so little about. General Relativity restricts possible values of the energy -momentum tensor by the so-called ''energy conditions'' (see Chapter 2). One of the simplest among them is the so-called Null Dominant Energy Condition (NDEC) $\rho+p\ge0$. The physical motivation of the latter is to avoid the vacuum instability. Applied to the dynamics of Universe, the NDEC requires that density of any allowed energy component cannot grow with the expansion of the Universe. The cosmological constant with $\dot\rho_\Lambda=0$, $\rho_\Lambda=const$ represents the limiting case. Because of our ignorance concerning the nature of dark energy it is reasonable to question whether this mysterious substance can differ from the already known ''good'' sources of energy and if it can violate the NDEC. Taking into account that dark energy must have positive density (it is necessary to make the Universe flat) and negative pressure (to provide the accelerated expansion of Universe), the violation of the NDEC must lead to $w<-1$. Such substance is called the phantom energy. The phantom field $\varphi$ minimally coupled to gravity has the following action:
\[S=\int d^4x \sqrt{-g}L=-\int d^4x \sqrt{-g}\left[\frac12g^{\mu\nu}\frac{\partial\varphi}{\partial x_\mu} \frac{\partial\varphi}{\partial x_\nu}+V(\varphi)\right],\]
which differs from the canonic action for the scalar field only by the sign of the kinetic term.}

\item \label{DE96} {  Show that the action of a scalar field minimally coupled to gravitation
\[S=\int d^4x\sqrt{-g}\left[\frac12(\nabla\varphi)^2-V(\varphi)\right]\]
leads, under the condition $\dot\varphi^2/2<V(\varphi)$, to $w_\varphi<-1$, i.e. the field is phantom.}

\item \label{DE97}   Obtain the equation of motion for the phantom scalar field described by the action of the previous problem.

\item \label{DE98}   Find the energy density and pressure of the phantom field.

\item \label{DE99}   Show that the phantom energy density grows with time. Find the dependence $\rho(a)$ for $w=-4/3$.

\item \label{DE100_0}   Show that the phantom scalar field violates all four energy conditions.

\item \label{DE100}   Show that in the phantom scalar field $(w<-1)$ dominated Universe the condition $\dot{H}>0$ always holds.

\item \label{DE101}   As we have seen in Chapter 3, the Friedman equations, describing spatially flat Universe, possess the duality, which connects the expanding and contracting Universe by appropriate transformation of the state equation. Consider the Universe where the weak energetic condition $\rho\ge0,\ \rho+p\ge0$ holds and show that the ideal liquid associated with the dual Universe is a phantom liquid or the cosmological constant.

\item \label{DE101_1}   Show that the Friedman equations for the Universe filled with dark energy in the form of cosmological constant and a substance with the state equation $p=w\rho$ can be presented in the form of nonlinear oscillator [see M.Dabrowski, arXiv0307128]
	\[\ddot X-\frac{D^2}{3}\Lambda X+D(D-1)kX^{1-2/D}=0\]
where
\[X=a^{D(w)},\quad D(w)=\frac32(1+w).\]

\item \label{DE102}   Show that the Universe dual to the one filled with a free scalar field, is described by the state equation $p=-3\rho$. 
\item \label{DE104}   Show that in the phantom component of dark energy the sound speed exceeds the light speed.

\item \label{DE105}  Construct the phantom energy model with negative kinetic term in the potential satisfying the slow-roll conditions \[\frac 1 V \frac{dV}{d\varphi}\ll1\] and \[\frac 1 V \frac{d^2V}{d\varphi^2}\ll1.\]

\subsection{Disintegration of Bound Structures}
{\it Historically the first criterion for decay of gravitationally bound systems due to the phantom dark energy was proposed by Caldwell, Kamionkowski and Weinberg (CKW) [see arXiv:astro-ph/0302506v1]. The authors argue that a satellite orbiting around a heavy attracting body becomes unbound when total repulsive action of the dark energy inside the orbit exceeds the attraction of the gravity center. Potential energy of gravitational attraction is determined by the mass $M$ of the attracting center, while the analogous quantity for repulsive potential equals to $\rho+3p$ integrated over the volume inside the orbit. It results in the following rough estimate for the disintegration condition
\begin{equation}\label{disintegration}
-\frac{4\pi}{3}(\rho+3p)R^3\simeq M.
\end{equation}
}
\item \label{DE108}   Show that for $w\ge-1$ a system gravitationally bound at some moment of time (Milky Way for example) remains bound forever.

\item \label{DE109}   Show that in the phantom energy dominated Universe any gravitationally bound system will dissociate with time.

\item \label{DE106}   Show that in a Universe filled with non--relativistic matter a hydrogen atom will remain a bound system forever.

\item \label{DE110}   \label{DE95_new} Demonstrate, that any gravitationally bound system with mass $M$ and radius (linear scale) $R$, immersed in the phantom background $\left( {w <  - 1} \right)$ will decay in time
\[t \simeq P\frac{|1+3w|}{|1+w|}\frac29\sqrt{\frac{3}{2\pi}}\]
before Big Rip. Here \[P=2\pi\sqrt{\frac{R^3}{GM}}\] is the period on the circular orbit of radius $R$ around the considered system.

\item \label{DE110_2}   Use the result of the previous problem to determine the time of disintegration for the following systems: galaxy clusters, Milky Way, Solar System, Earth, hydrogen atom. Consider the case $w=-3/2$.

\subsection{Big Rip, Pseudo Rip, Little Rip}
{\it The future finite-time singularity is an essential element of phantom cosmology [see S.Nojiri, S. Odintsov, arXiv:hep-th/0505215]. One may classify the future singularities as in the following way [see S.Nojiri, S. Odintsov and S.Tsajikava, arXiv:hep-th/0501025]:

1. For $t\to t_s$, $a\to\infty$, $\rho\to\infty$, $|p|\to\infty$ ("Big Rip").

The density of phantom dark energy and scale factor become infinite at some finite time $t_s$.

2. For $t\to t_s$, $a\to a_s$, $\rho\to\rho_s$ or $\rho\to0$, $|p|\to\infty$ (''sudden singularity'').

The condition $w<-1$ is necessary for future singularities, but it is not sufficient. If $w$ approaches to $-1$ sufficiently rapidly, then it is possible to have a model in which there are no future singularities. Models without future singularities in which $\rho_{DE}$ increases with time will eventually lead to dissolution of bound systems. This process received the name "Little Rip"[see P.Frampton, K.Ludwick and R.Scherrer, arXiv:1106.4996]. In the Big Rip the scale factor and energy density diverge at finite future time. As opposed to Big Rip in the $\Lambda$CDM, there is no such divergence. Little Rip represents an interpolation between these two limit cases.

3. For $t\to t_s$, $a\to a_s\ne0$, $\rho\to\infty$, $|p|\to\infty$.

4. For $t\to t_s$, $a\to a_s\ne0$, $\rho\to\rho_s$ (including $\rho_s=0$), while derivatives of $H$ diverge.

Here $t_s$, $a_s\ne0$ and $\rho_s$ are constants.}

\item \label{DE103}   For the flat Universe composed of matter $(\Omega_m\simeq0.3)$ and phantom energy $(w=-1.5)$ find the time interval left to the Big Rip.

{\it Immediate consequence of approaching the Big Rip is the dissociation of bound systems due to negative pressure inside them. }

\item \label{RIPS_1}   Show that all little-rip models can be described by condition $\ddot f>0$ where $f(t)$ is a nonsingular function such that $a(t)=\exp[f(t)]$.

\item \label{RIPS_2}   Consider the approach of the following authors [see S. Nojiri, S.D. Odintsov, and S. Tsujikawa, Phys. Rev. D 71, 063004 (2005); S. Nojiri and S.D. Odintsov, Phys. Rev. D 72, 023003 (2005);  H. Stefancic, Phys. Rev. D 71, 084024 (2005)], who expressed the pressure as a  function of  the density in the form
	\[p=-\rho-f(\rho).\]
Show that condition $f(\rho>0)$ ensures that the density increases with the scale factor.

\item \label{RIPS_3}   Find the dependencies $a(\rho)$ and $t(\rho)$ for the case of flat Universe filled by a substance with the following state equation \[p=-\rho-f(\rho).\]

\item \label{RIPS_4}   Solve the previous problem in the case of \(f(\rho)A\rho^\alpha,\ \alpha=const.\)

\item \label{RIPS_5}  Find the condition for big-rip singularity in the case $p=-\rho-f(\rho).$

\item \label{RIPS_6}   Show that taking a power law for $f(\rho)$, namely $f(\rho)=A\rho^\alpha$ a future singularity can be avoided for $\alpha\le1/2$.

\item \label{RIPS_7_1}   Solve the previous problem using the condition for absence of future singularities obtained in Problem \ref{RIPS_1}.

\item \label{RIPS_7_2}   Formulate the condition for the absence of a finite-time future (Big Rip) singularity in terms of function $\rho(a)$ .

{\it The problems below develop an alternative approach to investigate the singularities in the phantom Universe [see P-H. Chavanis, arXiv:1208.1195]}

\item \label{RIPS_7_3}   Consider the polytropic equation of state
\[p=\alpha\rho+k\rho^{1+1/n}\equiv-\rho+\rho\left(1+\alpha+k\rho^{1/n}\right)\]
under assumption $-1<\alpha\le1$. The case $\alpha=-1$ is treated separately in Problem \ref{RIPS_7_4}. The additional assumption $1+\alpha+k\rho^{1/n}\le0$ (and necessary condition $k<0$) guarantees that the density increases with the scale factor. This corresponds to phantom Universe. Find explicit dependence $\rho(a)$ and analyze limits $a\to0$ and $a\to\infty$.

\item \label{RIPS_7_4}   Consider the previous problem with $\alpha=-1$ and $k<0$. This equation of state was introduced by Nojiri and Odintsov (see problem \ref{RIPS_7_3}). Chavanis re-derives their results in a more transparent form.

\begin{flushright}
{\it {\bf "The little-rip dissociates all bound structures, but\\
 the strength of the dark energy is not enough to rip\\
 apart space-time as there is no finite-time singularity"}\\
P. Frampton, K. Ludwick1, and R. Scherrer}
\end{flushright}

[see A. Astashenok, S. Nojiri, S. Odintsov, and R. Scherrer, arXiv:1203.1976]

\item \label{RIPS_8}   Show that for any bound system the rip always occurs when either $H$ diverges or $\dot H$ diverges (assuming $\dot H>0$ ( expansion of Universe is accelerating)).

\item \label{RIPS_7}   Solve the previous problem in terms of function $f(\rho)$.

\item \label{RIPS_9}   Perform analysis of possible singularities in terms of characteristics of the scalar field $\varphi$ with the potential $V(\varphi)$.

{\it All the Big Rip, Little Rip and Pseudo-Rip arise from the assumption that the dark energy density $\rho(a)$ is monotonically increasing. Let us investigate what will happen if this assumption is broken and then propose a so-called "Quasi-Rip" scenario, which is driven by a type of quintom dark energy. In this work, we consider an explicit model of Quasi-Rip in details. We show that Quasi-Rip has an unique feature different from Big Rip, Little Rip and Pseudo-Rip. Our universe has a chance to be rebuilt in the ash after the terrible rip. This might be the last hope in the "hopeless" rip.}

\item \label{RIPS_65}   So-called soft singularities are characterized by a diverging $\ddot a$ whereas both the scale factor $a$ and $\dot a$ are finite. Analyze features of intersections between the soft singularities and geodesics.

\item \label{RIPS_66} {  (see F.Cannata, A. Kamenshchik, D.Regoli, arXiv:0801.2348) The power law cosmological evolution $a(t)\propto t^\beta$ leads to the Hubble parameter $H(t)\propto 1/t$. Consider a "softer" version of the cosmological evolution given by the law
	\[H(t)=\frac{S}{t^\alpha},\]
where $S$ is a positive constant and $0<\alpha<1$. Analyze the dynamics of such model at $t\to0$.}

\item \label{RIPS_67}   Reconstruct the potential of the scalar field model, producing the given cosmological evolution $H(t)$.

\item \label{RIPS_68}   Reconstruct the potential of the scalar field model, producing the cosmological evolution
\begin{equation}H(t)=\frac{S}{t^\alpha},\label{RIPS_68}\end{equation}
using the technique described in the previous problem.

\subsection{The Statefinder}
{\it In the models including dark energy in different forms it is useful to introduce a pair of cosmological parameters $\{r,s\}$, which is called the statefinder (see V.Sahni, T.Saini, A.Starobinsky, U.Alam astro-ph/0201498 ):
\[r\equiv\frac{\dddot a}{aH^3},\ s\equiv\frac{r-1}{3(q-1/2)}.\]
These dimensionless parameters are constructed from the scale factor and its derivatives. Parameter $r$ is the next member in the sequence of the kinematic characteristics describing the Universe's expansion after the Hubble parameter $H$ and the deceleration parameter $q$ (see Chapter ''Cosmography''). Parameter $s$ is the combination of $q$ and $r$ chosen in such a way that it is independent of the dark energy density. The values of these parameters can be reconstructed with high precision basing on the available cosmological data. After that the statefinder can be successfully used to identify different dark energy models.}

\item \label{DE111}   Explain the advantages for the description of the current Universe's dynamics brought by the  introduction of the statefinder.

\item \label{DE112}   Express the statefinder $\{r,s\}$ in terms of the total density, pressure and their time derivatives for a spatially flat Universe.

\item \label{DE113}   Show that for a flat Universe filled with a two-component liquid composed of non--relativistic matter (dark matter + baryons) and dark energy with relative density $\Omega _{DE}  = \rho _{DE} /\rho _{cr} $ the statefinder takes the form
$$
r = 1 + {\frac92}\Omega _{DE} w(1 + w) - {\frac32}\Omega _{DE}
{\frac{\dot w}{H}};
$$
$$
s = 1 + w - {\frac13}{\frac{\dot w}{wH}};\quad w \equiv
{\frac{p_{DE} } {\rho _{DE} }}.
$$

\item \label{DE114}   Express the statefinder in terms of Hubble parameter $H(z)$ and its derivatives.

\item \label{DE115}   Find the statefinders
\begin{description}

\item[a)] for dark energy in the form of cosmological constant;

\item[b)] for the case of time--independent state equation parameter $w$;

\item[c)] for dark energy in the form of quintessence.
\end{description}

\item \label{DE116}   Express the photometric distance $d_L(z)$ through the current values of parameters $q$ and $s$.

\subsection{Crossing the Phantom Divide}
{\it In the quintessence model of dark energy $-1<w<-1/3$.  In the phantom model with negative kinetic energy $w<-1$. Recent cosmological data seem to indicate that there occurred the crossing of the phantom divide line in the near past. This means that equation of state parameter $w_{DE}$ crosses the phantom divide line $w_{DE}=-1$. This crossing to the phantom region is possible neither for an ordinary minimally coupled scalar field nor for a phantom field. There are at least three ways to solve this problem. If dark energy behaves as quintessence at early stage, and evolves as phantom at the later stage, a natural suggestion would be to consider a 2-field model (quintom model): a quintessence and a phantom. The next possibility, discussed in the next Chapter, is to consider an interacting model, in which dark energy interacts with dark matter. Yet another possibility would be that General Relativity fails at cosmological scales. In this case quintessence or phantom energy can cross the phantom divide line in a modified gravity theory. We investigate this approach in Chapter 12.}

\item \label{DE117}   Show that at the point of  transition between the quintessence and the phantom phases $\dot H$ vanishes.

\item \label{DE117_1}   Show that the sound speed of a single perfect barotropic fluid is diverges when $w$ crosses the phantom divide line.

\item \label{DE117_11}   Find a dynamical law for the equation of state parameter $w=p/\rho$ in the barotropic cosmic fluid [see N.Caplar, H.Stefancic, arXiv:1208.0449].

\item \label{DE117_12}  Using the results of previous problem, find the functions $w(z)$, $\rho(z)$ and $p(z)$ for the simplest possibility $c_S=const$.

\item \label{DE117_13}   Realize the procedure described in the problem \ref{DE117_11} for the case of a minimally coupled scalar field $\varphi$ with potential $V(\varphi)$ in a spatially flat Universe.

\item \label{DE117_2} Consider the case of Universe filled with non-relativistic matter and quintessence and show that the condition to cross the phantom divide line $w=-1$ is equivalent to sign change in the following expression
\[\frac{dH^2(z)}{dz}-3\Omega_{m0}H_0^2(1+z)^2.\]

\item \label{DE118_}   Consider a model with the scale factor of the form
	\[a=a_c\left(\frac{t}{t-t_s}\right)^n,\]
where $a_c$ is a constant, $n>0$, $t_s$ is the time of a Big Rip singularity. Show that on the interval $0<t<t_s$ there is crossing of the phantom divide line $w=-1$.

\item \label{DE119_} Show, that for the model considered in the previous problem the parameter $H(t)$ and density $\rho(t)$ achieve their minimal values at the phantom divide point. (see K.Bamba, S.Capozziello, S.Nojiri, S.Odintsov,arXiv:1205.3421)

\item \label{DE120}   Find condition of intersection with the line $w=-1$ for the quintom Lagrangian
\[L=\frac12g^{\mu\nu}\left(\frac{\partial\varphi}{\partial x^\mu}\frac{\partial\varphi}{\partial x^\nu} - \frac{\partial\psi}{\partial x^\mu}\frac{\partial\psi}{\partial x^\nu} \right)-W(\varphi,\psi).\]

\section{Lost and Found}

\item \label{DE_new_1_1} Consider the case of spatially flat Universe dominated by non-relativistic matter and spatially homogeneous scalar complex field $\Phi$ and obtain the equations to describe the dynamics of such a Universe.
    
\item \label{DE_new_2_1}   Consider the case of the Universe composed of non-relativistic matter and quintessence and relate the quantities $\varphi ,\,\rho _\varphi  ,\,H,\,V(\varphi )$ with the redshift-dependent state equation parameter $w(z)$.

\item \label{DE_new_3_1}   When solving the equation for the scalar field one assumes that the time dependence of the scalar factor, which is necessary to calculate the Hubble parameter in the equation, is determined by the dominant component. Such approximation becomes invalid at some time moment, because the energy density for the scalar field decays slower than that of matter or radiation. Determine the value of the scalar field at that time moment for the potential of the problem \ref{DE73}

\section{Single Scalar Cosmology}

The discovery of the Higgs particle has confirmed that scalar fields play a fundamental
role in subatomic physics. Therefore they must also have been present in the early Universe and played a part
in its development. About scalar fields on present cosmological scales nothing is known, but in view of the observational evidence for accelerated expansion it is quite well possible that they take part
in shaping our Universe now and in the future. In this section we consider the evolution of a flat, isotropic and homogeneous Universe in the presence of a single cosmic
scalar field. Neglecting ordinary matter and radiation, the evolution of such a Universe is described by two
degrees of freedom, the homogeneous scalar field $\varphi(t)$ and the scale factor of the Universe $a(t)$. The
relevant evolution equations are the Friedmann and Klein-Gordon equations,
reading (in the units in which $c = \hbar = 8 \pi G = 1$)
\[
\frac{1}{2}\, \dot{\varphi}^2 + V = 3 H^2, \quad \ddot{\varphi} + 3 H \dot{\varphi} + V' = 0,
\]
where $V[\varphi]$ is the potential of the scalar fields, and $H = \dot{a}/a$ is the Hubble parameter.
Furthermore, an overdot denotes a derivative w.r.t.\ time, whilst a prime denotes a derivative w.r.t.\ the
scalar field $\varphi$.

\item\label{SSC_0}
Show that the Hubble parameter cannot increase with time in the single scalar cosmology.
\item\label{SSC_1}
Obtain first-order differential equation for the Hubble parameter $H$ as function of $\varphi$ and find its stationary points.
\item\label{SSC_2}
Consider eternally oscillating scalar field of the form $\varphi(t) = \varphi_0 \cos \omega t$ and analyze stationary points in such a model.
\item\label{SSC_3}
Obtain explicit solution for the Hubble parameter in the model considered in the previous problem.
\item\label{SSC_4}
Obtain explicit time dependence for the scale factor in the model of problem \ref{SSC_2}.
\item\label{SSC_5}
Reconstruct the scalar field potential $V(\varphi)$ needed to generate the model of problem \ref{SSC_2}.
\item\label{SSC_6_00}
Describe possible final states for the Universe governed by
a single scalar field at large times.
\item\label{SSC_6_0}
Formulate conditions for existence of end points of evolution in terms of the potential $V(\varphi)$.
\item\label{SSC_6_1}
Consider a single scalar cosmology described by the quadratic potential
\[
V = v_0 + \frac{m^2}{2}\, \varphi^2.
\] Describe all possible stationary points and final states of the Universe in this model.
\item\label{SSC_7}
Obtain actual solutions for the model of previous problem using the power series expansion
\[
H[\varphi] = h_0 + h_1 \varphi + h_2 \varphi^2 + h_3 \varphi^3 + ...
\] Consider the cases of $v_0 > 0$ and $v_0 < 0$.
\item\label{SSC_8}
Estimate main contribution to total expansion factor of the Universe.
\item\label{SSC_9_0}
Explain difference between end points and turning points of the scalar field evolution.
\item\label{SSC_9}
Show that the exponentially decaying scalar field
\[
\varphi(t) = \varphi_0 e^{-\omega t}
\]
can give rise to unstable end points of the evolution.
\item\label{SSC_10}
Analyze all possible final states in the model of previous problem.
\item\label{SSC_11}
Express initial energy density of the model of problem \ref{SSC_9} in terms of the $e$-folding number $N$.
\item\label{SSC_12}
Estimate mass of the particles corresponding to the exponential scalar field considered in problem \ref{SSC_9}.
\item\label{SSC_13}
Calculate the deceleration parameter for flat Universe filled with the scalar field in form of quintessence.
\item\label{SSC_14_}
When considering dynamics of scalar field $\varphi$ in flat Universe, let us define a function $f(\varphi)$ so that $\dot\varphi=\sqrt{f(\varphi)}$. Obtain the equation describing evolution of the function $f(\varphi)$. (T. Harko, F. Lobo  and M. K. Mak, Arbitrary scalar field and quintessence cosmological models, arXiv: 1310.7167)

\subsection{Exact Solutions for the Single Scalar Cosmology}

after Harko (arXiv:1310.7167v4)

\item\label{ES_0}
Rewrite the equations of the single scalar cosmology
\begin{equation}
3H^{2} =\rho _{\phi }=\frac{\dot{\phi}^{2}}{2}+V\left( \phi \right) ,
\label{H}
\end{equation}
\begin{equation}
2\dot{H}+3H^{2}=-p_{\phi }=-\frac{\dot{\phi}^{2}}{2}+V\left( \phi \right),
\label{H1}
\end{equation}
\begin{equation}
\ddot{\phi}+3H\dot{\phi}+V^{\prime }\left( \phi \right) = 0,  \label{phi}
\end{equation}
in terms of the parameter $G(\phi)$ introduced as
\[\dot\phi^2=2V(\phi)\sinh^2 G(\phi).\]
\item\label{ES_1}
Obtain equation to determine the parameter $G$ as function of time.
\item\label{ES_2}
Obtain equation to determine the parameter $G$ as function of scale factor.
\item\label{ES_3}
Obtain the deceleration parameter $q$ in terms of the parameter $G$.
\item\label{ES_4}
Obtain solution of the equation 
\begin{equation}
\frac{dG}{d\phi }+\frac{1}{2V}\frac{dV}{d\phi }\coth G+\sqrt{\frac{3}{2}}=0.
\label{fin}
\end{equation} with
\begin{equation}
V=V_{0}\exp \left( \sqrt{6}\alpha _{0}\phi \right).  \label{pp}
\end{equation}
in the case $\alpha _0\neq \pm 1$.
\item\label{ES_5}
Obtain explicit solution of the problem \ref{ES_4} in the case $\alpha _{0}=\pm \sqrt{2}$.
\item\label{ES_6}
Obtain explicit solution of the problem \ref{ES_4}
in the case $\alpha _{0}=\pm \sqrt{3/2}$.
\item\label{ES_7}
Obtain explicit solution of the problem \ref{ES_4}
in the case $\alpha _{0}=\pm 2/\sqrt{3}$.
\item\label{ES_8}
Obtain a particular solution of the problem \ref{ES_4}
in the case $G(\phi)=G_{0}=\mathrm{constant}$.
\item\label{ES_9} Obtain solution of the problem \ref{ES_4}
in the case $\alpha _0= \pm 1$.

\item\label{ES_10}
Obtain solution of the equation \begin{equation}
\frac{dG}{d\phi }+\frac{1}{2V}\frac{dV}{d\phi }\coth G+\sqrt{\frac{3}{2}}=0.
\label{fin}
\end{equation} with
\begin{equation}
\frac{1}{2V}\frac{dV}{d\phi }=\sqrt{\frac{3}{2}}\;\alpha _{1}\,\tanh G,
\end{equation}
where $\alpha _{1}$ is an arbitrary constant.

\item\label{ES_11}
Obtain solution of the equation (\ref{fin}) for the case
\begin{equation}
G=\mathrm{arccoth}\left( \sqrt{\frac{3}{2}}\frac{\phi }{\alpha _{2}}\right)
,\qquad \alpha _{2}=\mathrm{constant}.
\end{equation}

\item\label{ES_12}
Rewrite the equation (\ref{fin}) in form of the two linear differential equations for the variable $w=e^{-G}$.

\item\label{ES_13}
Obtain a consistency integral relation between the
separation function $M(\phi )$ and the self-interaction potential $V(\phi )$, corresponding to the equations for the variable $w$, obtained in the previous problem.

\item\label{ES_14}
Obtain exact solution of the equation (\ref{fin}) in the case $M\left( \phi \right) =\sqrt{V}$.

\item\label{ES_15}
Obtain exact solution of the equation (\ref{fin}) in the case $M\left( \phi \right) =V^{-3/2}$.

\section{Bianchi I Model}

(after Esra Russell, Can Battal Kilinc, Oktay K. Pashaev, Bianchi I Models: An Alternative Way To Model The Present-day Universe, arXiv:1312.3502)

Theoretical arguments and indications from recent observational data support the existence of an anisotropic phase that approaches an isotropic one. Therefore, it makes sense to consider models of a Universe with an initially anisotropic background. The anisotropic and homogeneous Bianchi models may provide adequate description of anisotropic phase in history of Universe. One particular type of such models is Bianchi type I (BI) homogeneous models whose spatial sections are flat, but the expansion rates are direction dependent,
\[ds^2={c^2}dt^2-a^{2}_{1}(t)dx^2-a^{2}_{2}(t)dy^2-a^{2}_{3}(t)dz^2\]
where $a_{1}$, $a_{2}$ and $a_{3}$ represent three different scale factors which are a function of time $t$.

\item\label{bianchi_01}
Find the field equations of the BI Universe.

\item\label{bi_2}
Reformulate the field equations of the BI Universe in terms of the directional Hubble parameters.
\[H_1\equiv\frac{\dot{a_1}}{a_1},\ H_2\equiv\frac{\dot{a_2}}{a_2},\ H_3\equiv\frac{\dot{a_3}}{a_3}.\]

\item\label{}
The BI Universe has a flat metric, which implies that its total density is equal to the critical density. Find the  critical density.

\item Obtain an analogue of the conservation equation $\dot\rho+3H(\rho+p)=0$ for the case of the BI Universe.

\item Obtain the evolution equation for the mean of the three directional Hubble parameters $\bar H$.

\item\label{}
Show that the system of equations for the BI Universe
\begin{align}
\nonumber
H_1H_2+H_1H_3+H_2H_3 & =\rho,\\
\nonumber
\dot H_1+ H_1^2 +\dot H_3+ H_3^2 +H_1H_3& =-p,\\
\nonumber
\dot H_1+ H_1^2 +\dot H_2+ H_2^2 +H_1H_2& =-p,\\
\nonumber
\dot H_2+ H_2^2 +\dot H_3+ H_3^2 +H_2H_3& =-p,
\end{align}
can be transformed to the following
\begin{align}
\nonumber
H_1H_2+H_1H_3+H_2H_3 & =\rho,\\
\nonumber
\dot H_1+ 3H_1\bar H & =\frac12(\rho-p),\\
\nonumber
\dot H_2+ 3H_2\bar H & =\frac12(\rho-p),\\
\nonumber
\dot H_3+ 3H_3\bar H & =\frac12(\rho-p).
\end{align}

\item Show that the mean of the three directional Hubble parameters $\bar H$ is related to the elementary volume of the BI Universe $V\equiv a_1a_2a_3$ as \[\bar H=\frac13\frac{\dot V}{V}.\]

\item\label{}
Obtain the volume evolution equation of the BI model.

\item\label{}
Find the generic solution of the directional Hubble parameters.

\item\label{}
Find the energy density of the radiation dominated BI Universe in terms of volume element $V_r$.

\item\label{}
Find the mean Hubble parameter of the radiation dominated case.

\item\label{}
Find the directional expansion rates of the radiation dominated model.

\item\label{}
Find time dependence for the scale factors $a_i$ in the radiation dominated BI Universe.

\item\label{bianchi_02}
Find the partial energy densities for the two components of the BI Universe dominated by radiation and matter in terms of volume element $V_{rm}$.

\item\label{bianchi_03}
Obtain time evolution equation for the total volume $V_{rm}$ in the BI Universe dominated by radiation and matter.

\item\label{bianchi_04}
Using result of the previous problem, obtain a relation between the mean Hubble parameter and the volume element.

\section{Hybrid Expansion Law}

In problems \ref{SSC_18} - \ref{SSC_19_0} we follow the paper of Ozgur Akarsu,  Suresh Kumar, R. Myrzakulov, M. Sami,  and Lixin Xu4, Cosmology with hybrid expansion law: scalar  field reconstruction of cosmic history and observational constraints (arXiv:1307.4911) to study expansion history of  Universe, using the hybrid expansion law---a product of power-law and exponential type of functions
\[a(t)=a_0\left(\frac{t}{t_0}\right)^\alpha\exp\left[\beta\left(\frac{t}{t_0}-1\right)\right],\]
where $\alpha$ and $\beta$ are non-negative constants. Further $a_0$ and $t_0$ respectively denote the scale factor and age of the Universe today.

\item\label{SSC_18}
Calculate Hubble parameter, deceleration parameter and jerk parameter for hybrid expansion law.

\item\label{SSC_18_2}
For hybrid expansion law find $a, H, q$ and $j$ in the cases of very early Universe $(t\to0)$ and for the late times $(t\to\infty)$.

\item\label{SSC_18_3}
In general relativity, one can always introduce an effective source that gives rise to a given expansion law. Using the ansatz of hybrid expansion law obtain the EoS parameter of the effective fluid.

\item\label{SSC_19}
We can always construct a scalar field Lagrangian which can mimic a given cosmic history. Consequently,  we can consider the quintessence realization of the hybrid expansion law. Find time dependence for the the quintessence field $\varphi(t)$ and potential $V(t)$, realizing the hybrid expansion law. Obtain the dependence $V(\varphi)$.

\item\label{SSC_19_1}
Quintessence paradigm relies on the potential energy of scalar fields to drive the late time acceleration of the Universe. On the other hand, it is also possible to relate the late time acceleration of the Universe with the kinetic term of the scalar field by relaxing its canonical kinetic term. In particular this idea can be realized with the help of so-called tachyon fields, for which
	\[\rho=\frac{V(\varphi)}{\sqrt{1-\dot\varphi^2}},\quad p=-V(\varphi)\sqrt{1-\dot\varphi^2}.\]
Find time dependence of the tachyon field $\varphi(t)$ and potential $V(t)$, realizing the hybrid expansion law. Construct the potential $V(\varphi)$.

\item\label{SSC_19_2}
Calculate Hubble parameter and deceleration parameter for the case of phantom field in which the energy density and pressure are respectively given by
\[\rho =-\frac{1}{2}\dot{\varphi}^2+V(\varphi),\quad p =-\frac{1}{2}\dot{\varphi}^2-V(\varphi).\]

\item\label{SSC_19_0}
Solve the problem \ref{SSC_19} for the case of phantom field.

\item\label{SSC_19_12}
Find EoS parameter for the case of phantom field.

\section{The Power-Law Cosmology}

\item\label{PWL_1}
Show that for power law $a(t)\propto t^n$ expansion  slow-roll inflation occurs when $n\gg1$.

\item\label{PWL_2}
Show that in the power-law cosmology the scale factor evolution $a\propto\eta^q$ in conformal time transforms into $a\propto t^p$ in physical (cosmic) time with \[p=\frac{q}{1+q}.\]

\item\label{PWL_3}
Show that if $a\propto\eta^q$ then the state parameter $w$ is related to the index $q$ by the following \[w=\frac{2-q}{3q}=const.\]

\end{enumerate}

\chapter{Dark Matter}
\begin{flushright}
\it Since the original suggestion of the\\
existence of dark matter,\\
the evidence  has become overwhelming.\\
The question has changed from \\
"Does dark matter exist?" to \\
"What is this most  common of substances?"\\
G.Jungman, M.Kamionkowski, K.Griest
\end{flushright}
{\it In the beginning of thirties of the last century a Swiss cosmologist F. Zwicky applied the virial theorem (in the gravitational field $2\langle E_{kin}\rangle+\langle E_{pot}\rangle=0$) in order to estimate the mass of the Coma cluster (Berenice's Hair). He was surprised to discover that in order to support the finite motion of the galaxies belonging to the cluster, its mass must be at least two orders of magnitude greater than the observed mass (in form of luminous galaxies). He was the first to introduce the term ''dark matter'' which strongly entered the vocabulary of modern cosmology. At present the term is understood as the non-baryon matter component which neither emits nor absorbs electromagnetic waves in any range.}
\section{Observational Evidence of the Dark matter Existence}
\begin{enumerate}

\item \label{DM01} {  Analyze the problem of a small satellite galaxy moving in the gravitational field of a large galaxy (the problem of non-decreasing behavior of the rotational curves.)}

\item\label{dm-p-1} {  What radial dependence of the density of a spherically symmetric galactic halo corresponds to the constant velocity of satellite galaxies?}

\item \label{dm-p-2}{  We have noted above that outside the optically observed range of a galactic disk the rotation curves of the dwarf satellite galaxies do not depend on the galacto-centric distance (the flatness of the rotational curves). External regions of the galaxies are mostly filled by cold neutral hydrogen. Evidently the rotation velocity of the gas gives important information about the mass distribution in the galaxy. How can one measure that velocity?}
\item \label{dm-p-3} {  Using the virial theorem, express the mass of a galaxy cluster through the observed quantities---the average velocity of galaxies in the cluster and its size. Estimate the mass of the Coma cluster (Berenice's hair) for $R\approx10^{23}m$, $\langle v^2\rangle^{1/2}\approx2\cdot10^6m/s$.}

\item \label{dm-p-4} {  Show that the age of the matter--dominated Universe contradicts observations.}

\item \label{dm-p-5} {  A galaxy has the visible mass of $10^{11}M_\odot$ and a horizontal rotation curve up to distance of $30kpc$ at velocity $250km/s$. What is the ration What is the dark to visible mass ratio in the galaxy?}

\end{enumerate}
\section{ANONIMOUS}
\begin{enumerate}[resume]

\item \label{dm-p-6} {  Find the lower limit for the mass of the dark matter particles in case they are:
\begin{description}

\item[a)] bosons;

\item[b)] fermions.
\end{description}}

\item \label{dm-p-7} {  Find the lower limit for the mass of fermion particles that constitute a compact spherical object of dark matter with radius $R$ and mass $M$.}

\item \label{dm-p-8} {  Show that particles that contitute a galaxy of mass $M=10^{10}M_\odot$ and radius $R=3kpc$ must be non-relativistic.}

\item \label{dm-p-9} {  In assumption that neutrinos have mass $m_\nu$ and temperature of decoupling  $T_D\simeq1MeV$, determine their contribution to the presently observed energy density.}

{\it If WIMPs represent dark matter, they will not just generate the background density of Universe, but they will also cluster together with the usual baryon matter in the galactic halo. In particular, they will be presented in our own galaxy --- the Milky Way. It gives hope to detect the relic WIMPs immediately on the Earth. What is these hope related to? By definition, WIMPs do not interact with photons. However, WIMPs could annihilate into usual matter (quarks and leptons) in the early Universe. Otherwise they would have too high relative abundance today. Due to the crossover symmetry, the temperature of annihilation of e.g. WIMPs into quarks is related to the amplitude of elastic scattering of WIMPs on quarks. Therefore WIMPs should interact, though weakly, with usual matter. Due to this interaction, WIMPs can elastically scatter on target nuclei and lead to a recoil of detector nuclei partially transferring energy to them. Therefore the search for the events of elastic scattering of WIMPs on detector nuclei is one of the prospective ways of galactic dark matter research.}

\item \label{dm-p-10} {  Show that preservation of thermal equilibrium for a certain component in the expanding Universe is possible only under the condition $\Gamma\ll H$, where $\Gamma$ is the rate of reaction needed to support the equilibrium.}

\item \label{dm-p-11} {  Show that at high temperatures $T\gg m_\chi$ ($m_\chi$ is the WIMPs mass) the ratio of equilibrium densities of WIMPs and photons is constant: $n_\chi^{eq}/n_\gamma^{eq}=const$.}

\item \label{dm-p-12} {  Show that the Boltzmann equation describing the evolution of WIMPs number density $n_\chi$ in particle number preserving interactions leads to the usual relation for non-relativistic matter $n_\chi\propto a^{-3}$.}
\item \label{dm-p-13} {  Show that in the early Universe for $T\ge m_\chi$, the WIMPs number density follows its equilibrium value at temperature decreasing.}
\item \label{DAMA_15} {  Show that decreasing of the annihilation cross-section leads to increasing of the relic density: contrary to Charles Darwin, the weakest wins.}
\item \label{dm-p-14} {  What temperature does neutrino's density "freezing" take place at?}

\item \label{dm-p-15} {  Considering the WIMPs as the thermal relics of the early Universe, estimate their current density.}

\section{Dark Matter Halo}

\item \label{dm-p-16} {  Estimate the local density of the dark halo in the vicinity of the Earth, assuming that its density decreases as $\rho_g=C/r^2$.}
\item {  \label{halo_model} Build the model of the spherically symmetric dark halo density
corresponding to the observed galactic rotation curves.}
\item \label{DM16} {  In frames of the halo model considered in the previous problem determine the local dark matter density $\rho_0$ basing on the given rotation velocities of satellite galaxies at the outer border of the halo $v_\infty\equiv v(r\rightarrow\infty)$ and in some point $r_0$.}
 \item \label{11_17} {  For the halo model considered in problem \ref{halo_model} obtain the dependencies $\rho(r)$ and $v(r)$ in terms of $\rho_0$ and $v_\infty$. Plot the dependencies $\rho(r)$ and $v(r)$.}
\item \label{dm-p-17} {  Many clusters are sources of X-ray radiation. It is emitted by the hot intergalactic gas filling the cluster volume. Assuming that the hot gas ($kT\approx10keV$) is in equilibrium in the cluster with linear size $R=2.5Mpc$ and core radius $r_c=0.25Mpc$, estimate the mass of the cluster.}

\section{Candidates for Dark Matter Particles}

\subsection{Standard Model Particles as Dark Matter Candidates}

\subsection{Supersymmetric Candidate Particles}

{\it Supersymmetry on weak scale ($\sim 100 {\rm GeV}$, it is the scale which lies in the center of the candidate particles search) represents the most motivated basis for new Particle Physics. It naturally provides the dark matter candidates with approximately correct relic density. This fact gives strongly fundamental and absolutely independent motivation for the supersymmetric theories. That is whay application of supersymmetry to cosmology and vice versa deserves the most attentive consideration. An indirect result of the dark matter research is to transfer the results of the supersymmetry theory into applicative domain where its predictions can be chacked in the nearest future.}

\item \label{dm-p-18} {  Why accelerator energies just exceeding the WIMP rest mass are insufficient to observe the WIMPs?}
\item \label{dm-p-19} {  What consequences follow from the $R$-parity conservation?}
\section{Dark Matter Detection}

\item \label{dm-p-20} {  Estimate the WIMPs' flow onto the Earth's surface.}
\item \label{dm-p-21} {  What are the main processes due to which the WIMPs can be detected?}
\item \label{11_20} {  Show that WIMPs of mass $\sim100GeV$ being elastically scattered on Xenon nuclei with mass $\sim130GeV$ lead to energy recoil $\le40keV$.}
\item \label{dm-p-22} {  Show that the WIMPs' elastic scattering experiments will be most efficient if the target nuclei's mass is comparable with the WIMPs' mass.}
\item \label{DM25}{  Determine the minimum velocity of WIMPs that can transmit energy $Q$ to a nucleus with mass $m_N$.}

\item \label{dm-p-23} {  Estimate the count rate for the detector registering elastic events of WIMPs.}
\item \label{dm-p-24} {  Obtain the expression for the count rate of the detector registering elastic events of WIMPs.}
\item \label{dm-p-25} {  Reconstruct the one--dimensional WIMPs' velocities distribution $f_1(v)$  (see the previous problem) basing on the given count rate of the elastic events.}

\item \label{dm-p-26} {  Construct a model-independent scheme for WIMPs' mass determination using the results of WIMPs' elastic scattering events detection for two or more sets of experimental data with detectors of different composition.}
\item \label{DM27} {  Show that if a WIMP has the mass of the order $100GeV$ and velocity of the order of $300km/s$, then it coherently interacts with nucleons of a detector nuclei.}
\item \label{dm-p-27} {  Find the total WIMP--nucleus cross--section determining the elastic events' count rate.}
\item \label{dm-p-28} {  How will the number of counts for the elastic events detector be affected by transition to heavier target nuclei at fixed detector mass?}
\item \label{dm-p-29} {  Show that if the WIMPs' mass is of the order of $100GeV$ then the elastic spin-independent cross-section for WIMPs on a nucleus with $A\sim100$ is eight orders of magnitude larger then the corresponding cross-section on a nucleon}
\item \label{dm-p-30} {  Estimate the annual modulations' amplitude of the WIMPs elastic scattering cross-section.}

\item \label{dm-p-31} {  Estimate the diurnal modulations' amplitude of the WIMPs elastic scattering cross-section.}

\section{The Dark Matter in the Solar System}
\item \label{dm-p-32} {  Estimate upper bound of the relative perturbation of the Earth orbit due to presence of the dark matter in the Solar system. Assume that the upper limit on the dark matter in the Solar system lies on the level $\rho_{DM}\approx1.5\cdot10^{-12}\,g/cm^3$ and it remains constant at the distance discussed.}
\item \label{dm-p-33} {  Find the velocity change of a spaceship rotating around the Earth with period $T$ as the result of dark matter particles scattering on the nuclei of particles that compose the spaceship.}

\item \label{dm-p-34} {  Assume that due to interaction with dark matter particles the spaceship's velocity changed by  $\Delta v$ in one period. Given its mass is $m$ and it moves around the Earth on a circular orbit, estimate the dark matter density in the Earth's neighborhood.}

\section{The Dark Stars}

\item \label{dm-p-35} {  Estimate how much will the period of rotation of the Earth around the Sun change in one year due to gravitational capture of dark matter particles.}

\item \label{dm-p-36} {  Estimate the rate of energy outcome due to the WIMP-annihilation process using the parameter values $m_{WIMP}=100GeV$ and $\langle \sigma v \rangle_{ann}=3\cdot 10^{-26} cm^3/sec$ for the annihilation cross section.}
\item \label{dm-p-37} {  It is theorized that the dark matter particles' annihilation processes could be a competitive energy source in the first stars. Why did those processes play an important role only in the early Universe and why are they not important nowadays?}

\end{enumerate}

\chapter{Interaction in the dark sector}
{\it The evolution of any broadly applied model is accompanied by multiple generalizations that aim to resolve conceptual problems, as well as to explain the ever-increasing array of observations. In the case of Standard Cosmological Model one of the most promising directions of generalization is the replacement of the cosmological constant with a more complicated, dynamic form of dark energy, as well as the inclusion of interaction between the dark components. Typically, dark energy (DE) models are based on the scalar fields minimally coupled to gravity, and do not implement the explicit coupling of the field to the background dark matter (DM). However there is no fundamental reason for this assumption in the absence of an underlying symmetry which would suppress the coupling. Given that we do not know the true nature of either DE or DM, one cannot exclude the possibility that there exists a coupling between them. Whereas new forces between DE and normal matter particles are heavily constrained by observations (e.g. in the solar system and gravitational experiments on Earth), this is not the case for DM particles. In other words, it is possible that the dark components interact with each other, while not being coupled to standard model particles. Therefore, the possibility of DE-DM interaction must be looked at with the utmost seriousness.}
\section{Physical mechanism of energy exchange}
\begin{enumerate}
\item \label{IDE_1}  Models with interaction between DM and the DE field can be realized if we make just an obvious assumption: the mass of the cold DM particles is a function of the DE field. Let the dark matter particles will be collisionless and nonrelativisic. Hence, the pressure of this fluid and its energy density are \(p_{dm}=0\) and \(\rho_{dm}=nm\) respectively, where $m$ is the rest mass and $n$ is the number density of DM particles. We define $m=\lambda\varphi$ where $\varphi$ is a scalar field and $\lambda$ is a dimensionless constant. Show how such assumption affects the scalar field dynamics (after \cite{0307350}).

\item \label{IDE_2}  Show that the DM on a inhomogeneous vacuum background can be treated as as interacting DE and DM. (after \cite{1209.0563}).

\item \label{IDE_3} Obtain general equations of motion for DE interacting with DM   \cite{1207.0250}.

\end{enumerate}

\section{Phenomenology of interacting models}
\begin{enumerate}[resume]
\item \label{IDE_4}  Find the effective state  parameters $w_{(de)eff}$ and $w_{(dm)eff}$ that would allow one to treat the interacting dark components as non-interacting.

\item \label{IDE_5} Using the effective state parameters obtained in the previous problem, analyze dynamics of dark matter and dark energy depending on sign of the rate of energy density exchange in  the dark sector.

\item \label{IDE_6}  Find the effective state  parameters $w_{(de)eff}$ and $w_{(dm)eff}$ for the case of the warm dark matter ($w_{dm}\ne0$) and analyze the features of dynamics in this case.

\item \label{IDE_7}  Show that the quintessence coupled to DM with certain sign of the coupling constant behaves like a phantom uncoupled model, but without negative kinetic energy.

\item \label{IDE_8}  In order to compare dynamics of a model with observational results it is useful to analyze all dynamic variables as functions of redshift rather than time. Obtain the corresponding transformation for the system of interacting dark components.

\item \label{IDE_9_0}  (after  \cite{0502034}) Show that energy exchange between dark components leads to time-dependent effective potential energy term in the first Friedman equation.

\item \label{IDE_9}  Show that the system of interacting components can be treated as the uncoupled one due to introduction of the partial effective pressure of the dark components
     \[\Pi_{de}\equiv\frac{Q}{3H},\quad \Pi_{dm}\equiv-\frac{Q}{3H}.\]

\item \label{IDE_10}  Assume that the mass $m_{dm}$ of dark matter particles depends on a scalar field $\varphi$. Construct the model of interacting dark energy and dark  matter in this case.

\item \label{IDE_12n}  Assume that the mass $m_{dm}$ of DM particles depends exponentially on the DE scalar field $m=m_*e^{-\lambda\varphi}$. Find the interaction term $Q$ in this case.

\item \label{IDE_11}  Find the equation of motion for the scalar field interacting with dark matter if its particles' mass depends on the scalar field.

\item \label{IDE_12}  Make the transformation from the variables $(\rho_{de}, \rho_{dm})$ to
\[\left(r=\frac{\rho_{dm}}{\rho_{de}}, \rho = \rho_{dm} + \rho_{de}\right)\] for the system of interacting dark components.

\item \label{IDE_13} Generalize the result of previous problem to the case of warm dark matter.

\item \label{IDE_16n} Calculate the derivatives $dr/dt$ and $dr/dH$ for the case of flat universe with the interaction $Q$.

\item \label{IDE_17n} It was shown in the previous problem that
\[\dot r=r\left(\frac{\dot\rho_{dm}}{\rho_{dm}}-\frac{\dot\rho_{de}}{\rho_{de}}\right) = 3Hr \left(w_{de} +\frac{1+r}{\rho_{dm}}\frac Q{3H}\right)=(1+r)\left[3Hw_{de}\frac{r}{1+r}+\Gamma\right],\quad \Gamma\equiv\frac Q {\rho_{de}}.\] Exclude the interaction $Q$ and reformulate the equation in terms of $\rho_{de}$, $H$ and its derivatives.

\item \label{IDE_18n}  Generalize the result, obtained in the previous problem, for the case of non-flat Universe \cite{0606555}.

\item \label{IDE_14}  Show that critical points in the system of equations obtained in problem \ref{IDE_12} exist only for the case of dark energy of the phantom type.

\item \label{IDE_15} Show that the result of previous problem holds also for warm dark matter.

\item \label{IDE_16}  Show that existence of critical points in the system of equations obtained in problem \ref{IDE_12} requires a transfer from dark energy to dark matter.

\item \label{IDE_17}  Show that the result of previous problem holds also for warm dark matter.

\item \label{IDE_18}  Assume that the ratio of the interacting dark components equals \[r\equiv\frac{\rho_{dm}}{\rho_{de}}\propto a^{-\xi}, \quad \xi\ge0.\] Analyze how the interaction $Q$ depends on $\xi$.

\item \label{IDE_19}  Show that the choice
\[r\equiv\frac{\rho_{dm}}{\rho_{de}}\propto a^{-\xi}, \quad (\xi\ge0)\] guarantees existence of an early matter-dominated epoch.

\item \label{IDE_20}  Find the interaction $Q$ for the Universe with interacting dark energy and dark matter, assuming that ratio of their densities takes the form
    \[r\equiv\frac{\rho_{dm}}{\rho_{de}}=f(a),\] where $f(a)$ is an arbitrary differentiable function of the scale factor.

\item \label{IDE_26n}  Let \[Q=\frac{\dot f(t)}{f(t}\rho_{dm}.\] Show that the sign of the deceleration parameter is defined by the ratio \[\frac{\dot f}{fH}.\]

\item \label{IDE_27n}  Show that in the model, considered in the previous problem, the transition from the accelerated expansion to the decelerated one can occur only due to time dependence of the interaction.

\end{enumerate}
\section{Simple linear models}
\begin{enumerate}[resume]
\item \label{IDE_21} Find the scale factor dependence for the dark matter density assuming that the interaction between the dark matter and the dark energy equals $Q=\delta(a) H\rho_{dm}$.

\item \label{IDE_22}  Obtain the equation for the evolution of the DE  energy density for   $Q=\delta(a) H\rho_{dm}$.

\item \label{IDE_23}  Find $\rho_{dm}$ and $\rho_{de}$ in the case  $Q=\delta H\rho_{dm}$, $\delta=const$, $w_{de}=const$.

\item \label{IDE_24}  As was shown above, interaction between dark matter and dark energy leads to non--conservation of matter, or equivalently, to scale dependence for the mass of particles that constitute the dark matter. Show that, within the framework of the model of previous problem ($Q=\delta H\rho_{dm}$, $\delta=const$, $w_{de}=const$) the relative change of particles mass per Hubble time equals to the interaction constant.

\item \label{IDE_25}  Find $\rho_{dm}$ and $\rho_{de}$ in the case  $Q=\delta H\rho_{de}$, $\delta=const$, $w_{de}=const$.

\item \label{IDE_26}  Find $\rho_{dm}$ and $\rho_{de}$ in the case  $Q=\delta(a) H\rho_{de}$, $\delta(a)=\beta_0a^\xi$, $w_{de}=const$.
    (after  \cite{1111.6743}.)

\item \label{IDE_27} Let's  look at a more general linear model for the expansion of a Universe that contains two interacting fluids with the equations of state
\[p_1 = (\gamma_1-1)\rho_1,\]	
\[p_2 = (\gamma_2-1)\rho_2,\]	
and energy exchange
	\[\dot\rho_1+3H\gamma_1\rho_1 = -\beta H\rho_1 + \alpha H\rho_2,\]
	\[\dot\rho_2+3H\gamma_2\rho_2 = \beta H\rho_1 - \alpha H\rho_2.\]
Here $\alpha$ and $\beta$ are constants describing the energy exchanges between the two fluids. Obtain the equation for $H(t)$ and find its solutions (After \cite{9702029,0604063}).

\item \label{IDE_28}  Show that the energy balance equations (modified conservation equations) for $Q\propto H$ do not depend on H.

\item \label{IDE_36n}  The Hubble parameter is present in the first Friedmann equation quadratically. This gives rise to a useful symmetry within a class of FLRW models. Because of this quadratic dependence, Friedmann's equation remains invariant under a transformation $H\to-H$ for the spatially flat case. This means it describes both expanding and contracting solutions. The transformation $H\to-H$ can be seen as a consequence of the change $a\to1/a$  of the scale factor of the FLRW metric. If, instead of just the first Friedmann equation, we want to make the whole system of Universe-describing equations invariant relative to this transformation, we must expand the set of values that undergo symmetry transformations. Then, when we refer to a duality transformation, we have in mind the following set of transformations
\[H\to\bar H=-H,\quad \rho\to\bar\rho=\rho,\quad p\to\bar p=-2\rho-p,\quad \gamma\equiv\frac{\rho+p}{\rho}\to\bar\gamma\equiv\frac{\bar\rho+\bar p}{\bar\rho}=-\gamma.\]

Generalize the duality transformation to the case of interacting components.(after \cite{0505096}.)

\end{enumerate}

\section{Cosmological models with a change of the direction of energy transfer}

{\it Let us consider one more type of interaction $Q$, whose sign (i.e., the direction of energy transfer) changes when the mode of decelerated expansion is replaced by the mode of accelerated expansion, and vice versa. The simplest interaction of this type is the one proportional to the deceleration parameter. An example of such interaction is
\[Q=q(\alpha\dot\rho+\beta H\rho),\]
where $\alpha$ and $\beta$ are dimensionless constants, and $\rho$ can be any of densities $\rho_{de}$, $\rho_{dm}$ or $\rho_{tot}$. In the following problems for simplicity we restrict ourselves to the decaying $\Lambda$ model, for which $\dot\rho_{de}=\dot\rho_\Lambda=-Q$ and $p_{de}=-\rho_{de}$. (after  \cite{1010.1074})}
\begin{enumerate}[resume]
\item \label{IDE_29}  Construct general procedure to the Hubble parameter and the deceleratio parameter for the case \(Q=q(\alpha\dot\rho_{dm}+\beta H\rho_{dm})\).

\item \label{IDE_30} Find deceleration parameter for the case $\alpha=0$ in the model considered in the previous problem.

\item \label{IDE_31}  Consider the model $Q=q(\alpha\dot\rho_\Lambda+3\beta H\rho_\Lambda)$. Obtain the Hubble parameter $H(a)$ and deceleration parameter for the case $\alpha=0$.

\end{enumerate}

\section{Non-linear interaction in the dark sector}
{\it The interaction studied so far are linear in the sense that the interaction term in the individual energy balance equations is proportional either to dark matter density or to dark energy density or to a linear combination of both densities. Also from a physical point of view an interaction proportional to the product of dark components seems preferred: an interaction between two components should depend on the product of the abundances of the individual components, as, e.g., in chemical reactions. Moreover, such type of interaction looks more preferable then the linear one when compared with the observations. Below we investigate the dynamics for a simple two-component model with a number of non-linear interactions.}

In problems 32-34 let a spatially flat FLRW Universe contain perfect fluids with densities $\rho_1$ and $\rho_2$. Consider a nonlinear interaction of the form $Q=\gamma\rho_1\rho_2$. (after \cite{1009.4942}).

\begin{enumerate}[resume]
\item \label{IDE_32}  Consider a model in which both fluids are dust. Find $r(t)\equiv\rho_1(t)/\rho_2(t)$.

\item \label{IDE_33}  Consider a Universe with more than two CDM components interacting with each other. What is the asymptotic behavior of the individual densities of the components in the limit $t\to\infty$?

\item \label{IDE_34}  Consider a Universe containing a cold dark matter  and a dark energy, in which the dark energy behaves like a cosmological constant.  Show that in such model dark energy is a perpetual component of the Universe.

\item \label{IDE_35}  Consider a two-component Universe with the interaction $Q=\gamma\rho_1\rho_2$. Let one component is CDM ($\rho_1=\rho_{dm}$, $w_1=0$), and the second is the dark energy with arbitrary state equation ($\rho_2=\rho_{de}$, $w_2=\gamma_{de}-1$). (The case considered in the previous problem corresponds to $\gamma_{de}=0$.) Find the relation between the dark energy and dark matter densities.

\item \label{IDE_36}  Let interaction term $Q$ be a non-linear function of the energy densities of the components and/or the total energy density. Motivated by the structure
\[\rho_{dm} = \frac r{1+r}\rho, \quad \rho_{de} = \frac r{1+r}\rho,\]
\[\rho\equiv\rho_{dm}+\rho_{de},\quad r\equiv\frac{\rho_{dm}}{\rho_{de}}\]
consider ansatz
	\[Q=3H\gamma\rho^m r^n(1+r)^s.\]
where $\gamma$ is a positive coupling constant. Show that for $s=-m$ interaction term is proportional to a power of products of the densities of the components. For $(m,n,s)=(1,1,-1)$ and $(m,n,s)=(1,0,-1)$ reproduce the linear case.(After   \cite{1112.5095})

\item \label{IDE_37}  Find analytical solution of non-linear interaction model covered by the ansatz of previous problem for $(m,n,s)=(1,1,-2)$, $Q=3H\gamma\rho_{de}\rho_{dm}/\rho$.

\item \label{IDE_38}  Find analytical solution of non-linear interaction model for $(m,n,s)=(1,2,-2)$, $Q=3H\gamma\rho_{dm}^2/\rho$.

\item \label{IDE_39}  Find analytical solution of non-linear interaction model for $(m,n,s)=(1,0,-2)$, $Q=3H\gamma\rho_{de}^2/\rho$.

\item \label{IDE_40}   Consider a flat Universe filled by CDM and DE with a polytropic equation of state
	\[p_{de}=K\rho_{de}^{1\frac1n}\]
where $K$ and $n$ are the polytropic constant and polytropic index, respectively. Find dependence of DE density on the scale factor under assumption that the interaction between the dark components is $Q=3\alpha H\rho_{de}$. (after \cite{1012.2692})

\item \label{IDE_41}  Show that under certain conditions the interacting polytropic dark energy with $Q=3\alpha H\rho_{de}$ behaves as the phantom energy.

\item \label{IDE_42} Find deceleration parameter for the system considered in the problem \ref{IDE_40}.
\end{enumerate}

\section{Phase space structure of models with interaction}
{\it The evolution of a Universe filled with interacting components can be effectively analyzed in terms of dynamical systems theory. Let us consider the following coupled differential equations for two variables
\begin{equation}
\label{IDE_s6_1}
\begin{array}{l}
\dot x=f(x,y,t),\\
\dot y=g(x,y,t).
\end{array}
\end{equation}	
We will be interested in the so-called autonomous systems, for which the functions $f$ and $g$ do not contain explicit time-dependent terms.
A point $(x_c,y_c)$ is said to be a fixed (a.k.a. critical) point of the autonomous system if
	\[f(x_c,y_c)=g(x_c,y_c)=0.\]
A critical point $(x_c,y_c)$ is called an attractor when it satisfies the condition \(\left(x(t),y(t)\right)\to(x_c,y_c)\) for $t\to\infty$.
Let's look at the behavior of the dynamical system (\ref{IDE_s6_1}) near the critical point. For this purpose, let us consider small perturbations around the critical point
	\[x=x_c+\delta x,\quad y=y_c+\delta y.\]
Substituting it into (\ref{IDE_s6_1}) leads to the first-order differential equations:
	\[\frac{d}{dN}\left(\begin{array}{c}\delta x\\ \delta y\end{array}\right) = \hat M \left(\begin{array}{c}\delta x\\ \delta y\end{array}\right).\]
Taking into account the specifics of the problem that we are solving, we made the change \[\frac{d}{dt}\to\frac{d}{dN},\]
where $N=\ln a$. The  matrix $\hat M$ is given by
\[\hat M =
\left(
\begin{array}{lr}
\frac{\partial f}{\partial x} & \frac{\partial f}{\partial y} \\
{} & {}\\
\frac{\partial g}{\partial x} & \frac{\partial g}{\partial y}
\end{array}
\right)
\]
The general solution for the linear perturbations reads
	\[\delta x=C_1e^{\lambda_1 N} + C_2e^{\lambda_2 N},\]
	\[\delta y=C_3e^{\lambda_1 N} + C_4e^{\lambda_2 N},\]
The stability around the fixed points depends on the nature of the eigenvalues.

Let us treat the interacting dark components as a dynamical system described by the equations
\[\rho'_{de}+3(1+w_{de})\rho_{de}=-Q\]
\[\rho'_{dm}+3(1+w_{dm})\rho_{dm}=Q\]
Here, the prime denotes the derivative with respect to $N=\ln a$. Note that although the interaction can significantly change the cosmological evolution, the system is still autonomous. We consider the following specific interaction forms, which were already analyzed above:
\[Q_1=3\gamma_{dm}\rho_{dm},\quad Q_1=3\gamma_{de}\rho_{de},\quad Q_1=3\gamma_{tot}\rho_{tot}\]}

\begin{enumerate}[resume]

\item \label{IDE_43} Find effective EoS parameters $w_{(dm)eff}$ and $w_{(de)eff}$ for the interactions $Q_1$, $Q_2$ and $Q_3$.

\item \label{IDE_52n} Find the critical points of equation for ratio $r=\rho_{dm}/\rho_{de}$ if $Q=3\alpha H(\rho_{dm}+\rho_{de})$, where the phenomenological parameter $\alpha$ is a dimensionless, positive constant, $w_{dm}=0$, $w_{de}=const$.

\item \label{IDE_53n} Show, that the remarkable property of the model, considered in the previous problem,
is that for the interaction parameter $\alpha$, consistent with the current observations $\alpha<2.3\times10^{-3}$ the ratio $r$ tends to a stationary but unstable value at early times, $r_s^+$, and to a stationary and stable value, $r_s^-$ (an attractor), at late times. Consequently, as the Universe expands, $r(a)$ smoothly evolves from $r_s^+$ to the attractor solution $r_s^-$.

\item \label{IDE_44}
Transform the system of equations
	\[\rho'_{de}+3(1+w_{de})\rho_{de}=-Q,\]
	\[\rho'_{dm}+3(1+w_{dm})\rho_{dm}=Q,\]
into the one for the fractional density energies.(after \cite{1003.2788})

\item \label{IDE_45} Analyze the critical points of the autonomous system, obtained in the previous problem
\[\Omega'_{dm}=3f_j \Omega_{dm}\Omega_{de},\]
\[\Omega'_{de}=-3f_j \Omega_{dm}\Omega_{de},\]
by imposing the conditions $\Omega'_{dm}=\Omega'_{de}=0$ and $\Omega_{dm}+\Omega_{de}=1$ (flatness of Universe).

\item \label{IDE_46} Construct the stability matrix for the dynamical system considered in the problem \ref{IDE_44} and determine its eigenvalues.

\item \label{IDE_47} Using result of the previous problem, determine eigenvalues of the stability matrix for the following cases: i) $\Omega_{dm} = 1$, $\Omega_{de} = 0$, $f_j \ne 0$; ii) $\Omega_{dm} = 0$, $\Omega_{de} = 1$, $f_j \ne 0$; iii) $f_j = 0$.

\item \label{IDE_48} Obtain position and type of the critical points obtained in the previous problem for the case of cosmological constant interacting with dark matter as $Q=3\gamma_{dm}\rho_{dm}$.

\item \label{IDE_49}
Construct the stability matrix for the following dynamical system
\begin{align}
\nonumber \rho' & = - \left(1+\frac{w_{de}}{1+r}\right)\rho,\\
\nonumber r' & = r \left[w_{de} - \frac{(1+r)^2}{r\rho}\Pi\right],
\end{align}
and determine its eigenvalues. (After \cite{1112.5095})
\end{enumerate}

\section{Peculiarity of  dynamics of scalar field coupled to dark matter}
\subsection{Interacting quintessence model}
{\it Given that the quintessence field and the dark matter have unknown physical natures, there seem to be no a priori reasons to exclude a coupling between the two components. Let us consider a two-component system (scalar field $\varphi$ + dark matter) with the energy density and pressure
\[\rho=\rho_\varphi+\rho_{dm},\quad p=p_\varphi+p_{dm}\]
(we do not exclude the possibility of warm DM ($p_{dm}\ne0$).)
If some interaction exists between the scalar field and DM, then
\[\dot\rho_{dm}+3H(\rho_{dm}+p_{dm})=Q\]
\[\dot\rho_\varphi+3H(\rho_\varphi+p_\varphi)=-Q.\]
Using the effective pressures $\Pi_\varphi$ and $\Pi_{dm}$,
\[Q=-3H\Pi_{dm}=3H\Pi_\varphi\]
one can transit to the system
\begin{align}
\nonumber
\dot\rho_{dm}+3H(\rho_{dm}+p_{dm}+\Pi_{dm}) & =0,\\
\nonumber
\dot\rho_\varphi+3H(\rho_\varphi+p_\varphi+\Pi_\varphi) & =0.
\end{align}}

\begin{enumerate}[resume]
\item \label{IDE_60}  Obtain the modified Klein-Gordon equation for the scalar field interacting with the dark matter.

\item \label{IDE_61} Consider a quintessence scalar field $\varphi$ which couples to the dark matter via, e.g., a Yukawa-like interaction $f(\varphi/M_{Pl})\bar\psi\psi$, where $f$ is an arbitrary function of $\varphi$ and $\psi$ is a dark matter Dirac spinor. Obtain the modified Klein-Gordon equation for the scalar field interacting with the dark matter in such way. (after \cite{0510628v2})

\item \label{IDE_62} (The problems \ref{IDE_62}-\ref{IDE_66} are inspired by \cite{0105479})

    Show that the Friedman equation with interacting scalar field and dark matter allow existence of stationary solution for the ratio $r\equiv\rho_{dm}/\rho_\varphi$.

\item \label{IDE_63}  Find the form of interaction $Q$ which provides the stationary relation $r$ for interacting cold dark matter and quintessence in spatially flat Universe.

\item \label{IDE_64} For the interaction $Q$ which provides the stationary relation $r$ for interacting cold dark matter and quintessence in spatially flat Universe (see the previous problem), find the dependence of $\rho_{dm}$ and $\rho_\varphi$ on the scale factor.

\item \label{IDE_65}  Show that in the case of interaction $Q$ obtained in the problem \ref{IDE_63}, the scalar field $\varphi$ evolves logarithmically with time.

\item \label{IDE_66}  Reconstruct the potential $V(\varphi)$, which realizes the solution $r=const$, obtained in the problem \ref{IDE_63}.

\item \label{IDE_67}  (after \cite{0503075})

Let the DM particle's mass $M$ depend exponentially on the DE scalar field as $M=M_*e^{-\lambda\varphi}$, where $\lambda$ is positive constant and the scalar field potential is
\[V(\varphi)=V_* e^{\eta\varphi}.\]
Obtain the modified Klein-Gordon equation for this case.

\item \label{IDE_68}  Let the DM particle's mass $M$ depend exponentially on the DE scalar field
as $M=M_*e^{-\alpha}$, and the scalar field potential is
\[V(\varphi)=V_* e^{\beta},\]
where $\alpha,\beta>0$. Obtain the modified Klein-Gordon equation for this case.
\end{enumerate}

\subsection{Interacting Phantom}
{\it Let the Universe contain only noninteracting cold dark matter ($w_{dm}=0$) and a phantom field ($w_{de}<-1$). The densities of these components evolve separately: $\rho_{dm}\propto a^{-3}$ and $\rho_{de}\propto a^{-3(1+w_{de})}$.  If matter domination ends at $t_m$, then at the moment of time \[t_{BR}=\frac{w_{de}}{1+w_{de}}t_m\] the scale factor, as well as a series of other cosmological characteristics of the Universe become infinite. This catastrophe has earned the name "Big Rip". One of the way to avoid the unwanted big rip singularity is to allow for a suitable interaction between the phantom energy and the background dark matter.}
\begin{enumerate}[resume]
\item \label{IDE_69}  Show that through a special choice of interaction, one can mitigate the rise of the phantom component and make it so that components decrease with time if there is a transfer of energy from the phantom field to the dark matter. Consider case of $Q=\delta(a)H\rho_{dm}$ and $w_{de}=const$.

\item \label{IDE_70} Calculate the deceleration parameter for the model considered in the previous problem.

\item \label{IDE_71} (after  \cite{0411524}.)

Let the interaction $Q$ of phantom field $\varphi$ with DM provide constant relation $r=\rho_{dm}/\rho_\varphi$. Assuming that $w_\varphi=const$, find $\rho_\varphi(a)$, $\rho_\varphi(\varphi)$ and $a(\varphi)$ for the case of cold dark matter (CDM).

\item \label{IDE_72}  Construct the scalar field potential, which realizes the given relation $r$ for the model considered in the previous problem.
\end{enumerate}

\subsection{Tachyonic Interacting Scalar Field}
{\it Let us consider a flat Friedmann Universe filled with a spatially homogeneous tachyon field $T$ evolving according to the Lagrangian
\[L=-V(T)\sqrt{1-g_{00}\dot T^2}.\]
The energy density and the pressure of this field are, respectively
\[\rho_T=\frac{V(T)}{\sqrt{1-\dot T^2}}\]
and
\[p_T=-V(T)\sqrt{1-\dot T^2}.\]
The equation of motion for the tachyon is
\[\frac{\ddot T}{1-\dot T^2}+3H\dot T+\frac{1}{V(T)}\frac{dV}{dT}.\]
(Problems \ref{IDE_73}-\ref{IDE_77} are after  \cite{0404086}.)}

\begin{enumerate}[resume]
\item \label{IDE_73} Find interaction of tachyon field with cold dark matter (CDM), which results in $r\equiv\rho_{dm}/\rho_T=const$.

\item \label{IDE_74}  Show that the stationary solution $\dot r=0$ exists only when the energy of the tachyon field  is transferred to the dark matter.

\item \label{IDE_75}  Find the modified Klein-Gordon equation for arbitrary interaction $Q$ of tachyon scalar field with dark matter.

\item \label{IDE_76} Find the modified Klein-Gordon equation for the interaction $Q$ obtained in the problem \ref{IDE_73} and obtain its solutions for the case $\dot\varphi=const$.

\item \label{IDE_77} Show that sufficiently small values of tachyon field provide the accelerated expansion of Universe.

\end{enumerate}

\section{Realization of interaction in the dark sector}
{\it Let us consider the Einstein field equations
\[R_{\mu\nu}-frac12Rg_{\mu\nu}=8\pi G\left(T_{\mu\nu}+\frac\Lambda{8\pi G}g_{\mu\nu}\right).\]	
According to the Bianchi identities, (i) vacuum decay is possible only from a previous existence of some sort of non-vanishing matter and/or radiation, and (ii) the presence of a time-varying cosmological term results in a coupling between $T_{\mu\nu}$ and $\Lambda$. We will assume (unless stated otherwise) coupling only between vacuum and CDM particles, so that
	\[u_\mu,T_{;\nu}^{(CDM)\mu\nu}=-u_\mu\left(\frac{\Lambda g^{\mu\nu}}{8\pi G}\right)_{;\nu}= -u_\mu\left(\rho_\Lambda g^{\mu\nu}\right)_{;\nu}\]
where $T_{\mu\nu}^{(CDM)}=\rho_{dm}u^\mu u^\nu$ is the energy-momentum tensor of the CDM matter and $\rho_\Lambda$ is the vacuum energy density. It immediately follows that
	\[\dot\rho_{dm}+3H\rho_{dm}=-\dot\rho_\Lambda.\]
Note that although the vacuum is decaying, $w_\Lambda=-1$ is still constant, the physical equation of state (EoS) of the vacuum $w_\Lambda\equiv=p_\Lambda/\rho_\Lambda$ is still equal to constant $-1$, which follows from the definition  of the cosmological constant.

(see  \cite{0408495,0507372} )}

\begin{enumerate}[resume]
\item \label{IDE_78}  Since vacuum energy is constantly decaying into CDM, CDM will dilute in a smaller rate compared with the standard relation $\rho_{dm}\propto a^{-3}$. Thus we assume that $\rho_{dm}=\rho_{dm0}a^{-3+\varepsilon}$, where $\varepsilon$ is a small positive small constant. Find the dependence $\rho_\Lambda(a)$ in this model.

\item \label{IDE_79} Solve the previous problem for the case when vacuum energy is constantly decaying into radiation.

\item \label{IDE_80} Show that existence of a radiation dominated stage is always guaranteed in scenarios, considered in the previous problem.

\item \label{IDE_81} Find how the new temperature law scales with redshift in the case of vacuum energy decaying into radiation.

\end{enumerate}

{\it Since the energy density of the cold dark matter is $\rho_{dm}=nm$, there are two possibilities for storage of the energy received from the vacuum decay process:

(i) the equation describing concentration, $n$, has a source term while the proper mass of CDM particles remains constant;

(ii) the mass $m$ of the CDM particles is itself a time-dependent quantity, while the total number of CDM particles, $N=na^3$, remains constant.

Let us consider both the possibilities.}
\begin{enumerate}[resume]
\item \label{IDE_82}  Find dependence of total particle number on the scale factor in the model considered in problem \ref{IDE_78}.

\item \label{IDE_83}  Find time dependence of CDM particle mass in the case when there is no creation of CDM particles in the model considered in problem \ref{IDE_78}.

\item \label{IDE_84} Consider a model where the cosmological constant $\Lambda$ depends on time as $\Lambda=\sigma H$. Let a flat Universe be filled by the time-dependent cosmological constant and a component with the state equation $p_\gamma= (\gamma-1)\rho_\gamma$. Find solutions of Friedman equations for this system \cite{0711.2686} .

\item \label{IDE_85} Show that the model considered in the previous problem correctly reproduces the scale factor evolution both in the radiation-dominated and non-relativistic matter (dust) dominated cases.

\item \label{IDE_86} Find the dependencies $\rho_\gamma(a)$ and $\Lambda(a)$ both in the radiation-dominated and non-relativistic matter dominated cases in the model considered in problem \ref{IDE_84}.

\item \label{IDE_87}  Show that for the $\Lambda(t)$ models \[T\frac{dS}{dt}=-\dot\rho_\Lambda a^3.\]

\subsection{Time-dependent cosmological ''constant''}

\item Consider a two-component Universe filled by matter with the state equation $p=w\rho$ and cosmological constant and rewrite the second Friedman equation in the following form
      \begin{equation}\label{mainDE}
\frac{\ddot{a}}{a} = \frac{1}{2}\left( 1 + 3w\right)
                     \left( \frac{\dot{a}^2}{a^2} + \frac{k}{a^2} \right)
                   + \frac{1-3w}{6} \Lambda .
\end{equation}

\item Consider a two-component Universe filled by matter with the state equation $p=w\rho$ and cosmological constant with quadratic time dependence $\Lambda(\tau)=\mathcal{A}\tau^2,$ and find the time dependence for of the scale factor.

\item \label{tauell} Consider a flat two-component Universe filled by matter with the state equation $p=w\rho$ and cosmological constant with quadratic time dependence $\Lambda(\tau)=\mathcal{A}\tau^{\ell}.$ Obtain the differential equation for Hubble parameter in this model and classify it.

\item Find solution of the equation obtained in the previous problem \ref{tauell} in the case  ${\ell =1}.$   Analyze the obtained solution.

\item Solve the equation obtained in the problem \ref{tauell} for ${\ell =2}.$ Consider the following cases
\begin{description}
 \item[a)] $\lambda_0 > -1/(3\gamma\tau_0)^2,$
 \item[b)] $\lambda_0 = -1/(3\gamma\tau_0)^2,$
 \item[c)] $\lambda_0 < -1/(3\gamma\tau_0)^2$
\end{description}
 (see the previous problem). Analyze the obtained solution.

\item Consider a flat two-component Universe filled by matter with the state equation $p=w\rho$ and cosmological constant with the following scale factor dependence
\begin{equation}
\Lambda = {\cal B} \, a^{-m},
\label{Bam}
\end{equation}
Find dependence of energy density of matter on the scale factor in this model.

\item Find dependence of deceleration parameter on the scale factor for the model of previous problem.

\end{enumerate}
\section{Statefinder parameters for interacting dark energy and cold dark matter.}
\begin{enumerate}[resume]
\item \label{IDE_88} (after  \cite{0311067})

    Show that in flat Universe both the Hubble parameter and deceleration parameter do not depend on whether or not dark components are interacting. Become convinced the second derivative $\ddot H$ does depend on the interaction between the components.

\item \label{IDE_89} Find  statefinder parameters for interacting dark energy and cold dark matter.

\item \label{IDE_90} Show that the statefinder parameter $r$ is generally necessary to characterize any variation in the overall equation of state of the cosmic medium.

\item \label{IDE_91} Find relation between the statefinder parameters in the flat Universe.

\item \label{IDE_92} Express the statefinder parameters in terms of effective state parameter $w_{(de)eff}$, for which \[\dot\rho_{de}+3H(1+w_{(de)eff})\rho_{de}=0.\]

\item \label{IDE_93} Find the statefinder parameters for $Q=3\delta H\rho_{dm}$, assuming that $w_{de}=const$.

\item \label{IDE_94} Find statefinder parameters for the case $\rho_{dm}/\rho_{de}=a^{-\xi}$, where $\xi$ is a constant parameter in the range $[0,3]$ and $w_{de}=const$.

\item \label{IDE_95} Show that in the case $\rho_{dm}/\rho_{de}=a^{-\xi}$ the current value of the statefinder parameter $s=s_0$ can be used to measure the deviation of cosmological models from the SCM.

\item \label{IDE_96} Find how the statefinder parameters enter the expression for the luminosity distance.
\end{enumerate}

\section{Interacting holographic dark energy}
{\it The traditional point of view assumed that dominating part of
degrees of freedom in our World are attributed to physical fields.
However it became clear soon that such concept complicates
the construction of Quantum Gravity: it is necessary to introduce small
distance cutoffs for all integrals in the theory in order to make it
sensible. As a consequence, our World should be described on a
three-dimensional discrete lattice with the period of the order of Planck
length. Lately some physicists share an even more radical
point of view: instead of the three-dimensional lattice, complete
description of Nature requires only a two-dimensional one, situated on
the space boundary of our World. This approach is based on the
so-called ''holographic principle''. The name is related to the optical
hologram, which is essentially a two-dimensional record of a
three-dimensional object. The holographic principle consists of two
main statements:
\begin{enumerate}
\item All information contained in some region of space can be
''recorded'' (represented) on the boundary of that region.
\item The theory, formulated on the boundaries of the considered
region of space, must have no more than one degree of freedom per
Planck area:
\begin{equation}
\label{Hol_f:1}
    N\le \frac{A}{A_{pl}},\quad A_{pl}=\frac{G\hbar}{c^3}.
\end{equation}
\end{enumerate}
Thus, the key piece in the holographic principle is
the assumption that all the information about the Universe can be
encoded on some two-dimensional surface --- the holographic
screen. Such approach leads to a new interpretation of cosmological
acceleration and to an absolutely unusual understanding of Gravity. The
Gravity is understood as an entropy force, caused by variation of
information connected to positions of material bodies. More
precisely, the quantity of information related to matter and its
position is measured in terms of entropy. Relation between the
entropy and the information states that the information change is
exactly the negative entropy change $\Delta I=-\Delta S$. Entropy change
due to matter displacement leads to the so-called entropy force, which,
as will be proven below, has the form of gravity. Its origin
therefore lies in the universal tendency of any macroscopic theory
to maximize the entropy. The dynamics can be constructed in terms
of entropy variation and it does not depend on the details of
microscopic theory. In particular, there is no fundamental field
associated with the entropy force. The entropy forces are typical
for macroscopic systems like colloids and biophysical systems. Big
colloid molecules, placed in thermal environment of smaller
particles, feel the entropy forces. Osmose is another phenomenon
governed by the entropy forces.

Probably the best known example of the entropy force is the elasticity of a polymer
molecule. A single polymer molecule can be modeled as a
composition of many monomers of fixed length. Each monomer can
freely rotate around the fixation point and choose any spacial
direction. Each of such configurations has the same energy. When the
polymer molecule is placed into a thermal bath, it prefers to form a
ring as the entropically most preferable configuration: there are
many more such configurations when the polymer molecule is short,
than those when it is stretched. The statistical tendency to transit
into the maximum entropy state transforms into the macroscopic
force, in the considered case---into the elastic force.

Let us consider a small piece of holographic screen and a particle
of mass $m$ approaching it. According to the holographic principle,
the particle affects the amount of the information (and therefore of the
entropy) stored on the screen. It is natural to assume that
entropy variation near the screen is linear on the displacement
$\Delta x$:
\begin{equation}
\Delta S = 2\pi k_B \frac{mc}{\hbar} \Delta x. \label{delta_s}
\end{equation}
The factor $2\pi$ is introduced for convenience, which the reader
will appreciate solving the problems of this section. In order to
understand why this quantity should be proportional to mass,
let us imagine that the particle has split into two or more particles
of smaller mass. Each of those particles produces its own entropy
change when displaced by $\Delta x$. As entropy and mass are both
additive, then it is natural that the former is proportional to the
latter. According to the first law of thermodynamics, the entropy
force related to information variation satisfies the equation
\begin{equation}
F\Delta x = T\Delta S. \label{delta_x}
\end{equation}
If we know the entropy gradient, which can be found from (\ref{delta_s}),
and the screen temperature, we can calculate the entropy
force.

An observer moving with acceleration $a$, feels the
temperature (the Unruh temperature)
\begin{equation}
\label{Hol_f_Unruh:4}
k_B T_U=\frac{1}{2\pi}\frac\hbar c a.
\end{equation}
Let us assume that the total energy of the system equals $E$. Let us
make a simple assumption that the energy is uniformly distributed
over all $N$ bits of information on the holographic screen. The
temperature is then defined as the average energy per bit:
\begin{equation}
E =\frac12 N k_B T. \label{average_e}
\end{equation}
Equations (\ref{delta_s})--(\ref{average_e}) allow one to describe the holographic
dynamics, and as a particular case---the dynamics of the Universe, and
all that without the notion of Gravity.}
\begin{enumerate}[resume]

\item\label{IDE_97_1} For the interacting holographic dark energy $Q = 3\alpha H \rho_L,$ with the Hubble radius as the IR cutoff, find the depending on the time for the scale factor, the Hubble parameter and the deceleration parameter.

\item \label{IDE_97}  Show that for the choice $\rho_{hde}\propto H^2$ ($\rho_{hde}=\beta H^2$, $\beta=const$)an interaction is the only way to have an equation of state different from that of the dust.

\item \label{IDE_98} Calculate the derivative \[\frac{d\rho_{de}}{d\ln a}\] for the holographic dark energy model, where IR cut-off $L$ is chosen to be equal to the future event horizon. (after \cite{0711.1641})

\item \label{IDE_99}  Find the effective state parameter value $w_{eff}$, such that \[\rho'_{de}+3(1+w_{eff})\rho_{de}=0\] for the holographic dark energy model, considered in the previous problem, with the interaction of the form $Q=3\alpha H\rho_{de}$.

\item \label{IDE_100}  Analyze how fate of the Universe depends on the parameter $c$ in the holographic dark energy model, where IR cut-off $L$ is chosen to be equal to the future event horizon.

\item \label{IDE_101}  In the case of interacting holographic Ricci dark energy with interaction is given by
 \begin{eqnarray}
 \label{intrate}
   Q=\gamma H \rho_{_{\cal R}},
  \end{eqnarray}
  where $\gamma$ is a dimensionless parameter,
find the dependence of the density of dark energy and dark matter on the scale factor.

\item \label{IDE_102}  Find the exact solutions for linear interactions between Ricci DE and DM, if the energy density of Ricci DE is given by $ \rho_x =\left(2\dot H + 3\alpha H^2\right)/\Delta,$  where $\Delta=\alpha -\beta$ and $\alpha,\,\beta$ are constants.

\item\label{IDE_103}  Find  the equation of motion for the relative density
   \begin{equation}\label{eq7}
 \Omega_q=\frac{n^2}{H^2T^2},
\end{equation}
were
\begin{equation}\label{AGE_U}
T=\int_{0}^{a}{\frac{d{a}'}{H{a}'}}.
\end{equation}
 of interacting agegraphic dark energy and the deceleration parameter, for sets of   interaction term $Q=3\alpha H\rho_q;\; 3\beta H\rho_m;\; 3\gamma H\rho_{tot}.$
\end{enumerate}

\section{Transient acceleration}
\begin{enumerate}[resume]

\item Consider a simple model of transient acceleration with decaying cosmological constant
\begin{equation} \label{ec}
\dot{\rho}_{m} + 3\frac{\dot{a}}{a}\rho_{m} = - \dot{\rho}_\Lambda\;,
\end{equation}
where $\rho_{m}$ and $\rho_\Lambda$ energy density DE and cosmological constant  $\Lambda$. At the early stages of the expansion of the Universe, when $\rho_\Lambda$ is quite small, such a decay does not influence cosmological evolution in any way. At later stages, as the DE contribution increases, its decay has an ever increasing effect on the standard dependence of the DM energy density $\rho_{m} \propto a^{-3}$ on the scale factor $a$. We consider the deviation to be described by a function  of the scale factor - $\epsilon(a)$.
\begin{equation} \label{dm}
\rho_{m} = \rho_{m, 0}a^{-3 + \epsilon(a)}\;,
\end{equation}
where $a_0 = 1$ in the present epoch.
Other fields of matter (radiation, baryons) evolve independently and are conserved. Hence, the DE density has the form
\begin{equation}\label{decayv}
\rho_{\Lambda} =  \rho_{m0} \int\limits_{a}^{1}\frac{\epsilon(\tilde{a}) + \tilde{a}\epsilon' \ln(\tilde{a})}{\tilde{a}^{4 - \epsilon(\tilde a)}} d\tilde{a} + {\rm{X}}\;,
\end{equation}
where the prime denotes the derivative with respect to the scale factor, and ${\rm{X}}$ is the integration constant. If radiation is neglected, the first Friedmann equation takes the form
\begin{equation}
\label{friedmann} {{H}}= H_0\left[\Omega_{b,0}{a}^{-3} + \Omega_{m0}\varphi(a) + {\Omega}_{{\rm{X,0}}}\right]^{1/2},
\end{equation}
Using the assumption that the function $\epsilon (a)$ has the following simple form
\begin{equation}
\label{Parametrization_a}
\epsilon(a) = \epsilon_0a^\xi\ = \epsilon_0(1+z)^{-\xi},
\end{equation}
where $\epsilon_0$ and $\xi$ can take both positive and negative values, find function $\varphi(a)$ and relative energy density $\Omega_b(a)$, $\Omega_{m}(a)$ and $\Omega_{\Lambda}(a)$.

\item Using the results of the previous problem, find the deceleration parameter for this model is $q(a)$. Draw the graph deceleration parameter as a function of $\log(a)$ for various values of $\epsilon_0$ and $\xi$: $\xi = 1.0$ and $\epsilon_0 = 0.1$, $\xi = -1.0$ and $\epsilon_0 = 0.1$, $\xi = 0.8$ and $\epsilon_0 = 0.5$, $\xi = -0.5$ and $\epsilon_0 = -0.1.$.

\item Consider the possibility of an accelerating transient regime within the interacting scalar field model using potential of the form
\begin{equation}
\label{V_fransient}
V(\phi)=\rho_{\phi\,0}[1-\frac{\lambda}{6}(1+\alpha \sqrt{\sigma}\phi)^2)]
\exp{[-\lambda\sqrt{\sigma}(\phi+ \frac{\alpha \sqrt{\sigma}}{2}\phi^2)]},
\end{equation}
where $\rho_{\phi\,0}$ is a constant energy density, $\sigma=8\pi G/\lambda$, and $\alpha$ and $\lambda$ are two dimensionless, positive parameters of the model, that the deceleration parameter is non-monotonically dependent on the scale factor.
Plot the deceleration parameter as a function of the scale factor.

\item Consider a simple parameterization:
\begin{equation}
\label{Qsimple}
Q=3\beta(a)H \rho_{de}
\end{equation}
with a simple power-law ansatz for $\beta(a)$, namely:
\begin{equation}
\label{cas32}
\beta(a)=\beta_0 a^\xi.
\end{equation}
Substituting this interaction form into conservation equations for DM and DE:
\begin{eqnarray}\label{eom1}
\dot{\rho}_{dm}+3H\rho_{dm}=Q,
\end{eqnarray}
\begin{eqnarray}\label{eom2}
\dot{\rho}_{de}+3H(\rho_{de}+p_{de})=-Q,
\end{eqnarray}
we get
 \begin{equation}
\label{rhophi2}
\rho_{de}=\rho_{de0}\, a^{-3(1+w_0)}\cdot
\exp{\left[\frac{3\beta_0(1-a^\xi)}{\xi}\right]},
\end{equation}
where the integration constant $\rho_{de0}$ is value of the dark energy at present,
and the dark energy EoS parameter $w\equiv p_{de}/\rho_{de}$ is a constant-$w_0$.
Substituting Eq. (\ref{rhophi2}) into Eq. (\ref{eom2}), we get
the dark matter energy density,
 \begin{equation}\label{rhom2}
\rho_{dm}=f(a)\rho_{dm0},
\end{equation}
where
\begin{equation}
\label{f1}
f(a)\equiv \frac{1}{a^3}\left\{1-\frac{\Omega_{de0}}{\Omega_{dm0}}\frac{3\beta_0 a^{-3w_0}e^{\frac{3\beta_0}{\xi}}}{\xi}\cdot\left[a^\xi E_{\frac{3w_0}{\xi}}\left(\frac{3\beta_0 a^\xi}{\xi}\right)-a^{3w_0} E_{\frac{3w_0}{\xi}} \left(\frac{3\beta_0}{\xi}\right)\right]\right\},
\end{equation}
where $\rho_{dm0}$ is dark matter density at present day, and $E_n(z)=\int_1^\infty t^{-n}e^{-xt}dt$ the
usual exponential integral function.
Note however that Eq. (\ref{rhom2}) is an
analytical expression, while in the corresponding expressions were left as
integrals and were calculated numerically. Obviously, in the case of
non-interaction (that is, for $\beta_0=0$), Eq. (\ref{rhom2}) recovers the standard
result $\rho_{dm}=\rho_{dm0}/a^3$.
For the special case $\xi=0$ find dimensionless Hubble parameter $E^2(z)\equiv \frac{H^{2}}{H^{2}_0}$, the evolution of the density parameters $\Omega_b(a)$, $\Omega_{dm}(a)$ and $\Omega_{de}(a)$ and $q(a)$. For what values of $\beta_0$ the cosmic acceleration is transient?

\item Consider the flat FLRW cosmology with two coupled
homogeneous scalar fields $\Phi$ and $\Psi$:
\begin{eqnarray}
\dot{\rho_b} &=& -3 H \gamma_b \rho_b \\
\ddot{\Phi} &=& -3H\dot{\Phi} - \partial_\Phi V \\
\ddot{\Psi} &=& -3H\dot{\Psi} - \partial_\Psi V \\
\dot{H} &=& -4\pi G(\gamma_m\rho_m+\gamma_r\rho_r+\gamma_Q\rho_Q)~,
\end{eqnarray}
\begin{equation}
H^2 = \frac{8\pi G}{3}(\rho_m+\rho_r+\rho_Q)-\frac{k}{a^2}~.
\end{equation}
Here a dot denotes a derivative with respect to the cosmic time $t$, the subscript $b$
refers to the dominant background quantity, either dust (m) or radiation (r) while $Q$
refers to the Dark Energy sector, here the two quintessence scalar fields.

The quintessence fields with potential $V$ have the following
energy density and pressure:
\begin{eqnarray}
  \rho_Q =\frac{1}{2}\dot{\Phi}^2 + \frac{1}{2}\dot{\Psi}^2 + V(\Phi,\Psi)\\
  p_Q =\frac{1}{2}\dot{\Phi}^2 + \frac{1}{2}\dot{\Psi}^2 - V(\Phi,\Psi)
\end{eqnarray}
with $p_Q = (\gamma_Q-1)\rho_Q$.
It is convenient to define the following new variables :
\begin{equation}
X_{\Phi}=\sqrt{\frac{8\pi G}{3H^2}}~\frac{\dot{\Phi}}{\sqrt{2}},~~~
X_{\Psi}=\sqrt{\frac{8\pi G}{3H^2}}~\frac{\dot{\Psi}}{\sqrt{2}},~~~
X_V=\sqrt{\frac{8\pi G}{3H^2}}~\sqrt{V}.
\end{equation}
Find expressions for the $\Phi'$, $\Psi'$, $X_{\Phi}'$, $X_{\Psi}'$, $X_V'$ where a prime denotes a derivative with respect to the quantity $N$, the number of e-folds with respect to the present time,
\begin{equation}
N\equiv {\rm ln} \frac{a}{a_0}~,\label{N}
\end{equation}
and we have also $H=\dot{N}$.

\item Using result from the previous problem find the relative energy density for matter, radiation and quintessence,  $\Omega_m$, $\Omega_r$ and $\Omega_Q$, the deceleration parameter $q$, the Hubble-parameter-free luminosity distance $D_L$ and the age of the Universe $t_0$.
\end{enumerate}

\chapter{Standard Cosmological Model}
\begin{flushright}
\it "The history of cosmology shows that in every age\\ devout
people believe that they have at last discovered\\ the true nature
of the Universe"\\ E.R.~Harrison
\end{flushright}
{\it On the frontier between XX and XXI centuries the Standard
Cosmological Model (SCM) became the dominant model of Universe. It
is based on two most important observational results:\\
1. Accelerated expansion of the Universe\\
2. Euclidean geometry of space.\\
The theoretical basis of the SCM is the theory of General
Relativity. Also it is assumed that early Universe is adequately
described by the inflation theory (see Chapter
\ref{chapter_inflation}). SCM fixes a set of parameters of the
Universe and, in particular, its energy composition: $\Omega_{k0}=0$, $\Omega_{m0}=0.27$, $\Omega_{r0}=8.1 \times 10^{-5}$, $\Omega_{\Lambda}=0.73$. According to
SCM, two components dominate in the present Universe---the dark
energy (in form of cosmological constant $\Lambda$) and the cold
dark matter (CDM). Therefore the model was named $\Lambda CDM$.

In the problems below, unless specified otherwise, the standard values for the SCM parameters are to be used.}
\section{Characteristic Parameters and Scales}
\begin{enumerate}
\item \label{scm-p-1} {  Calculate the dark energy density and the cosmological constant value.}

\item \label{scm-p-2} {  Estimate total number of baryons in the Universe.}

\item \label{SCM_2} {  Find dependence of relative density of dark energy $\Omega_\Lambda$ on the redshift. Plot $\Omega_\Lambda(z)$.}

\item \label{scm-p-3} {  Estimate total number of stars in the Universe.
}

\item \label{scm-p-4} {  Find the ratio of dark energy density to the energy density of
electric field of intensity $1\,V/m$. Compare the dark energy
density with gravitational field energy density on the Earth
surface.}

\item \label{scm-p-5} {  Estimate the distance between two neutral hydrogen atoms at which the gravitational force of their attraction is balanced by the repulsion force generated by dark energy in the
form of cosmological constant. Make the same estimates for the
Sun-Earth system.}

\item \label{scm-p-6} {  Can an open Universe recollapse or a closed Universe expand forever?}

\item \label{scm-p-7} {  Current observations show that the Universe is flat with high precision: $|\Omega_{curv}|<0.02$. Consider a hypothetic situation: all dark energy in the Universe instantly disappears, but the curvature remains as small as it is, and $k=+1$ ($\Omega_{curv}<0$) (closed Universe). What ratio of scale factors $a(t)/a_0$ corresponds to the moment when expansion change to contraction in such a Universe?}

\item \label{scm-p-8} {  Calculate magnitude of physical acceleration.}

\item \label{scm-p-9} {  How far can one see in the Universe?}

\item \label{scm-p-10} {  Find age of the Universe.
}

\item \label{scm-p-11} {  Give a qualitative explanation why the age of Universe in SCM is considerably greater than the age of matter dominated Universe (Einstein-de Sitter model).}


\section{Evolution of Universe}
\item \label{scm-p-12} {  Rewrite the first Friedman equation in terms of redshift and analyze the contributions of separate components at different stages of Universe evolution.
}

\item \label{scm-p-12} {  Find the time dependence for the scale factor and analyze asymptotes of the dependence. Plot $a(t)$.
}

\item \label{scm-p-14} {  Determine the redshift value corresponding to equality of radiation and matter densities.}

\item \label{scm-p-15} {  Show that the following holds: $\dot H= -4\pi G\rho_m$
and $\ddot H= 12\pi G\rho_m H$.
}

\item {  \label{SCM_18}Expand the scale factor in Taylor series near the time moment
$t_0$:
\[\frac{a(t)}{a(t_0)}=1+\sum\limits_{n=1}^\infty\frac{A_n(t_0)}{n!}[H_0(t-t_0)]^n;\quad A_n\equiv\frac{1}{aH^n}\frac{d^na}{dt^n}\]
and calculate values for few first coefficients $A_n$.}

\item \label{scm-p-16} {  Show that all the coefficients $A_n$ can be expressed through
elementary functions of the deceleration parameter $q$ or the
density parameter \[\Omega_m=\frac23(1+q).\]}

\item \label{scm-p-17} {  Consider the case of flat Universe filled by non-relativistic matter and dark
energy with state equation $p_X=w\rho_X$, where the state parameter $w$ is parameterized as the following \[w=w_0+ w_a(1-a)=w_0 + w_a \frac{z}{1+z}.\]
Express current values of cosmographic parameters through $w_0$ and
$w_a$. 
}

\item \label{scm-p-18} {  Show that the results of the previous problem applied to SCM coincide with
the ones obtained in the problem \ref{SCM_18}.
}

\item \label{SCM10} {  Photons with $z=0.1,\ 1,\ 100,\
1000$ are registered. What was the Universe age $t_U$ in the moment of
their emission? What period of time $t_{ph}$ were the photons on the
way? Plot $t_U(z)$ and $t_{ph}(z)$.
}

\item \label{scm-p-19} {  Determine the present physical distance to the object that emitted light with current redshift $z$.
}

\item \label{SCM_15} {  A photon was emitted at time $t$ and registered at time $t_0$ with red
shift $z$. Find and plot the dependence of emission time on the redshift $t(z)$.
}

\item \label{scm-p-20} {  Find the time dependence for the scale factor and analyze its asymptotes. Plot $a(t)$.
}

\item \label{scm-p-21} {  Using the explicit solution for scale factor $a(t)$, obtained in the previous problem, find the cosmic horizon $R_h$ (see Chapter 3).}

\item \label{scm-p-22} {  Show that the Universe becomes $R_h$-delimited only after the cosmological constant starts to dominate.}

\item \label{scm-p-23} {  Analyze the stability of Friedmann equations.}

\item \label{scm-p-24} {  Determine solutions for the perturbations $\delta\rho_m$ and $\delta H$. Make sure that the solutions are stable.}

\item \label{scm-p-25} {  Rewrite the first Friedman equation in terms of conformal time.}

\item \label{scm-p-26} {  Find relation between the scale factor and conformal time}

\item \label{scm-p-27} {  Find explicit dependence of the scale factor on the conformal
time.}

\item \label{scm-p-28} {  Find relative density of dark energy $10^9$ years later.}

\item \label{SCM_17} {  Find variation rate of relative density of dark energy. What are its asymptotic values? Plot its time dependence.}

\item \label{scm-p-29} {  Estimate size of the cosmological horizon.}

\item \label{SCM_19} {  Find time dependence of Hubble parameter and plot it.}

\item \label{scm-p-30} {  Find time dependence of dark matter density.}
	
\item \label{scm-p-30_} {  Find the asymptotic (in time) value of the Hubble parameter.}

\item \label{scm-p-31} {  At present the age of the Universe $t_0\simeq13.7\cdot10^9$ years is close to the Hubble time $t_H=H_0^{-1}\simeq14\cdot10^9$ years.
Does the relation $t^*\simeq t_H(t^*)=H^{-1}(t^*)$ for the age $t^*$
of the Universe hold for any moment of its evolution?}

\item \label{scm-p-32} {  Determine values of state parameter for the dark energy, that provide accelerated expansion of Universe in the present time.}

\item \label{scm-p-33} {  Find current value of the deceleration parameter.}

\item \label{SCM_24} {  Find the redshift dependence of
the deceleration parameter. Analyze the limiting cases.}

\item \label{SCM_25} {  Find and plot the time dependence of
the deceleration parameter.}

\item \label{scm-p-34} {  Find the moment of time when the dark energy started to dominate over dark matter. What redshift did it correspond to?}

\item \label{scm-p-35} {  Determine the moment of time and redshift value corresponding to the transition from decelerated
expansion of the Universe to the accelerated one.}

\item \label{scm-p-36} {  Solve the previous problem using the derivative $d\eta/d\ln a$.}
\item \label{scm-p-37} {  Is dark energy domination necessary for transition to the accelerated expansion?}

\item \label{scm-p-38} {  Consider flat Universe composed of matter and dark energy in form of cosmological constant. Find the redshift value corresponding to equality of densities of
the both components $\rho_m(z_{eq})=\rho_\Lambda(z_{eq})$ and the one corresponding to beginning of the accelerated expansion $q\left(z_{accel} \right) = 0$. Obtain relation between $z_{eq}$ and $z_{accel}$.}

\item \label{scm-p-39} {  Show that density perturbations stop to grow after the transition
from dust to $\Lambda$-dominated era.}

\item \label{scm-p-40} {  What happens to the velocity fluctuations of non-relativistic
matter and radiation with respect to Hubble flow in the epoch of
cosmological constant domination?}

\item \label{scm-p-41} {  Find the ratio of baryon to non-baryon components in the galactic halo.}

\item \label{scm-p-42} {  Imagine that in the Universe described by SCM the dark energy was
instantly switched off. Analyze further dynamics of the Universe.}

\item \label{local group} Estimate density of dark energy in form of cosmological
constant using the Hubble diagram (see fig.\ref{hubble_diagram})
for the neighborhood of the Local group.
\begin{figure}
\includegraphics[width=\textwidth]{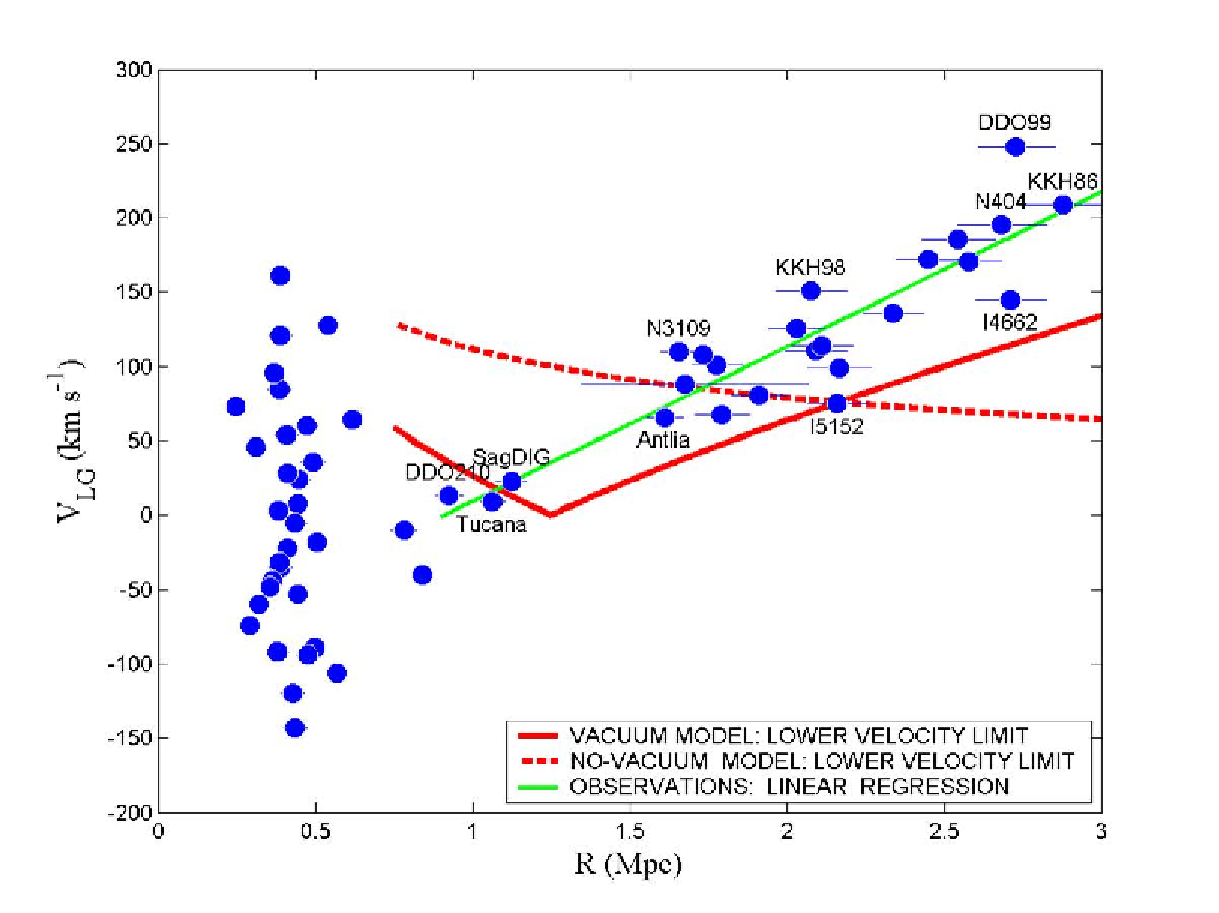}
\caption{To problem \ref{local group}: Hubble diagram for the
neighborhood of the Local group. \label{hubble_diagram}}
\end{figure}

\item \label{scm-p-44} {  Estimate the Local group mass by methods used in the previous
problem.
}

\item \label{scm-p-45} {  Find ''weak'' points in the argumentation of the two preceding
problems.
}

\item \label{scm-p-46} {  The product of the age of the Universe and current Hubble parameter (the Hubble's
constant) is a very important test (Sandage consistency test) of
internal consistency for any model of Universe. Analyze the
parameters on which the product $H_0 t_0$ depends in the Big Bang
model and in the SCM.
}

\item \label{SCM_38} {  Show that for a fixed source of radiation the luminosity distance for high redshift values in flat Universe is greater for the dark energy dominated case compared to the non-relativistic matter dominated one.}

\item \label{scm-p-47} {  In the observations that discovered the accelerated expansion of
the Universe the researchers in particular detected two $Ia$ type
supernovae: $1992P$, $z=0.026$, $m=16.08$ and $1997ap$, $z=0.83$,
$m=24.32$. Show that these observed parameters are in accordance with the
SCM.}

\item \label{scm-p-48} {  Find the redshift value, at
which a source of linear dimension $d$ has minimum visible angular
size.}
%

\item \label{scm-p-49} {  Compare the observed value of the dark energy density with the
one expected from the dimensionality considerations (the
cosmological constant problem).}

\item \label{scm-p-50} {  Determine the density of vacuum energy using the Planck scale as cutoff parameter.
}

\item \label{scm-p-51} {  Identifying the vacuum fluctuations density with the observable dark energy value in SCM, find the required frequency cutoff magnitude in the fluctuation spectrum.}
\[\nu_{max}=10^{12}Hz.\]

\item \label{scm-p-52} {  What purely cosmological problem originates from the divergence of the zero-point energy density?}

\item \label{scm-p-53} {  With exact supersymmetry, the bosonic contribution to cosmological constant is canceled by its fermionic counterpart. However, we know that our world looks like not supersymmetric. Supersymmetry, if exists, has to be broken above or around $100GeV$ scale. Compare the observed value of the dark energy density with the one expected from broken supersymmetry.}

\item \label{scm-p-54} {  Determine duration of the inflation period.}

\item \label{scm-p-55} {  Plot the dependence of luminosity distance $d_L$ (in units of
$H_0^{-1}$) on the redshift $z$ for the two-component flat Universe
with non-relativistic liquid ($w=0$) and cosmological constant
($w=-1$). Consider the following cases:
\begin{description}
  \item [a)] $\Omega_\Lambda^0=0$;
  \item [b)] $\Omega_\Lambda^0=0.3$;
  \item [c)] $\Omega_\Lambda^0=0.7$;
  \item [d)] $\Omega_\Lambda^0=1$.
\end{description}
}

\item \label{scm-p-56} {  For a Universe filled by dark energy with state equation
$p_{DE}=w_{DE}\rho_{DE}$ and non-relativistic matter obtain the
Taylor series for $d_L$ in terms of $z$ near the observation point
$z_0=0$. Explain the obtained result.}

\item \label{SCM_45} {  Determine position of the first acoustic peak in the CMB power
spectrum produced by baryon oscillations on the surface of the last
scattering.}
\item \label{scm-p-57} {  Compare the asymptotes of time dependence of the scale factor
$a(t)$ for the SCM and de Sitter models. Explain physical reasons of
their distinction.
}

\item \label{scm-p-58} {  Redshift for any object slowly changes due to the
acceleration (or deceleration) of the Universe expansion. Estimate
change of velocity in one year.
}

\item \label{scm-p-59} {  Determine the lower limit
of ratio of the total volume of the Universe to the observed one?
}
\item \label{scm-p-60} {  What is the difference between the inflationary expansion in the early
Universe and the present accelerated expansion?
}
%

\item \label{scm-p-61} {  Compare the values of Hubble parameter at the beginning of the
inflation period and at the beginning of the present accelerated
expansion of Universe.
}

\item \label{scm-p-62} {  Consider the quantity
\[O(x)=\frac{h^2(x)-1}{x^3-1},\] where $x=1+z$, $h(x)=H(x)/H_0(x)$.
Show that if the state parameter $w=const$ then $O(x)=\Omega_{0m}$
for cosmological constant, $O(x)>\Omega_{0m}$ for quintessence and
$O(x)<\Omega_{0m}$ for phantom energy, and therefore $O(x)$ can be
used to probe the dark energy state equation.}
\item \label{SCM_52} {  Plot the dependencies $h(x)=H(x)/H_0(x)$, where $x=1+z$, in the SCM,
for quintessence and for phantom energy cases.}
\item \label{scm-p-63} {  Find the constraints imposed by the Weak Energy Condition (WEC) for the dark energy on the redshift-dependent Hubble parameter. (Inspired by A.Sen, R.Scherrer, arXiv:astro-ph/ 0703416) }
\item \label{scm-p-64} {  Using the statefinders, show that the power law cosmology mimics SCM model at   (see Chapters 3 and 9).}
\end{enumerate}

\chapter{Weak field limit and gravitational waves}

\section{Motivation and symmetries / Introduction}

\begin{enumerate}
\item\label{gw1}
Show that relic neutrinos can provide us with information on the Universe at temperatures less that 1 MeV (i.e. from the age of the order of seconds). What is the corresponding limit for gravitons?

\item\label{gw2}
The basic theory of gravitational waves, discussed below, deals mostly with small perturbations on flat Minkowski background. Can this approximation be useful for studying waves on a non-trivial non-flat background, and if yes then in which cases?

\item\label{gw3}
Show that generation of either electromagnetic or gravitational monopole radiation is impossible.

\item\label{gw4}
Show that dipole gravitational radiation is prohibited by the momentum conservation law.

\item\label{gw5}
Obtain, using dimensional analysis, the quadrupole formula for the energy loss by a system due to emission of gravitational waves
\[\frac{dE}{dt}\sim
	\frac{G}{c^5}\cdot
	{\dddot{Q}}_{\alpha\beta}
	{\dddot{Q}}^{\alpha\beta},\]
where
\[Q_{\alpha\beta}=\int d^3 x \;
	( x_\alpha x_\beta -\tfrac13 r^2 \delta_{\alpha\beta})
	\rho(\mathbf{x}).\]
is the reduced quadrupole moment of the system.

\item\label{gw6}
Using the quadrupole formula, find the upper limit for gravitational luminosity of a source.

\item\label{gw7}
Construct the limiting luminosity from dimensional analysis

\item\label{gw8}
Why the upper limit on luminosity, obtained in General Relativity, should not change in the future quantum theory of gravitation?

\item\label{gw9}
Estimate the gravitational luminosity for gravitationally bound systems

\item\label{gw10}
Estimate the upper bound of the gravitational wave frequency generated by a compact source with size $R$ and mass $M$.

\section{Linearized Einstein equations}

Let us consider small perturbations on Minkowski background, such that in some frame the metric can be presented in the form
\begin{equation}
\label{WFL}
g_{\mu\nu}(x)=\eta_{\mu\nu}+h_{\mu\nu}(x),
\qquad |h_{\mu\nu}(x)|\ll 1.
\end{equation}

We can also consider perturbations on the background of other exact solutions of the Einstein equations by replacing $\eta_{\mu\nu}$ with the corresponding $g_{\mu\nu}^{(0)}$. Thus cosmological perturbations are naturally studied in the Friedmanninan background.

The linearized Einstein equations are obtained in the first order by $h_{\mu\nu}$, discarding quadratic terms. On Minkowski background the zero-order terms for the curvature tensor and its contractions vanish, so from the Einstein's equation the stress-energy tensor in the considered region must also be small (if non-zero) and $\sim h$. The constraints this places on matter will be considered in more detail in the next section.

\item\label{gw11}
Show that on Minkowski background the inverse metric is
\[g^{\mu\nu}(x)=\eta^{\mu\nu}-h^{\mu\nu}(x)+O(h^2),
\quad\text{where}\quad
h^{\mu\nu}\equiv\eta^{\mu\rho}\eta^{\nu\sigma}
h_{\rho\sigma},\]
and we agree to use $\eta$ for raising and lowering of the indices.

\item\label{gw12}
Show$^*$ that using the background metric $g_{\mu\nu}^{(0)}$ to raise and lower indices instead of the true metric $g_{\mu\nu}$ only makes difference in the next order by $h$.

Consider for definiteness a second rank tensor $A_{\mu\nu}$:
\begin{align*}A_{\mu\nu}
=g_{\mu\rho}g_{\nu\sigma}A^{\rho\sigma}
=g_{\mu\rho}^{(0)}g_{\nu\sigma}^{(0)}A^{\rho\sigma}
	+O(hA).
\end{align*}

$^*$ Is it really a problem at all?

\item\label{gw13}
Derive the curvature, Ricci and Einstein tensors in the first order by $h_{\mu\nu}$.

\item\label{gw14}
Write the Einstein's tensor in terms of the trace-reversed metric perturbation
\[\bar{h}_{\mu\nu}
=h_{\mu\nu}-\frac{1}{2}h\;\eta_{\mu\nu}.\]

\section{Gauge transformations and degrees of freedom}

The general equations
\[G_{\mu\nu}=\frac{8\pi G}{c^4}T_{\mu\nu}\]
are valid in any coordinate frame, in which the metric obeys (\ref{WFL}), so$^*$ we have the freedom to make coordinate transformation
\[x^{\mu}\to {x'}^{\mu}=x^{\mu}+\xi^{\mu}(x),\]
with four arbitrary functions $\xi^\mu$, which are of the first order by $h_{\mu\nu}$.

$^*$In addition to global Lorentz transformations, which are symmetries of the Minkowski background, or in general the isometries of the background spacetime.

\item\label{gw15}
Find $h_{\mu\nu}$ in the new (primed) coordinates; show that curvature tensor and its contractions are gauge invariant and do not change their functional form.

In a given frame the metric perturbation $h_{\mu\nu}$ can be decomposed into pieces which transform under spatial rotations as scalars, vectors and tensors (the irreducible representations of the rotation group $SO(3)$) in the following way (spatial components are denoted by Greek indices from the beginning of the alphabet $\alpha,\beta,\gamma\ldots=1,2,3$):
\begin{align}
h_{00}&=2\Phi;\\
h_{0\alpha}&=-w_{\alpha};\\
h_{\alpha\beta}&=2\big( s_{\alpha\beta}
	+\Psi\eta_{\alpha\beta}\big),
\end{align}
where $h_{\alpha\beta}$ is further decomposed in such a way that $s_{ij}$ is traceless and $\Psi$ encodes the trace:
\begin{align}
 h\equiv h_{\alpha}^{\alpha}
	&=\eta^{\alpha\beta}h_{\alpha\beta}
	=0+2\Psi \delta^{\alpha}_{\alpha}=6\Psi;\\
\Psi&=\tfrac{1}{6}h;\\
s_{\alpha\beta}&=\tfrac{1}{2}\big(h_{\alpha\beta}
	-\tfrac{1}{6}h\; \eta_{\alpha\beta}\big).
\end{align}
Thus the metric takes the form
\[ds^{2}=(1+2\Phi)dt^2 -2w_{\alpha}dt\,dx^{\alpha}
	-\big[(1-2\Psi)\eta_{\alpha\beta}
		-2s_{\alpha\beta}\big]dx^\alpha dx^\beta\]

\item\label{gw16}
Write down geodesic equations for a particle in the weak field limit in terms of fields  $\Phi$, $w_\alpha$, $h_{\alpha\beta}$.  What are the first terms of expansion by $v/c$ in the non-relativistic limit?

\item\label{gw17}
Derive the Einstein equations for the scalar $\Phi,\Psi$, vector $w^\alpha$ and tensor $s_{\alpha\beta}$ perturbations. Which of them are dynamical?

\item\label{gw18}
Find the gauge transformations for the scalar, vector and tensor perturbations.

\item\label{gw19}
This one is equivalent to Gaussian normal coordinates and is fixed by setting
\begin{equation}
\Phi=0,\qquad w^\alpha=0.
\end{equation}
Write the explicit coordinate transformations and the metric in this gauge.

\item\label{gw20}
This is a generalization of the conformal Newtonian or Poisson gauge sometimes used in cosmology, which is fixed by demanding that
\begin{equation}
\partial_\alpha s^{\alpha\beta}=0,\qquad
\partial_\alpha w^\alpha =0.
\end{equation}
Find the equations for $\xi^\mu$ that fix the transverse gauge.

\section{Wave equation}

\item\label{gw21}
The Lorenz${}^{*}$ gauge conditions are
\begin{equation}
\label{LorenzGauge}
\partial_{\mu}\bar{h}^{\mu\nu}=0.
\end{equation}
Write down the Einstein's equations in the Lorenz gauge.

${}^{*}$By analogy with electromagnetism; note the spelling: Lorenz, not Lorentz.

\item\label{gw22}
Lorenz frame is the coordinate frame in which the Lorenz gauge conditions (\ref{LorenzGauge}) are satisfied.

1) find the coordinate transformation  $x^\mu \to {x'}^{\mu}=x^\mu +\xi^\mu$ from a given frame to the Lorenz frame;

2) is the Lorenz frame unique? what is the remaining freedom for the choice of $\xi^\mu$?

\item\label{gw23}
Obtain the Einstein's equation in the Lorenz gauge in the first order by $h$ directly from the action

\item\label{gw24}
Simplify the vacuum Einstein's equations in the transverse gauge (without the Lorenz gauge conditions), assuming vanishing boundary conditions.

The only difference of the wave equation for gravitational perturbations from the one for the electromagnetic field is in its tensorial nature. The Green's function for the wave equation is known 
and the retarded solution, which is usually thought of as the one physically relevant, is
\begin{equation}
\bar{h}_{\mu\nu}(t,\mathbf{x})=\frac{1}{4\pi}\int
	\frac{d^{3}x'}{|\mathbf{x}-\mathbf{x'}|}\cdot
	\frac{16\pi G}{c^4}T_{\mu\nu}
	\Big(t-\frac{|\mathbf{x}-\mathbf{x'}|}{c},
		\mathbf{x'}\Big).
\end{equation}

\section{Transverse traceless gauge}

From now on we consider only vacuum solutions. Suppose we use the Lorenz gauge. As shown above, we still have the freedom of coordinate transformations with $\square \xi^\mu =0$, which preserve the gauge.  So, let us choose some arbitrary (timelike) field $u^\mu$ and, in addition to the Lorenz gauge conditions, demand that the perturbation is also transverse $u^{\mu}\bar{h}_{\mu\nu}=0$ with regard to it plus that it is traceless $\bar{h}^{\mu}_{\mu}=0$. Then $h_{\mu\nu}=\bar{h}_{\mu\nu}$ and we can omit the bars. In the frame of observers with 4-velocity $u^\mu$ the full set of conditions that fix ''the transverse traceless (TT) gauge'' is then
\begin{equation}
\partial_\mu {h^{\mu}}_{\nu}=0,\quad
h_{0\mu}=0,\quad {h^{\mu}}_{\mu}=0.
\end{equation}

Simplest solutions of the vacuum wave equation are plane waves
\[h_{\mu\nu}=h_{\mu\nu}e^{ik_{\lambda}x^\lambda}.\]
Indeed, substitution into $\square h_{\mu\nu}=0$ yields
\[k^{\lambda}k_{\lambda}\cdot h_{\mu\nu}=0.\]
Thus:

1) either the wave vector is null $k^{\lambda}k_{\lambda}=0$, which roughly translates as that gravitational waves propagate with the speed of light,

2) or $h_{\mu\nu}=0$, which means that in any other (non-TT) coordinate frame, in which metric perturbation is non-zero, it is due to the oscillating coordinate system, while the true gravitational field vanishes.

\item\label{gw25}
Rewrite the gauge conditions for the TT and Lorenz gauge

1) in terms of scalar, vector and tensor decomposition;
2) in terms of the metric perturbation for the plane wave solution
\[h_{\mu\nu}=h_{\mu\nu}e^{ik_{\lambda}x^\lambda}
	=h_{\mu\nu}e^{i\omega t-ikz}\]
with wave vector $k^{\mu}=(\omega,0,0,k)$ directed along the $z$-axis.

So any plane-wave solution with $k^{\mu}=(\omega,0,0,k)$ in the $z$-direction in the TT gauge has the form (in coordinates $t,x,y,z$)
\begin{equation}
h_{\mu\nu}=\begin{pmatrix}
	0&0&0&0\\
	0&h_{+}&h_{\times}&0\\
	0&h_{\times}&-h_{+}&0\\
	0&0&0&0\end{pmatrix}
e^{ikz-i\omega t},
\end{equation}
or more generally any wave solution propagating in the $z$ direction can be presented as
\begin{align}
h_{\mu\nu}(t,z)&=
	\begin{pmatrix}
	0&0&0&0\\
	0&1&0&0\\
	0&0&-1&0\\
	0&0&0&0\end{pmatrix} h_{+}(t-z)
	+\begin{pmatrix}
	0&0&0&0\\
	0&0&1&0\\
	0&1&0&0\\
	0&0&0&0\end{pmatrix}h_{\times}(t-z)=\\
	&=e^{(+)}_{\mu\nu}h_{+}(t-z)
		+e^{(\times)}_{\mu\nu}h_{\times}(t-z).
\end{align}
Here $h_{+}$ and $h_\times$ are the amplitudes of the two independent components with linear polarization, and $e^{(\times)}_{\mu\nu},e^{(+)}_{\mu\nu}$ are the corresponding polarization tensors.

\item\label{gw26}
Show that $e^{(\times)}_{\mu\nu}$ and $e^{(+)}_{\mu\nu}$ transform into each other under rotation by $\pi/8$

\item\label{gw27}
Consider the plane wave solution of the wave equation in the Lorenz gauge:
\[\bar{h}_{\mu\nu}
	=A_{\mu\nu}e^{i\,k_{\lambda}x^\lambda},
	\quad k^\mu k_\mu =0.\]

1) Show that the TT gauge is fixed by the coordinate transformation $x\to x+\xi$ with
\begin{align}
\xi_{\mu}&=B_{\mu}e^{ik_\lambda x^\lambda};\\
B_{\lambda}
	&=-\frac{A_{\mu\nu}l^\mu l^\nu}
			{8i \omega^4}k_\lambda
	-\frac{A^{\mu}_{\mu}}{4i\omega^2}l_\lambda
	+\frac{1}{2i\omega^2}A_{\lambda\mu}l^\mu;\\
&\text{where}\quad
k^\mu=(\omega,\mathbf{k}),\quad
	l^{\mu}=(\omega,-\mathbf{k}).
\end{align}

2) What is the transformation to the Lorenz gauge for arbitrary gravitational wave in vacuum?

\item\label{gw28}
Consider the plane-wave solution, in which
\[R_{\mu\nu\rho\sigma}
	=C_{\mu\nu\rho\sigma}e^{ik_{\lambda}x^{\lambda}}.\]
1) Using the Bianchi identity, show that all components of the curvature tensor can be expressed through $R_{0\alpha0\beta}$;

2) Show that in the coordinate frame such that $k^\mu =(k,0,0,k)$ is directed along the $z$-axis the only possible nonzero components are $R_{0x0x}$, $R_{0y0y}$ and $R_{0x0y}$, obeying $R_{0x0x}=-R_{0y0y}$, leaving only two independent non-zero components;

3) in the TT gauge (denoted by the superscript $TT$)
\[R_{0\alpha 0\beta}
	=-\tfrac{1}{2}\partial_0^2 g_{\alpha\beta}^{TT};\]

\section{Gravitational waves and matter}

\item\label{gwm1}
Consider a plane wave with "$+$" polarization.
1) Write down the equations of motion for a non-relativistic particle; what happens to particles at rest?

2) Find the geodesic deviation from the Raychaudhuri equation and show that the result is the same;

\item\label{gwm2}
Find the variation of proper distance between two particles at rest in the TT frame in the presence of a plane gravitational wave with the "$+$" polarization. Show that a ring of test particles placed initially at rest in the plane orthogonal to the wave vector will be distorted into an ellipse with its axes directed along the polarization tensor's main axes and  oscillating with the wave's frequency.

\item\label{gwm3}
Derive the equation of motion and geodesic deviation in the proper frame of one of the particles.

\item\label{gwm4}
Consider the plain gravitational wave with "$+$" polarization, propagating in the $z$ direction
\[h_{\alpha \beta }(t,z)=e_{\alpha\beta}^{(+)}f(t-z)\]
and introduce new coordinates in the $(x,y)$ plane at $z=0$:
\[X=\left( 1+\frac{1}{2}f \right)x,
\quad Y=\left( 1-\frac{1}{2}f \right)y\]
The coordinates $(x,y)$ of test particles do not change with time, but $(X,Y)$ do.

Show, that the distance between the particles can be calculated in the first order by the wave's amplitude in the $(X,Y)$ coordinates using the Euclidean metric

\item\label{gwm5}
Show that far from an isolated non-relativistic system of small enough mass the spacial components of the metric perturbation in the long-wave approximation are
\[\bar{h}_{\alpha\beta}(t,\mathbf{r})
	=\frac{2}{r}\frac{d^2 I_{\alpha\beta}(t_r)}{dt^2},\]
where $t_{r}=t-r$ is retarded time and
\[I_{\alpha\beta}=\int d^3 x\;
	 x^\alpha x^\beta T_{00}(t,\mathbf{x})\]
is the second mass moment of the system.

\section{Energy of gravitational waves}

\item
Derive the quadrupole formula for the energy loss by a system due to emission of gravitational waves

\section{Binary systems}

\item
Find the strain tensor $\bar{h}_{\alpha\beta}$ of the metric perturbation far from a binary system with components of equal masses $M$. The radius of the orbit is $R$, orbital frequency $\omega$.

\item
Calculate the gravitational luminosity for the binary system with equal masses of the previous problem.

\item
Calculate the gravitational luminosity for a binary system with components of arbitrary masses $M_1$ and $M_2$.

\item
Find the rate of the two components of a binary with masses $M_1$ and $M_2$ approaching due to emission of gravitational waves. Obtain the distance between them $R$ as a function of time.

\item
Calculate the lifetime of a binary system of mass $M$ and orbital radius $R$ due to emission of gravitational waves alone. Make estimates for the Hulse-Taylor binary pulsar PSR1913+16 with the following characteristics: $M\approx 2.8 M_\odot$, $R\approx 4.5 R_\odot$, $R_\odot \approx 7\times 10^8$m.

\item
Show that the mass of a binary can be determined through the frequency of gravitational waves it generates.

\section{Gravitational Waves: scale of the phenomenon}

\item
A world champion in sprint starts the race. Estimate the portion of energy he spends that goes into production of gravitational waves.

\item
Imagine a dumb-bell  consisting of two 1-ton compact masses with their centers separated by 2 meters and spinning at 1 kHz about a line bisecting and orthogonal to their symmetry axis.Estimate the amplitude of gravitational waves at the distance of $r=300km$ from this source.

\item
Estimate the amplitude of the gravitational wave produced by a supernova explosion in our Galaxy.

\item
Consider a pair of $1.4{M}_{\odot}$ neutron stars $15Mpc$ away (e.g., near the center of the Virgo galactic cluster) on a circular orbit of $20km$ radius and orbital frequency of $400Hz$. Estimate the amplitude of gravitational wave.

\item
Estimate the energy flux of gravitational waves, registered on Earth, from a pair of merging neutron stars in the Virgo cluster with the same parameters. Compare with the energy flux from the Sun.

\item
In the Solar system the most considerable source of gravitational waves is the subsystem of Sun and Jupiter. Estimate its power.

\item
Estimate the lifetime of the hydrogen atom in the $3d$ state with respect to decay into $1s$ due to gravitational (as opposed to electromagnetic) interaction, and emission of a graviton.

\section{Generation and detection of gravitational waves}

\item
Show that effectiveness of gravitational wave production increases with the increase of mass.

\item
What is the reason for the very low efficiency of gravitational waves' production, i.e. conversion of mechanical energy into that of gravitational waves?

\item
What happens to a single particle as a gravitational wave passes through?

\item
One may wonder, how it is possible to infer the presence of an astronomical body by the gravitational waves that it emits, when it is clearly not possible to sense its much larger stationary (essentially Newtonian) gravitational potential. What gives us hope to overcome this problem?

The next three problems are inspired by S.Hughes, Listening to the Universe with gravitational-wave astronomy, arXiv:astro-ph/0210481

\item
What could be the origin of this poetic terminology?

\item
The most promising sources (neutron binaries, supernovae) of gravitational waves should give the amplitudes of the order of $h\sim 10^{-21}$. For every kilometer of baseline $L$ we need to be able to measure a distance shift of $\Delta L$ better than $10^{-16}cm$. How can we possibly hope to measure an effect that is $\sim 10^{12}$ times smaller than the wavelength of visible light (all interferometers use optical lasers)?

\item
The atoms on the surface of the interferometers’ test mass mirrors oscillate with an amplitude
\[\delta l_{atom}
	=\sqrt{\frac{kT}{m\omega^{2}}}
	\sim {10}^{-10}cm\]
at room temperature $T$, with $m$ the atomic mass, and with a vibrational frequency $\omega \sim {10}^{14}s^{-1}$. This amplitude is huge relative to the effect of the gravitational waves. Why doesn't it wash out the gravitational wave?

\end{enumerate}

\chapter{Observational Cosmology}

\section{Point gravitational lenses}
\begin{enumerate}
\item \label{ocgl1}
Compare the dependence of refraction angle $\hat{\alpha}$ on impact parameter $p$ for optical and gravitational lenses.

\item \label{ocgl2} Obtain the formula $\hat{\alpha}=r_g/p$ for refraction of light ray using the Newtonian theory.

\item \label{ocgl3} Calculate the angle of refraction of light in the gravitational field of the Sun.

\item \label{ocgl4} Propagation of light in gravitational field could be considered as propagation in a ''medium''. Calculate the effective refractive index for such a medium.

\item \label{ocgl5}Determine the dependence of a ray's shifting from the axis of symmetry after the refraction on a nontransparent lens. Find the region of shadow and estimate its size, considering the Sun as a lens.

\item \label{ocgl6}
What scales of angles and distances determine the position of the images of the light source after the passage through the gravitational lens? Consider two cases: 1) the source and the lens are at cosmological distances from the observer; 2) the distance from the observer to the lens is much smaller than the distance to the source.

\item \label{ocgl7} Show that when the gravitational lens is placed between source and observer in the general case the two images of the source would be observed. How are the images placed relative to the lens and observer?

\item \label{ocgl8} How should source, gravitational lens and observer be placed relative to each other in order to observe the Einstein ring? Calculate the radius of the ring.

\item \label{ocgl9} How would the Einstein ring change if we take into account the finite size of the source? Estimate the space characteristics of the observed image assuming that the radius of the lens is much smaller than the raius of the lens.

\item \label{ocgl10} Qualitatively consider the general situation, when a source of finite size, lens and observer are not on one line. Estimate the angular sizes of the observed images.

\item \label{ocgl11} Calculate the angular shift of the Einstein ring from the circle of the gravitational lens. Estimate it, considering the Sun as a lens. Is this value observable?

\item \label{ocgl12} Recently the exceptional phenomenon was observed using the Hubble space telescope: the double Einstein ring, formed by the influence of the gravitational field of the galaxy on the light from two other more distant galaxies. What conditions are necessary for the observation of this phenomenon?

\item \label{ocgl13} Calculate the energy amplification coefficient for images produced by the gravitational lens. Determine its peculiarities. Compare with an optical lens.

\section{Microlensing and Weak Lensing}

\item \label{ocgl14}The microlensing occurs when a lens passes in front of a distant source. Consider the conditions of existence of such an effect and its main features.

\item \label{ocgl15}In order to describe the probability of microlensing the notion of optical depth is used. Consider its physical sense. How many sources should one observe in order to register a few events per year?

\item \label{ocgl16}How can microlensing help to detect and characterize dark matter in a galactic halo?

\item \label{ocgl17}Obtain the general relation for reflection angle in the lens' field, if the surface density of the lens is given.

\item \label{ocgl18}Introduce the Jacobian of the transformation from the coordinates without lensing to the coordinates after lensing. What is its physical sense?

\item \label{ocgl19}Propose a method to reconstruct the mass distribution basing on the measured surface brightness of the images.

\end{enumerate}

\chapter{Holographic Universe}
{\it The traditional point of view assumed that dominating part of
degrees of freedom in our World are attributed to physical fields.
However it became clear soon that such concept complicates
the construction of Quantum Gravity: it is necessary to introduce small
distance cutoffs for all integrals in the theory in order to make it
sensible. As a consequence, our world should be described on a
three-dimensional discrete lattice with the period of the order of Planck
length. Lately some physicists share an even more radical
point of view: instead of the three-dimensional lattice, complete
description of Nature requires only a two-dimensional one, situated on
the space boundary of our World. This approach is based on the
so-called ''holographic principle''. The name is related to the optical
hologram, which is essentially a two-dimensional record of a
three-dimensional object. The holographic principle consists of two
main statements:
\begin{enumerate}
\item All information contained in some region of space can be
''recorded'' (represented) on the boundary of that region.
\item The theory, formulated on the boundaries of the considered
region of space, must have no more than one degree of freedom per
Planck area:
\begin{equation}
\label{Hol_f:1}
    N\le \frac{A}{A_{pl}},\quad A_{pl}=\frac{G\hbar}{c^3}.
\end{equation}
\end{enumerate}
Thus, the key piece in the holographic principle is
the assumption that all the information about the Universe can be
encoded on some two-dimensional surface --- the holographic
screen. Such approach leads to a new interpretation of cosmological
acceleration and to an absolutely unusual understanding of Gravity. The
gravity is understood as an entropy force, caused by variation of
information connected to positions of material bodies. More
precisely, the quantity of information related to matter and its
position is measured in terms of entropy. Relation between the
entropy and the information states that the information change is
exactly the negative entropy change $\Delta I=-\Delta S$. Entropy change
due to matter displacement leads to the so-called entropy force, which,
as will be proven below, has the form of gravity. Its origin
therefore lies in the universal tendency of any macroscopic theory
to maximize the entropy. The dynamics can be constructed in terms
of entropy variation and it does not depend on the details of
microscopic theory. In particular, there is no fundamental field
associated with the entropy force. The entropy forces are typical
for macroscopic systems like colloids and biophysical systems. Big
colloid molecules, placed in thermal environment of smaller
particles, feel the entropy forces. Osmose is another phenomenon
governed by the entropy forces.

Probably the best known example of the entropy force is the elasticity of a polymer
molecule. A single polymer molecule can be modeled as a
composition of many monomers of fixed length. Each monomer can
freely rotate around the fixation point and choose any spacial
direction. Each of such configurations has the same energy. When the
polymer molecule is placed into a thermal bath, it prefers to form a
ring as the entropically most preferable configuration: there are
many more such configurations when the polymer molecule is short,
than those when it is stretched. The statistical tendency to transit
into the maximum entropy state transforms into the macroscopic
force, in the considered case---into the elastic force.

Let us consider a small piece of holographic screen and a particle
of mass $m$ approaching it. According to the holographic principle,
the particle affects the amount of the information (and therefore of the
entropy) stored on the screen. It is natural to assume that
entropy variation near the screen is linear on the displacement
$\Delta x$:
\begin{equation}
\Delta S = 2\pi k_B \frac{mc}{\hbar} \Delta x. \label{delta_s}
\end{equation}
The factor $2\pi$ is introduced for convenience, which the reader
will appreciate solving the problems of this Chapter. In order to
understand why this quantity should be proportional to mass,
let us imagine that the particle has split into two or more particles
of smaller mass. Each of those particles produces its own entropy
change when displaced by $\Delta x$. As entropy and mass are both
additive, then it is natural that the former is proportional to the
latter. According to the first law of thermodynamics, the entropy
force related to information variation satisfies the equation
\begin{equation}
F\Delta x = T\Delta S. \label{delta_x}
\end{equation}
If we know the entropy gradient, which can be found from (\ref{delta_s}),
and the screen temperature, we can calculate the entropy
force.

An observer moving with acceleration $a$, feels the
temperature (the Unruh temperature)
\begin{equation}
\label{Hol_f_Unruh:4}
k_B T_U=\frac{1}{2\pi}\frac\hbar c a.
\end{equation}
Let us assume that the total energy of the system equals $E$. Let us
make a simple assumption that the energy is uniformly distributed
over all $N$ bits of information on the holographic screen. The
temperature is then defined as the average energy per bit:
\begin{equation}
E =\frac12 N k_B T. \label{average_e}
\end{equation}
Equations (\ref{delta_s})--(\ref{average_e}) allow one to describe the holographic
dynamics, and as a particular case---the dynamics of the Universe, and
all that without the notion of Gravity.}

\section{Information, Entropy and Holographic Screen}
\begin{enumerate}
\item \label{HU01}  Show that information density on holographic screen is limited by the value
 $\sim 10^{69} \: \mbox{bit/m}^2.$

\item \label{HU02} Choosing the Hubble sphere as a holographic screen, find its area in the de Sitter model (recall that in this model the Universe dynamics is determined by the cosmological constant
 $\Lambda >0$).

\item \label{HU03}  Find the entropy of a quantum system composed of $N$ spin-$1/2$ particles.

\item \label{HU04}   Consider a 3D lattice of spin-$1/2$ particles and prove that entropy of such a system defined in a standard way is proportional to its volume.

\item \label{HU05}   Find the change of the holographic screen area when it is crossed by a spin-$1/2$ particle.

\item \label{HU06}   What entropy change corresponds to the process described in the previous problem?

\item \label{HU07}   According to Hawking, black holes emit photons (the black hole evaporation) with thermal  spectrum and effective temperature
$$
T_{BH}=\frac{\hbar c}{4\pi k_{_B}}r_{g},
$$
where $r_{g}={2GM}/{c^2}$ is the black hole Schwarzschild radius. Neglecting
accretion and CMB absorption, determine the life time of a black hole
with initial mass $M_{0}.$  Calculate the lifetime for black holes
with the mass equal to Planck mass, mass of the Earth and Solar mass.

\item \label{HU08}   Show that black holes have negative thermal capacity.

\item \label{HU09}  Show that the black hole evaporation process is accompanied by increase of its temperature. Find the relative variation of the black hole temperature if its mass gets twice smaller due to evaporation.
    
\item \label{HU10}   Show that in order to produce an entropy force when a particle approaches the holographic screen, its temperature must be finite.

\item \label{HU11}  Determine the entropic acceleration of a particle crossing the holographic screen with temperature ${{T}_{b}}$.

\item \label{HU12}  Show that the Unruh temperature is the Hawking radiation temperature $$T_{BH}=\frac{\hbar c^3}{8\pi k_{_B}GM}$$ with substitution $a\to g,$ where $g$ is the surface gravity of the black hole.

\item \label{HU13}  In order to verify experimentally the Unruh effect, it is planned to accelerate particles with acceleration of the order of  $10^{26} \mbox{\it m/}\sec^2$. What vacuum temperature does this acceleration correspond to?

\item \label{HU14}   Derive the second Newton's law from the holographic principle.

\item \label{HU15}  Show that the inertia law can be obtained from the holographic principle.

\item \label{HU16}   Obtain the Newton's law of universal gravitation using the holographic principle.

\item \label{HU17}   Compare how close to a black hole are Earth, Sun and observable Universe.

\item \label{HU18}   Estimate the temperature of the Hubble sphere, considering it as a holographic screen.

\item \label{HU19}  Show that equilibrium between the relic radiation and holographic screen is possible only at Planck temperature.

\item \label{HU20}   According to the holographic ideology, all physical phenomena can be described by the boundary layer theory. Therefore a conclusion comes that one should account for contribution of such surface terms while deriving the equations of General Relativity. Show that consideration of the boundary terms in the Einstein-Hilbert action is equivalent to introduction of non-zero energy-momentum tensor into the standard Einstein equations.

\item \label{HU21}  Derive the Friedman equations from the holographic principle.

\item \label{HU22}   Derive the Friedman equations from the holographic principle for a non-flat Universe.

\item \label{HU23}   Consider an effective field theory with the ultraviolet cutoff parameter equal to $\Lambda$ and entropy satisfying the inequality
\[S\equiv L^3\Lambda^3\le S_{BH}\simeq L^2 M_{Pl}^2,\]
where $S_{BH}$ is the Beckenstein-Hawking entropy, and show that such a theory necessarily contains the states with Schwarzschild radius $R_S$ much greater than the linear dimensions $L$ of the system. (see Zimdahl, Pavon, 0606555)

\item \label{HU24}   Estimate the dark energy density assuming that the total energy in a region with linear size $L$ cannot exceed the black hole mass of the same size.

\item \label{HU25}  Find the correspondence between the ultraviolet and infrared cutoff scales.

\item \label{HU26}   Show that the entropy force produces negative pressure.

\item \label{HU27}   Taking the Hubble sphere for the holographic screen, and using the SCM parameters, find the entropy, the force acting on the screen and the corresponding pressure.

\item \label{HU28}   The most popular approach to explain the observed accelerated expansion of the Universe assumes introduction of dark energy in the form of cosmological constant into the Friedman equations. As we have seen in the corresponding Chapter, this approach is successfully realized in SCM. Unfortunately, it leaves aside the question of the nature of the dark energy. An alternative approach can be developed in the frame of holographic dynamics. In this case it is possible to explain the observations without the dark energy. It is replaced by the entropy force, acting on the cosmological horizon (Hubble sphere) and leading to the accelerated Universe's expansion.

    Show that the Hubble sphere acceleration obtained this way agrees with the result obtained in SCM.

\item \label{HU29}   Plot the dependence of the deceleration parameter on the red shift in the Universe composed of non-relativistic matter. Take into account the negative pressure generated by the entropy force (see Problem \ref{HU26}). Compare the result with the SCM.

\item \label{HU30}   Show that the coincidence problem does not arise in the models with the holographic dark energy.

\item \label{HU31}  Show that the holographic screen in the form of Hubble sphere cannot explain the accelerated expansion of Universe.

\item \label{HU33}  Obtain the equation of motion for the relative density of the holographic dark energy in the case when the particle horizon serves as the holographic screen.

\item \label{HU34}  Solve the equation of motion obtained in the previous problem.

\item \label{HU35}   Find dependence the of holographic dark energy density on the scale factor taking the cosmological event horizon $R_h$ as the holographic screen. Find the equation of state parameter for such a dark energy.

\item \label{HU36}  Obtain the equation of motion for the relative density of the holographic dark energy in the case when the event horizon serves as a holographic screen.

\item \label{HU37}   Solve the equation of motion obtained in the previous problem.

\item \label{HU38}  Find the dependence of cosmological parameters (Hubble parameter $H$, state equation parameter $w$ and deceleration parameter $q$) on red shift in the model of
the Universe composed of radiation, non-relativistic matter and
holographic dark energy, taking as the holographic screen the Ricci
scalar's $R$ characteristic length
$$
\rho_{_{RDE}}=-\frac{\alpha}{16\pi} R
=\frac{3\alpha}{8\pi}\left(\dot{H}+2H^2+\frac{ k}{a^2}\right),
$$
where $\alpha $ is a positive constant, $k $ is the sign of space
curvature.

\item \label{HU39}  Find the dependence of holographic dark energy density on the scale factor for the model of the previous problem.

\item \label{HU40}   Find the pressure of holographic dark energy for the model of Problem \ref{HU38}.

\item \label{HU41}   In the frame of the holographic dark energy model considered in Problem \ref{HU38}, find the dependence $w(z)$ using the SCM parameters:
$\Omega_{k0}=0$, $\Omega_{m0}=0.27$, $\Omega_{r0}=8.1 \times 10^{-5}$,$\Omega_{X0}=0.73$, $w_0=-1$.

\item \label{HU42}   Find the state equation parameter for the holographic dark energy in the agegraphic dark energy model, in which
the holographic screen is taken in the form of the surface situated at the
distance traversed by light during the Universe's lifetime $T$.

\item \label{HU43}  Find the equation of motion for the relative density of agegraphic
dark energy in the model considered in the previous problem.

\item \label{HU44}   Using the equation of motion for the relative density of agegraphic
dark energy, obtained in the previous problem, find its dependence on scale factor in the Universe dominated by non-relativistic matter.

\item \label{HU45}  Using the equation of motion for the relative density of agegraphic
dark energy, obtained in Problem \ref{HU43}, find its dependence on scale factor in the case of the dark energy domination.

\item \label{HU46}   Derive the exact solution of the equation of motion obtained in Problem \ref{HU43}.

\item \label{HU47}   Find the state equation parameter for the agegraphic dark energy
in the case when this form of dark energy interacts with dark
matter: \[  \dot{\rho}_m+3H\rho_m=Q,\] \[
\dot{\rho}_q+3H(1+w_q)\rho_q=-Q.\]

\item \label{HU48}  In the model of interacting agegraphic dark energy and dark matter find the equation of motion for the relative density of agegraphic dark energy.

{\it Isolated systems are known to evolve towards the equilibrium state in such a way that entropy $S(x)$ never decreases
\[\frac{dS(x)}{dx}\ge0,\]
while
\[\frac{d^2S(x)}{dx^2}<0.\]
In the context of eternally expanding FRW-cosmology the above mentioned conditions imply that the entropy available to comoving observer plus the entropy on the cosmological horizon satisfy the conditions (generalized second law of thermodynamics)
\[S'(a)\ge0,\ S''(a)\le0\]
where the prime stands for differentiation with respect to scale factor $d/da$ and $a\to\infty.$ The latter condition means that we deal below with late stages of the Universe evolution.}

\item \label{HU49}   Show that in the Friedmannian Universe with energy density $\rho$ the cosmological horizon area equals
\[A=\frac{3}{2G}\frac 1 \rho.\]

\item \label{HU50}   Show that in the Friedmannian Universe filled with a substance with state equation $p=w\rho$ the cosmological horizon area grows
 with the Universe's expansion under the condition $1+w>0$.

 \item \label{HU51}  Using the results of Problem \ref{HU49}, find the entropy of the Universe.

\item \label{HU52}   Find the state equation for the substance, filling the Universe with $A''\le0$.

\end{enumerate}

\chapter{Horizons}
In this chapter we assemble the problems that concern one of the most distinctive features of General Relativity and Cosmology --- the horizons. The first part gives an elementary introduction into the concept in the cosmological context, following and borrowing heavily from the most comprehensible text by E. Harrison \cite{Harrison}. Then we move to more formal exposition of the subject, making use of the seminal works of W. Rindler \cite{Rindler} and G.F.R. Ellis, T. Rothman \cite{Ellis}, and consider first simple, and then composite models, such as $\Lambda$CDM. The fourth section elevates the rigor one more step and explores the causal structure of different simple cosmological models in terms of conformal diagrams, following mostly the efficient approach of V. Mukhanov \cite{Muchanov}. The section on black holes relates the general scheme of constructing conformal diagrams for stationary black hole spacetimes, following mostly the excellent textbook of K. Bronnikov and S. Rubin \cite{BrRubin}. The consequent sections focus on more specific topics, such as the various problems regarding the Hubble sphere, inflation and holography.

\section{Simple English}
A vague definition of a horizon can be accepted as the following: it is a frontier between things observable and things unobservable.

\textit{Particle horizon.} If the Universe has a finite age, then light travels only a finite distance in that time and the volume of space from which we can receive information at a given moment of time is limited. The boundary of this volume is called the particle horizon.

\textit{Event horizon.} The event horizon is the complement of the particle horizon. The event horizon encloses the set of points from which signals sent at a given moment of time will never be received by an observer in the future.

Space-time diagram is a representation of space-time on a two-dimensional plane, with one timelike and one spacelike coordinate. It is typically used for spherically symmetric spacetimes (such as all homogeneous cosmological models), in which angular coordinates are suppressed.

\begin{enumerate}
\item Draw a space-time diagram that shows behaviour of worldlines of comoving observers in a
\begin{enumerate}
\item stationary universe with beginning
\item expanding universe in comoving coordinates
\item expanding universe in proper coordinates
\end{enumerate}

\item Suppose there is a static universe with homogeneously distributed galaxies, which came into being at some finite moment of time. Draw graphically the particle horizon for some static observer.

\item How does the horizon for the given observer change with time?

\item Is there an event horizon in the static Universe? What if the Universe ends at some finite time?

\item Consider two widely separated observers, A and B (see Fig. \ref{Conf-Harr-Hor}). Suppose  they have overlapping horizons, but each can apparently see things that the other cannot. We ask: Can B communicate to A information that extends A's knowledge of things beyond his horizon? If so, then a third observer C may communicate to B information that extends her horizon, which can then be communicated to A. Hence, an unlimited sequence of observers B, C, D, E,\ldots may extend A's knowledge of the Universe to indefinite  limits. According to this argument A has no true horizon. This is the horizon riddle. Try to resolve it for the static Universe.
\begin{figure}[!htb]
\center
\includegraphics*[width=0.6\textwidth]{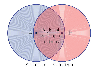}
\caption{\label{Conf-Harr-Hor} The horizon riddle: can two observers with overlapping horizons pass information to each other regarding things outside of the other's horizon?}
\end{figure}

\item Suppose observer O in a stationary universe with beginning sees A in some direction at distance $L$ and B in the opposite direction, also at distance $L$. How large must $L$ be in order for A and B to be unaware of each other's existence at the time when they are seen by O?

\item Draw spacetime diagrams in terms of comoving coordinate and conformal time and determine whether event or particle horizons exist for:
\begin{enumerate}
\item the universe which has a beginning and an end in conformal time. The closed Friedman universe that begins with Big Bang and ends with Big Crunch belongs to this class.

\item the universe which has a beginning but no end in conformal time.  The Einstein--de Sitter universe and the Friedman universe of negative curvature, which begin with a Big Bang and expand forever, belong to this class.

\item the universe which has an end but no beginning in conformal time. The de Sitter and steady-state universes belong to this class.

\item the universe which has no beginning and no ending in conformal time, as in the last figure of Fig. \ref{Conf-Harr15-18}. The Einstein static and the Milne universes are members of this class.
\end{enumerate}
Conformal time is the altered time coordinate $\eta=\eta (t)$, defined in such a way that lightcones on the spacetime diagram in terms of $\eta$ and comoving spatial coordinate are always straight diagonal lines, even when the universe is not stationary.

\item Draw the spacetime diagram in terms of comoving space and ordinary time or the universe with an end but no beginning in conformal time.

\item Formulate the necessary conditions in terms of conformal time for a universe to provide a comoving observer with
\begin{enumerate}
\item a particle horizon
\item an event horizon
\end{enumerate}

\item Consider two galaxies, observable at present time, $A$ and $B$. Suppose at the moment of detection of light signals from them (now) the distances to them are such that $L_{det}^{A}<L^{B}_{det}$. In other words, if those galaxies had equal absolute luminosities, the galaxy $B$ would seem to be dimmer. Is it possible for galaxy $B$ (the dimmer one) to be closer to us at the moment of its signal's emission than galaxy $A$ (the brighter one) at the moment of $A$'s signal's emission?

\item Show on a spacetime diagram the difference in geometry of light cones in universes with and without particle horizons.

\end{enumerate}

\section{Simple Math}
The problems of this section need basic understanding of Friedman equations, definitions of proper, comoving and conformal coordinates, the cosmological redshift formula and simple cosmological models (see Chapters 2 and 3).

Let us make our definitions a little more strict.

A \textbf{particle horizon}, for a given observer $A$ and cosmic instant $t_0$ is a surface in the instantaneous three-dimensional section $t=t_0$ of space-time, which divides all comoving particles\footnote{Rindler uses the term ``fundamental observers''} into two classes: those that have already been observable by $A$ up to time $t_0$ and those that have not. 

An \textbf{event horizon}, for a given observer $A$, is a hyper-surface in space-time, which divides all events into two non-empty classes: those that have been, are, or will be observable by $A$, and those that are forever outside of $A$'s possible powers of observation. It follows from definition that event horizon, and its existence, depend crucially on the observer's (and the whole Universe's) future as well as the past: thus it is said to be an essentially \emph{global concept}. It is formed by null geodesics.

The following notation is used hereafter: $L_p (t_0)$ is the proper distance from observer $A$ to its particle horizon, measured along the slice $t=t_0$. For brevity, we will also call this distance simply ``the particle horizon in proper coordinates'', or just ``particle horizon''. The corresponding comoving distance $l_p$ is the particle horizon in comoving coordinates.

Likewise, $L_e$ is the proper distance from an observer to its event horizon (or, rather, its section with the hypersurface $t=t_0$), measured also along the slice $t=t_0$. It is called ``space event horizon at time $t_0$'', or just the event horizon, for brevity. The respective comoving distance is denoted $l_e$.

\begin{enumerate}[resume]
\item The proper distance $D_p (t_0)$ between two comoving observers is the distance measured between them at some given moment of cosmological time $t=t_0$. It is the quantity that would be obtained if all the comoving observers between the given two measure the distances between each other at one moment $t=t_0$ and then sum all of them up. Suppose one observer detects at time $t_{0}$ the light signal that was emitted by the other observer at time $t_e$. Find the proper distance between the two observers at $t_0$ in terms of $a(t)$.

\item Show, that the proper distance $L_p$ to the particle horizon at time $t_0$ is
\begin{equation}
L_p (t_0)=\lim\limits_{t_e \to  0}D_p (t_e ,t_0). \label{LpDp}
\end{equation}

\item The past light cone of an observer at some time $t_0$ consists of events, such that light emitted in each of them reaches the selected observer at $t_0$. Find the past light cone's equation in terms of proper distance vs. emission time $D_{plc} (t_e)$. What is its relation to the particle horizon?

\item The simplest cosmological model is the one of \emph{Einstein-de Sitter}, in which the Universe is spatially flat and filled with only dust, with $a(t)\sim t^{2/3}$. Find the past light cone distance $D_{plc}$ (\ref{Dplc}) for Einstein-de Sitter.

\item Demonstrate that in general $D_{plc}$ can be non-monotonic. For the case of Einstein-de Sitter show that its maximum -- the maximum emission distance -- is equal to $8/27 L_H$, while the corresponding redshift is $z=1.25$.

\item In a matter dominated Universe we see now, at time $t_0$, some galaxy, which is now on the Hubble sphere. At what time in the past was the photon we are registering emitted?

\item Show that the particle horizon in the Einstein-de Sitter model recedes at three times the speed of light.

\item Does the number of observed galaxies in an open Universe filled with dust increase or decrease with time?

\item Draw the past light cones $D_{plc}(t_e)$ for Einstein-de Sitter and a Universe with dominating radiation on one figure; explain their relative position.

\item Find the maximum emission distance and the corresponding redshift for power law expansion $a(t)\sim (t/t_0 )^n$.

\item Show that the most distant point on the past light cone was exactly at the Hubble sphere at the moment of emission of the light signal that is presently registered.

\item Show that the comoving particle horizon is the age of the Universe in conformal time

\item Show that
\begin{align}
\frac{dL_p}{dt}&=L_p (z)H(z)+1;\\
\frac{dL_e}{dt}&=L_e (z)H(z)-1.
\end{align}

\item Find $\ddot{L_p}$ and $\ddot{L_e}$.

\item Show that observable part of the Universe expands faster than the Universe itself. In other words, the observed fraction of the Universe always increases.

\item Show that the Milne Universe has no particle horizon.

\item Consider a universe which started with the Big Bang, filled with one matter component. How fast must $\rho(a)$ decrease with $a$ for the particle horizon to exist in this universe?

\item Calculate the particle horizon for a universe with dominating
\begin{enumerate}
\item radiation;
\item matter.
\end{enumerate}

\item Consider a flat universe with one component with state equation  $p =w \rho$. Find the particle horizon at present time $t_0$.

\item Show that in a flat universe in case of domination of one matter component with equation of state $p=w\rho$, $w>-1/3$
\begin{equation}
L_{p} (z)=\frac{2}{H(z) (1+3w)},\qquad \dot{L_p}(z)=\frac{3(1+w)}{(1+3w)}.
\end{equation}

\item Show that in a flat universe in case of domination of one matter component with equation of state $p=w\rho$, $w<-1/3$
\begin{equation}
L_{e} (z)=-\frac{2}{H(z) (1+3w)},\qquad \dot{L_p}(z)=-\frac{3(1+w)}{(1+3w)}.
\end{equation}

\end{enumerate}

\begin{enumerate}[resume]
\item Estimate the particle horizon size at matter-radiation equality.
\end{enumerate}

\section{Composite models}

\begin{enumerate}[resume]

\item Consider a flat universe with several components $\rho=\sum_i \rho_i$, each with density $\rho_i$ and partial pressure $p_i$ being related by the linear state equation  $p_i =w_i \rho_i$. Find the particle horizon at present time $t_0$.

\item Suppose we know the current material composition of the Universe $\Omega_{i0}$, $w_i$ and its expansion rate as function of redshift $H(z)$. Find the particle horizon $L_p (z)$ and the event horizon $L_p (z)$ (i.e the distances to the respective surfaces along the surface $t=const$) at the time that corresponds to current observations with redshift $z$.

\item When are $L_p$ and $L_e$ equal? It is interesting to know whether both horizons might have or not the same values, and if so, how often this could happen.

\item Find the particle and event horizons for any redshift $z$ in the standard cosmological model -- $\Lambda$CDM.

\item Express the particle $L_p (z)$ and event $L_e (z)$ horizons in $\Lambda$CDM through the hyper-geometric function.
\end{enumerate}

\section{Causal structure}
The causal structure is determined by propagation of light and is best understood in terms of \emph{conformal diagrams}. In this section we construct and analyze those for a number of important model cosmological solutions (which are assumed to be already known), following mostly the exposition of \cite{Muchanov}.

In terms of comoving distance $\tilde{\chi}$ and conformal time $\tilde{\eta}$ (in this section they are denoted by tildes) the two-dimensional radial part of the FLRW metric takes form
\begin{equation}
	ds_{2}^{2}=a^{2}(\tilde{\eta})\big[d\tilde{\eta}^2 -d\tilde{\chi}^2\big].
	\label{ds2Dconf}
\end{equation}
In the brackets here stands the line element of two-dimensional Minkowski flat spacetime. Coordinate transformations that preserve the \emph{conformal} form of the metric
\begin{equation*}
ds_2^2 =\Omega^{2}(\eta,\chi)\big[d\eta^2 -d\chi^2 \big],
\end{equation*}
are called conformal transformations, and the corresponding coordinates $(\eta,\chi)$ -- conformal coordinates.

\begin{enumerate}[resume]
\item Show that it is always possible to construct $\eta(\tilde{\eta},\tilde{\chi})$, $\chi(\tilde{\eta},\tilde{\chi})$, such that the conformal form of metric (\ref{ds2Dconf}) is preserved, but $\eta$ and $\chi$ are bounded and take values in some finite intervals. Is the choice of $(\eta,\chi)$ unique?
\end{enumerate}
In this section we will reserve notation $\eta$ and $\chi$ and names ``conformal coordinates'' and ``conformal variables'' to such variables that can only take values in a bounded region on $\mathbb{R}^2$; $\tilde{\eta}$ and $\tilde{\chi}$ can span infinite or semi-infinite intervals. Spacetime diagram in terms of conformal variables $(\eta,\chi)$ is called conformal diagram. Null geodesics $\eta=\pm \chi + const$ are diagonal straight lines on conformal diagrams.

\begin{enumerate}[resume]
\item Construct the conformal diagram for the closed Universe filled with
\begin{enumerate}
\item radiation;
\item dust;
\item mixture of dust and radiation.
\end{enumerate}
Show the particle and event horizons for the observer at the origin $\chi=0$ (it will be assumed hereafter that the horizons are always constructed with respect to this chosen observer).

\item Construct the conformal diagram for the de Sitter space in the closed sections coordinates. Provide reasoning that this space is (null) geodesically complete, i.e. every (null) geodesic extends to infinite values of affine parameters at both ends.

\item  Rewrite the metric of de Sitter space (\ref{dSclosed}) in terms of ``static coordinates'' $T,R$:
\begin{equation}
\tanh (H_\Lambda T)=-\frac{\cos\eta}{\cos\chi},\qquad
H_\Lambda R =\frac{\sin\chi}{\sin\eta}.
\end{equation}
\begin{enumerate}
\item What part of the conformal diagram in terms of $(\eta,\chi)$ is covered by the static coordinate chart $(T,R)$?
\item Express the horizon's equations in terms of $T$ and $R$
\item Draw the surfaces of constant $T$ and $R$ on the conformal diagram.
\item Write out the coordinate transformation between $(\eta,\chi)$ and $(T,R)$ in the regions where $|\cos\eta|>|\cos\chi|$. Explain the meaning of $T$ and $R$.
\end{enumerate}

\item  The scale factor in flat de Sitter is $a(t)=H_\Lambda^{-1} e^{H_\Lambda t}$.
\begin{enumerate}
\item Find the range of values spanned by conformal time $\tilde{\eta}$ and comoving distance $\tilde{\chi}$ in the flat de Sitter space
\item Verify that coordinate transformation
\begin{equation}
\tilde{\eta}=\frac{-\sin\eta}{\cos\chi-\cos\eta},\qquad
\tilde{\chi}=\frac{\sin\chi}{\cos\chi-\cos\eta}
\end{equation}
bring the metric to the form of that of de Sitter in closed slicing (it is assumed that $\tilde{\eta}=0$ is chosen to correspond to infinite future).
\item Which part of the conformal diagram is covered by the coordinate chart $(\tilde{\eta},\tilde{\chi})$? Is the flat de Sitter space geodesically complete?
\item Where are the particle and event horizons in these coordinates?
\end{enumerate}

\item  What parts of the spacetime's boundary on the conformal diagram of flat de Sitter space corresponds to
\begin{enumerate}
\item spacelike infinity $i^0$, where $\tilde{\chi}\to +\infty$;
\item past timelike infinity $i^-$, where $\tilde{\eta}\to -\infty$ and from which all timelike worldlines emanate
\item past lightlike infinity $J^-$, from which all null geodesics emanate?
\end{enumerate}

\item  Consider the de Sitter space in open slicing, in which $a(t)=H_\Lambda \sinh (H_\Lambda t)$, so conformal time is
\begin{equation}
\tilde{\eta} =\int\limits_{+\infty}^{t} \frac{dt}{a(t)},
\end{equation}
where again the lower limit is chosen so that the integral is bounded.
\begin{enumerate}
\item Find $\tilde{\eta}(t)$ and verify that coordinate transformation from $(\tilde{\eta},\tilde{\chi})$ to $\eta,\chi$, such that
\begin{equation}
\tanh\tilde{\eta}=\frac{-\sin\eta}{\cos\cos\chi},\qquad
\tanh\tilde{\chi}=\frac{\sin\chi}{\cos\eta}
\end{equation}
brings the metric to the form of de Sitter in closed slicing.
\item What are the ranges spanned by $(\tilde{\eta},\tilde{\chi})$ and $(\eta,\chi)$? Which part of the conformal diagram do they cover?
\end{enumerate}

\item Rewrite the Minkowski metric in terms of coordinates $(\eta,\chi)$, which are related to $(t,r)$ by the relation
\begin{equation}
\tanh \tilde{\eta}=\frac{\sin\eta}{\cos\chi},\qquad
\tanh \tilde{\chi}=\frac{\sin\chi}{\cos\eta}
\end{equation}
that mirrors the one between the open and closed coordinates of de Sitter. Construct the conformal diagram and determine different types of infinities. Are there new ones compared to the flat de Sitter space?

\item The choice of conformal coordinates is not unique. Construct the conformal diagram for Minkowski using the universal scheme: first pass to null coordinates, then bring their span to finite intervals with $\arctan$ (one of the possible choices), then pass again to timelike and spacelike coordinates.

\item Draw the conformal diagram for the Milne Universe and show which part of Min\-kow\-ski space's diagram it covers.

\item Consider open or flat Universe filled with matter that satisfies strong energy condition $\varepsilon +3p>0$. What are the coordinate ranges spanned by the comoving coordinate $\tilde{\chi}$ and conformal time $\tilde{\eta}$? Compare with the Minkowski metric and construct the diagram. Identify the types of infinities and the initial Big Bang singularity.

\item Draw the conformal diagram for open and flat Universes with power-law scale factor $a(t)\sim t^{n}$, with $n>1$. This is the model for the power-law inflation. Check whether the strong energy condition is satisfied.
\end{enumerate}

\section{Conformal diagrams: stationary black holes}
In the context of black hole spacetimes there are many subtly distinct notions of horizons,
 with the most useful being different from those used in cosmology. In particular, particle horizons do not play any role. The event horizon is defined not with respect to some selected observer, but with respect to \emph{all} external observers: in an asymptotically flat spacetime\footnote{Meaning the spacetime possesses the infinity with the same structure as that of Minkowski, which is important.} a future event horizon is the hypersurface which separates the events causally connected to future infinity and those that are not. Likewise the past event horizon delimits the events that are causally connected with past infinity or not. Another simple but powerful concept is the  \emph{Killing horizon}: in a spacetime with a Killing vector field $\xi^\mu$ it is a (hyper-)surface, on which $\xi^\mu$ becomes lightlike. We will see explicitly for the considered examples that the Killing horizons are in fact event horizons by constructing the corresponding conformal diagrams. In general, in the frame of GR a Killing horizon in a stationary spacetime is (almost) always an event horizon and also coincides with most other notions of horizons there are.

This section elaborates on the techniques of constructing conformal diagrams for stationary black hole solutions. The construction for the Schwarzschild black hole, or its variation, can be found in most textbooks on GR; for the general receipt see \cite{BrRubin}.

\subsection{Schwarzschild-Kruskal black hole solution}
\begin{enumerate}[resume]
\item  The simplest black hole solution is that of Schwarzschild, given by
\begin{equation}
ds^{2}=f(r)dt^2 -\frac{dr^2}{f(r)}-r^2 d\Omega^2,\qquad f(r)=1-\frac{r_g}{r},
	\label{SchwBH}
\end{equation}
where $r_g$ is the gravitational radius, and $d\Omega^2$ is the angular part of the metric, which we will not be concerned with. The surface $r=r_g$ is the horizon. Focus for now only on the external part of the solution,
\[\big\{-\infty<t<+\infty,\;\;
	 r_g<r<+\infty)\big\}.\]
The general procedure of building a conformal diagram for the $(t,r)$ slice, as discussed in the cosmological context earlier, works here perfectly well, but needs one additional step in the beginning:
\begin{enumerate}
\item use a new radial coordinate to bring the metric to conformally flat form;
\item pass to null coordinates;
\item shrink the ranges of coordinate values to finite intervals with the help of $\arctan$;
\item return to timelike and spacelike coordinates.
\end{enumerate}
Identify the boundaries of Schwarzschild's exterior region on the conformal diagram and compare it with Minkowski spacetime's.

\item  The region $r\in (0,r_g)$ represents the black hole's interior, between the horizon $r=r_g$ and the singularity $r=0$. Construct the conformal diagram for this region following the same scheme as before.
\begin{enumerate}
\item Which of the coordinates $(t,r)$ are timelike and which are spacelike?
\item Is the singularity spacelike or timelike?
\item Is the interior solution static?
\end{enumerate}
The Schwarzschild black hole's interior is an example of the \emph{T-region}, where $f(r)<0$, as opposed to the \emph{R-region}, where  $f(r)>0$.

\item  Consider radial motion of a massive particle and show that the exterior and interior parts of the Schwarzschild are not by themselves geodesically complete, i.e. particle's worldlines are terminated at the horizon at finite values of affine parameter. Show that, on the other hand, the horizon is not a singularity, by constructing the null coordinate frame, in which the metric on the horizon is explicitly regular.

\item  Geodesic incompleteness means the full conformal diagram must be assembled from the parts corresponding to external and internal solutions by gluing them together along same values of $r$ (remember that each point of the diagram corresponds to a sphere). Piece the puzzle.

Note that a) there are two variants of both external and internal solutions' diagrams, differing with orientation and b) the boundaries of the full diagram must go along either infinities or singularities.

\end{enumerate}

\subsection{Other spherically symmetric black holes}
The Schwarzschild metric (\ref{SchwBH}) is the vacuum spherical symmetric solution of Einstein's equation. If we add some matter, we will obtain a different solution, e.g. Reissner-Nordstr\"{o}m for an electrically charged black hole. In all cases, spherical symmetry means that in appropriately chosen coordinate frame the metric takes the same form (\ref{SchwBH}),
\begin{equation}
 ds^2 =f(\rho)dt^2 -\frac{d\rho^2}{f(\rho)}-r^2 (\rho)d\Omega^2 ,
\end{equation}
but with some different function $f(\rho)$: one can always ensure that the metric functions have this form by choosing the appropriate radial coordinate\footnote{It is sometimes called the ``quasiglobal coordinate''} $\rho$. The angular part $\sim d\Omega^2$ does not affect causal structure and conformal diagrams. The zeros $\rho=\rho^\star$ of $f(\rho)$ define the surfaces, which split the full spacetime into R- and T-regions and are the horizon candidates.

\begin{enumerate}[resume]
\item  Consider radial geodesic motion of a massive or massless particle. Make use of the integral of motion $E=-u^\mu \xi_\mu$ due to the Killing vector and find $\rho (t)$  and $\rho(\lambda)$, where $\lambda$ is the affine parameter (proper time $\tau$ for massive particles)
\begin{enumerate}
\item When is the proper time of reaching the horizon $\tau^\star =\tau (\rho^\star)$ finite?
\item Verify that surface $\rho=\rho^\star$ is a Killing horizon
\end{enumerate}

\item  Let us shift the $\rho$ coordinate so that $\rho^\star =0$, and assume that
\begin{equation}
f(\rho)=\rho^{q}F(\rho),\qquad q\in\mathbb{N},
\end{equation}
where $F$ is some analytic function, with $F(0)\neq 0$. Suppose we want to introduce a new radial ``tortoise'' coordinate $x$, such that the two-dimensional part of the metric in terms of $(t,x)$ has the conformally flat form:
\begin{equation}
ds^2_2 =f(\rho(x))\big[dt^2 -dx^2 \big].
\end{equation}
\begin{enumerate}
\item What is the asymptotic form of relation $\rho (x)$?
\item Rewrite the metric in terms of null coordinates
\begin{equation}
V=t+x,\qquad W=t-x.
\end{equation}
Where is the horizon in terms of these coordinates? Is there only one?
\item Suppose we pass to new null coordinates $V=V(v)$ and $W=W(w)$. What conditions must be imposed on functions $V(v)$ and $W(w)$ in order for the mixed map $(v,W)$ to cover the past horizon and the map $(w,V)$ to cover the future horizon without singularities?
\end{enumerate}

\end{enumerate}
Thus a horizon candidate $\rho=\rho^\star$ is defined by condition $f(\rho^\star)=0$. If $\rho^\star =\pm\infty$, then the proper time of reaching it diverges, so it is not a horizon, but an infinitely remote boundary of spacetime.  It can be a remote horizon if $x^\star =\pm \infty$. In case $x^\star =O(1)$ the candidate is not a horizon, but a singularity, so again there is no continuation. An asymptotically flat spacetime by definition has the same structure of infinity as that of Minkowski spacetime; in general this is not always the case---recall the de Sitter and other cosmological spacetimes.

\begin{enumerate}[resume]
\item  Draw the parts of conformal diagrams near the boundary that correspond to the limiting process
\begin{equation}
\rho\to\rho_0 ,\qquad |\rho_0|<\infty,
\end{equation}
under the following conditions:
\begin{enumerate}
\item spacelike singularity: $f(\rho_0)>0$, $|x_0 |<\infty$;
\item timelike singularity: $f(\rho_0)<0$, $|x_0 |<\infty$;
\item asymptotically flat infinity: $\rho_0 =\pm \infty$, $f(\rho)\to f_0>0$;
\item a horizon in the R region: $f(\rho)\to +0$, $|x_0 |\to \infty$;
\item a horizon in the T region: $f(\rho)\to -0$, $|x_0 |\to \infty$;
\item remote horizon in a T-region: $\rho_0 =\pm \infty$, $f(\rho)\to -<0$;
\end{enumerate}
Remember that any spacelike line can be made ``horizontal'' and any timelike one can be made ``vertical'' by appropriate choice of coordinates.
\end{enumerate}
For arbitrary $f(\rho)$ we can
\begin{itemize}
\item split the full spacetime into regions between the zeros $\rho_i$ of $f(\rho)$;
\item draw for each region the conformal diagram, which is a square $\Diamond$ or half of it $\triangle$, $\nabla$, $\triangleleft$, $\triangleright$;
\item glue the pieces together along the horizons $\rho=\rho_i$, while leaving singularities and infinities as the boundary.
\end{itemize}
The resulting diagram can turn out to be either finite, as for Schwarzshild, when singularities and infinities form a closed curve enclosing the whole spacetime, or not.

\begin{enumerate}[resume]
\item  Draw the conformal diagrams for the following spacetimes:
\begin{enumerate}
\item Reissner-Nordstr\"{o}m charged black hole:
\begin{equation}
f(r)=1-\frac{r_g}{r}+\frac{q^2}{r^2}, \qquad 0<q<r_{g},\qquad r>0;
\end{equation}
\item Extremal Reissner-Nordstr\"{o}m charged black hole
\begin{equation}
f(r)=\Big(1-\frac{q}{r}\Big)^2, \qquad q>0,\qquad r>0;
\end{equation}
\item Reissner-Nordstr\"{o}m-de Sitter charged black hole with cosmological constant (it is not asymptotically flat, as $f(\infty)\neq 1$)
\begin{equation}
f(r)=1-\frac{r_g}{r}+\frac{q^2}{r^2}-\frac{\Lambda r^2}{3}\qquad q,\Lambda>0;
\end{equation}
analyze the special cases of degenerate roots.
\end{enumerate}

\end{enumerate}

\section{Hubble sphere}
The Hubble radius is the proper distance $R_H (t)=c H^{-1}(t)$. The sphere of this radius is called the Hubble sphere. From definition, the Hubble recession ``velocity'' of a comoving observer on the Hubble sphere is $v(R_H)=H R_H =c$ and equal to the speed of light $c$. This is true, of course, at the same moment of time $t$, for which $H (t)$ is taken.

\begin{enumerate}[resume]
\item All galaxies inside the Hubble sphere recede subluminally (slower than light) and all galaxies outside recede superluminally (faster than light). This is why the Hubble sphere is sometimes called the ``photon horizon''.  Does this mean that galaxies and their events outside the photon horizon are permanently hidden from the observer's view? If that were so, the photon horizon would also be an event horizon. Is this correct? 

\item Show, by the example of static universe, that the Hubble sphere does not coincide with the boundary of the observable Universe.

\item Estimate the ratio of the volume enclosed by the Hubble sphere to the full volume of a closed Universe.

\item Show that in a spatially flat Universe ($k=0$), in which radiation is dominating, the particle horizon coincides with the Hubble radius.

\item Find the dependence of comoving Hubble radius $R_H /a$ on scale factor in a flat Universe filled with one component with the state equation $\rho=w p$.

\item Express the comoving particle horizon through the comoving Hubble radius for the case of domination of a matter component with state parameter $w$.

\item Show that
\begin{equation}
\frac{dR_H }{dt}=c(1+q),\label{dRHdt}
\end{equation}
where
\begin{equation}
q=-\frac{\ddot{a}/a}{H^{2}}
\end{equation}
 is the deceleration parameter.

\item Show that
\begin{equation}
\frac{dL_p}{dt}=1+\frac{L_p}{R_H}.
\end{equation}

\item Show that in the Einstein-de Sitter Universe the relative velocity of the Hubble sphere and galaxies on it is equal to $c/2$.

\item Find $a(t)$ in a universe with constant positive deceleration parameter $q$.

\item Show that in universes of constant positive deceleration $q$, the the ratio of distances to the particle and photon horizons is $1/q$.

\item Show that the Hubble sphere becomes degenerate with the particle horizon at $q=1$ and with the event horizon at $q=-1$.

\item Show that if $q$ is not constant, comoving bodies can be inside and outside of the Hubble sphere at different times. But not so for the observable universe; once inside, always inside.

\end{enumerate}

\section{Proper horizons}
The following five problems are based on work by F. Melia \cite{Melia}.

Standard cosmology is based on the FLRW metric for a spatially homogeneous and isotropic three-dimensional space, expanding or contracting with time. In the coordinates used for this metric, $t$ is the cosmic time, measured by a comoving observer (and is the same everywhere), $a(t)$ is the expansion factor, and $r$ is an appropriately scaled radial coordinate in the comoving frame.

F. Melia  demonstrated the usefulness of expressing the FRLW metric in terms of an observer-dependent coordinate $R=a(t)r$, which explicitly reveals the dependence of the observed intervals of distance, $dR$, and time on the curvature induced by the mass-energy content between the observer and $R$; in the metric, this effect is represented by the proximity of the physical radius $R$ to the cosmic horizon $R_{h}$, defined by the relation
\[R_{h}=2G\,M(R_h).\]
In this expression, $M(R_h)$ is the mass enclosed within $R_h$ (which terns out to be the Hubble sphere). This is the radius at which a sphere encloses sufficient mass-energy to create divergent time dilation for an observer at the surface relative to the origin of the coordinates.
\begin{enumerate}[resume]
\item\label{dyn-Melia1} Show that in a flat Universe $R_{h}=H^{-1}(t)$.

\item\label{dyn-Melia2} Represent the FLRW metric in terms of the observer-dependent coordinate $R=a(t)r$.

\item\label{dyn-Melia3} Show, that if we were to make a measurement at a fixed distance $R$ away from us, the time interval $dt$ corresponding to any measurable (non-zero) value of $ds$ must go to infinity as $r\to R_h$.

\item\label{dyn-Melia4} Show that $R_h$ is an increasing function of cosmic time $t$ for any cosmology with $w>-1$.

\item\label{dyn-Melia5} Using FLRW metric in terms of the observer-dependent coordinate $R=a(t)r$, find $\Phi(R,t)$ for the specific cosmologies:
\begin{enumerate}
\item[a)] the De Sitter Universe ;
\item[b)] a cosmology with  $R_h =t$, ($w=-1/3$);
\item[c)] radiation dominated Universe  ($w=1/3$);
\item[d)] matter dominated Universe ($w=0$).
\end{enumerate}

\end{enumerate}

\section{Inflation}
\begin{enumerate}[resume]
\item Is spatial curvature important in the early Universe? Compare the curvature radius with the particle horizon.

\item Comoving Hubble radius
\begin{equation}
	r_H =\frac{R_H}{a}=\frac{1}{aH}=\frac{1}{\dot{a}}
\end{equation}
plays crucial role in inflation. Express the comoving particle horizon $l_e$ in terms of $r_H$. 

\item Show that for the conventional Big Bang expansion (with $w\geq 0$) the comoving particle horizon and Hubble radius grow monotonically with time.

\end{enumerate}

\textbf{The flatness problem}
\begin{enumerate}[resume]
\item The ``flatness problem'' can be stated in the following way: spacetime in General Relativity is dynamical, curving in response to matter in the Universe. Why then is the Universe so closely approximated by Euclidean space? Formulate the ``flatness problem'' in terms of the comoving Hubble radius.

\item Inflation is defined as any epoch, in which scale factor grows with acceleration\footnote{Technically, this includes also the current epoch of cosmological history -- late-time accelerated cosmological expansion.}, i.e. $\ddot{a}>0$. Show that this condition is equivalent to the comoving Hubble radius decreasing with time.

\item Show how inflation solves the flatness problem.

\end{enumerate}

\textbf{The horizon problem}
\begin{enumerate}[resume]
\item What should be the scale of homogeneity in the early Universe, as function of time, for the observed cosmological background (CMB) to be almost isotropic? Compare this with the functional dependence of comoving particle horizon on time. How can this be compatible with the causal evolution of the Universe?

\item If CMB was strictly isotropic, in what number of causally independent regions temperature had had to be kept constant at Planck time?

\item Consider the case of dominating radiation. Show, that at any moment in the past, within the matter comprising today the observable Universe, one can find regions that are out of causal contact.

\item Illustrate graphically the solution of the horizon problem by the inflation scenario.

\end{enumerate}

\textbf{Growth of perturbations}
\begin{enumerate}[resume]

\item Show that any mechanism of generation of the primary inhomogeneities in the Big Bang model violates the causality principle.

\item How should the early Universe evolve in order to make the characteristic size $\lambda_P$ of primary perturbations decrease faster than the Hubble radius, if one moves backward in time?

\end{enumerate}

\section{Holography}
In the context of holographic description of the Universe (see the minimal introduction on the subject in the corresponding Chapter)
 the Hubble sphere is often treated as  the holographic screen, and consequently called a horizon, although technically it is not.

\begin{enumerate}[resume]
\item Formulate the problem of the cosmological constant (see chapter on Dark Energy) in terms of the Hubble radius.

\item Choosing the Hubble sphere as the holographic screen, find its area in the de Sitter model (recall that in this model the Universe's dynamics is determined by the cosmological constant $\Lambda >0$).

\item Find the Hubble sphere's area in the Friedman's Universe with energy density $\rho$.

\item Show that in the flat Friedman's Universe filled with a substance with state equation $p=w\rho$ the Hubble sphere's area grows with the Universe's expansion under the condition $1+w >0$.

\item Estimate the temperature of the Hubble sphere $T_H$ considering it as the holographic screen.

\item Taking the Hubble sphere for the holographic screen and using the SCM parameters, find the entropy, force acting on the screen and the corresponding pressure.

\item The most popular approach to explain the observed accelerated expansion of the Universe assumes introduction of dark energy in the form of cosmological constant into the Friedman equations. As seen in the corresponding Chapter, this approach is successfully realized in SCM. Unfortunately, it leaves aside the question of the nature of the dark energy. An alternative approach can be developed in the frame of holographic dynamics. In this case it is possible to explain the observations without the dark energy. It is replaced by the entropy force, which acts on the cosmological horizon (in this case it is the Hubble sphere) and leads to the accelerated expansion of the Universe. Show that the Hubble sphere's acceleration obtained this way agrees with the result obtained in SCM.

\item Show that 
 a pure de Sitter Universe obeys the holographic principle in the form
\begin{equation}
N_{sur}=N_{bulk}.
\end{equation}
Here $N_{sur}$ is the number of degrees of freedom on the Hubble sphere, and $N_{bulk}$ is the effective number of degrees of freedom, which are in equipartition at the horizon temperature $T_H =\hbar H /2\pi k_B$.

\item If it is granted that the expansion of the Universe is equivalent to the emergence of space (in the form of availability of greater and greater volumes of space), then the law governing this process must relate the emergence of space to the difference
\begin{equation}
 N_{sur}-N_{bulk}
\end{equation}
(see previous problem). The simplest form of such a law is
\begin{equation}
\Delta V\sim ( N_{sur}-N_{bulk}) \Delta t .
\end{equation}
We could imagine this relation as a Taylor series expansion truncated at the first order. Show that this assumption is equivalent to the second Friedman equation.

\end{enumerate}

\chapter{Quantum Cosmology}

\begin{enumerate}

\item \label{QC_1}
Consider two neutral particles of equal mass $M$ in flat space, whose motion obeys a Schrodinger equation. The stationary ground state of the system is the analog of an atom, but with gravitational binding instead of electricity. Estimate the size $R$ of the ground state wave function. (Craig J. Hogan, Quantum indeterminacy in local measurement of cosmic expansion, 1312.7797)

\item \label{QC_2}
Obtain the condition for the acceleration due to the gravitation of the bodies to be smaller than the cosmic acceleration, and their gravitational binding energy to be less than their cosmic expansion kinetic energy. Give an interpretation of the obtained inequality in terms of the gravitational free-fall time and the space-time curvature.

\item \label{QC_3}
The standard quantum uncertainty in position $x$ of a body of mass $M$ measured at two times separated by an interval $\tau$ is
\begin{equation}\label{quantum}
\Delta x_q(\tau)^2\equiv \langle (x(t)-x(t+\tau))^2\rangle> 2\hbar \tau/ M.
\end{equation}
Consider two bodies of identical mass $M$ in  a state of minimal relative displacement uncertainty $\Delta x$. What value of their masses should be in order to make the uncertainty in position to be less than the change in their separation due to cosmic expansion.

\item \label{QC_4}
Estimate minimum size of a system needed for the uncertainty in position of its parts to be less than the change in their separation due to cosmic expansion.

\item \label{QC_5}
(see M.Gasperini, String theory and primordial cosmology, 1402.0101)

Like all classical theories, GR has a limited validity range. Because of those limits the standard cosmological model cannot be extrapolated to physical regimes where the energy and the space-time curvature are too high. Show that SCM is not applicable in the vicinity of the  initial  singularity (Big Bang).

\item \label{QC_6}
Construct Lagrangian and hamiltonian corresponding to the action
\[S=\int\sqrt{-g}\left(\frac R{16\pi G}-\rho\right),\quad \rho=\sum\limits_i\rho_i.\]

\item \label{QC_7}
Using the lagrangian obtained in the previous problem, obtain the equation of motion for scale factor in closed FRLW Universe, filled with cosmological constant $\rho=\rho_\Lambda$ and show that it coincides with the first Friedmann equation (see Vilenkin, 9507018)

\item \label{QC_8}
Obtain and analyze the solution $a(t)$ of the equation obtained in the previous equation.

\item \label{QC_9}
Construct an effective Hamiltonian for the de Sitter model.

\item \label{QC_10}
Using result of the previous problem, obtain the Wheeler-DeWitt equation for the de Sitter model.

\item \label{QC_11}
Obtain  Wheeler-de Witt equation for a Universe filled by cosmological constant and radiation. (1301.4569)

\end{enumerate}

\section{Introduction to vacuum fluctuations}

(V. F. Mukhanov and S. Winitzki, Quantum Fields in Classical Backgrounds, Lecture notes --- 2004)

A vacuum fluctuations are fluctuation of an "empty" space. The word empty quoted because the quantum vacuum represents a collection of zero-point oscillations of quantum fields. We shall consider only the scalar fields. A free real massive classical scalar field $\varphi(x,t)$ (in a stationary Universe) satisfies the Klein-Gordon equation
\[\ddot\varphi-\Delta\varphi+m^2\varphi=0.\]
It is convenient to use the spatial Fourier decomposition,
\[\varphi(\vec{x},t)=\frac1{(2\pi)^{3/2}}\int d^3k e^{-i\vec{k}\vec{x}}\varphi_{\vec{k}}(t).\]
It is useful to consider a field $\varphi(\vec{x},t)$ in a box (a cube with sides $L$) of finite volume $V$ with the periodic boundary conditions imposed on the field $\varphi$ at the box boundary. In this case the Fourier decomposition can be written as
\[\varphi(\vec{x},t)=\frac1{\sqrt V}\sum\limits_k \varphi_{\vec{k}}(t)e^{i\vec{k}\vec{x}},\]
\[\varphi_{\vec{k}}(t)=\frac1{\sqrt V}\int d^3x \varphi(\vec{x},t)e^{i\vec{k}\vec{x}},\]
where the sum goes over three-dimensional wave numbers $k$ with components of the form
\[k_{(x,y,z)}=\frac{2\pi n_{(x,y,z)}}{L},\quad n_{(x,y,z)}=0,\pm1,\pm2,\ldots\]
The Klein-Gordon equation for each $k$-component is
\[\ddot\varphi_{\vec{k}}+(k^2+m^2)\varphi_{\vec{k}}=0.\]
Each (in general) complex function $\varphi_{\vec{k}}(t)$ satisfies the harmonic oscillator equation with the frequency $\omega_k=(k^2+m^2)^{1/2}$. The functions $\varphi_{\vec{k}}(t)$ are called the modes of the field $\varphi$.

To quantize the field, each mode $\varphi_{\vec{k}}(t)$ is quantized as a separate harmonic oscillator. We replace the classical "coordinates" $\varphi_{\vec{k}}$ and momenta $\pi_{\vec{k}}\equiv\dot\varphi^*_{\vec{k}}$ by operators $\hat\varphi_{\vec{k}}$ and $\hat\pi_{\vec{k}}$ with the equal-time commutation relations
\[[\hat\varphi_{\vec{k}},\hat\pi_{\vec{k}}]=i\delta(\vec{k}+\vec{k}').\]

\begin{enumerate}[resume]
\item \label{QC_12}
Construct the wave function of the vacuum state of a  scalar field.

\item \label{QC_13}
Find the wave function, obtained in the previous problem, in the limit of quantization inside an infinitely large box ($V\to\infty$).

\item Estimate the typical amplitude $\delta\varphi_{\vec{k}}$ of fluctuations in the mode $\varphi_{\vec{k}}$.

\item \label{QC_15}
Show, that if $\varphi_L$ is the average of $\varphi(t)$ over a volume $V=L^3$, the typical fluctuation $\delta\varphi_L$ of $\varphi_L$ is \[\langle\varphi_L^2\rangle\sim k^3\left(\delta\varphi_{\vec{k}}\right)_{k=L^{-1}}\].

\item \label{QC_16}
Analyze the expression for amplitude of the scalar field fluctuations, obtained in the previous problem, as a function of linear size $L$ of the region of the averaging.

\item Obtain the modified Raychaudhuri equation in LQC.

\item \label{QC_18}
Consider a model of Universe filled with dust-like matter, i.e. $p=0$, to demonstrate the main distinction between the standard (Fridmannian) and the LQC (see A. Barrau1, T. Cailleteau, J. Grain4, and J. Mielczarek, Observational issues in loop quantum cosmology, arXiv:1309.6896 ).

\item \label{QC_19}
Estimate the typical wavelength of photons radiated by a black hole of mass M and
compare it with the size of the black hole (the Schwarzschild radius).

\item \label{QC_20}
The temperature of a sufficiently small black hole can be high enough to efficiently produce baryons (e.g. protons) as components of the Hawking radiation. Estimate the
required mass M of such black holes and compare their Schwarzschild radius with the
size of the proton (its Compton length).

\item \label{QC_21}
(see V.Frolov, A.Zelnikov, Introduction to black hole physics, Oxford University Press, 2011)

GR allows the existence of black holes of arbitrary mass. Why do we not observe formation of small mass black holes in the laboratory or our everyday life?

\item \label{QC_22}
Black holes of small masses can be created in the early Universe when the matter density was high. Such black holes are called the primordial black holes (PBHs). The mass spectrum of the PBHs could span an enormous mass range. Determine the mass range of the PBHs, created during the radiation dominated epoch $10^{-43}sec<t<1sec$.

\item \label{QC_23}
(see V. F. Mukhanov and S. Winitzki, Introduction to Quantum Fields in Classical Backgrounds, Lecture notes  - 2004)

A glass of water is moving with constant acceleration. Determine the smallest acceleration that would make the water boil due to the Unruh effect.
\end{enumerate}

In the de Broglie-Bohm causal interpretation of quantum mechanics  the Schrodinger equation for a single nonrelativistic particle
\[i\hbar\frac{\partial\psi(x,t)}{\partial t}=\left[-\frac{\hbar^2}{2m}\nabla^2+V(x)\right]\psi(x,t)\]	
with the substitution $\Psi=Re^{iS/\hbar}$ is transformed to the system of equations
\begin{align}
\nonumber \frac{\partial S}{\partial t} + \frac{(\nabla S)^2}{2m} + V-\frac{\hbar^2}{2m}\frac{\Delta R}R &=0,\\
\nonumber \frac{\partial R^2}{\partial t} + \nabla\left(R^2\frac{\nabla S}{m}\right) &=0.
\end{align}
The first equation is a Hamilton-Jacobi type equation for a particle submitted to an external potential which is the classical potential plus a new quantum potential
\[Q=-\frac{\hbar^2}{2m}\frac{\Delta R}R\]
The second equation can be treated as the continuity equation for a fluid with the density $\rho=R^2$. In a series of the problems below let us confine ourselves to the simple case when the Universe is supposed to be filled with only one component, namely, the nonrelativistic gas of point-like particles of the equal mass $m$. Then the energy-momentum tensor components have the following  form:
\[T^{\mu\nu}=(\rho_m+\rho_q+p_q)-p_qg^{\mu\nu}\]
where $\rho_m$ is the classical energy density of nonrelativistic component (the corresponding classical pressure $p_m$ is equal to zero), while $\rho_q$ and $p_q$ represent quantum admixtures, generated by interaction with the quantum potential.

\begin{enumerate}[resume]
\item \label{QC_24}
Obtain relativistic generalization for the quantum potential.

\item \label{QC_25}
Find energy density $\rho_q$ of quantum admixture for the case when the background density $\rho_m$ represents a gas of point-like particles of the equal masses $m$.

\item \label{QC_26}
Calculate energy density and pressure generated by the quantum potential in Friedmannian Universe filled with non-relativistic matter.

\item \label{QC_27}
Show that in the limit $a\to\infty$ the de Broglie-Bohm model disregards quantum effects.

\end{enumerate}



\begin{thebibliography}{99}
\bibitem{siv1} Sivaram, C.: 1993a, Mod. Phys. Lett. 8,321.; 14. Sivaram, C.: 1993b, Astrophys. Spc. Sci. 207, 317.; 15. Sivaram, C.: 1993c, Astron. Astrophys. 275, 37.; 16. Sivaram, C.: 1994a, Astrophysics. Spc. Sci., 215, 185.; 17. Sivaram, C.: 1994b. Astrphysics .Spc .Sci., 215,191.; 18. Sivaram, C.: 1994c. Int. J. Theor. Phys. 33, 2407.

\bibitem{siv2} Sivaram, C.: 1982, Astrophysics. Spc. Sci. 88,507.; 5. Sivaram, C.: 1982, Amer. J. Phy. 50, 279.; 6. Sivaram, C.: 1983, Amer. J. Phys. 51, 277.; 7. Sivaram, C.: 1983, Phys. Lett. 60B, 181.

\bibitem{siv3} C Sivaram, Astr. Spc. Sci, 219, 135; IJTP, 33, 2407, 1994, 17. C Sivaram, Mod. Phys. Lett., 34, 2463, 1999

\bibitem{Schw} K. Schwarzschild, On the gravitational field of a mass point according to Einstein's theory, \textit{Sitzungsber. Preuss. Akad. Wiss. Phys. Math. Kl.},  p.189 (1916) arXiv:physics/9905030v1.

\bibitem{Birkhoff} G.D. Birkhoff, Relativity and Modern Physics, p.253, Harvard University Press, Cambridge (1923).

\bibitem{Jebsen} J.T. Jebsen, \textit{Ark. Mat. Ast. Fys.} (Stockholm) \textbf{15}, nr.18 (1921), see also arXiv:physics/0508163v2.

\bibitem{Kerr63} R.P. Kerr, Gravitational field of a spinning mass as an example of algebraically special metrics. \textit{Phys. Rev. Lett.} \textbf{11} (5), 237 (1963).

\bibitem{BoyerLindquist67} R.H. Boyer, R.W. Lindquist. Maximal Analytic Extension of the Kerr Metric. \textit{J. Math. Phys} \textbf{8}, (1967).

\bibitem{Wald} Wald R.M, General relativity. U. Chicago, 1984, 505p (ISBN 0226870332).

\bibitem{Carroll} Carroll S., Spacetime and geometry: an introduction to General Relativity. AW, 2003, 525p (ISBN 0805387323).

\bibitem{LL} Landau L.D., Lifshitz E.M. Vol. 2. The classical theory of fields [4ed., Butterworth-Heinemann, 1994].

\bibitem{t'Hooft} G. `t Hooft, Introduction to General Relativity, Caputcollege 1998.

\bibitem{MTW} Charles W. Misner, Kip S. Thorne. John Archibald Wheeler,  Gravitation. W.H. Freeman and Company, 1973.


\bibitem{Hobson2006} Hobson M., Efstathiou G., Lasenby A. General relativity: an introduction for physicists. CUP, 2006 (ISBN 0521536391).

\bibitem{Padmanabhan2010} Padmanabhan T. Gravitation: Foundations and Frontiers. CUP, 2010 (ISBN 9780521882231).

\bibitem{li} A. P. Lightman, W. H. Press, R. H. Price, and S. A. Teukolsky, Problem book in Relativity and Gravitation (Princeton University Press, Princeton, New Jersey, 1975).

\bibitem{mb} Black Holes: The Membrane Paradigm. Edited by Kip S. Thorne, Richard H. Price, Douglas A. Macdonald. Yale University Press New Haven and London, 1986.

\bibitem{0307350} M. Hoffman, Cosmological constraints on a dark matter---dark energy interaction, arXiv:astro-ph/0307350
\bibitem{1209.0563} Josue De-Santiago, David Wands, Yuting Wang, Inhomogeneous and interacting vacuum energy,  arXiv:astro-ph/1209.0563
\bibitem{1207.0250} Solano, U.Nucamendi, Reconstruction of the interaction term between dark matter and dark energy using SNe Ia, BAO, CMB, H(z) and X-ray gas mass fraction arXiv:1207.0250
\bibitem{0502034} M. Szydlowski, Cosmological model with energy transfer,  arXiv:astro-ph/0502034v2
\bibitem{0606555} Winfried Zimdahl,  D. Pavon, Interacting holographic dark energy, arXiv:astro-ph/0606555v3
\bibitem{1111.6743} Xi-ming Chen, Yungui Gong, E. Saridakis, Time-dependent interacting dark energy and transient acceleration, arXiv:1111.6743v4 [astro-ph.CO]
\bibitem{9702029} L. P. Chimento, Form invariance of differential equations in general relativity, arXiv:physics/9702029v1 [math-ph]
\bibitem{0604063} J.Barrow, T. Clifton, Cosmologies with energy exchange, arXiv:gr-qc/0604063v2
\bibitem{0505096} L.Chimento, D.Pavon, Dual interacting cosmologies and late accelerated
expansion, arXiv:gr-qc/0505096v3
\bibitem{1010.1074}  Hao Wei, Cosmological Constraints on the Sign-Changeable Interactions, arXiv:1010.1074v2 [gr-qc]
\bibitem{1009.4942} S. Z.W. Lip, Interacting Cosmological Fluids and the Coincidence Problem, arXiv:1009.4942v2 [gr-qc]
\bibitem{1112.5095} F.Arevalo, A.Bacalhau and W. Zimdahl,  Cosmological dynamics with non-linear interactions, arXiv:1112.5095v2 [astro-ph.CO]
\bibitem{1012.2692}  M. Malekjani, A. Khodam-Mohammadi,  M. Taji, Cosmological implications of interacting polytropic gas dark energy model in non-flat universe, arXiv:1012.2692v2 [gr-qc]
\bibitem{1003.2788} Yi Zhang, ,Hui Li, A New Type of Dark Energy Model, arXiv:1003.2788v2 [astro-ph.CO]
\bibitem{0510628v2} S. Das,  P. Corasaniti,  J. Khoury, Super-acceleration as Signature of Dark Sector Interaction, arXiv:astro-ph/0510628v2
\bibitem{0105479} W. Zimdahl, D. Pavon, L. Chimento, Interacting Quintessence, arXiv:astro-ph/0105479v2
\bibitem{0503075} Xin Zhang, Statefinder diagnostic for coupled quintessence, arXiv:astro-ph/0503075v1
\bibitem{0411524} Zong-Kuan Guo, Yuan-Zhong Zhang, Interacting Phantom Energy,  arXiv:astro-ph/0411524v1
\bibitem{0404086} R. Herrera, D. Pavon,  W. Zimdahl, Exact solutions for the interacting tachyonic-dark matter system, arXiv:astro-ph/0404086v1
\bibitem{0408495}  Peng Wang,  Xin-He Meng, Can vacuum decay in our Universe?, arXiv:astro-ph/0408495v3
\bibitem{0507372} J. S. Alcaniz, J. A. S. Lima, Interpreting Cosmological Vacuum Decay, arXiv:astro-ph/0507372v2
\bibitem{0711.2686} S. Carneiro, M. A. Dantas, C. Pigozzo,  J. S. Alcaniz, Observational constraints on late-time $\Lambda(t)$ cosmology, arXiv:0711.2686v2 [astro-ph]
\bibitem{0311067} W. Zimdahl, D. Pavon, Statefinder parameters for interacting dark energy, arXiv:gr-qc/0311067v1
\bibitem{0711.1641}  Yin-Zhe Ma,  Yan Gong,  Xuelei Chen, Features of holographic dark energy under the combined cosmological constraints, arXiv:0711.1641v4 [astro-ph]
\bibitem{arXiv:0911.5687} Luis P. Chimento, Linear and nonlinear interactions in the dark sector,  Phys.Rev.\textbf{D 81}:043525,2010 arXiv:0911.5687v2 [astro-ph.CO]
\bibitem{Harrison} E. Harrison. Cosmology: the science of the Universe. CUP (1981)

\bibitem{Rindler} W. Rindler. Visual horizons in world-models. MNRAS 116 (6), 662--677 (1956)

\bibitem{Ellis} G.F.R. Ellis and T. Rothman. Lost horizons. Am. J. Phys. 61, 883 (1993)

\bibitem{Muchanov} V.F. Muchanov. Physical foundations of cosmology (CUP, 2005) ISBN~0521563984

\bibitem{BrRubin} K.A. Bronnikov and S.G. Rubin. Black Holes, Cosmology and Extra Dimensions. (WSPC, 2012), ISBN~978-9814374200

\bibitem{Melia} F. Melia. The cosmic horizon.  MNRAS 382 (4), 1917--1921 (2007);  F.~Melia and M.~Abdelqader, The Cosmological Spacetime,  Int. J. Mod. Phys. D 18, 1889 (2009) (arXiv:0907.5394)

\bibitem{Baumann} D. Baumann, TASI Lectures on Inflation, (arXiv: 0907.5424)]
\end{thebibliography}
\end{document}